\documentclass[openright,12pt]{report}

\usepackage[paper=a4paper,dvips,top=3.0cm,left=2.7cm, right=2.7cm,bottom=2.5cm]{geometry}
\usepackage{amsmath,amssymb}  
\usepackage{graphicx}  
\usepackage{color}  
\usepackage{float}  
\usepackage{setspace} 
\usepackage{hyperref}   
\hypersetup{colorlinks=true,linkcolor=beamer@mediumblue, citecolor=beamer@new}

\usepackage{fancyhdr}  
\definecolor{beamer@blue}{RGB}{0,0,255}
\definecolor{beamer@mediumblue}{RGB}{0,0,190}
\definecolor{beamer@midnightblue}{RGB}{25,25,112}
\definecolor{beamer@navy}{RGB}{0,0,128}
\definecolor{beamer@darkblue}{RGB}{0,0,139}
\definecolor{beamer@purple}{RGB}{128,0,128}
\definecolor{beamer@levander}{RGB}{100.,149.,237.}
\definecolor{beamer@green}{RGB}{0,128,0}
\definecolor{beamer@darkgreen}{RGB}{0,150,0}
\definecolor{beamer@olive}{RGB}{128,128,0}
\definecolor{beamer@darkolivegreen}{RGB}{85,107,47}
\definecolor{beamer@gray}{RGB}{190,190,190}
\definecolor{beamer@ivry}{RGB}{220,220,220}
\definecolor{beamer@new}{RGB}{40,120,50}
\definecolor{shadecolor}{RGB}{220,220,220}
\definecolor{beamer@darkslategray}{RGB}{47,79,79}
\definecolor{beamer@chocolate}{RGB}{210,105,30}
\definecolor{beamer@orangered}{RGB}{255,69,0}
\definecolor{beamer@maroon}{RGB}{128,0,0}
\definecolor{beamer@white}{RGB}{234,242,243}
\definecolor{beamer@silver}{RGB}{0.5,0.45,0.37}

\begin{document}

\thispagestyle{empty}
\begin{center}  
\Large{\textbf{Solvable Models on Noncommutative Spaces with Minimal Length Uncertainty Relations}}\\
\vspace{2.0cm}
\Large{\textbf{Sanjib Dey}} \\
\vspace{1.0cm}
\large{\textbf{Ph.D. THESIS}}\\
\vspace{1.2cm}
\includegraphics[scale=0.5]{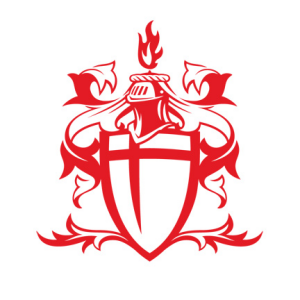}\\
\vspace{1.2cm}
\large{\textbf{DEPARTMENT OF MATHEMATICS}}\\
\large{\textbf{CITY UNIVERSITY LONDON}}\\
\vspace{2.0cm}
\large{\textbf{SEPTEMBER 2014}} \\
\vspace{2.0cm}
\large{\textbf{SUPERVISOR : Prof. Dr. Andreas Fring}}
\end{center}


\newpage
\thispagestyle{empty}
\begin{flushright}
\large{To my Parents}
\end{flushright}


\tableofcontents
\rhead{Contents}
\lhead{}
\chead{}



\chapter*{Acknowledgements}
\addcontentsline{toc}{chapter}{Acknowledgements}
\pagenumbering{roman}
\normalsize{First of all I would like to take the opportunity to thank the person with whom I was most directly involved in this work, Prof. Andreas Fring, supervisor and friend, who is an inspiring mentor, and initiated me into this fascinating field of research. His insight, quest for perfection, and passion for science has always inspired me. He has always been very patient, and ensured the fact that I understood every subtle point. I am very grateful for your enormous friendly support. It has been a real pleasure working with you and I hope to collaborate in future. \\
     \indent I would like to thank the members of the panel who kindly accepted the laborious task of accessing this thesis, Dr. Alessandro de Martino, Prof. Fabio Bagarello and Prof. Joseph Chuang. I feel extremely honoured by their assistance.\\
     \indent I am also very grateful for the financial support, the academic infrastructure and the research environment provided by City University London. I am obliged to my co-supervisor Dr. Olalla Castro Alvaredo and I would like to thank her for her support during my research days.\\
     \indent It has been a great experience to work with Dr. Laure Gouba, Dr. Paulo G. Castro, Boubakeur Khantoul and Thilagarajah Mathanaranjan. Thanks to all of you for your collaboration. Special thanks to Prof. Ali Mostafazadeh and the other organisers of PHHQP-XII, for giving me an opportunity to present my work in such a nice international conference. Special thanks also to Prof. Bijan Bagchi. I owe so much to his help. I am deeply indebted to you for your support and scientific discussions.\\
     \indent I thank all the people from the department, and my officemates, In particular Andrea, Cheng, Emanuele, Zoe, Oliver, Elizabeth, David, Veronika, Tom, Karan, Davide, Klodeta, Niamh, Lleonard, Raza, Jan, William for sharing good times with me, and many days and nights working together.


\chapter*{Declaration}
\addcontentsline{toc}{chapter}{Declaration}
The work presented in this thesis is based on investigations believed to be  original and carried out at the Department of Mathematics, City University London in collaboration with Prof. Andreas Fring. It has not been presented elsewhere for a degree, diploma, or similar qualification at any university or similar institution. I have clearly stated my contributions, in jointly-authored works, as well as referenced the contributions of other people working in this area.\\ 
    \indent Powers of discretion are granted to the University Librarian to allow the thesis to be copied in whole or in part without further reference to the author. This permission covers only single copies made for study purposes, subject to normal conditions of acknowledgement.
    

\chapter*{Thesis Commission}
\large{\textbf{Dr. Alessandro de Martino}}\\
\normalsize{Department of Mathematics, City University London, London, UK}
\vspace{1.0cm}\\
\large{\textbf{Prof. Fabio Bagarello}}\\
\normalsize{Dipartimento di Energia, Ingegneria dell'Informazione e Modelli Matematici, Facolt$\grave{a}$ di Ingegneria, Universit$\grave{a}$ di Palermo, Palermo, Italy and \\
INFN, Sezione di Torino, Italy}
\vspace{1.0cm}\\
\large{\textbf{Prof. Joseph Chuang}}\\
\normalsize{Department of Mathematics, City University London, London, UK}
}
\small{
    

\newpage
\thispagestyle{empty}
\huge{\textbf{Publications}} \\
\small{
\begin{enumerate}
\item $\mathcal{PT}$-symmetric noncommutative spaces with minimal volume uncertainty relations\\
S. Dey, A. Fring and L. Gouba\\
\textbf{J. Phys. A: Math. Theor. 45, 385302 (2012)}\\
arXiv:1205.2291. Chapter 4
\item Squeezed coherent states for noncommutative spaces with minimal length uncertainty relations​\\
S.Dey and A.Fring\\
\textbf{Phys. Rev. D 86, 064038 (2012)}\\
arXiv:1207.3297. Chapter 6
\item Time-dependent q-deformed coherent states for generalized uncertainty relations\\
S. Dey, A. Fring, L. Gouba and P. G. Castro\\
\textbf{Phys. Rev. D 87, 084033 (2013)}\\
arXiv:1211.4791. Chapter 6
\item Hermitian versus non-Hermitian representations for minimal length uncertainty relations\\
S. Dey, A. Fring and B. Khantoul\\
\textbf{J. Phys. A: Math. Theor. 46, 335304 (2013)}\\
arXiv:1302.4571. Chapter 5 
\item Bohmian quantum trajectories from coherent states\\
S. Dey and A.Fring\\
\textbf{Phys. Rev. A 88, 022116 (2013)}\\
arXiv:1305.4619. Chapter 7

\rhead{Publications}
\lhead{}
\chead{}

\item The two dimensional harmonic oscillator on a noncommutative space with minimal uncertainties\\
S. Dey and A. Fring\\
\textbf{Acta Polytechnica 53, 268$-$270 (2013)}\\
arXiv:1207.3303. Chapter 4
\item Non-Hermitian systems of Euclidean Lie algebraic type with real energy spectra \\
S. Dey, A. Fring and T. Mathanaranjan\\
\textbf{Ann. Phys. 346, 28-41 (2014)}\\
arXiv:1401.4426. Chapter 3 
\item Quantum versus classical $\mathcal{PT}$-symmetry\\
S. Dey and A. Fring\\
in preparation, Chapter 7
\end{enumerate}
}
\normalsize{


\chapter*{Abstract}
\addcontentsline{toc}{chapter}{Abstract}
Intuitive arguments involving standard quantum mechanical uncertainty relations suggest that at length scales close to the Planck length, strong gravity effects limit the spatial as well as temporal resolution smaller than fundamental length scale ($l_p=$ Planck Length $=\sqrt{G\hbar/c^3}\approx 1.616 \times 10^{-35}~$m), leading to space-space as well as space-time uncertainties. Space-time cannot be probed with a resolution beyond this scale i.e. space-time becomes "fuzzy" below this scale, resulting into noncommutative space-time. Hence it becomes important and interesting to study in detail the structure of such noncommutative spacetimes and their properties, because it not only helps us to improve our understanding of the Planck scale physics but also helps in bridging standard particle physics with physics at Planck scale.

Our main focus in this thesis is to explore different methods of constructing models in these kind of spaces in higher dimensions. In particular, we provide a systematic procedure to relate a three dimensional $q$-deformed oscillator algebra to the corresponding algebra satisfied by canonical variables describing  non-commutative spaces. The representations for the corresponding operators obey algebras whose uncertainty relations lead to minimal length, areas and volumes in phase space, which are in principle natural candidates of many different approaches of quantum gravity. We study some explicit models on these types of noncommutative spaces, in particular, we provide solutions of three dimensional harmonic oscillator as well as its decomposed versions into lower dimensions. Because the solutions are computed in these cases by utilising the standard Rayleigh-Schr{\"o}dinger perturbation theory, we investigate a method afterwards to construct models in an exact manner. We demonstrate three characteristically different solvable models on these spaces, the harmonic oscillator, the manifestly non-Hermitian Swanson model and an intrinsically noncommutative model with P{\"o}schl-Teller type potential. In many cases the operators are not Hermitian with regard to the standard inner products and that is the reason why we use $\mathcal{PT}$-symmetry and pseudo-Hermiticity property, wherever applicable, to make them self-consistent well defined physical observables. We construct an exact form of the metric operator, which is rare in the literature, and provide Hermitian versions of the non-Hermitian Euclidean Lie algebraic type Hamiltonian systems. We also indicate the region of broken and unbroken $\mathcal{PT}$-symmetry and provide a theoretical treatment of the gain loss behaviour of these types of systems in the unbroken $\mathcal{PT}$-regime, which draws more attention to the experimental physicists in recent days.

\rhead{Abstract}
\lhead{}
\chead{}

Apart from building mathematical models, we focus on the physical implications of noncommutative theories too. We construct Klauder coherent states for the perturbative and nonperturbative noncommutative harmonic oscillator associated with uncertainty relations implying minimal lengths. In both cases, the uncertainty relations for the constructed states are shown to be saturated and thus imply to the squeezed coherent states. They are also shown to satisfy the Ehrenfest theorem dictating the classical like nature of the coherent wavepacket. The quality of those states are further underpinned by the fractional revival structure which compares the quality of the coherent states with that of the classical particle directly. More investigations into the comparison are carried out by a qualitative comparison between the dynamics of the classical particle and that of the coherent states based on numerical techniques. We find the qualitative behaviour to be governed by the Mandel parameter determining the regime in which the wavefunctions evolve as soliton like structures. We demonstrate these features explicitly for the harmonic oscillator, the P{\"o}schl-Teller potential and a Calogero type potential having singularity at the origin, we argue on the fact that the effects are less visible from the mathematical analysis and stress that the method is quite useful for the precession measurement required for the experimental purpose. In the context of complex classical mechanics we also find the claim that "the trajectories of classical particles in complex potential are always closed and periodic when its energy is real, and open when the energy is complex", which is demanded in the literature, is not in general true and we show that particles with complex energies can possess a closed and periodic orbit and particles with real energies can produce open trajectories. 


\chapter{Introduction} \label{introduction}

\pagenumbering{arabic}

Physics has gone through many ups and downs in its way and a number of conceptual changes occurred in the history of the early twentieth century. In particular, experiments in atomic physics revealed the quantum structure of nature at microscopic distances and Einstein's theory of general relativity provided a deeper understanding of space and time at a macroscopic scale. Quantum mechanics
was later successfully combined with special relativity, leading to the theory of quantum fields and eventually to the standard model of particle physics. However, when one tries to probe small distances with high energies, in particular around the Planck scale, the effects of gravity become so important that they would significantly disturb the space-time structure and result in discreteness of the space-time manifold. Therefore the understanding of the unification of general relativity with the laws of quantum mechanics becomes very  much necessary, albeit it has not been successfully understood yet and quantum gravity remains the most challenging and fascinating field of research in 21st century, in the areas of mathematical physics, mathematics and phenomenology.

There are by now a multitude of possibilities to describe the quantum behaviour of the gravitational field. The best candidate so far in the approach is probably string theory \cite{polchinski_volI,polchinski_volII,green_schwarz_witten}, which tries to unify all fundamental interactions in a single mathematical framework, whereas the idea of loop quantum gravity \cite{rovelli} is to quantise the gravitational field only, with all other fundamental forces being kept separated. In addition, to these two major frameworks, there are also other influential approaches like asymptotic safety \cite{reuter,niedermaier_reuter}, causal dynamical triangulations \cite{ambjorn_loll}, doubly special relativity \cite{amelino1} and many more, which have already led to interesting insights into the quantum nature of space-time.

At the very early days of quantum gravity, people picked up the two well known, but old proposals of quantisation of field, namely, the canonical quantisation and the path integral quantisation and essentially attempted \cite{dewitt,hartle_hawking,hawking} to utilise both of them to quantise the gravitational field too. However, from the field theoretical point of view, the theory of relativity is not renormalizable and leads to ultraviolet divergences, which effectively stops any further progress. Nevertheless, if a minimal observable length is introduced as an effective cut off in the ultraviolet domain, it is indeed possible to make the theory renormalizable.

It was suggested very early by the founding fathers of quantum mechanics, most notably Heisenberg, in the pioneering days of quantum field theory that one could use a noncommutative structure of space-time coordinates at very small length scales to introduce the effective ultraviolet cut-off. It was Snyder \cite{snyder} who first formalized this idea in an article entirely devoted to the subject. His idea was further extended later by Yang \cite{yang} who replaced the algebra of noncommutating linear operators by the algebra of functions to describe a general geometrical structure. However, these suggestions were largely ignored at that time, perhaps due to the failure of making accurate experimental predictions of the theory, but mostly because of its timing. At around the same time, the renormalization group program of quantum field theory finally proved to be successful at accurately predicting numerical values for physical observables in quantum electrodynamics and therefore the quantum theory in noncommutative space-time went through a long period of ostracism.

However, the interest in this kind of theory was resurfaced with Seiberg and Witten \cite{seiberg_witten}, who showed that string theory, at a certain low-energy limit, can be realized as an effective quantum field theory in a noncommutative space-time. Some important mathematical developments of the 1980s have also contributed to this rebirth, for instance Connes \cite{connes1985} and Woronowicz \cite{woronowicz} revived the notion by introducing a differential structure in the noncommutative framework and noncommutative theories have been an area of intense research since then. For a concrete review of the topic, see for instance \cite{garay,connes,madore,douglas_nekrasov,szabo}.

\rhead{Introduction}
\lhead{Chapter 1}
\chead{}

Therefore, the noncommutative theories are believed to be suggested possible pathways of many physical problems from time to time. The reason is because the commutation relations emerging out of these kind of theories quite naturally lead to a special kind of uncertainty relations which deliberately give rise to the existence of a minimal observable length. This is amazing enough that almost all the attempts towards quantum gravity indicate the existence of a minimum measurable length scale which is expected to be close, or identical to the Planck length \cite{doplicher_fredenhagen_roberts,doplicher_fredenhagen_roberts2}
\begin{equation}
l_P=\sqrt{\frac{G\hbar}{c^3}} \approx 1.616 \times 10^{-35}~\text{m},
\end{equation}
that combines the fundamental constants of nature (Newton's constant $G$, reduced Planck's constant $\hbar$ and the speed of light $c$) in a dimensionally appropriate way. Because it is a constant of nature, and not an artificially imposed cut-off, such an ultraviolet regulator would be extremely welcome and somewhat natural. Therefore, in addition to the expected quantum uncertainty, there is another uncertainty which arises for the space-time fluctuations at the Planck scale and that is the so-called minimal length. A minimal length can be found in string theory \cite{veneziano,gross_mende,yoneya,amati_cliafaloni_veneziano,guida_konishi_provero}, path integral quantum gravity \cite{padmanabhan1,padmanabhan2,padmanabhan3,padmanabhan4,greensite}, loop quantum gravity \cite{rovelli}, doubly special relativity \cite{magueijo_smolin,magueijo_smolin2,girelli_livine_oriti,cortes_gamboa,ghosh_subir,ali_das_vagenas,nozari_etemadi} and many more. Within a string theoretical argument, one claims that a string cannot probe distances smaller than its length as can be viewed from simple arguments \cite{kato,konishi_paffuti_provero}. In 1993, Maggiore \cite{maggiore} has also found that an uncertainty relation that gives rise to a minimum length can be derived from the measurement of the radius of a black hole. Moreover, some Gedankenexperiments or thought experiments \cite{scardigli,ng_dam,mead}, in the spirit of black hole physics have also supported the idea. For review purpose, one may follow \cite{garay,ng,hossenfelder}.

Note that, the existence of a minimal length is an obvious contradiction with the conventional version of Heisenberg Uncertainty Principle (HUP) which puts no lower or upper bound on the non-simultaneous measurements of the position or the momentum of a particle. In fact, in ordinary quantum mechanics the position uncertainty $\Delta x$, can be made arbitrarily small by letting the momentum uncertainty $\Delta p$, to grow correspondingly. However, for energies close to the Planck energy $(E_p=\sqrt{\hbar c^5/G}\approx 1.22 \times 10^{19}~$GeV), the particle's Schwarzschild radius $(r_s=2Gm/c^2)$ and its Compton wavelength $(\lambda_c=h/mc)$ become approximately of the order of the Planck length. So, in order to merge the idea of the minimal length into quantum mechanics, one needs to modify the ordinary uncertainty principle to the so-called generalized uncertainty principle (GUP), which can be done immediately from the noncommutative space time structure. Indeed, the notion of minimal length should quantum mechanically be described as a minimal uncertainty in position measurements. The introduction of this idea has drawn much attention in recent years and many papers have appeared in the literature \cite{kempf,kempf_mangano_mann,sun_fu,macfarlane,biedenharn,brodimas_jannussis_mignani, quesne_tkachuk,bagchi_fring,fring_gouba_scholtz,fring_gouba_bagchi_area,hossenfelder1, chang_lewis_minic_takeuchi,lewis_takeuchi,pedram_nozari_taheri,dorsch_nogueira,nozari_etemadi,gavrilik_kachurik} to address the effects of GUP on various quantum mechanical systems and phenomena.

In our discussion our main focus is to describe such kind of behaviour of the noncommutative space-time, which modifies the Heisenberg uncertainty relation to introduce the generalised uncertainty relation and provides the existence of the minimal observable length. In chapter \ref{chapter2}, we introduce the simplest possible example of the noncommutative space, namely the flat noncommutative spaces and thereafter the Snyder's Lorentz invariant version of this and discuss various physical consequences of the theory. Later the $q$-deformed space is introduced, on which we place more attention in our investigations. The most important thing that we try to explain in this discussion is how to construct the models explicitly from the $q$-deformed noncommutative spaces and how the commutation relations in these spaces lead to the existence of minimal length. 

In chapter \ref{chapterPT}, we introduce the basic notions of $\mathcal{PT}$-symmetric quantum mechanics. This is important to understand for us because, when one deals with the systems in noncommutative spaces, it is quite natural that one ends up with systems where the Hamiltonians are not Hermitian with respect to the standard inner products and thus can not be interpreted as observables. The virtue of the $\mathcal{PT}$-symmetric theory is that, it gives us some insights of how to deal with the non-Hermitian operators and construct well-defined physical systems out of that. On the other hand, the theory of pseudo-Hermiticity (see \ref{subsection322}) provides more accurate expressions of the Hermitian versions of the corresponding non-Hermitian Hamiltonians, which are sometimes easier to handle for more complicated problems. Finally we present some examples of Euclidean Lie algebraic type $\mathcal{PT}$-symmetric non-Hermitian Hamiltonians and obtain their Hermitian versions, which have been worked out in one of our recent publications \cite{dey_fring_mathanaranjan}. Whereas most of the works provide perturbative expressions, in our work we obtain an exact form of the metric operator which can be found rarely in the literature. We also point out the region of broken and unbroken symmetry and discuss the physical implications of these kind of Hamiltonians in the unbroken $\mathcal{PT}$ regime and provide a theoretical explanation of the gain/loss behaviour of the system in the context of $\mathcal{PT}$-symmetric wave-guide, which draws much more attention to the experimental physicists in recent days.

With the introduction of basic necessary tools in chapter \ref{chapter2} and \ref{chapterPT}, in chapter \ref{noncommutativemodelsin3D} \cite{dey_fring_gouba,dey_fring_2DHO} we present the principal segments of our investigations. We start with the most general Ansatz of the generators of the flat noncommutative spaces in three dimensions and explain the fact of how the number of free parameters can be reduced using the $\mathcal{PT}$-symmetric behaviour of the system. Afterwards, the generators of the $q$-deformed noncommutative spaces, rather than that of the flat noncommutative spaces, are assumed to be linear in three dimensions and subsequently the existence of minimal volume is worked out following the outlines of \cite{bagchi_fring,fring_gouba_scholtz,fring_gouba_bagchi_area}, which has not been explored notably in the literature. We also show the explicit construction procedure of different kind of models in higher dimensions. As for example, we present the harmonic oscillator models in one, two and three dimensions and solve them finally by utilising the usual perturbation technique. 

In chapter \ref{chapter_solvable} \cite{dey_fring_khantoul}, we probe the procedure of how the models, or more appropriately, a class of models in the noncommutative spaces could be solved exactly, rather than using the standard Rayleigh-Schr{\"o}dinger perturbation theory. As a first example we choose the one dimensional harmonic oscillator in the noncommutative space in different representations which has been constructed in the previous chapter and thereafter a couple of other models, namely the popular model of Swanson and a version of the P{\"o}schl-Teller model in disguise.

Instead of constructing complicated mathematical models in noncommutative spaces, in chapter \ref{chapter_coherent} \cite{dey_fring_squeezed,dey_fring_gouba_castro}, our quest is to explore the physical implications of these kind of theories, which are indeed more exciting to realise the connection to the physical world. We construct the coherent states of a couple of models, at first, the perturbative harmonic oscillator in the noncommutative space and later that of the exact case. In both the cases we pick up the Klauder coherent states instead of the Glauber coherent states (as Klauder coherent states are applicable to generalised models, whereas Glauber coherent states are suitable for harmonic oscillator only) and analyse the expectation values of the observables systematically to verify the generalised uncertainty relation and show that the product of the position and momentum uncertainties are indeed saturated which must hold for the case of qualitatively good coherent states. We also check the compatibility of the results with the Ehrenfest's theorem and the fractional revival structure (see \ref{section65}) of the systems to test the qualities further.

In chapter \ref{chapterBohmian} \cite{dey_fring_bohmian,dey_fring_bohmian_2}, we build up an absolutely different procedure of inspecting the qualitative behaviour of the coherent states, mainly based on numerical techniques. Certainly this procedure provides an impressive insight into the theory, because, with this method one can compute the dynamics of the coherent states and the classical cases separately and then compare them together to judge the quality of the coherent states. For this, we utilize the usual formulations of Bohmian mechanics and apply them to calculate the dynamics of the coherent states, whereas the dynamics of the classical cases are computed by solving the standard canonical equations of motion. To provide some concrete examples, we choose the harmonic oscillator model first, as it is always very instructive to understand the basic foundation and then we choose more complicated models such as P{\"o}schl-Teller model and a Calogero type potential having singularity at the origin. We probe the method for both the real and complex Hamiltonians, with the emphasis on the complex side. The later case is more interesting from the classical point of view and in addition it is itself an active field of research. However, we will see that the pathway will not be so spontaneous as we expect, because the coherent states do not always behave like a classical particle but exhibit the exact classical behaviour under certain circumstances. We introduce the Mandel parameter $Q$ (\ref{mandel}) to be the additional condition and impose the restriction that $Q$ must be zero or very close to zero which is not quite visible from the mathematical analysis. Initially we investigate the above procedure for the models in the usual space, however, our actual motivation is to examine the quality of the coherent states for the models of noncommutative space. We noticed that most of the models explored so far in noncommutative space are in momentum space and no one has ever solved them in position space. On the other hand, Bohmian mechanics has not been formulated in the momentum space yet and are subject to construction which is very contradictory and of course not satisfactory. However, we do keep the interpretational issues aside and focus on the construction of position space wave function of the noncommutative harmonic oscillator. We provide a Fourier transformed version of this, which we have not applied to the context yet, but one can surely utilize those wave functions to investigate further. 

While investigating the Calogero type potential we find an interesting observation in the context of complex classical mechanics. People usually claim \cite{bender_holm_hook, bender_holm_hook_2} that the trajectories of classical particles moving in complex potentials, whose energies are real, are always closed and periodic, whereas those having complex energies are always open and therefore demand that the complex but $\mathcal{PT}$-symmetric classical Hamiltonians always produce close and periodic trajectories, whereas non-$\mathcal{PT}$-symmetric Hamiltonians possess open trajectories. We contradict with this statement and claim that this is no longer true in general and thus dispute the connection of complex classical trajectories with the $\mathcal{PT}$-symmetry of the Hamiltonians.


\chapter{Noncommutative Space-Time and Minimal Length} \label{chapter2}
Before starting rigorous discussion on various different notions of noncommutative theory, let us first quickly recall some of the fundamental results of the theory of quantum mechanics. A quantum mechanical phase-space is defined by replacing the canonical position and momentum variables $x$, $p$ with the quantum mechanical operators $\hat{x}$, $\hat{p}$, which are by definition Hermitian and obey the Heisenberg's canonical commutation relation $\left[\hat{x}_i,\hat{p}_j\right]=i \hbar~\delta_{ij}$. The ordinary phase space is recovered in the classical limit $\hbar \rightarrow 0$ (In this energy scale, this often corresponds to large quantum numbers $n$ for discrete energy levels). The birth of the Heisenberg uncertainty principle essentially takes place with the introduction of the canonical commutation relation, which can be recognized easily, by assuming $\vert \psi \rangle$ to be the eigenvector of position as well as the momentum operator with the eigenvalues $x_0$ and $p_0$ respectively, such that
\begin{equation}
\left[\hat{x},\hat{p}\right]\vert\psi\rangle=\left(x_0 p_0-p_0 x_0\right) \vert\psi\rangle=0.
\end{equation}
However, the Heisenberg's canonical commutation relation requires 
\begin{equation}
\left[\hat{x},\hat{p}\right]\vert\psi\rangle=i\hbar \vert\psi\rangle\neq 0,
\end{equation}
which implies that no quantum state can be simultaneous position and momentum eigenstate, or in other words, position and momentum observables can not be measured simultaneously with a high precision. The more precisely the position of a quantum particle is measured, the less precise the measurement of momentum becomes and vice-versa. The product of the uncertainties is bound by a constant, represented by the reduced Planck's constant $\hbar$
\begin{equation} \label{commrelation}
\Delta x\Delta p  \geq \frac{1}{2}\left\vert \left\langle [x,p]\right\rangle\right\vert = \frac{\hbar}{2}~~,
\end{equation}
where $\Delta x=\sqrt{\langle x^2 \rangle-\langle x \rangle ^2}$ and $\Delta p=\sqrt{\langle p^2 \rangle-\langle p \rangle ^2}$ are the standard deviations of position and momentum observables. Thus, with the introduction of Heisenberg's uncertainty principle, the concept of quantum mechanical phase space is changed drastically and the notion of a point is replaced with that of the Planck cell. It was John Von Neumann \cite{birkhoff_neumann}, who first came forward to describe the quantum phase space rigorously, referring to the fact that the notion of a point in a quantum phase space is meaningless because of the Heisenberg uncertainty principle of quantum mechanics and dubbed his study as "pointless geometry". This led to the theory of Von Neumann algebras and essentially to the birth of "noncommutative geometry", referring to the study of topological spaces whose commutative $C^\ast$-algebras \cite{connes} of functions are replaced by noncommutative algebras \cite{connes,madore,landi}.

\rhead{Noncommutative Spacetime $\&$ Minimal Length}
\lhead{Chapter 2}
\chead{}

\section{Flat Noncommutative Space} \label{sectionflat}
Just like the quantisation of classical phase space, the simplest and most commonly studied version of non-commutative spaces replace the standard set of commuting coordinates  by the Hermitian generators $x^\mu$ of a
noncommutative $C^\ast$-algebra of "functions on spacetime" which obey the commutation relations
\begin{equation} \label{nccommutator}
\left[x^\mu,x^\nu\right]=i \theta^{\mu\nu}, \qquad \left[x^\mu,p_\nu\right]=i\hbar \delta^\mu_\nu,\quad \text{and} \quad \left[p_\mu,p_\nu\right]=0,
\end{equation}
with the deformation parameter $\theta^{\mu\nu}$ being a constant antisymmetric tensor, where $\mu,\nu=1,2,3,4$. Because the space-time coordinates no longer commute in this scenario, the underlying space disappears and the noncommutative space is introduced.

However, there are some crucial issues related with the application of (\ref{nccommutator}) to physical problems. In standard relativistic theory, one can immediately observe that a nonvanishing $\theta^{\mu\nu}$ can and does break the Lorentz-Poincar{\'e} symmetries \cite{carroll_harvey_kostelecky_lane_okamoto,banerjee_chakraborty_kumar}. Indeed, the coordinates $x^\mu$ transform as vectors, while $\theta^{\mu\nu}$ is constant in all reference frames. Nevertheless, in spite of this well recognised problem, all fundamental issues like unitarity \cite{gomis_mehen}, causality \cite{seiberg_susskind_toumbas}, UV/IR divergences \cite{minwalla_van_seiberg} and anomalies \cite{ardalan_sadooghi,banerjee_ghosh,banerjee1} have been discussed in a formally Lorentz invariant manner, using the representation of the usual Poincar{\'e} algebra. These results have been achieved employing the Weyl-Moyal correspondence, which assigns an established ordinary theory to a noncommutative one by replacing ordinary fields with noncommutative fields and ordinary products with Moyal $\star$ products
\begin{equation}
\phi(x)\psi(x)\rightarrow\phi(x)\star\psi(x),
\end{equation}
where the Moyal $\star$ product was originally introduced long time back, by Moyal \cite{moyal} to facilitate Wigner's phase space formulation of quantum mechanics, which has been applied more recently in a wide field of research, such as noncommutative M-theory \cite{fairlie}, string theory \cite{seiberg_witten}, integrable field theories \cite{dimakis,grisaru,moriconi,lechtenfeld} etc. It is defined as
\begin{equation}\label{moyalast}
\phi(x)\star\psi(x)=\phi(x)e^{\frac{i}{2}\theta^{\mu\nu}\overleftarrow{\partial}_\mu \overrightarrow{\partial}_\nu}\psi(y)\vert_{x=y}~,
\end{equation}
more useful and symmetric version of which \cite{moyal,fairlie1,carroll} is
\begin{equation}
\phi(x)\star\psi(x)=\phi(x)e^{\frac{i}{4}\theta^{\mu\nu}\left(\overleftarrow{\partial}_\mu \overrightarrow{\partial}_\nu-\overleftarrow{\partial}_\nu \overrightarrow{\partial}_\mu\right)}\psi(y)\vert_{x=y}~.
\end{equation}
Consequently, the commutators of operators are replaced by Moyal brackets, for instance in (\ref{nccommutator})
\begin{equation}
\left[x^{\mu},x^\nu\right] \rightarrow \left[x^{\mu},x^\nu\right]_\star\equiv x^\mu\star x^\nu-x^\nu\star x^\mu=i \theta^{\mu\nu}.
\end{equation}
In fact, admitting noncommutativity to be relevant only at very short distances, it has been often treated as a perturbation and only the corrections to first order in $\theta$ were computed. As a result, the NC QFT was practically considered Lorentz invariant in zeroth order in $\theta^{\mu\nu}$, with the first order corrections coming only from the $\star$ product.

However, long before these considerations, Snyder \cite{snyder} in 1947 proposed the noncommutativity in a slightly different manner so that the commutation relations inherit the Lorentz covariance by its construction 
\begin{alignat}{1} \label{snyderalgebra}
\left[x^\mu,x^\nu\right] &=i \theta\left(x^\mu p^\nu-x^\nu p^\mu\right) \nonumber \\
\left[x^\mu,p_\nu\right] &=i\hbar\left(\delta^\mu_\nu+\theta p^\mu p_\nu\right) \\
\left[p_\mu,p_\nu\right] &=0 \nonumber,
\end{alignat}
with $\theta$ being the constant deformation parameter. This is checked comfortably by taking the standard Lorentz transformation $x^\mu\rightarrow x^\mu+\delta x^\mu$ and $p^\mu\rightarrow p^\mu+\delta p^\mu$ with
\begin{eqnarray}
\delta x^\mu &=& \omega^{\mu\alpha}x_\alpha \notag\\
\delta p^\mu &=& \omega^{\mu\alpha}p_\alpha~,
\end{eqnarray}
on both sides of the relations (\ref{snyderalgebra}), where, $\omega^{\mu\alpha}=-\omega^{\alpha\mu}$.  First consider, the transformation on the left hand side of the first equation of (\ref{snyderalgebra})
\begin{eqnarray}
\delta\left[x^\mu,x^\nu\right] &=& \left[\delta x^\mu,x^\nu\right]+\left[x^\mu,\delta x^\nu\right] \notag \\
&=& \omega^{\mu\alpha}\left[x_\alpha,x^\nu\right]+\omega^{\nu\alpha}\left[x^\mu,x_\alpha\right] \notag \\
&=& i\theta\omega^{\mu\alpha}\left(x_\alpha p^\nu-x^\nu p_\alpha\right)-i\theta\omega^{\nu\alpha}\left(x_\alpha p^\mu-x^\mu p_\alpha\right),
\end{eqnarray}
and then the similar expression is obtained by transforming the right hand side of the same relation
\begin{eqnarray}
i\theta\delta\left(x^\mu p^\nu-x^\nu p^\mu\right)=i\theta\omega^{\mu\alpha}\left(x_\alpha p^\nu-x^\nu p_\alpha\right)-i\theta\omega^{\nu\alpha}\left(x_\alpha p^\mu-x^\mu p_\alpha\right).
\end{eqnarray}
An identical treatment follows for the other two relations. Whereas, the usual Lorentz transformations leaves the Snyder algebra (\ref{snyderalgebra}) covariant, it can be easily checked that they are not invariant under the translation symmetry, so that the original Poincar{\'e} symmetry is violated. In \cite{banerjee_kulkarni_samanta}, the authors found the deformation of the original Poincar{\'e} symmetries so that the deformed version becomes compatible with the Snyder algebra (\ref{snyderalgebra}). Subsequently the deformation of the conformal symmetries as well as the action were obtained which immediately tells that one could easily formulate some well defined physical models with the  coordinates corresponding to the noncommutative space-time. Apart from the particular case of Snyder noncommutativity (\ref{snyderalgebra}), many authors have also argued that in general it might be possible to introduce quantum deformations of Poincar{\'e} symmetries such that the particular form of the commutators (\ref{nccommutator}) remain covariant while the deformed Poincar{\'e} generators preserve the original Poincar{\'e} algebra. This has been discussed in great detail, for a constant $\theta^{\mu\nu}$ using either higher order differential operators \cite{wess,aschieri_blohmann_dimitrijevic_meyer_schupp_wess,koch_tsouchnika,banerjee2} or twist functions following from quantum group arguments \cite{chaichian_kulish_nishijima_tureanu,chaichian_presnajder_tureanu,aschieri_dimitrijevic_meyer_wess, castro_kullock_toppan}.  For Lie algebraic and quadratic deformations in (\ref{nccommutator}), such an analysis was carried out in \cite{lukierski_woronowicz}.

However, in our discussion, we will follow a slightly different approach. We start with the deformed commutation relation between the creation and annihilation operators, which is roughly outlined in \cite{kempf,kempf_mangano_mann,gomes_kupriyanov,gomes_kupriyanov_silva}, and construct the noncommutative models out of that. Note that the procedure is slightly different from the flat noncommutative space and is quite often known as the q-deformed noncommutative space in the literature. This approach has many advantages, for instance, it allows for the explicit construction of the entire Fock space \cite{biedenharn,macfarlane,sun_fu}. Moreover, this theory is enriched with other physical implications as indicated in the introduction that this immediately leads to the generalised version of the uncertainty relation giving rise to the minimal measurable length that becomes meaningful in almost all the approaches of quantum gravity. Besides, the theory has also been proved to be compatible with the theory of doubly special relativity (DSR), which we will discuss in detail later in this chapter. However, before that, we first look at the procedure of constructing noncommutative models from the $q$-deformed noncommutative space.

\section{q-Deformed Noncommutative Space}\label{section22}\label{sectionqdef}
The concept of the q-deformation essentially starts with the study of the quantum inverse problem method \cite{takhtadzhan_faddeev} and the solutions to the Yang-Baxter equation \cite{kulish_sklyanin}. The growing interest on the subject connects to the fact that the properties of quantum groups and algebras are quite similar to those of Lie algebras in connection with the representation theory and its applications. The q-deformation of Lie algebras has been discovered by Jimbo \cite{jimbo1,jimbo2} and Drinfeld \cite{drinfeld}. Subsequently, these ideas were implemented to the construction of the q-deformed harmonic oscillator by many authors, for instance one may look at the references \cite{sun_fu,macfarlane,biedenharn,kulish_damaskinsky,arik_coon}.

The natural question that one could ask afterwords is that, why the quantum groups are necessary to be deformed in this picture. The answer is very simple and lies on the fact that the space-time structures at the Planck scale do not produce the usual scenario, rather, they are deformed in these cases which gives rise to the noncommutative structure of the space-time and that is the reason that the deformation of the structure of the Lie groups and Lie algebras become necessary in order to describe various different features of these kind of spaces. Quite naturally, there are numerous ways of deforming the standard commutation relations between the dynamical variables $X$ and $P$. One might for instance deform the Heisenberg's canonical commutation relations \cite{schwenk,brodimas_jannussis_mignani,kempf_mangano_mann,quesne_tkachuk} which is of the form
\begin{equation}\label{simpdeform}
PX-qXP=i\hbar,
\end{equation}
for other possibilities one may look at \cite{sun_fu,macfarlane,biedenharn}. However, we will assume instead the $q$-deformation \cite{bagchi_fring} of the corresponding commutation relation between the creation and annihilation operator of so-called $q$-bosons $A^\dagger$ and $A$, respectively, to be deformed in the form 
\begin{equation} \label{q2deformation}
AA^\dagger-q^2 A^\dagger A=1
\end{equation}
and analyse systematically how the deformed relation (\ref{q2deformation}) leads to the non-commutative space-time structure. We will assume further that the representations for the position $X$ and momentum $P$ operator to be linear in $A$ and $A^\dagger$
\begin{equation}\label{qXP}
X=\alpha \left(A^\dagger+ A\right), \qquad P=i\beta\left(A^\dagger-A\right), \qquad \alpha, \beta \in \mathbb{R}.
\end{equation}
Thereafter using equation (\ref{q2deformation}) and (\ref{qXP}), the commutation relation of the position $X$ and momentum $P$ operators is computed to be of the form
\begin{equation}\label{XPcom}
[X,P]=2i\alpha\beta\left[1+\left(q^2-1\right)A^\dagger A\right].
\end{equation}
It is worth mentioning at this point, that the generators of the q-deformed noncommutative spaces will always be represented from now and onwards, by the capital letters of the corresponding observables. 

Using equation (\ref{qXP}), one can express the product of the creation and annihilation operator as
\begin{equation}
A^\dagger A=2i\alpha\beta\left[1+\frac{q^2-1}{4}\left(\frac{X^2}{\alpha^2}+\frac{P^2}{\beta^2}+\frac{i}{\alpha\beta}\left[X,P\right]\right)\right],
\end{equation}
which when replaced in (\ref{XPcom}) and solved subsequently for $\left[X,P\right]$, one obtains the $q$-deformed commutation relation
\begin{equation}\label{XPcompartial}
\left[X,P\right]=\frac{4i\alpha\beta}{1+q^2}\left[1+\frac{q^2-1}{4}\left(\frac{X^2}{\alpha^2}+\frac{P^2}{\beta^2}\right)\right].
\end{equation}
The relation (\ref{XPcompartial}) can be simplified successively by choosing constraint on the parameter
\begin{equation}
\alpha \beta=\frac{\hbar}{2} \qquad\text{i.e.}\qquad\alpha=\frac{\hbar}{2 \beta},
\end{equation}
and assuming further that the deformation parameter to be of the form $q=e^{2 \tau \beta^2}$ ($\tau \in \mathbb{R}^+,[\tau]=[P]^{-2}$), with a subsequent non-trivial limit $\beta \rightarrow 0$, one obtains the simple deformed canonical commutation relation
\begin{equation} \label{ncsimplecommutator}
[X,P]=i \hbar \left(1+\tau P^2\right).
\end{equation}
It is customary to look at the trivial limit $\tau \rightarrow 0$, i.e. $q \rightarrow 1$, for which the commutation relation becomes the usual canonical commutation relation. However, having obtained a simple version of the noncommutative commutation relation (\ref{ncsimplecommutator}), it is quite straight forward to find a representation for $X$ and $P$, which will reproduce the commutator (\ref{ncsimplecommutator}). For instance we may select the position $X$ and momentum $P$ operators in terms of the undeformed observables $x$ and $p$, i.e. $[x,p]=i\hbar$, as
\begin{equation}
X=\left(1+\tau p^2\right) x \qquad \text{and} \qquad P=p,
\end{equation}
which one may utilize in the formulation of different kinds of models. In \cite{bagchi_fring}, such an example is explicitly computed for the case of one dimensional harmonic oscillator. Examples in higher dimensions have also been shown, such as the free particle \cite{fring_gouba_scholtz} and the harmonic oscillator \cite{fring_gouba_bagchi_area} in two dimension. We have extended the ideas in chapter \ref{noncommutativemodelsin3D} to the three dimensional harmonic oscillator and subsequently reduced to the lower dimensional cases, which are clearly a challenge from the computational point of view. One should note that the construction of models in these spaces would not be an easy task as the operators $X$ and $P$ in the deformed commutation relation (\ref{ncsimplecommutator}) are in general not Hermitian $X^\dagger=X+2 i \tau \hbar P$ and $P^\dagger=P$, whereas that of the operators in the flat noncommutative spaces are naturally resulting to be Hermitian. We explore the procedure rigorously in Chapter \ref{chapterPT} of how to tackle with the difficulties involving the non-Hermitian operators and how a self-consistent physical theory can be constructed out of that. 

\section{Minimal Length}
With the discussion of the previous section, we are now convinced that it is indeed possible to build different types of models based on the noncommutative spaces. Besides the computational challenge of constructing complex mathematical models, as already been mentioned in the introduction, the noncommutative theory possess many compelling physical implications too. To understand this, let us look at the uncertainty relations induced by the flat noncommutative space (\ref{nccommutator}), which requires to replace the "Heisenberg's Uncertainty Relations" by the "Generalised Uncertainty Relations"
\begin{equation} \label{GUP1}
\Delta x^\mu \Delta x^\nu \geq \frac{1}{2}\left\vert\theta^{\mu\nu}\right\vert,
\end{equation}
and consequently the space-time point is replaced by the Planck cell of dimension given by the Planck area. In these kind of situations, one may encounter many interesting phenomena, in particular, when the commutation relations are modified in such a way that their structure constants $\theta^{\mu\nu}$ involve higher powers of coordinates or momenta. The picture is quite clearly visible from the $q$-deformed noncommutative spaces. For instance, selecting the simplest version of the $q$-deformed commutation relation (\ref{ncsimplecommutator}), the generalised version of the uncertainty relation appears to be
\begin{alignat}{1} \label{genuncert}
\Delta X \Delta P & \geq \frac{1}{2}\left\vert\langle [X,P]\rangle\right\vert \nonumber \\ & \geq \frac{\hbar}{2}\left(1+\tau \langle P^2 \rangle\right) \nonumber \\ & \geq \frac{\hbar}{2}\left[1+\tau \left(\Delta P\right)^2+\alpha\right],
\end{alignat} 
with $\alpha=\tau \langle P\rangle^2$. While in ordinary quantum mechanics, as expressed by the equation (\ref{commrelation}), which simply have reduced Planck's constant $\hbar$ as structure constant, $\Delta x$ can be made arbitrarily small by letting $\Delta p$ grow correspondingly, this is no longer the case if the equation (\ref{genuncert}) is taken into account. In this case, if $\Delta X$ is decreased and $\Delta P$ is increased correspondingly, the new term $\tau (\Delta P)^2$ on the right hand side will eventually grow faster than the left hand side. Hence $\Delta X$ can no longer be made arbitrarily small, which essentially gives rise to the so called "minimal length" $\Delta X_0$, beyond which one can not probe and provide any physically measurable information about the system.

Snyder's idea \cite{snyder} was that, if one could find a coherent description of the space-time structure such that the space-time point is replaced by a small length scale, then the ultraviolet divergences of quantum field theory could be eliminated, and that would be equivalent to use an ultraviolet cut-off in momentum space integrations to compute the Feynman diagrams, which inevitably lead to a minimum observable length scale below which all the phenomena are ignored. The old belief was therefore that the simplest and most elegant way of introducing the cut-off is in terms of the space-time coordinates $x^\mu$. Many other possibilities have been explored from time to time, which give rise to other interesting physical interpretations. However, before analysing those kind of situations, we will first investigate the generalised uncertainty relation (\ref{genuncert}) associated with the noncommutative algebras and see how the minimal length can be calculated explicitly and thus the compatibility with Snyder's idea. We start with considering the relation, $f\left(\Delta X,\Delta P\right)=\Delta X \Delta P-\frac{1}{2}\left\vert\langle [X,P]\rangle\right\vert$, which needs to be minimised as a function of $\Delta P$, in order to determine the minimum value for $\Delta X=\Delta X_0$. This means, we need to solve the two equations 
\begin{alignat}{1}
\partial_{\Delta P}f\left(\Delta X,\Delta P\right) & =0 \\ \nonumber
f\left(\Delta X,\Delta P\right) & =0,
\end{alignat}
and subsequently compute the smallest value for $\Delta X_{\text{min}}$ in order to obtain the absolute minimal length $\Delta X_0$ for which $\langle X \rangle~ \text{and}~\langle P \rangle$ are taken to be zero. In the standard scenario, i.e. when $x$ and $p$ commute up to a constant, the result is therefore zero. For the case at hand, the commutation relation involves higher powers of $X$ or $P$ and therefore the minimal length takes the form
\begin{equation}
\Delta X_{\text{min}}=\hbar \sqrt{\tau}\sqrt{1+\tau \langle P \rangle^2},
\end{equation}
so that taking $\langle P \rangle=0$, the absolutely smallest uncertainty in position has the value
\begin{equation}
\Delta X_0=\hbar \sqrt{\tau}.
\end{equation}
The result can be visualised from figure \ref{fig0}.
 
\begin{figure}[h]
\centering   \includegraphics[scale=0.6]{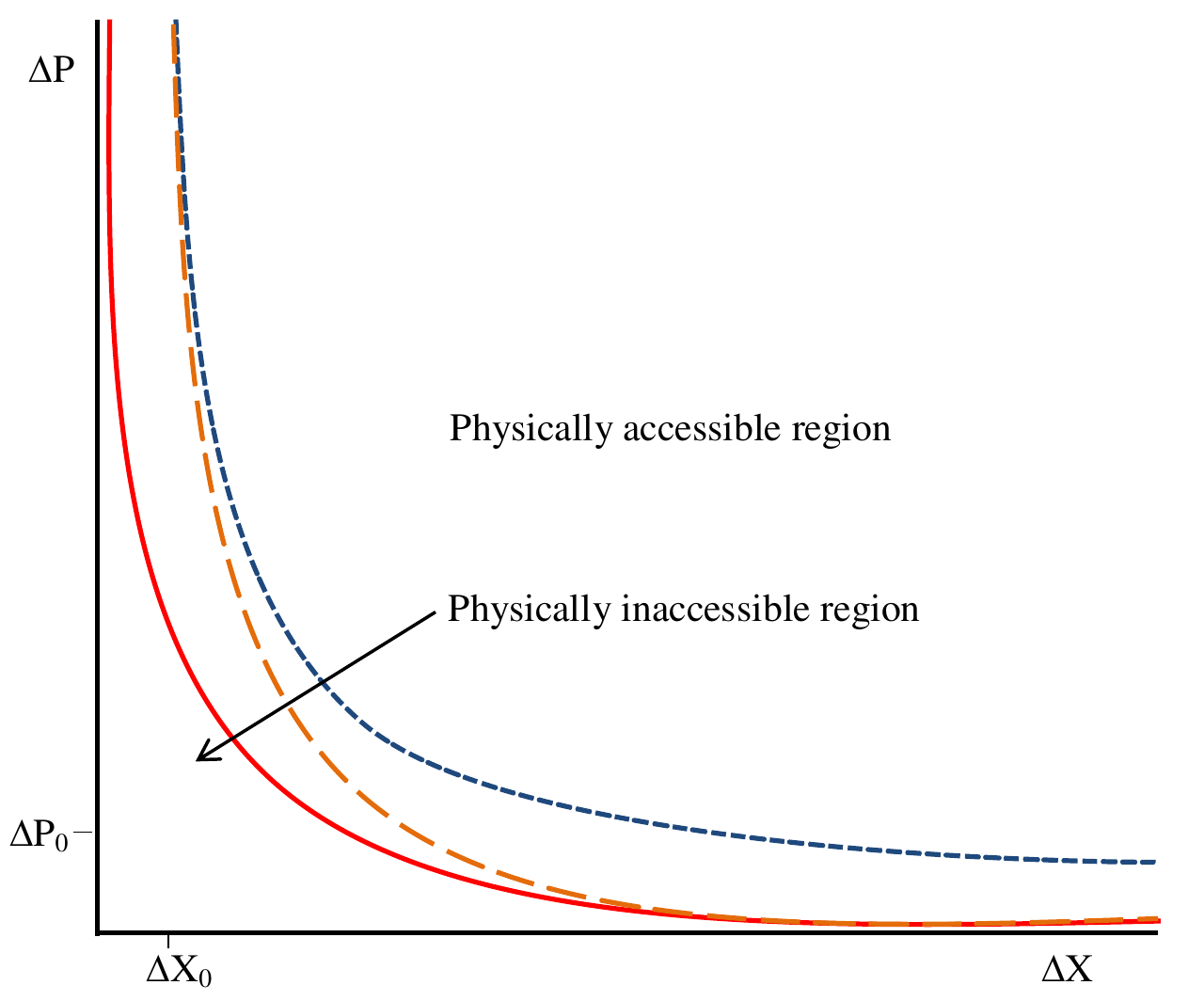} \centering   
\caption{\small{Solid red line corresponds to the Heisenberg's uncertainty relation which shows neither minimal length nor minimal momentum, Orange dashed line corresponding to Generalised uncertainty relation (\ref{genuncert}) leads to minimal length and Blue dotted line represents equation (\ref{genuncert227}) leading to both minimal length and minimal momentum.}}
\label{fig0}
\end{figure}  

Note that there is no non-vanishing minimal uncertainty in momentum in this case (\ref{genuncert}). This circumstance always arises when the commutation relation is extended in such a way that the uncertainty relation involves the higher power of both the position and momentum uncertainty, 
\begin{alignat}{1} \label{genuncert227}
[X,P] & = i\hbar \left(1+\tau X^2+\rho P^2\right) \nonumber \\
\Delta X \Delta P & \geq \frac{\hbar}{2}\left[1+\tau \left(\Delta X\right)^2+\rho \left(\Delta P\right)^2+\alpha\right],
\end{alignat}
with $\alpha=\tau \langle X\rangle^2+\rho \langle P\rangle^2$, so that it turns out a non-zero minimal uncertainty in both the position $\Delta X_0$ and momentum $\Delta P_0$. 

Recently, it has been argued that Einstein's special relativity predictions appear to be violated by certain observations of ultra-high-energy cosmic rays \cite{bird} and at the same time quantum gravity arguments also encourage the idea. The minimal length which is a natural candidate of the quantum theory of gravity, by definition, should not undergo a Lorentz contraction when it is boosted. That means a modification of the Lorentz transformations becomes necessary. The new transformations should not only leave the speed of light $c$ to be invariant, but also keep the minimal length as a second invariant. Such a modification has been achieved recently by many authors \cite{amelino3,magueijo_smolin,amelino1,amelino2,kowalski}, which has been dubbed as the doubly special relativity (or deformed special relativity) in the literature. These transformations can be described through the generators of the $\kappa$-Poincar{\'e} Hopf algebra \cite{lukierski_ruegg_nowicki_tolstoy,majid_ruegg,lukierski_ruegg_zakrzewski,kowalski2,kowalski_nowak, kowalski_nowak2}, and therefore it is evident that the doubly special relativity also suggests a similar kind of modification of the usual canonical commutation relations and essentially the noncommutative structure of the space-time.  

It has also been suggested in this context that the minimal measurable length would restrict a test particle's momentum to take any arbitrary values which deliberately leads to an upper bound $P_{\text{max}}$ of the momentum. This effect is also visible from the considerations of generalised uncertainty relation, for instance, if we start with the commutation relation with a linear term in momentum, $[X,P]=i \hbar \left(1-\alpha P+ \beta P^2\right)$, and pursue the same calculation as done before, one could obtain the maximal value of $P=P_{\text{max}}$ by replacing the minimal value of $X$ into the equation \cite{magueijo_smolin,magueijo_smolin2,cortes_gamboa,ghosh_subir,ali_das_vagenas,nozari_etemadi}.

The crucial point that one should note at this point is that the calculations which are carried out here are just few easy examples in one dimension but the same analysis could also be performed in higher dimensions. In that case one would eventually follow up more exciting situations, for example in two dimensions one could obtain many similar type of non-zero commutation relations involving the operators corresponding to position and momentum in $X$ and $Y$ direction and achieve a membrane type "minimal area" \cite{fring_gouba_bagchi_area,fring_gouba_scholtz}. By now many studies on the structure of such type of generalised canonical relations have been carried out \cite{kempf,kempf_mangano_mann,sun_fu,macfarlane,biedenharn,brodimas_jannussis_mignani,quesne_tkachuk, bagchi_fring,fring_gouba_scholtz,fring_gouba_bagchi_area,hossenfelder1,chang_lewis_minic_takeuchi, lewis_takeuchi,pedram_nozari_taheri,dorsch_nogueira,nozari_etemadi,gavrilik_kachurik} , albeit mostly in dimensions less than three. In chapter \ref{noncommutativemodelsin3D}, we explore how the situation can be dealt with in three dimension and play with the commutators to obtain a fuzzy "minimal volume" \cite{dey_fring_gouba}.


\chapter{Non-Hermitian Hamiltonians $\&$ Noncommutativity} \label{chapterPT}
Before moving to the main part of the work, in this chapter we will briefly review some necessary parts of the theory of $\mathcal{PT}$-symmetric non-Hermitian Hamiltonian systems, in particular the pseudo-Hermitian Hamiltonians. As we have already discussed in the introduction why it would be likely to start the discussion with the topic, that is because, whenever we consider the systems of noncommutative spaces, it is always more probable that we end up with the Hamiltonian systems which are non-Hermitian with respect to the standard inner products. With the knowledge of how to deal with the non-Hermitian Hamiltonians and how to develop a consistent quantum mechanical theory out of the non-Hermitian dissipative systems, it would always be easy to tackle the difficulties arising in systems of non-commutative space. To start our discussion, let us first introduce few basic notions of quantum theory. 

\section{Hermiticity Leads to Reality of Spectrum}
The theory of quantum mechanics is nearly one hundred years old and has become an acceptable theory in modern science because of the wide range of success in verification of its theoretical predictions in form of experimental results. In an introductory course of quantum physics, one learns the basic axioms that define and characterize the theory. For example the energy spectrum is required to be real such that all measurements of the energy of a system yield real results, eigenvalues are bounded below so that there exists a ground state and most importantly the time evolution of a quantum state must be unitary because of the fact that the expected result of a probability measurement of a state cannot grow or decay in time. A quantum theory of elementary particles must also satisfy the physical axioms of Lorentz covariance and causality. \\
    \indent However, there is another axiom which is rather more mathematical in character, that is the requirement that the Hamiltonian operator $H$, which governs the dynamics of the quantum system, must be Hermitian. The Hermiticity of $H$ is expressed by the equation
\begin{equation}
\langle\psi\vert H\psi\rangle=\langle H\psi\vert\psi\rangle,\quad \text{i.e.}\quad H=H^\dagger \label{dirac}
\end{equation}
where the Dirac Hermitian conjugation symbol '$\dagger$' represents the combined operations of matrix transposition and complex conjugation. We will adopt here the conventions used in the physics literature and avoid technicalities of domain issues leading to the well-known distinction between Hermitian and self-adjoint operators. The mathematical condition (\ref{dirac}) is very convenient to use as a machinery to derive all the above axioms using this single requirement. For instance, we consider a diagonalisable operator,  Hamiltonian $H$ with discrete eigenvalues $a^\prime$ and $a^{\prime\prime}$ with right and left eigenvectors, $\vert a^\prime\rangle$ and $\langle a^{\prime\prime}\vert$ respectively 
\begin{alignat}{1}
H\vert a^\prime\rangle &=a^\prime\vert a^\prime\rangle \\
\langle a^{\prime\prime}\vert H^\dagger &=a^{\prime\prime\ast} \langle a^{\prime\prime} \vert.
\end{alignat}
Projecting the state on the left to the first equation by $\langle a^{\prime\prime} \vert$ and second equation on the right by $\vert a^\prime\rangle$ and subtracting, we obtain
\begin{equation}\label{eqn34}
\langle a^{\prime\prime}\vert H \vert a^\prime\rangle-\langle a^{\prime\prime}\vert H^\dagger\vert a^\prime\rangle=\left(a^\prime-a^{\prime\prime\ast}\right)\langle a^{\prime\prime}\vert a^\prime\rangle.
\end{equation}
Considering the Hamiltonian operator Hermitian, the above expression (\ref{eqn34}) reduces to 
\begin{equation}\label{eqn35}
\left(a^\prime-a^{\prime\prime\ast}\right)\langle a^{\prime\prime}\vert a^\prime\rangle=0.
\end{equation}
Now $a^{\prime}$ and $a^{\prime\prime}$ can be taken to be either same or different. Let us first choose them  to be same. We consider the fact that $\vert a^\prime\rangle$ is not a null ket, we then deduce

\rhead{Non-Hermitian Hamiltonians $\&$ Noncommutativity}
\lhead{Chapter 3}
\chead{}

\begin{equation}
a^\prime=a^{\prime\ast},
\end{equation}
which is to say that the eigenvalues are real, if and only if the Hamiltonian operator is Hermitian. Let us now assume $a^\prime$ and $a^{\prime\prime}$ to be different. Because of the just-proved reality condition, the difference $\left(a^\prime-a^{\prime\prime\ast}\right)$ that appears in (\ref{eqn35}) is equal to $\left(a^\prime-a^{\prime\prime}\right)$, which can not vanish by assumption. The inner product $\langle a^{\prime\prime}\vert a^\prime\rangle$ must then vanish:
\begin{equation}
\langle a^{\prime\prime}\vert a^\prime\rangle=0,~~~\left(a^\prime\neq a^{\prime\prime}\right)
\end{equation}
which says the eigenstates of $H$ corresponding to different eigenvalues are orthogonal. It is conventional to normalize $\vert a^\prime\rangle$ so that $\{\vert a^\prime\rangle\}$ form an orthonormal set:
\begin{equation}
\langle a^{\prime\prime}\vert a^\prime\rangle=\delta_{a^{\prime\prime}a^\prime}~~.
\end{equation}
It is easy to check that the set of eigenkets are complete:
\begin{equation}
\displaystyle\sum\limits_{a^\prime}\vert a^\prime\rangle\langle a^\prime\vert=\bf{1}.
\end{equation}
The dynamics of a quantum state can be obtained by operating the time evolution operator on the state as:
\begin{equation}
\vert a^\prime\left(t\right)\rangle=e^{-iHt}\vert a^\prime\left(0\right)\rangle,
\end{equation}
such that the probability density is conserved
\begin{equation}
\langle a^\prime(t)\vert a^\prime(t)\rangle=\langle a^\prime(0)\vert e^{iH^\dagger t}e^{-iHt}\vert a^\prime(0)\rangle=\langle a^\prime(0)\vert a^\prime(0)\rangle,~~\text{iff}~~H=H^\dagger
\end{equation}
Thus the Hermiticity property ensures the reality of the spectrum.

On the other hand, the Hamiltonians that are non-Hermitian, appear quite often in the literature to describe dissipative systems, such as the phenomenon of radioactive decay. However, these non-Hermitian Hamiltonians are only approximate, phenomenological descriptions of physical processes. They cannot be regarded as fundamental because they violate the requirement of unitarity, thus describing open systems rather than closed self-consistent ones. A non-Hermitian Hamiltonian whose purpose is to describe a particle that undergoes radioactive decay predicts that the probability of finding the particle gradually decreases in time. Of course, a particle cannot just disappear because this would violate the conservation of probability; rather, the particle transforms into other particles. Thus, a non-Hermitian Hamiltonian that describes radioactive decay can at best be a simplified, phenomenological, and non-fundamental description of the decay process because it ignores the precise nature of the decay products. In his book on quantum field theory Barton \cite{barton} gives the standard reasons for why a non-Hermitian Hamiltonian cannot provide a fundamental description of nature: “A non-Hermitian Hamiltonian is unacceptable partly because it may lead to complex energy eigenvalues, but chiefly because it implies a non-unitary S matrix, which fails to conserve probability and makes a hash of the physical interpretation.   
    
\section{Quantum Mechanics of Non-Hermitian Hamiltonians} \label{section2.2}
However, one natural question that one may ask at this point is, whether the non-Hermitian Hamiltonians could play an important role in the formulation of complete and fundamental quantum theories. The possibility that those systems can possess discrete eigenstates with real positive energies has already been indicated by von Neumann and Wigner \cite{neumann} almost eighty five years ago. Later, this type of systems were under more intense scrutiny and nowadays the properties of these so-called BICs (bound states in the continuum) are fairly well understood for many concrete examples \cite{friedrich1,magunov,rotter1} together with their bi-orthonormal eigenstates \cite{persson,moiseyev}.

Whereas the above type of Hamiltonians only possess single states with these "strange properties" \cite{neumann}, it was observed very recently in a ground-breaking numerical study by Bender and Boettcher \cite{bender_boettcher} that the Hamiltonians with potential terms $V = x^2\left(ix\right)^\nu$ for $\nu \geq 0$ possess an entirely real and positive spectrum. With this revolutionary discovery indeed a new era has been opened up in the area of non-Hermitian systems. It was only the introduction of $\mathcal{PT}$ symmetry into the Hamiltonian which guarantees the reality of spectrum. The idea has been enriched afterwards by many authors as we will discuss in the subsequent sections.
    
\subsection{Role of $\mathcal{PT}$-Symmetry in Physics}
The idea of the $\mathcal{PT}$ symmetry was introduced to the subject in the desire of describing the non-Hermitian quantum theories into a consistent framework. The symmetry is associated to systems which are invariant under simultaneous operations of the Parity $\mathcal{P}$ and time reversal $\mathcal{T}$ operators. The role of which can be understood by operating those operators on position and momentum operators
\begin{alignat}{2}\label{PTsym}
\mathcal{P} x \mathcal{P}^{-1} &= -x  &\qquad~~ \mathcal{P} p \mathcal{P}^{-1} &= -p, \nonumber \\  
\mathcal{T} x \mathcal{T}^{-1} &= x & \mathcal{T} p \mathcal{T}^{-1} &= -p.
\end{alignat}
$\mathcal{P}$ and $\mathcal{T}$ are reflection operators, i.e. the square of them yields a unit operator
\begin{alignat}{2}
\mathcal{P}^2 &= \bf{1}  &\qquad~~~ \mathcal{P}^{-1} &= \mathcal{P},\nonumber \\  
\mathcal{T}^2 &= \bf{1} & \mathcal{T}^{-1} &= \mathcal{T}.
\end{alignat}
Note that the fundamental commutation relation of quantum mechanics (Heisenberg algebra) is left invariant under the operation of parity, which ensures the fact that the parity operator $\mathcal{P}$ is linear
\begin{eqnarray}
\mathcal{P}xp\mathcal{P}-\mathcal{P}px\mathcal{P} &=& i\hbar\mathcal{P}^2 \notag \\
xp-px &=& i \hbar~\bf{1}.~~\label{com}
\end{eqnarray} 
On the other hand, the time reversal operator $\mathcal{T}$ leaves the commutation relation (\ref{com}) invariant as well, but this requires that the sign of the complex number $i$ to be changed, such that
\begin{eqnarray}
\mathcal{T} xp \mathcal{T}-\mathcal{T} px \mathcal{T} &=& \mathcal{T} i \mathcal{T}\hbar \notag \\
-xp+px &=& \mathcal{T} i \mathcal{T}\hbar \notag \\
\therefore~~~\mathcal{T} i \mathcal{T}^{-1} &=& -i,
\end{eqnarray}
which establishes the fact that $\mathcal{T}$ is antilinear \cite{wigner}. Therefore the combined $\mathcal{PT}$ operation can be defined as an anti-linear operation,
\begin{equation}\label{PTantilinear}
\mathcal{PT}~:\qquad p\rightarrow p, \quad x\rightarrow -x, \quad i \rightarrow -i.
\end{equation}
The importance of $\mathcal{PT}$-symmetry follows from the work of Bender and Boettcher \cite{bender_boettcher}, where a whole new class of non-Hermitian, but $\mathcal{PT}$ symmetric Hamiltonians,
\begin{equation}
H=p^2+m^2 x^2-\left(ix\right)^N \qquad \text{with} \quad N \in \mathbb{R},
\end{equation}
produce completely real eigenvalues, which was actually motivated by the conjecture by Bessis and Zinn-Justin \cite{bessis} who claimed that the spectrum of the Hamiltonian $H=p^2+x^2+ix^3$ might be real. One should note that all one requires for the reality of the spectrum is an anti-linear symmetry \cite{wigner} and $\mathcal{PT}$ is only one example of anti-linear symmetry. However, with the motivations of $\mathcal{PT}$, the reality of the spectrum was attributed initially to the $\mathcal{PT}$-symmetry of the Hamiltonian. In fact, when the wavefunctions are simultaneous eigenstates of the Hamiltonian and the $\mathcal{PT}$-operator, one can easily argue that the spectrum has to be real \cite{wigner,bender_making_sense,bender_brody_jones}. However despite the fact that $\left[\mathcal{PT},H\right]=0$, this is not always guaranteed, because the $\mathcal{PT}$ operator is an anti-linear operator \cite{weigert}. As a consequence one may also encounter conjugate pair of eigenvalues for broken $\mathcal{PT}$ symmetry \cite{bender_brody_jones}, that is when $[\mathcal{PT},H]=0$ but $\mathcal{PT}\phi\neq \phi$. One may use various techniques \cite{dorey,weigert_detecting} to verify case-by-case, whether the $\mathcal{PT}$ symmetry is broken or not.

Therefore a $\mathcal{PT}$-symmetric Hamiltonian, though they are non-Hermitian, in principle in the $\mathcal{PT}$ unbroken regime can also produce a quantum theory as a Hermitian Hamiltonian. However, what is essential is to have a fully consistent quantum theory whose dynamics is described by a non-Hermitian Hamiltonian. In order to achieve this, one needs to modify the inner product for the corresponding Hilbert space. The natural choice of the inner product suitable for $\mathcal{PT}$-symmetric quantum mechanics is the $\mathcal{PT}$-inner product which can be defined as
\begin{equation}
\langle \phi \vert \psi \rangle ^{\mathcal{PT}}=\int \left[\phi\left(x\right)\right]^{\mathcal{PT}}\psi\left(x\right) dx=\int \left[\phi\left(-x\right)\right]^\ast \psi \left(x\right)dx.
\end{equation}
Note that, the boundary conditions (vanishing $\phi$, $\psi$ at $x\rightarrow\pm\infty$) must be imposed properly at this point to solve the eigenfunctions of the time independent Schr{\"o}dinger equation, which are located in this context within the wedges bounded by Stokes lines in the complex x-plane \cite{bender_boettcher} and that is the reason that one must integrate the system within this specified region. However, the inner product is not yet acceptable to formulate a valid quantum theory, because the norm of a state is not always positive definite, which is once again due to the fact that the wavefunctions may not be simultaneous eigenfunctions of $H$ and $\mathcal{PT}$ due to the antilinearity property of the $\mathcal{PT}$ operator. Bender, Brody and Jones \cite{bender_brody_jones} overcame this problem consistently by introducing a new type of inner product, namely the $\mathcal{CPT}$-inner product:
\begin{equation} \label{CPT}
\langle \phi \vert \psi \rangle ^{\mathcal{CPT}}=\int_{\Gamma}\left[\phi\left(x\right)\right]^{\mathcal{CPT}} \psi \left(x\right) dx,
\end{equation}
where 
\begin{equation}
\phi^{\mathcal{CPT}}(x)=\int \mathcal{C}(x,y)\phi^\ast(-y)~dy,
\end{equation}
with $\Gamma$ being the contour in the Stokes wedges and $\mathcal{C}$ being the charge conjugation operator, which commutes with both $H$ and $\mathcal{PT}$. The $\mathcal{C}$ operator is defined in coordinate space as a sum over the normalized $\mathcal{PT}$ eigenfunctions $\phi_n$ of the Hamiltonian: 
\begin{equation}\label{cop}
\mathcal{C}\left(x,y\right)=\sum_{n=0}^{\infty}\phi_n\left(x\right)\phi_n\left(y\right),
\end{equation}
which has eigenvalues $\pm 1$ and measures the sign of the $\mathcal{PT}$-norm of an eigenstate. As a result, states with negative norms are multiplied by $-1$ when acted on by the $\mathcal{C}$ operator and therefore the inner products become positive definite, $\langle \phi_n \vert\phi_m \rangle_\mathcal{CPT}=\delta_{nm}$ and the completeness relation reads \cite{weigert_completeness}:
\begin{equation}
\displaystyle\sum\limits_{n=0}^\infty \phi_n \left(x\right)\left[\mathcal{CPT} \phi_n\left(y\right)\right]=\delta\left(x-y\right).
\end{equation}
However, it is not an easy work to obtain a closed expression of $\mathcal{C}\left(x, y\right)$, one often has to rely on various approximation techniques \cite{bender_wang,bender_jones,bender_tan}. Nevertheless, the series (\ref{cop}) might also be non-convergent, if one does not choose the topology properly. Therefore, in these kind of computations perturbative series might have no meaning at all. Apart from examples of exactly solvable $\mathcal{PT}$-symmetric systems, such as Scarf I \cite{roychoudhury_roy} and matrix models \cite{mostafazadeh4,mostafazadeh6,mostafazadeh_time_dependent}, it is almost impossible to obtain a closed form of the operator. Therefore the initial drawback of this formulation was that the $\mathcal{C}$ operator needed to be determined dynamically, which requires in principle the knowledge of all eigen-functions.

Meanwhile also alternative methods have been developed to compute the $\mathcal{C}$ operator and the procedure circumvents the difficult problem of evaluating the infinite sum (\ref{cop}). For instance, noting that $\mathcal{C}$ is a symmetry of the Hamiltonian and in addition an involution, such that one may compute it alternatively by solving the algebraic equations \cite{bender_brody_jones_prd}
\begin{equation} \label{CPTsolution}
\left[\mathcal{C},H\right]=0, \qquad \left[\mathcal{C},\mathcal{PT}\right]=0 \qquad \text{and} \qquad \mathcal{C}^2=1.
\end{equation}
It might not be possible to solve the equations (\ref{CPTsolution}) in an exact and unique fashion, however, one can invoke the usual perturbation technique \cite{bender_brody_jones_prd} instead.

Thus studies on $\mathcal{PT}$-symmetric quantum mechanics certainly make one point clear, namely the requirement of Dirac Hermiticity ($H = H^\dagger$) for a Hamiltonian to possess real eigenvalues might be relaxed and the Dirac Hermiticity condition could be replaced by $H=H^{\mathcal{PT}}$, provided that one works in the unbroken $\mathcal{PT}$ regime. However, it is worth mentioning that the $\mathcal{PT}$-symmetry is only a sufficient condition but not necessary for the reality of spectrum. Meanwhile alternative possibility arose at around the same time which is recognised by the name "pseudo-Hermiticity of the Hamiltonian".

\subsection{Pseudo Hermiticity: an Alternative Approach to the Reality of Spectrum}\label{subsection322}
The concept of pseudo-Hermiticity was introduced very early in 1940s by Dirac and Pauli \cite{pauli}, and was discussed later by Lee, Wick, and Sudarshan \cite{gupta,bleuler,sudarshan,lee}, who were trying to resolve the problems that arose in the context of quantizing electrodynamics and other quantum field theories in which negative norm states appear as a consequence of renormalization. Even before the discovery of $\mathcal{PT}$-symmetry \cite{bender_boettcher} and the introduction of the $\mathcal{CPT}$-inner product (\ref{CPT}), there have been very general considerations \cite{dieudonne,scholtz_geyer_hahne} addressing the question of how a consistent quantum mechanical framework can be constructed from the non-Hermitian Hamiltonian systems. It was understood at that time that quasi-Hermitian systems \cite{dieudonne,scholtz_geyer_hahne} would lead to positive inner products. The concept was illustrated later by Mostafazadeh \cite{mostafazadeh1,mostafazadeh2,mostafazadeh3,mostafazadeh4}, who proposed that instead of considering quasi-Hermitian Hamiltonians one may investigate pseudo-Hermitian Hamiltonians satisfying
\begin{equation}\label{hermiticity}
h=\eta H \eta^{-1}=h^\dagger=\eta^{-1}H^\dagger \eta \quad \Leftrightarrow \quad H^\dagger=\rho H \rho^{-1}~\text{with}~\rho=\eta^\dagger\eta,
\end{equation}
where $\rho$ is a linear, invertible, Hermitian and positive operator acting in the Hilbert space, such that $H$ becomes a self-adjoint operator with regard to this metric $\rho$, as explained in more detail below. $\eta$ is often called the Dyson map \cite{dyson}. Note that the usual Hermiticity condition is recovered with the choice of $\eta$ to be $\bf{1}$. Since the Hermitian Hamiltonian $h$ and non-Hermitian Hamiltonian $H$ are related by a similarity transformation, they belong to the same similarity class and therefore have the same eigenvalues.

We should stress that this is the most frequently used terminology and at times it is mixed up and people imply different properties by using the same names. Making no assumption on the positivity of the $\rho$ in (\ref{hermiticity}), the relation on the right-hand side constitutes the well known "pseudo-Hermiticity" condition \cite{sudarshan,mostafazadeh2,mostafazadeh5}, when the operator $\rho$ is linear, invertible and Hermitian. In case the operator $\rho$ is positive but not invertible, this condition is usually referred to as "quasi-Hermiticity" \cite{scholtz_geyer_hahne,dieudonne,williams}. With regard to the properties of discrete spectra of $H$, the difference is irrelevant as both conditions may be used to establish its reality. However, in the case of pseudo-Hermiticity this is guaranteed, whereas in the case of quasi-Hermiticity one merely knows that it could be real. With regard to the construction of a metric operator the difference becomes also important, since pseudo-Hermiticity may lead to an indefinite metric, whereas quasi-Hermiticity will guarantee the existence of a positive definite metric.
\begin{center}
  \begin{tabular}{| l | c | c | c |}
    \hline
    &$H^\dagger=\eta^\dagger\eta H\left(\eta^\dagger\eta\right)^{-1}$ & $H^\dagger \rho =\rho H$ & $H^\dagger =\rho H \rho^{-1}$ \\ \hline
    Positivity of $\rho$ & \checkmark & \checkmark & $\times$ \\ \hline
    Hermiticity of $\rho$ & \checkmark & \checkmark & \checkmark \\ \hline
    Invertibility of $\rho$ & \checkmark & $\times$ & \checkmark \\ \hline
    Spectrum of $H$ & Real & Could be real & Real \\ \hline
    Definite metric & Guaranteed & Guaranteed & Not conclusive \\ \hline
    Terminology for $H$ & (\ref{hermiticity}) & Quasi-Hermiticity & Pseudo-Hermiticity \\
    \hline
  \end{tabular}
\end{center}

Coming back to the discussion of equation (\ref{hermiticity}), the time-independent Schr{\"o}dinger equations corresponding to the Hermitian and non-Hermitian Hamiltonian can be written down as
\begin{equation}
h\phi=\epsilon \phi \qquad \text{and} \qquad H\Phi=\epsilon \Phi,
\end{equation}
where the wavefunctions are related as
\begin{equation}
\Phi=\eta^{-1} \phi.
\end{equation}
Therefore, the inner products for the wavefunctions $\Phi$ related to the non-Hermitian $H$ may now simply taken to be
\begin{equation}\label{innerproduct}
\langle\Phi\vert\Phi'\rangle_\eta:=\langle\Phi\vert\eta^2\Phi'\rangle,
\end{equation}
where the inner product on the right hand side of (\ref{innerproduct}) is the conventional inner product associated to the Hermitian Hamiltonian $h$. Crucially we have $\langle\Phi\vert H \Phi'\rangle_\eta=\langle H \Phi\vert\Phi'\rangle_\eta$.

To summarize, it is relatively straight forward to compute the Hermitian counterpart $h$ of the non-Hermitian Hamiltonian $H$, for which one needs to construct the metric operator followed by the equation (\ref{hermiticity}). Thus a key task that remains to calculate in this approach is to find $\rho$ and $\eta$. In practical terms, however, there are very few examples \cite{bagchi_quesne_roychoudhury,znojil_geyer,musumbu_geyer_heiss,assis_fring_lie} where one can compute them in an exact manner, as for example we have worked out \cite{dey_fring_mathanaranjan} an exact form of the metric in the context of Euclidean Lie algebraic Hamiltonians which will be discussed later in this chapter. However there are many other methods such as spectral theory, perturbation technique  \cite{bender_brody_jones_prd,mostafazadeh_batal,jones,mostafazadeh8,krejvcivrik_bila_znojil,mostafazadeh9, ghatak_mandal} etc., which one may follow up for the construction of the metric operator. We provide a quick review on the perturbation method here. As a starting point, one can choose the metric to be of the form $\eta=e^{q/2}$ and employ the Baker-Campbell-Hausdorff formula
\begin{equation} \label{hausdorff}
e^A B e^{-A}=B+\left[A,B\right]+\frac{1}{2!}\left[A,\left[A,B\right]\right]+\frac{1}{3!}\left[A,\left[A,\left[A,B\right]\right]\right]+......
\end{equation}
to calculate the similarity transformation (\ref{hermiticity})
\begin{equation} \label{transformed_ham}
H^\dagger=\eta^2 H \eta^{-2}=H+\left[q,H\right]+\frac{1}{2!}\left[q,\left[q,H\right]\right]+\frac{1}{3!}\left[q,\left[q,\left[q,H\right]\right]\right]+......~~.
\end{equation}
Assuming the non-Hermitian Hamiltonian $H$ to be decomposed into a Hermitian part $h_0$, perturbed by another Hermitian part $h_1$ whose coupling is imaginary
\begin{equation}\label{hoih1}
H=h_0+i \epsilon h_1 \qquad \text{with} \qquad h_i^\dagger=h_i \qquad \text{and} \qquad \epsilon \in \mathbb{R},
\end{equation}    
the relation (\ref{transformed_ham}) acquires the form
\begin{equation}\label{h0h1}
i \left[q,h_0\right]+\frac{i}{2}\left[q,\left[q,h_0\right]\right]+\frac{i}{3!}\left[q,\left[q,\left[q,h_0\right]\right]\right]...=\epsilon \left(2 h_1+\left[q,h_1\right]+\frac{1}{2}\left[q,\left[q,h_1\right]\right]+...\right).
\end{equation}
One can furthermore expand $q$ in powers of $\epsilon$ as $q=\sum_{n=0}^{\infty} \epsilon^n q_n$. Note that the coefficients of even powers of $\epsilon$ contain no additional information because the equation arising from the coefficient of even powers can be derived from the equations arising from the coefficients of odd powers. Therefore for our convenience, $q$ can be expanded as a series in odd powers of $\epsilon$
\begin{equation}\label{qexpansion}
q=\epsilon q_1+\epsilon^3 q_3+\epsilon^5 q_5+.....~~.
\end{equation}
Now replacing the equation (\ref{qexpansion}) into (\ref{h0h1}) and identifying the terms that are of the same order in powers of $\epsilon$, we obtain the first three equations
\begin{alignat}{1} \label{purturbation}
\left[h_0,q_1\right] & = 2 i h_1 \nonumber \\  
\left[h_0,q_3\right] & = \frac{i}{6} \left[q_1,\left[q_1,h_1\right]\right], \\
\left[h_0,q_5\right] & = \frac{i}{6} \left(\left[q_1,\left[q_3,h_1\right]\right]+\left[q_3,\left[q_1,h_1\right]\right]-\frac{1}{60}\left[q_1,\left[q_1,\left[q_1,\left[q_1,h_1\right]\right]\right]\right]\right), \nonumber
\end{alignat}
which can be used to determine the unknown quantities $q_i$ recursively. Having obtained the metric operator, it is now easy to compute the Hermitian counterpart of the Hamiltonian once again by using the formula (\ref{hausdorff})
\begin{equation}
h=\eta H \eta^{-1}=H+\frac{1}{2}\left[q,H\right]+\frac{1}{2!2^2}\left[q,\left[q,H\right]\right]+\frac{1}{3!2^3}\left[q,\left[q,\left[q,H\right]\right]\right]+......~~.
\end{equation}
In \cite{faria_fring}, the authors have obtained a closed formula for both the metric and the Hermitian Hamiltonian, which can of course be used for convenience. One should however, note, that in practice, the above-stated procedure may lead to rather cumbersome relations involving commutators, which can be overcome by using the Moyal product (\ref{moyalast}). In the present context of studying non-Hermitian Hamiltonians such possibilities have been explored in \cite{scholtz_geyer1,scholtz_geyer2,faria_fring1}, which one may look at for more detailed explanations.

\subsection{$\mathcal{CPT}$-Symmetry Versus Pseudo-Hermiticity}
In the previous subsections we have learnt enough about how to formulate the usual quantum mechanical framework from the non-Hermitian Hamiltonian systems. Certainly there is many more things left in the discussion, which we will elaborate on whenever necessary. However, let us quickly discuss about the superiority and/or necessity of the two procedures which we have discussed already. \\
    \indent Taking the similarity transformation (\ref{hermiticity}) into account and considering the Hamiltonian $H$ to be $\mathcal{PT}$-symmetric and then in addition picking up the solution of equation (\ref{CPTsolution}) to be $\mathcal{C}=\eta^{-2}\mathcal{P}$, the $\mathcal{CPT}$ inner product (\ref{CPT}), the $\eta$-inner product (\ref{innerproduct}) and the conventional inner product related to the Hermitian Hamiltonian coincide
\begin{equation}\label{innerequality}
\langle\Phi\vert\Phi'\rangle_\mathcal{CPT}=\langle\Phi\vert\Phi'\rangle_\eta=
\langle\phi\vert\phi'\rangle.
\end{equation}
With regard to (\ref{innerequality}), one may wonder why one requires the $\mathcal{CPT}$-inner products when one may in fact use the $\eta$-inner products, or even more radically why one needs the non-Hermitian formulations at all when they can always be related to the standard inner products. In fact, these issues are quite controversially discussed \cite{mostafazadeh6,bender_chen_milton,jones,mostafazadeh7}. Despite of the fact that the power of $\mathcal{PT}$-symmetry is limited, in particular the fact that it does not guarantee a positive spectrum, it is a very good guiding principle to select potentially interesting non-Hermitian Hamiltonians on the classical level, e.g. for many-particle systems \cite{mallick_kundu,fring1}. This property can be read off directly from a classical Hamiltonian, whereas even when one has identified such Hamiltonians, a proper analysis requires the construction of the similarity transformation $\eta$ or the $\mathcal{CPT}$-operator, which is usually not evident a priori. With regard to the inner products, it appears far easier to construct $\eta$ rather than the $\mathcal{CPT}$-operator, but in principle, if one has $\eta$ in hand, one also has the other. However, the metric operator $\eta$ is not unique and one can construct different versions of it and thus different Hermitian versions of the non-Hermitian Hamiltonian. The fact is slightly confusing and one does not know which version one should accept. Therefore, it remains an open problem to unify the metric operator and thus the Hamiltonian.    \\
    \indent One apparent virtue of the non-Hermitian formulation, using $\mathcal{CPT}$ or $\eta$-inner products, is that one may relate simple non-Hermitian Hamiltonians to fairly complicated Hermitian Hamiltonians. It is sometimes argued that the computations in the non-Hermitian framework are simpler to perform \cite{bender_chen_milton}, but this statement has been challenged \cite{mostafazadeh7}. Certainly, this feature can not be elevated to a general principle \cite{faria_fring}. Even when the non-Hermitian Hamiltonian looks simpler than its Hermitian counterpart, this is not true for the corresponding wavefunctions, which still take on a simpler form in the Hermitian formulation. 
    
\subsection{Applications of $\mathcal{PT}$-Symmetry}\label{sectionopticallattice}
The idea of $\mathcal{PT}$-symmetry has expanded into a wide range of fundamental and applied sciences, most notably photonics \cite{el_makris,musslimani,makris_el_chris,guo_salamo,makris,ruter, midya_roy_roychoudhury,jones1,graefe_jones,longhi_valle}, plasmonics \cite{benisty_degiron,lupu_benisty}, quantum optics of atomic gases \cite{hang_huang_konotop,sheng_miri}, studies of the Bose-Einstein condensation \cite{graefe_korsch_niederle,klaiman_gunther_moiseyev, graefe_korsch_niederle_2,cartarius_wunner,graefe_JPA_2012, dast_haag,heiss_cartarius}, and physics of electronic circuits \cite{schindler_li,schindler_lin}. However, we are going to discuss here the applications of $\mathcal{PT}$-symmetric Hamiltonian systems in the context of optical lattices, which is of great interest to experimental physicists in recent days. It has been suggested \cite{el_makris,musslimani,makris_el_chris,guo_salamo} that optical lattices can be studied in the context of $\mathcal{PT}$-symmetric non-Hermitian Hamiltonian systems when the potentials are taken to be complex. The action of the standard $\mathcal{PT}$ operator (\ref{PTsym}) on a non-Hermitian $\mathcal{PT}$-symmetric Hamiltonian $H=p^2/2m+V(x)$, whose potential $V(x)$ is complex can be understood as
\begin{eqnarray}
\mathcal{T}H \mathcal{T}^{-1} &=& \frac{p^2}{2}+V^\ast(x) \nonumber \\
(\mathcal{PT})H (\mathcal{PT})^{-1} &=& \frac{p^2}{2}+V^\ast(-x).
\end{eqnarray}
In general, we state that the $\mathcal{PT}$-symmetry of a system is unbroken provided that the eigenfunctions of a complex Hamiltonian $H$ are also eigenfunctions of the $\mathcal{PT}$ operator. In the present case $[\mathcal{PT},H]\neq 0$. Therefore, it follows that a necessary condition for a Hamiltonian to be $\mathcal{PT}$-symmetric is $V(x)=V^\ast(-x)$ (but not sufficient). In other words, $\mathcal{PT}$-symmetry requires that the real part of the potential $V$ is an even function of position $x$, whereas the imaginary part is odd, that is, the Hamiltonian must be of the form $H=p^2/2m+V_R(x)+i \epsilon V_I(x)$, where $V_{R,I}$ are the symmetric and antisymmetric components of $V$ respectively. Clearly, if $\epsilon=0$, the Hamiltonian becomes Hermitian. It turns out that, even if the imaginary component is finite, these class of potentials can possess entirely real spectra as long as $\epsilon$ stays below some threshold value, $\epsilon<\epsilon_{th}$. When the limit is crossed ($\epsilon>\epsilon_{th}$), the spectrum ceases to be real and starts producing complex eigenvalues. Thus phase transition occurs and $\mathcal{PT}$-symmetry is spontaneously broken. Broken $\mathcal{PT}$-symmetry typically involves the unfolding of an eigenfunction into complementary eigenfunctions at the so called exceptional points \cite{kato_book,cartarius_main_wunner,cejnar_heinze_macek}.

However, the fact that the complex $\mathcal{PT}$-symmetric potentials can be realized in the framework of optics has been proposed both theoretically \cite{el_makris,musslimani,makris_el_chris,midya_roy_roychoudhury, jones1,graefe_jones,longhi_valle} and experimentally \cite{guo_salamo,ruter}. In the optical analogy the complex refractive index distribution plays the role of the optical potential. The refractive index is then considered as
\begin{equation}
n(x)=n_0+n_R(x)+i n_I(x),
\end{equation}
where $n_0$ represents constant background index, $n_R$ is the real index profile
of the structure, and $n_I$ stands for the gain or loss component. With these considerations the quantum mechanical Schr{\"o}dinger equation of the $\mathcal{PT}$-symmetric Hamiltonian with complex potentials can be identified with the paraxial approximation of the equation of propagation of an electromagnetic wave in a medium, but with different interpretations for the symbols appearing therein. The equation of propagation then takes the form
\begin{equation}
i\frac{\partial \psi(x,z)}{\partial z}=-\left(\frac{\partial^2}{\partial x^2}+V(x)\right)\psi(x,z),
\end{equation}
where, $\psi(x,z)$ represents the envelope function of the amplitude of the electric field, $z$ is a scaled propagation distance, and $V(x)$ is the optical potential, proportional to the variation in the refractive index of the material through which the wave is passing. Propagation through such a medium exhibits many new and interesting properties, such as non-reciprocal diffraction \cite{berry} and birefringence \cite{makris}. One of the main features of complex optical lattices is the non-conservation of the total power. In the PT-symmetric case this can lead to effects such as power oscillations \cite{makris}.

In a recent work Chong and his collaborators \cite{chong} have introduced the concept of coherent perfect absorber (CPA) as the time reversed counterpart of a laser, where the incoming coherent light is completely absorbed. Soon after this discovery, Longhi \cite{longhi} showed that an optical medium that satisfies the parity-time ($\mathcal{PT}$)-symmetry condition $\epsilon(-r)=\epsilon^\ast (r)$ for the dielectric constant can behave simultaneously as a laser oscillator (i.e. it can emit outgoing coherent waves) and as a coherent perfect absorber, fully absorbing incoming coherent waves with the right amplitudes and phases. Owing to such a special property, they refer to such an optical device as a PT CPA-laser.

In the subsequent section we consider $\mathcal{PT}$-symmetric systems of Euclidean Lie algebraic type and point out the region of broken and unbroken symmetry as well as the exceptional points. We demonstrate theoretical gain/loss behaviour of the systems from purely theoretical background which could be realised experimentally after further investigations. 

\section{$\mathcal{PT}$-Invariant Non-Hermitian Hamiltonians of Euclidean Lie Algebraic Type} \label{section33}
Having obtained the general formalities of the $\mathcal{PT}$-symmetric non-Hermitian Hamiltonian systems and accepting the fact that a non-Hermitian formulation of quantum mechanics is more straight-forward in a pseudo-Hermitian formulation procedure, rather than a $\mathcal{CPT}$ scheme, in this section, let us try to find the Hermitian counterparts of some special type of non-Hermitian Hamiltonians, namely the Euclidean Lie algebraic type Hamiltonians.

Quasi-exactly solvable models \cite{turbiner,kamran} of Lie algebraic type are believed to be almost all related to $sl_2 \left(\mathbb{C}\right)$ with their compact and non-compact real forms $su(2)$ and $su(1, 1)$, respectively \cite{humphreys}. The nature of those models dictates that essentially all the wavefunctions related to solutions for the time-independent Schr{\"o}dinger equation of these type of models may be expressed in terms of hypergeometric functions. Non-Hermitian variants of these models expressed generically in terms of $su(2)$ or $su(1, 1)$ generators have been investigated systematically in \cite{assis_fring_lie,assis} and large classes of models were found to possess real or partially real spectra despite their non-Hermitian nature. Under certain constraints on the coupling constants the models could be mapped to Hermitian isospectral counterparts. Positive Hermitian metric operators were shown to exist, such that a consistent quantum mechanical description for these type of models is possible.

It is, however, also well known that there exists an interesting subclass of solvable models related to Mathieu functions which are known to possess solutions, which are not expressible in terms of hypergeometric functions. In a more generic setting these types of models are known to be related to specific representations of the Euclidean algebra rather than to its subalgebra $sl_2\left(\mathbb{C}\right)$. This feature makes models based on them interesting objects of investigation from a mathematical point of view. In a more applied setting it is also well known that the Mathieu equation arises in optics as a reduction from the Helmholtz equation as explained in section \ref{sectionopticallattice} in detail. This analogue setting of complex quantum mechanics is currently under intense investigation. Concrete versions of complex potentials leading to real Mathieu potentials have recently been studied from a theoretical as well as experimental point of view in \cite{musslimani,makris,guo_salamo,midya_roy_roychoudhury,jones1,graefe_jones}. Further applications are found for instance in the investigation of complex crystals \cite{longhi_valle}.

It was recently shown that for Euclidean-algebra in two, $E_2$ \cite{bender_kalveks} and three dimensions, $E_3$ \cite{jones_kalveks}, some simple non-Hermitian versions also possess real spectra. Here \cite{dey_fring_mathanaranjan} we will follow the line of thoughts of \cite{assis_fring_lie} and investigate systematically the analogous of quasi-exactly solvable models of Lie algebraic type, that in those models, which can be written as bilinear combinations in terms of the Euclidean algebra generators.
    
\subsection{$\mathcal{PT}$-Symmetric $E_2$-Invariant Non-Hermitian Hamiltonians}
The commutation relations of the $E_2$-algebra, obeyed by the three generators $u$, $v$ and $J$ are known as
\begin{equation}\label{EEE2}
\left[u,J\right]=iv \qquad \left[v,J\right]=-iu \qquad \text{and} \qquad \left[u,v\right]=0.
\end{equation}
Naturally, there are many possible representations for this algebra, as for instance one used in the context of quantizing strings on tori \cite{isham} acting on square integrable wavefunctions on a circle $L^2(\mathbf{S}^1,d \theta)$ with
\begin{equation}
J:=-i\partial _{\theta },\qquad u:=\sin \theta ,\qquad \text{and\qquad }
v:=\cos \theta ,  \label{rep11}
\end{equation}
or in two-dimension in terms of generators of the Heisenberg canonical commutators $q_j$, $p_j$ satisfying $\left[q_j,p_k\right]=i\delta_{jk}$ for $j$, $k=1,2$
\begin{alignat}{3}\label{rep22}
J: &=q_{1}p_{2}-p_{1}q_{2} & ,\qquad u: &=p_{2} & ,\qquad \text{and}\qquad v: &=p_{1}, \nonumber \\ 
J: &=q_{1}p_{2}-p_{1}q_{2} & ,\qquad u: &=q_{2} & ,\qquad \text{and}\qquad v: &=q_{1}, \\
J: &=p_{1}q_{2}-q_{1}p_{2} & ,\qquad u: &=q_{1} & ,\qquad \text{and}\qquad v: &=q_{2}, \nonumber 
\end{alignat}
and many more. It is important to note that the $E_2$-algebra is left invariant with regard to an anti-linear symmetry \cite{wigner}. As previously noted \cite{bender_berry_mandilara,fring_smith,dey_fring_gouba} in dimensions larger than one there are in general various types of anti-linear symmetries, which by a slight abuse of language we all refer to as $\mathcal{PT}$-symmetries. For instance, it is easy to see that the algebra (\ref{EEE2}) is left invariant under the following anti-linear maps
\begin{equation}
\begin{array}{lllll}\label{PT1to5}
\mathcal{PT}_{1}:~~~~~ & J\rightarrow -J,~~ & u\rightarrow -u,~~ & 
v\rightarrow -v,~~~ & i\rightarrow -i, \\ 
\mathcal{PT}_{2}: & J\rightarrow -J, & u\rightarrow u, & v\rightarrow v, & 
i\rightarrow -i, \\ 
\mathcal{PT}_{3}: & J\rightarrow J, & u\rightarrow v, & v\rightarrow u, & 
i\rightarrow -i, \\ 
\mathcal{PT}_{4}: & J\rightarrow J, & u\rightarrow -u, & v\rightarrow v, & 
i\rightarrow -i, \\ 
\mathcal{PT}_{5}: & J\rightarrow J, & u\rightarrow u, & v\rightarrow -v, & 
i\rightarrow -i.
\end{array}
\end{equation}
Each of these symmetries may be utilized to describe different types of
physical scenarios. For instance, $\mathcal{PT}_{1}$ was considered in \cite{bender_kalveks} with $\mathcal{P}_{1}:\theta \rightarrow \theta +\pi $
corresponding to a reflection of the particle to the opposite side of the
circle for the representation (\ref{rep11}). For the same representation we
can identify the remaining symmetries as $\mathcal{P}_{2}:\theta \rightarrow
\theta +2\pi $, $\mathcal{P}_{3}:\theta \rightarrow \pi /2-\theta $, $\mathcal{P}_{4}:\theta \rightarrow \pi -\theta $ and $\mathcal{P}_{5}:\theta
\rightarrow -\theta $. Of course other representations allow for different
interpretations. For instance, in the two dimensional representations (\ref{rep22}) the symmetry $\mathcal{PT}_{3}$ can be used when describing systems with two particle species as it may be viewed as a particle exchange, or an annihilation of a particle of one species accompanied by the creation of a
particle of another species, together with a simultaneous reflection $\mathcal{PT}_{3}:p_{1}\leftrightarrow p_{2},$ $q_{1}\leftrightarrow -q_{2}$.

$\mathcal{PT}_{i}$-invariant Hamiltonians $H$ in term of bilinear
combinations of $E_{2}$-generators are then easily written down. Crucially,
this very general symmetry allows for non-Hermitian Hamiltonians to be
considered since it is anti-linear \cite{wigner}. Following the general
techniques developed over the last years \cite{bender_boettcher,bender_making_sense,mostafazadeh5} in the context of $\mathcal{PT}$-symmetric non-Hermitian quantum mechanics as discussed in section \ref{section2.2}, we attempt to map these non-Hermitian Hamiltonians $H\neq H^{\dagger }$ to isospectral Hermitian counterparts $h=h^{\dagger }$ by means of a similarity transformation $h=\eta H\eta ^{-1}$ as explained in (\ref{hermiticity}). When $\eta $, often referred to as the Dyson map \cite{dyson}, is Hermitian, the latter equation is equivalent to $H^{\dagger}=\eta ^{2}H\eta ^{-2}$, which is another equation one might utilize to determine $\eta $. Taking here the Dyson map to be of the general form 
\begin{equation}
\eta =e^{\lambda J+\rho u+\tau v},\qquad \text{\ \ \ \ \ \ \ \ for }\lambda
,\tau ,\rho \in \mathbb{R},  \label{eta}
\end{equation}
we can easily compute the adjoint action of this operator on the $E_{2}$-generators. We find
\begin{eqnarray}
\eta J\eta ^{-1} &=&J+i(\rho v-\tau u)\frac{\sinh \lambda }{\lambda }+(\rho
u+\tau v)\frac{1-\cosh \lambda }{\lambda },  \label{ad1} \\
\eta u\eta ^{-1} &=&u\cosh \lambda -iv\sinh \lambda ,  \label{ad2} \\
\eta v\eta ^{-1} &=&v\cosh \lambda +iu\sinh \lambda .  \label{ad3}
\end{eqnarray}
As explained in (\ref{hermiticity}), once $\eta $ is identified the metric operators needed for a consistent
quantum mechanical formulation can in general be taken to be $\rho =\eta
^{\dagger }\eta $. Let us now construct isospectral counterparts, if they
exist, for non-Hermitian Hamiltonians symmetric with regard to the various
different types of $\mathcal{PT}$-symmetries. It should be noted that exact
computations of this type remain a rare exception for instance look at \cite{musumbu_geyer_heiss,assis_fring_lie,fring_smith, bagarello,bagarello_lattuca,bagarello_fring} and even for some of the
simplest potentials the answer is only known perturbatively as we have already discussed in (\ref{purturbation}). Even for the simple prototype non-Hermitian potential $V=i\varepsilon x^{3}$ \cite{bender_brody_jones_prd,mostafazadeh8,faria_fring}, the Dyson map is only known perturbatively.

\subsubsection{$\mathcal{PT}_1$-Invariant Hamiltonians of $E_2$-Lie Algebraic Type}
The most general $\mathcal{PT}_{1}$-invariant Hamiltonian expressed in terms
of bilinear combinations of the $E_{2}$-generators is 
\begin{equation}
H_{\mathcal{PT}_{1}}=\mu _{1}J^{2}+i\mu _{2}J+i\mu _{3}u+i\mu _{4}v+\mu
_{5}uJ+\mu _{6}vJ+\mu _{7}u^{2}+\mu _{8}v^{2}+\mu _{9}uv,  \label{HPT1}
\end{equation}
with $\mu _{i}\in \mathbb{R}$ for $i=1,\ldots ,9$. Clearly the Hamiltonian $H_{\mathcal{PT}_{1}}$ is non-Hermitian with regard to the standard inner
product when considering it for a Hermitian representation with $J^{\dagger
}=J$, $v^{\dagger }=v$ and $u^{\dagger }=u$, unless $\mu _{2}=0$, $\mu
_{5}=-2\mu _{4}$, $\mu _{6}=2\mu _{3}$. The specific case $H_{BK}=J^{2}+igv$
when $\mu _{i}=0$ for $i\neq 1,4$ was studied in \cite{bender_kalveks}, where
partially real spectra were found but no isospectral counterparts were
constructed. Using the relations (\ref{ad1})-(\ref{ad3}), we compute the
adjoint action of $\eta $ on $H$ and subsequently demand the result to be
Hermitian. This requirement will constrain our 12 free parameters $\mu
_{i},\lambda ,\tau ,\rho $. A priori it is unclear whether solutions to the
resulting set of equations exist. For $H_{\mathcal{PT}_{1}}$ we find the
manifestly Hermitian isospectral counterpart
\begin{equation}
h_{\mathcal{PT}_{1}}=\mu _{1}J^{2}+\mu _{3}\{v,J\}-\mu _{4}\{u,J\}-\frac{2\mu _{3}\mu _{4}}{\mu _{1}}uv+\frac{\mu _{4}^{2}-\mu _{3}^{2}}{\mu _{1}}u^{2}+\mu _{8}(u^{2}+v^{2}).
\end{equation}
As usual, we denote by $\{A,B\}:=AB+BA$ the anti-commutator. Without loss of
generality we may set $\mu _{8}=0$ since $C=u^{2}+v^{2}$ is a Casimir
operator for the $E_{2}$-algebra and can therefore always be added to $H$ having simply the effect of shifting the ground state energy. The remaining constants $\mu _{i}$ have been constrained to 
\begin{equation}
\tau =\frac{\lambda \mu _{3}}{\mu _{1}},~\rho =-\frac{\lambda \mu _{4}}{\mu
_{1}},~\mu _{2}=0,~\mu _{5}=-2\mu _{4},~\mu _{6}=2\mu _{3},~\mu _{7}=\mu
_{8}+\frac{\mu _{4}^{2}-\mu _{3}^{2}}{\mu _{1}},~\mu _{9}=-\frac{2\mu
_{3}\mu _{4}}{\mu _{1}},~  \label{const}
\end{equation}
by the requirement that $h_{\mathcal{PT}_{1}}$ ought to be Hermitian,
whereas $\lambda ,\mu _{1},\mu _{3},\mu _{4}$ are chosen to be free. We
observe that we have been led to the constraints (\ref{const}), of which a
subset stated that $H_{\mathcal{PT}_{1}}$ is already Hermitian before the
transformation. We also note that the constraints (\ref{const}) do not allow a
reduction to the Hamiltonian $H_{BK}$, dealt with in \cite{bender_kalveks}, as for instance $\mu _{5}=0$ implies $\mu _{4}=0$.

Having guaranteed that $H_{\mathcal{PT}_{1}}$ possess real eigenvalues under
certain constraints we may now also compute the corresponding solutions to
the time-independent Schr\"{o}dinger equation $h_{\mathcal{PT}_{1}}\phi
=E\phi $ or equivalently to $H_{\mathcal{PT}_{1}}\psi =E\psi $ with $\psi
=\eta ^{-1}\phi $. We find 
\begin{equation}
\phi (\theta )=e^{-\frac{i\mu _{4}\cos \theta }{\mu _{1}}-i\frac{\sin \theta 
}{\mu _{1}}\mu _{3}}\left[ c_{1}\exp \left( -i\theta \sqrt{\frac{E}{\mu _{1}}+\frac{\mu _{3}^{2}}{\mu _{1}^{2}}}\right) +\frac{i}{2\sqrt{\frac{E}{\mu _{1}}+\frac{\mu _{3}^{2}}{\mu _{1}^{2}}}}c_{2}\exp \left( i\theta \sqrt{\frac{E}{\mu _{1}}+\frac{\mu _{3}^{2}}{\mu _{1}^{2}}}\right) \right] ,
\end{equation}
with normalization constants $c_{1}$, $c_{2}$. Imposing either bosonic or
fermionic boundary conditions, i.e. $\psi (\theta +2\pi )=\pm \psi (\theta )$, we obtain the discrete real energy eigenvalues
\begin{equation}
\text{bosonic:~~}E_{n}=\mu _{1}\left( n^{2}-\frac{\mu _{3}^{2}}{\mu _{1}^{2}}\right) ,\qquad \text{fermionic:~~}E_{n}=\mu _{1}\left( n^{2}+n+\frac{1}{4}-\frac{\mu _{3}^{2}}{\mu _{1}^{2}}\right) ,~~n\in \mathbb{Z}\text{.}
\end{equation}
As expected, the wavefunctions are eigenstates of the $\mathcal{PT}$-operator, selecting different behaviours for the two linearly independent
parts of $\phi (\theta )$, acting as $\mathcal{PT}_{1}\phi
_{n}(c_{1})=(-1)^{n}\phi _{n}(c_{1})$ and $\mathcal{PT}_{1}\phi
_{n}(c_{2})=(-1)^{n+1}\phi _{n}(c_{2})$.

\subsubsection{$\mathcal{PT}_2$-Invariant Hamiltonians of $E_2$-Lie Algebraic Type}
Similarly as in the previous subsection we use the adjoint action of $\eta $
as specified in (\ref{eta}) to map the general $\mathcal{PT}_{2}$-symmetric
and for $\mu _{2}\neq 0$, $\mu _{5}\neq 2\mu _{4}$, $\mu _{6}=-2\mu _{3}$
non-Hermitian Hamiltonian
\begin{equation}
H_{\mathcal{PT}_{2}}=\mu _{1}J^{2}+i\mu _{2}J+\mu _{3}u+\mu _{4}v+i\mu
_{5}u~\!\!J+i\mu _{6}v~\!\!J+\mu _{7}u^{2}+\mu _{8}v^{2}+\mu _{9}uv,
\end{equation}
to the Hermitian isospectral counterpart
\begin{eqnarray}
h_{\mathcal{PT}_{2}} &=&\mu _{1}J^{2}+\mu _{3}\tanh \frac{\lambda }{2}\{u,J\}+\mu _{4}\tanh \frac{\lambda }{2}\{u,J\}+\frac{2\mu _{3}\mu _{4}}{\mu
_{1}}\tanh ^{2}\frac{\lambda }{2}uv \\
&&+\frac{\mu _{3}^{2}}{\mu _{1}}\frac{\cosh \lambda }{\cosh ^{2}\frac{\lambda }{2}}u^{2}+\left( \frac{\mu _{3}^{2}}{\mu _{1}}+\frac{\mu _{4}^{2}}{\mu _{1}}\tanh ^{2}\frac{\lambda }{2}\right) v^{2}+\mu _{8}(u^{2}+v^{2}). 
\notag
\end{eqnarray}
In this case the coupling constants are constraint to
\begin{equation}
\rho =\tau \frac{\mu _{3}}{\mu _{4}}=\frac{\mu _{3}\lambda \coth \lambda }{\mu _{1}},~~\mu _{2}=0,~~\mu _{5}=2\mu _{4},~\mu _{6}=-2\mu _{3},~~\mu
_{7}=\mu _{8}+\frac{\mu _{3}^{2}-\mu _{4}^{2}}{\mu _{1}},~\mu _{9}=\frac{2\mu _{3}\mu _{4}}{\mu _{1}},
\end{equation}
We note that once again we have only the four free parameters $\lambda ,\mu
_{1},\mu _{3},\mu _{4}$ left at our disposal, as $\mu _{8}$ may be set to
zero for the above mentioned reason. As in the previous case these
conditions imply also that the original Hamiltonian $H_{\mathcal{PT}_{2}}$
is already Hermitian when these type of constraints are imposed.

\subsubsection{$\mathcal{PT}_{3}$-Invariant Hamiltonians of $E_{2}$-Lie Algebraic Type}
As the general $\mathcal{PT}_{3}$-invariant Hamiltonian of Lie algebraic
type we consider
\begin{eqnarray}
H_{\mathcal{PT}_{3}} &=&\mu _{1}J^{2}+\mu _{2}J+\mu _{3}(u+v)+i\mu
_{4}(u-v)+\mu _{5}(u+v)~\!\!J+i\mu _{6}(u-v)~\!\!J \notag \\
&&+i\mu _{7}(v^{2}-u^{2})+\mu _{8}(v^{2}+u^{2})+\mu _{9}uv.
\end{eqnarray}
For Hermitian representations of the $E_{2}$-generators this Hamiltonian is
non-Hermitian unless $\mu _{6}=\mu _{7}=0$ and $\mu _{5}=2\mu _{4}$. As
isospectral Hermitian counterpart we find in this case 
\begin{eqnarray}
h_{\mathcal{PT}_{3}} &=&\mu _{1}J^{2}+\mu _{2}J+\frac{1}{2}\left( \mu
_{5}+\mu _{6}\tanh \frac{\lambda }{2}\right) \{u+v,J\} \\
&&\!\!\!\!\!\!\!\!\!\!\!\!+\left\{ \frac{1}{2\mu _{1}}\left[ \mu
_{5}^{2}+\mu _{6}^{2}\tanh ^{2}\frac{\lambda }{2}+\mu _{6}\mu _{5}\frac{4+4\cosh \lambda -2\cosh (2\lambda )}{\sinh (2\lambda )}\right] +\frac{2\mu
_{7}}{\sinh (2\lambda )}\right\} uv  \notag \\
&&\!\!\!\!\!\!\!\!\!\!\!\!+\left[ \mu _{3}-\frac{\mu _{6}}{2}+\left( \mu
_{4}-\frac{\mu _{5}}{2}\right) \tanh \frac{\lambda }{2}\right] (u+v)+\left[
\mu _{8}+\frac{\mu _{5}\mu _{6}\sinh \lambda +\mu _{6}^{2}\cosh \lambda }{2\mu _{1}(1+\cosh \lambda )}\right] (u^{2}+v^{2})  \notag
\end{eqnarray}
with only four constraining equations
\begin{eqnarray}
\rho &=&\tau =\frac{\lambda \left( \mu _{5}+\mu _{6}\coth \lambda \right) }{%
2\mu _{1}},~~~\coth \lambda =\frac{\mu _{2}\mu _{5}+\mu _{1}\left( \mu
_{6}-2\mu _{3}\right) }{\mu _{1}\left( 2\mu _{4}-\mu _{5}\right) -\mu
_{2}\mu _{6}},~  \label{co1} \\
\mu _{9} &=&\frac{\mu _{5}^{2}+\mu _{6}^{2}+2\mu _{6}\mu _{5}\coth (2\lambda
)}{2\mu _{1}}+2\mu _{7}\coth (2\lambda ).
\end{eqnarray}
Thus, in this case we have eight free parameters left. We also note that
unlike as for the $\mathcal{PT}_{1}$ and $\mathcal{PT}_{2}$ symmetric cases
we are not led to constraints which render the original Hamiltonian $H_{\mathcal{PT}_{3}}$ Hermitian. For $\mu _{1}=1$, $\mu _{7}=2q$ and all other
coupling constants vanishing the Schr\"{o}dinger equation with representation (\ref{rep11}) converts into the standard Mathieu differential equation, see e.g. \cite{gradshteyn},
\begin{equation}
-\phi ^{\prime \prime }(\theta )+2iq\cos (2\theta )\phi (\theta )=E\phi
(\theta ).  \label{mat}
\end{equation}
with purely complex coupling constant. Unfortunately for this choice of the
coupling constants the Dyson map is no longer well-defined, because of the
last equation in (\ref{co1}), such that it remains an open problem to find
the corresponding isospectral counterpart for this scenario.

\subsubsection{$\mathcal{PT}_{4}$-Invariant Hamiltonians of $E_{2}$-Lie Algebraic Type}
The general $\mathcal{PT}_{4}$-invariant Hamiltonian we consider is
\begin{equation}
H_{\mathcal{PT}_{4}}=\mu _{1}J^{2}+\mu _{2}J+i\mu _{3}u+\mu _{4}v+i\mu
_{5}u~\!\!J+\mu _{6}v~\!\!J+\mu _{7}u^{2}+\mu _{8}v^{2}+i\mu _{9}uv.
\end{equation}
This Hamiltonian is non-Hermitian unless $\mu _{5}=\mu _{9}=0$ and $\mu
_{6}=2\mu _{3}$. Constraining now the parameters as 
\begin{eqnarray}
&&\rho =0,\quad \tau =\frac{\lambda \left( \mu _{5}\coth \lambda +\mu
_{6}\right) }{2\mu _{1}},\quad \coth (2\lambda )=\frac{4\mu _{1}(\mu
_{8}-\mu _{7})-\mu _{5}^{2}-\mu _{6}^{2}}{2\mu _{5}\mu _{6}}, \\
&&\mu _{3}=\frac{\mu _{1}\mu _{5}+\mu _{2}\mu _{6}-2\mu _{1}\mu _{4}}{2\mu
_{1}}\tanh \lambda +\frac{\mu _{2}\mu _{5}}{2\mu _{1}}+\frac{\mu _{6}}{2},\qquad \mu _{9}=0,\quad
\end{eqnarray}
we map this to the isospectral counterpart 
\begin{eqnarray}
h_{\mathcal{PT}_{4}} &=&\mu _{1}J^{2}+\mu _{2}J+\frac{1}{2}\left( \mu
_{6}+\mu _{5}\tanh \frac{\lambda }{2}\right) \{v,J\} \\
&&+\left[ \frac{\mu _{2}\tanh \left( \frac{\lambda }{2}\right) \left( \mu
_{5}+\mu _{6}\tanh \lambda \right) }{2\mu _{1}}+\left( \mu _{4}-\frac{\mu
_{5}}{2}\right) \text{sech}\lambda \right] v  \notag \\
&&+\left[ \frac{\mu _{5}^{2}\left( \tanh ^{2}\frac{\lambda }{2}-\cosh
(2\lambda )\right) -2\mu _{6}^{2}\sinh ^{2}\lambda +2\mu _{5}\mu _{6}\left(
\tanh \frac{\lambda }{2}-\sinh (2\lambda )\right) }{8\mu _{1}}\right.  \notag
\\
&&+\left. \frac{\mu _{8}-\mu _{7}}{2}\cosh (2\lambda )\right] \left(
v^{2}-u^{2}\right) +\frac{\mu _{5}^{2}\cosh \lambda +\mu _{5}\mu _{6}\sinh
\lambda }{4\mu _{1}(1+\cosh \lambda )}+\frac{1}{2}\left( \mu _{7}+\mu
_{8}\right) .  \notag
\end{eqnarray}
Thus, in this case we have seven free parameters left to our disposal. Also
in this case we obtained a genuine non-Hermitian/Hermitian isospectral pair
of Hamiltonians.

\subsubsection{$\mathcal{PT}_{5}$-Invariant Hamiltonians of $E_{2}$-Lie Algebraic Type}
As general $\mathcal{PT}_{5}$-invariant Hamiltonian we consider 
\begin{equation}
H_{\mathcal{PT}5}=\mu _{1}J^{2}+\mu _{2}J+\mu _{3}u+i\mu _{4}v+\mu
_{5}u~\!\!J+i\mu _{6}v~\!\!J+\mu _{7}u^{2}+\mu _{8}v^{2}+i\mu _{9}uv.~~
\end{equation}%
This Hamiltonian is non-Hermitian unless $\mu _{6}=\mu _{9}=0$ and $\mu
_{5}=-2\mu _{4}$. In the same manner as in the previous subsections we
construct the isospectral counterpart 
\begin{eqnarray}
h_{\mathcal{PT}_{5}} &=&\mu _{1}J^{2}+\mu _{2}J+\frac{1}{2}\left( \mu_{5}-\mu_{6}\tanh \frac{\lambda }{2}\right) \{u,J\} \\
&&+\left[ \frac{2\mu _{5}^{2}\sinh ^{2}\lambda +\mu _{6}^{2}(\text{sech}^{2}
\frac{\lambda }{2}+\cosh (2\lambda )-1)+2(\tanh \frac{\lambda }{2}-\sinh
(2\lambda ))\mu _{5}\mu _{6}}{8\mu _{1}}\right.  \notag \\
&&+\left. \frac{\mu _{8}-\mu _{7}}{2}\cosh (2\lambda )\right] (v^{2}-u^{2})+
\left[ \text{csch}\lambda \left( \mu _{4}+\frac{1}{2}\mu _{5}\right) +\frac{
\mu _{2}}{2\mu _{1}}(\mu _{5}-\coth \lambda \mu _{6})\right] u  \notag \\
&&+\frac{\mu _{6}^{2}\cosh \lambda -\mu _{5}\mu _{6}\sinh \lambda }{4\mu
_{1}(1+\cosh \lambda )}+\frac{1}{2}\left( \mu _{7}+\mu _{8}\right) ,  \notag
\end{eqnarray}
where the constants are constraint to$\allowbreak $
\begin{eqnarray}
&&\tau =0,\quad \rho =\frac{\lambda \left( \mu _{5}-\mu _{6}\coth \lambda
\right) }{2\mu _{1}},\quad \coth (2\lambda )=\frac{\mu _{5}^{2}+\mu
_{6}^{2}-4\mu _{1}\mu _{7}+4\mu _{1}\mu _{8}}{2\mu _{5}\mu _{6}}, \\
&&\mu _{3}=\frac{\left( 2\mu _{1}\mu _{4}+\mu _{1}\mu _{5}-\mu _{2}\mu
_{6}\right) \coth (\lambda )}{2\mu _{1}}+\frac{\mu _{2}\mu _{5}}{2\mu _{1}}-
\frac{\mu _{6}}{2},\quad \mu _{9}=0.\quad
\end{eqnarray}
Thus, in this case we have also seven free parameters left to our disposal.

Having obtained the Hermitian counterpart, let us construct in this case some solutions to the time-independent Schr\"{o}dinger equation. The
discussion of the entire parameter space is a formidable task, but as we
shall see it will be sufficient to focus on some special parameter choices
in order to extract different types of qualitative behaviour. We will also
make contact to some special cases previously treated in the literature,
notably in the area of complex optical lattices.

\subsubsection{Maps to a Three Parameter Real Mathieu Equation}
First we specify our parameters further such that only three are left free 
\begin{eqnarray}
\mu _{1} &=&1,\quad \mu _{2}=0,\quad \mu _{5}=-2\mu _{4},\quad \mu
_{6}=-2\mu _{3},\quad \mu _{8}=\mu _{9}=0,\quad  \label{cons1} \\
\tau &=&0,\quad \rho =\lambda \left( \mu _{3}\coth \lambda -\mu _{4}\right)
,\quad \coth (2\lambda )=\frac{\mu _{3}^{2}+\mu _{4}^{2}-\mu _{7}}{2\mu
_{3}\mu _{4}}.  \label{cons2}
\end{eqnarray}
The corresponding isospectral pair of Hamiltonians simplifies in this case to 
\begin{eqnarray}
H_{\mathcal{PT}_{5}}^{(3)} &=&J^{2}-i\mu _{3}\{v,~\!\!J\}-\mu
_{4}\{u,~\!\!J\}+\mu _{7}u^{2},~~ \\
h_{\mathcal{PT}_{5}}^{(3)} &=&J^{2}+\alpha \{u,J\}+\beta u^{2}+\gamma ,
\end{eqnarray}
where $\alpha $, $\beta $, $\gamma $ are functions of $\mu _{3}$, $\mu _{4}$, $\mu _{7}$
\begin{eqnarray}
\alpha &=&\mu _{3}\tanh \frac{\lambda }{2}-\mu _{4}, \\
\beta &=&\frac{2\mu _{3}}{1+\cosh \lambda }(\mu _{3}\cosh \lambda -\mu
_{4}\sinh \lambda )+\mu _{7}-2\gamma , \\
\gamma &=&(\mu _{3}\cosh \lambda -\mu _{4}\sinh \lambda )^{2}-\mu _{7}\sinh
^{2}\lambda .
\end{eqnarray}
For the representation (\ref{rep11}) the standard Mathieu differential
equation (\ref{mat}) with real coupling constant is easily converted into
the time-independent Schr\"{o}dinger equation 
\begin{equation}
h_{\mathcal{PT}_{5}}^{(3)}\psi (\theta )=E\psi (\theta )  \label{S4}
\end{equation}
with the transformations $\phi (\theta )\rightarrow e^{-i\alpha \cos \theta
}\psi (\theta )$, $q\rightarrow (\alpha ^{2}-\beta )/4$ and $E\rightarrow
E+(\alpha ^{2}-\beta )/2-\gamma $. Therefore (\ref{S4}) is solved by
\begin{equation}
\psi (\theta )=e^{i\alpha \cos \theta }\left[ c_{1}C\left( E+\frac{\alpha
^{2}-\beta }{2}-\gamma ,\frac{\alpha ^{2}-\beta }{4},\theta \right)
+c_{2}S\left( E+\frac{\alpha ^{2}-\beta }{2}-\gamma ,\frac{\alpha ^{2}-\beta 
}{4},\theta \right) \right]
\end{equation}
where $C$ and $S$ denote the even and odd Mathieu function \cite{gradshteyn}, respectively. A discrete energy spectrum is extracted in the usual way by imposing periodic boundaries $\psi (\theta +2\pi )=e^{i\pi s}\psi (\theta )$ as quantization condition. While in general anyonic conditions are possible in dimensions lower than 4, we present here only the bosonic and fermionic case, that is $s=0$ and $s=1$, respectively. As the Mathieu equation is known to possess infinitely many periodic solutions, the boundary condition as such is not sufficient to obtain a unique solution. However, the latter is achieved by demanding in addition the continuity of the energy levels at $q=0$. The inclusion of all values for $s$ will naturally lead to band structures.

\begin{figure}[H]
\centering   \includegraphics[width=7.5cm,height=6.0cm]{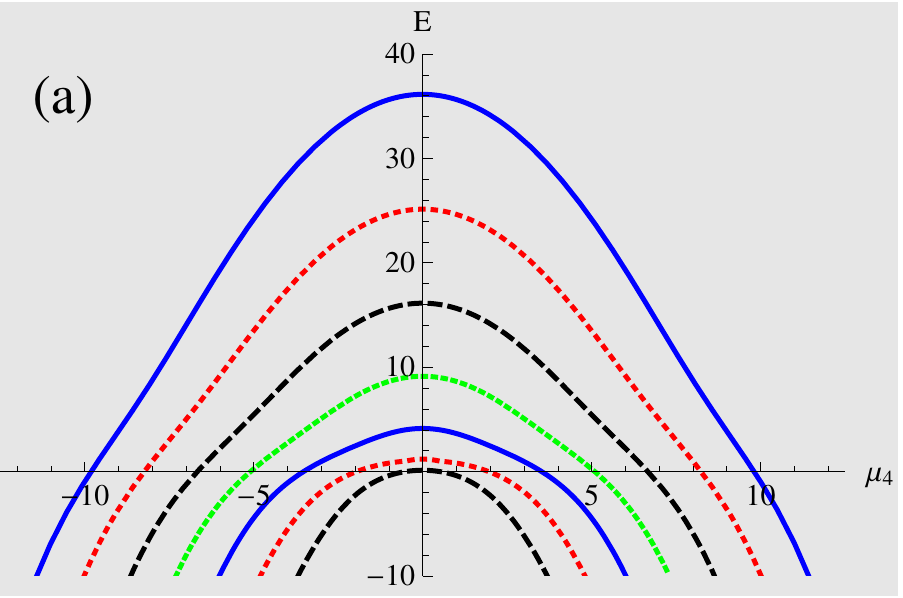}
\includegraphics[width=7.5cm,height=6.0cm]{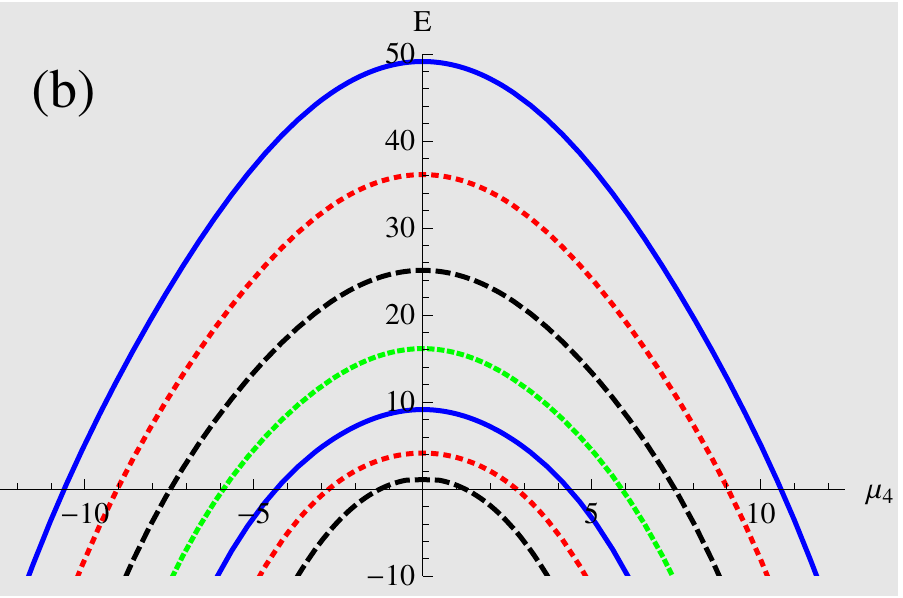}
\includegraphics[width=7.5cm,height=6.0cm]{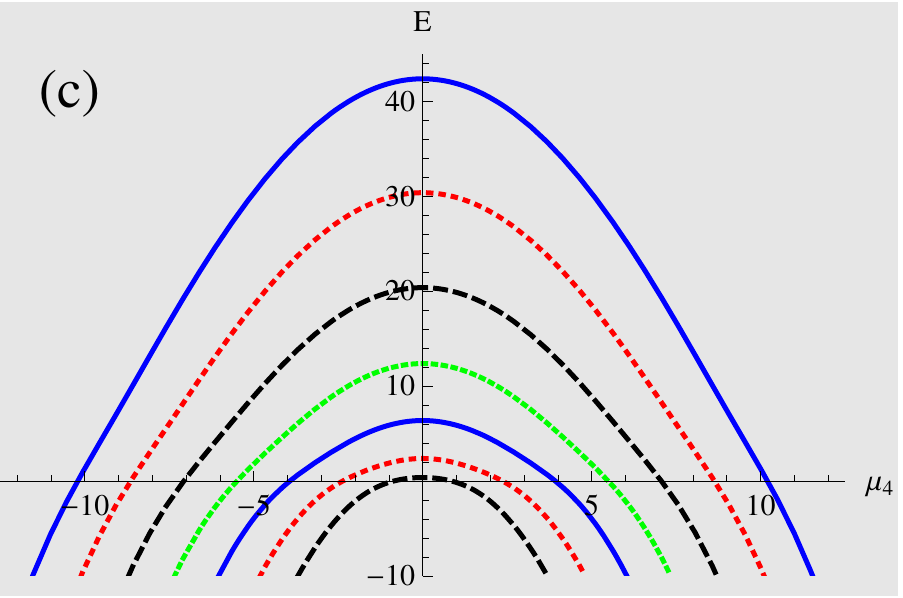}
\includegraphics[width=7.5cm,height=6.0cm]{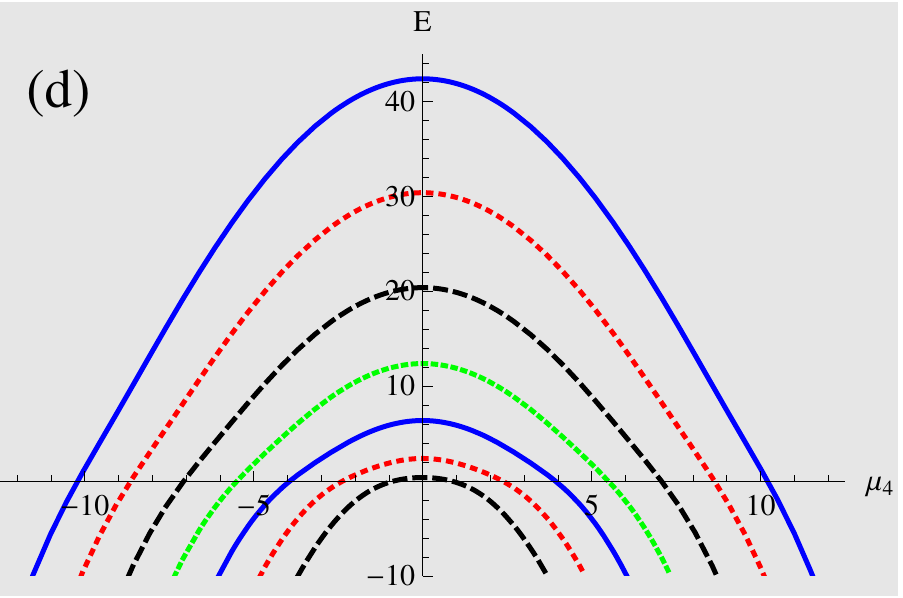} \centering   
\caption{\small{Entirely real energy eigenvalue spectrum for the non-Hermitian
Hamiltonian $H_{\mathcal{PT}_{5}}^{(3)}$ as a function of $\protect\mu _{4}$
with $\protect\mu _{3}=1/2$ and $\protect\mu _{7}=0$. The values for even
(odd) eigenfunctions with bosonic and fermionic boundary conditions are
displayed in the panels a and c (b and d), respectively.}}
\label{fig1}
\end{figure}

We commence our numerical analysis by taking $\mu _{7}=0$. In this case the map $\eta $ is well-defined, except when $\mu _{3}=\mu _{4}$ for which $\lambda \rightarrow \infty $ by (\ref{cons2}). Thus we expect an entirely
real energy spectrum. In figure \ref{fig1} we present the results of our
numerical solutions for the computation of the lowest seven energy levels,
demonstrating that this is indeed the case for the even and odd solutions
for bosonic as well as fermionic boundary conditions.

\begin{figure}[h]
\centering   \includegraphics[width=7.5cm,height=6.0cm]{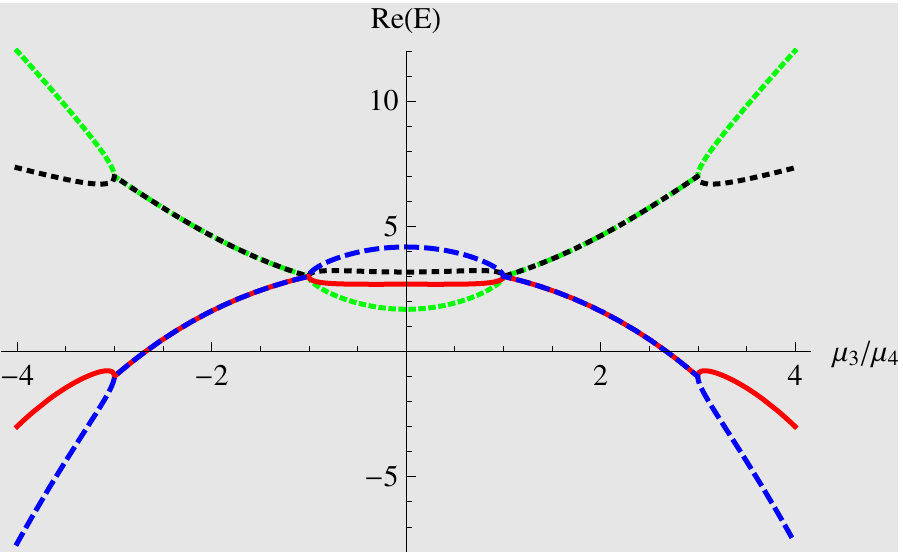}
\includegraphics[width=7.5cm,height=6.0cm]{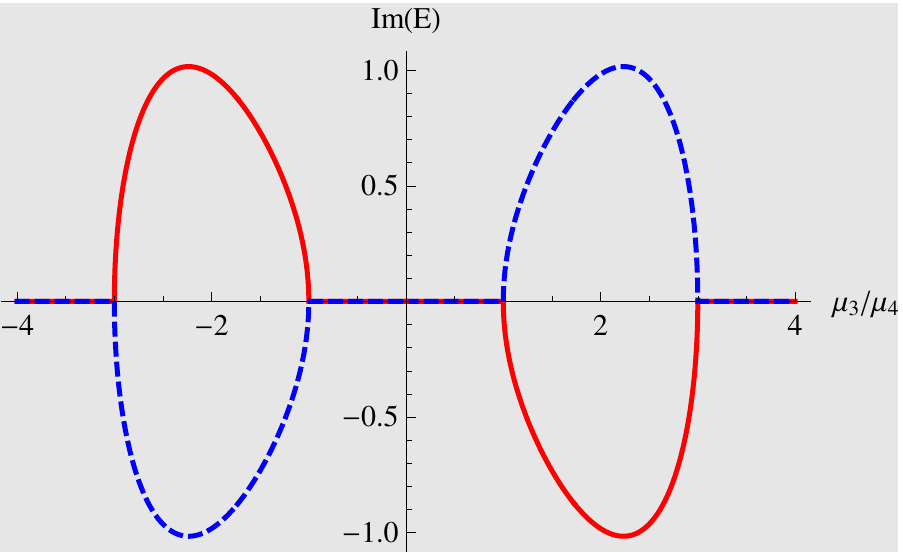} \centering   
\caption{\small{Energy eigenvalue spectra in the spontaneously broken regime for $H_{\mathcal{PT}
_{5}}^{(3)}$ as a function of $\protect\mu_3$ with fixed values $\protect\mu
_4=1$ and $\protect\mu_7=4$ with even (green, short dashed) and odd (black,
dotted) eigenfunctions for bosonic boundary conditions and as a function of $
\protect\mu_4$ with fixed values $\protect\mu_3=1$ and $\protect\mu_7=4$
with even (red, solid) and odd (blue, dashed) eigenfunctions for bosonic
boundary conditions. The exceptional points are located at ($\protect\mu
_{3/4}=\pm 1, E=3$), ($\protect\mu_{3}=\pm 3, E=7$) and ($\protect\mu%
_{4}=\pm 3, E=-1$).}}
\label{fig2}
\end{figure} 

For nonzero values of $\mu _{7}$ we can enter the ill-defined region for the Dyson map as for the last constraint in (\ref{cons2}) we may encounter values on the right hand side between $-1$ and $1$. Viewing the energy eigenvalues as functions of $\mu _{3/4}$ we expect therefore to find four exceptional points at $\mu _{3/4}=\pm \mu _{4/3}\pm \sqrt{\mu _{7}}$. As an example we fix $\mu _{3/4}=1$ and $\mu _{7}=4$, such that $\eta (\mu _{4/3})$ is only well defined for $|\mu _{4/3}|<1$ or $|\mu _{4/3}|>3$. Indeed our numerical solutions for this choice presented in figure \ref{fig2} confirm this prediction. We observe that the eigenvalues acquire a complex part when  $1<\mu _{3/4}<3$ and $-3<\mu _{3/4}<-1$ and is real otherwise. We present here only the spectrum for bosonic boundary condition with an even and odd wavefunction since the qualitative behaviour for the other cases and levels are very similar as already noted in the previous example. 

We clearly observe the typical behaviour of spontaneously broken $\mathcal{PT}$-symmetry in form of two of the real eigenvalues merging into complex conjugate pairs at exceptional points. We further note that there are three disconnected regions $|\mu _{3/4}|<1$ or $|\mu _{3/4}|>3$ in which all the eigenvalues are real.

\begin{figure}[h]
\centering   \includegraphics[width=7.5cm,height=6.0cm]{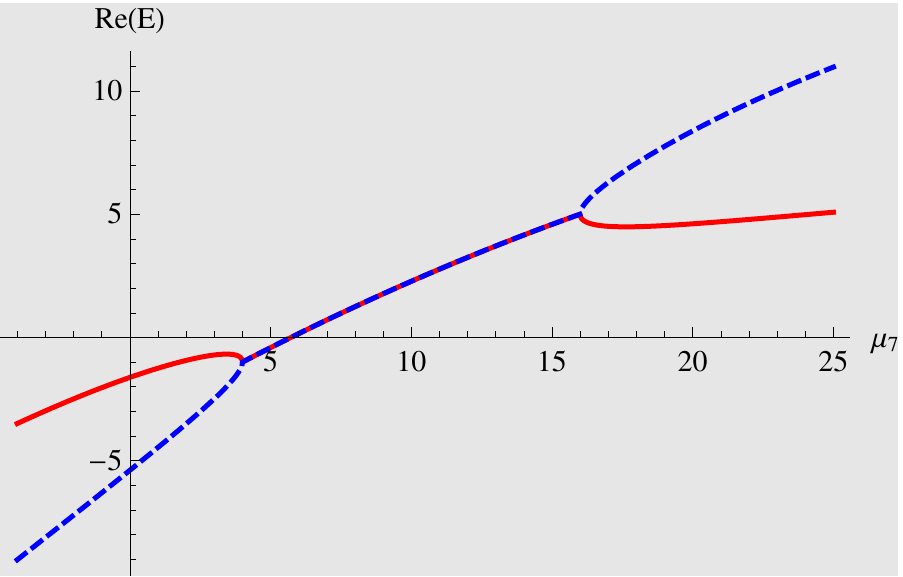}
\includegraphics[width=7.5cm,height=6.0cm]{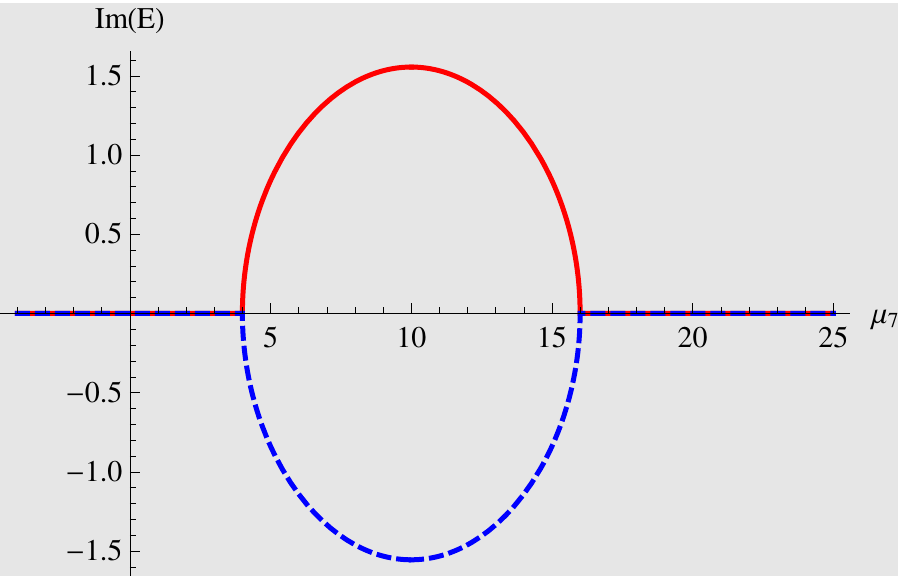} \centering   
\caption{\small{Energy eigenvalue spectra in the spontaneously broken regime for $H_{\mathcal{PT}
_{5}}^{(3)}$ as a function of $\protect\mu_7$ with fixed values $\protect\mu
_3=1$ and $\protect\mu_4=3$ with even (red, solid) and odd (blue, dashed)
eigenfunctions. The exceptional points are located at ($\protect\mu_{7}=4 ,
E=-1$) and ($\protect\mu_{7}=16, E=5$).}}
\label{fig3}
\end{figure}

Alternatively we may also view the energy spectra as functions of $\mu _{7}$, in which case we expect just two exceptional points at $(\mu _{3}\pm \mu_{4})$ $^{2}$. Our numerical solutions for this choice are presented in figure \ref{fig3}, which clearly confirms these values and the predicted
qualitative behaviour.

\begin{figure}[h]
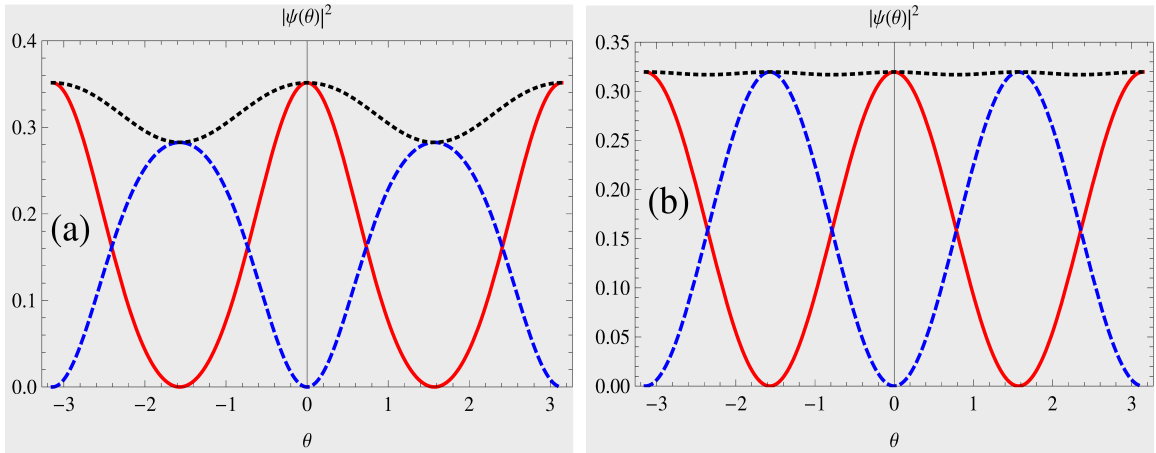

\centering
\includegraphics[width=7.5cm,height=6.0cm]{Intensityunbroken.pdf}
\includegraphics[width=7.5cm,height=6.0cm]{Intensitybroken.pdf} 
\centering   
\caption{\small{Intensities of an even (red, solid) and odd (blue, dashed) wavefunction for two consecutive quantum numbers together with their sum (black, dotted) (a) in the unbroken $\mathcal{PT}$-regime with $\protect\mu _{3}=0.8$, $\protect\mu _{4}=1$, $\protect\mu _{7}=4$ and (b) in the broken $\mathcal{PT}$-regime with $\protect\mu _{3}=1.2$, $\protect\mu_{4}=1 $, $\protect\mu _{7}=4$.}}
\label{figInt}
\end{figure}

We conclude this subsection by considering the intensities, as in principle
these quantities are experimentally accessible. In figure \ref{figInt} we display the intensity $I(\theta )=|\psi (\theta )|^{2}$ for an odd and even wavefunction merging at the exceptional points whose energy spectrum is displayed in figure \ref{fig2}. In the spontaneously broken $\mathcal{PT}$-regime we clearly observe the loss/gain symmetry around the line $I_{\max}(\theta )/2$, which is absent in the unbroken $\mathcal{PT}$-regime. Searching for a measurable quantity that can be used to identify the symmetry breaking we observe that
\begin{equation}\label{inten}
I(\theta ):=|\psi _{\text{even}}(\theta )|^{2}+|\psi _{\text{odd}}(\theta )|^{2}-|\psi _{\text{even}}(0)|^{2} \left\{\begin{array}{cl}=0 & \text{for broken}~\mathcal{PT}~\text{symmetry} \\ \neq 0 &  \text{for unbroken}~\mathcal{PT}~\text{symmetry} \end{array} \right. .
\end{equation}
We note that the change from one regime to the other is very abrupt and sharp. This effect is very strongly displayed in figure \ref{figp}, where we scan over a larger range for the coupling constant $\mu _{3}$ entering and leaving the broken $\mathcal{PT}$-regime. We depict $I(\theta )$ as defined in (\ref{inten}) and clearly observe an oscillatory behaviour in the unbroken $\mathcal{PT}$-regime ($\mu _{3}<1$ and $\mu _{3}>3$) and complete annihilation in the region where the symmetry is spontaneously broken ($1$ $<\mu _{3}<3$). This qualitative behaviour is reminiscent of the symmetric gain/loss behaviour observed in complex optical potentials \cite{guo_salamo}.

\begin{figure}[h]
\centering   \includegraphics[width=12.5cm]{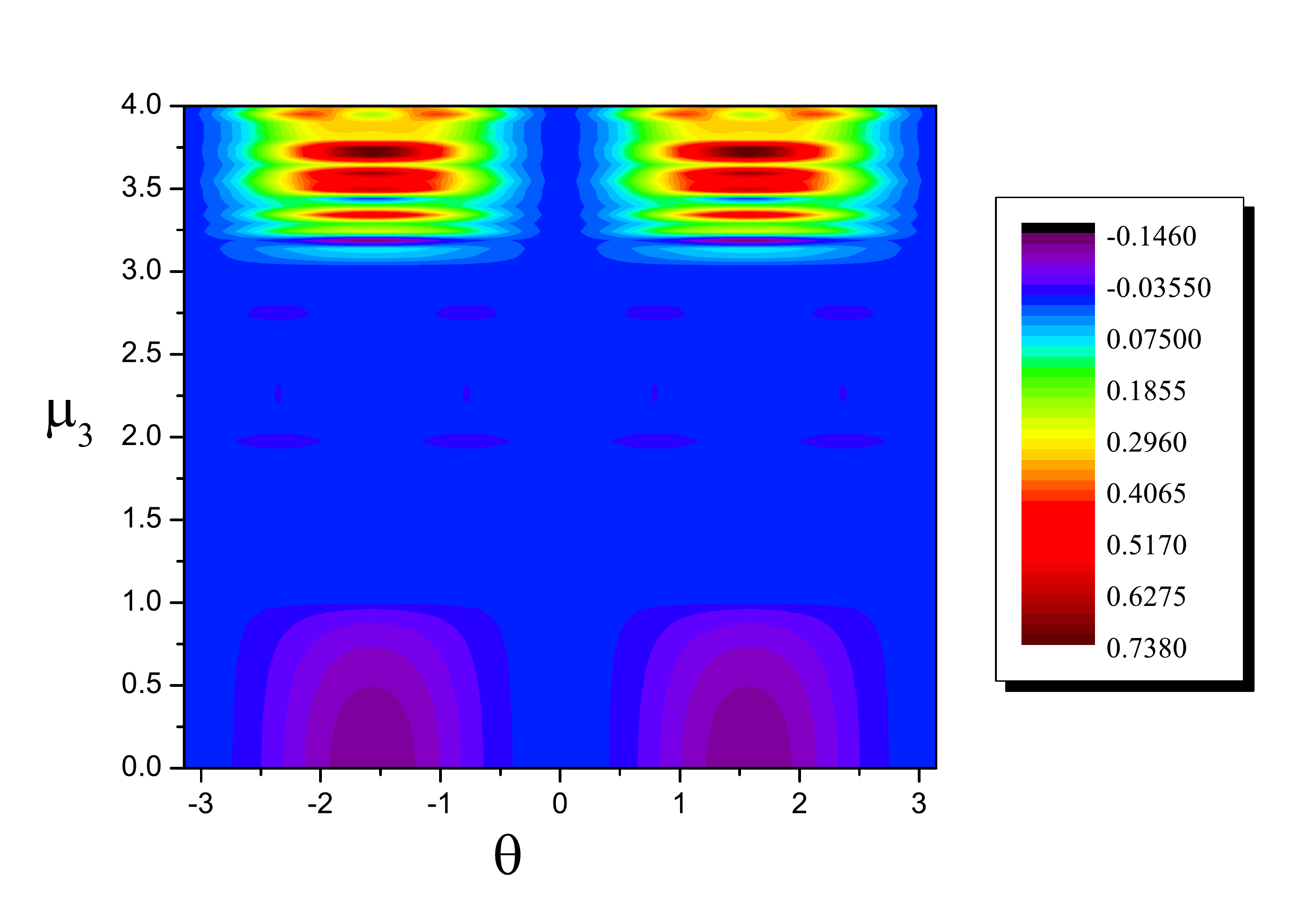} \centering   
\caption{\small{Intensity sum $I(\protect\theta )=|\protect\psi _{\text{even}}(
\protect\theta )|^{2}+|\protect\psi _{\text{odd}}(\protect\theta )|^{2}-|
\protect\psi _{\text{even}}(0)|^{2}$ as a function of $\protect\mu _{3}$
with fixed values $\protect\mu _{4}=1$ and $\protect\mu _{7}=4$. }}
\label{figp}
\end{figure}

Based on our observation we propose (\ref{inten}) as a measurable quantity that can be used as a criterium to distinguish between unbroken $\mathcal{PT}$-symmetric and spontaneously broken $\mathcal{PT}$-symmetric regimes. At this point this behaviour remains an observation for which we have no rigorous explanation.

\subsubsection{Sinusoidal Optical Lattices}
For different choices of parameters we can also make contact with a simpler example currently of great interest, since it can be realized experimentally in form of optical lattices. Making the simple choice
\begin{equation}
\mu _{1}=1,\quad \mu _{2}=\mu _{3}=\mu _{4}=\mu _{5}=\mu _{6}=0\quad \tau
=\rho ,\quad \coth (2\lambda )=\frac{\mu _{7}-\mu _{8}}{\mu _{9}},
\label{conop}
\end{equation}
we obtain the isospectral Hermitian counterpart
\begin{equation}
h_{\mathcal{PT}_{4/5}}^{(ol)}=J^{2}+\frac{1}{2}\sqrt{(\mu _{7}-\mu
_{8})^{2}-\mu _{9}^{2}}(v^{2}-u^{2})+\frac{1}{2}(\mu _{7}+\mu _{8}).
\label{OP}
\end{equation}
Taking the representation (\ref{rep11}) in (\ref{OP}), the further special
choices $\mu _{7}=0$, $\mu _{8}=-4$, $\mu _{9}=-8V_{0}$ or $\mu _{7}=-\mu
_{8}=A/2$, $\mu _{9}=-2AV_{0}$ reduce the potential to the sinusoidal
optical lattice potential
\begin{equation}
V(x)=4\cos^2 x+4 i V_0 \sin 2x,
\end{equation}
which has been dealt with in \cite{midya_roy_roychoudhury} and \cite{jones1}. In both cases the requirement for the validity of the Dyson map $\left\vert (\mu _{7}-\mu _{8})/\mu _{9}\right\vert <1$, implied by the last equation in (\ref{conop}), boils down to $\left\vert V_{0}\right\vert <1/2$ confirming the finding in \cite{midya_roy_roychoudhury} and \cite{jones1} that only in this regime the corresponding potential leads to a real energy eigenvalue spectrum.

\subsubsection{Complex Mathieu Equation}
We conclude by discussing the parameter choice
\begin{equation}\label{eqn376}
\mu _{1}=1,\quad \mu _{2}=0,\quad \mu _{3}=-\frac{\mu _{6}}{2},\quad \mu
_{5}=-\mu _{4},\quad \mu _{7}=\frac{\mu _{4}^{2}}{2},\quad \mu _{8}=-\frac{\mu_{6}^{2}}{4},\quad \mu _{9}=-\frac{\mu _{4}\mu _{6}}{2}.
\end{equation}
In that case the reported similarity transformation is invalid. However,
similarly as in the previous case we may solve the corresponding Schr\"{o}dinger equation exactly by mapping it to the Mathieu equation, which is however complex in this case. We then find the solution 
\begin{equation}
\psi (\theta )=e^{-i\mu _{4}/2\cos \theta +\mu _{6}/2\sin \theta }\left[
c_{1}C\left( 4E,i\mu _{4},\theta /2\right) +c_{2}S\left( 4E,i\mu _{4},\theta
/2\right) \right] .  \label{sol3}
\end{equation}
As in the previous case we impose bosonic or fermionic boundary conditions
to determine the spectrum. Our results are depicted in figure \ref{fig4}.

\begin{figure}
\centering   \includegraphics[width=7.5cm,height=6.0cm]{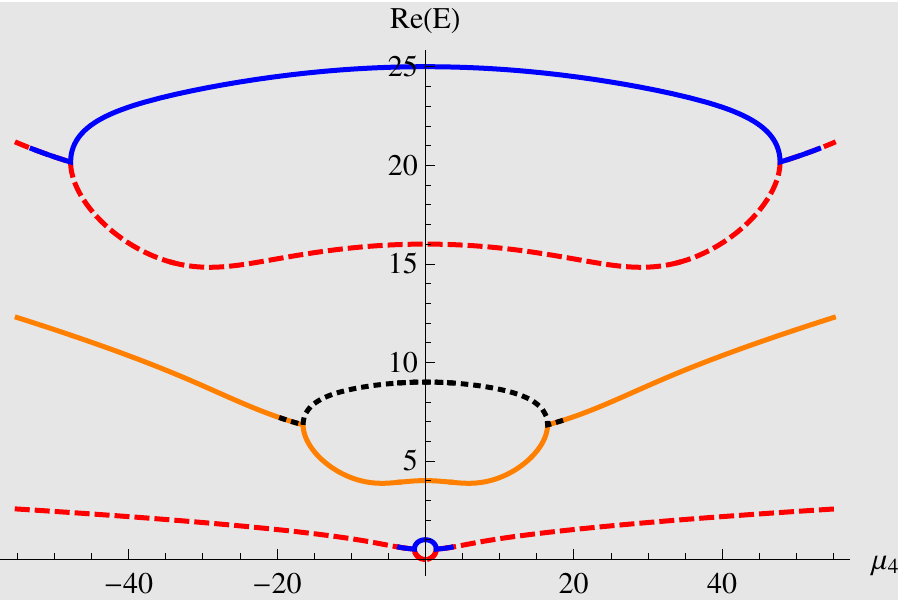}
\includegraphics[width=7.5cm,height=6.0cm]{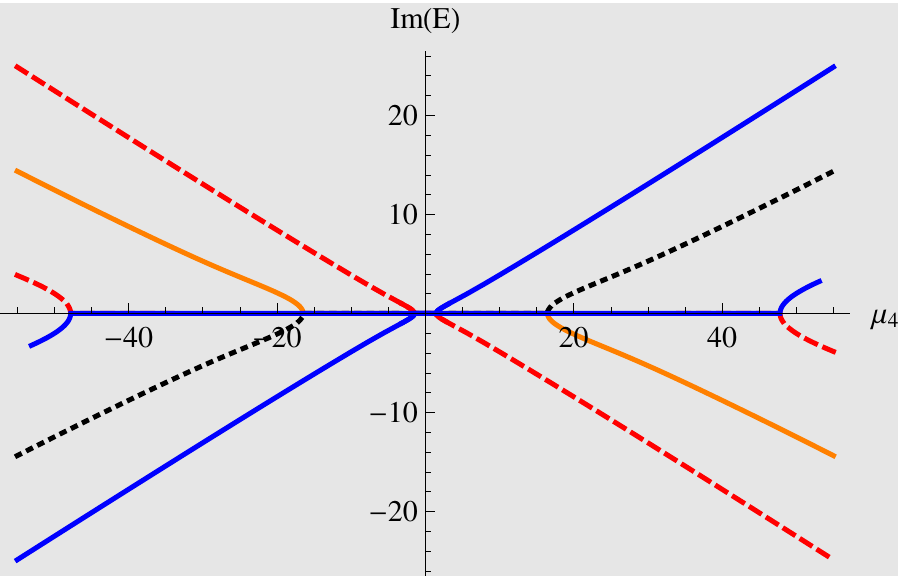} \centering   
\caption{\small{Energy eigenvalue spectra in the spontaneously broken regime for the parameter choice (\ref{eqn376}) as a function of $\protect\mu_4$ with even (blue and orange, solid lines) and odd (black and red, dashed lines) eigenfunctions for bosonic boundary conditions. The exceptional points are located at $(\protect
\mu_4=\pm 1.4687, E=0.5205)$, $(\protect\mu_4=\pm 16.47116, E=6.8323)$ and $(
\protect\mu_4=\pm 47.80596, E=20.1677)$.}}
\label{fig4}
\end{figure}

We clearly observe the usual merging of two energy levels at the exceptional
points where they split into complex conjugate pairs. Since the real part of
the energy eigenvalues is monotonically increasing we note that the spectrum
is entirely real for $\left\vert \mu _{4}\right\vert \leq 1.46876$. It
remains an open challenge to explain the origin of this value for instance
by finding an exact similarity transformation. As we expect, this behaviour
is similar to the one reported in \cite{bender_kalveks}.

\subsection{$\mathcal{PT}$-Symmetric E$_{3}$-Invariant Systems of Lie Algebraic Type}
The $E_{3}$-algebra is the rank 3 extension of the $E_{2}$-algebra, spanned
by six generators $J_{i}$, $P_{i}$ for $i=1,2,3$ satisfying the algebra
\begin{equation}
\left[ J_{j},J_{k}\right] =i\varepsilon _{jkl}J_{l},\qquad \left[ J_{j},P_{k}
\right] =i\varepsilon _{jkl}P_{l},\qquad \text{and\qquad }\left[ P_{j},P_{k}
\right] =0.
\end{equation}
Evidently every subset $\{J_{j},P_{k},P_{l}\}$ with $j\neq k\neq l$
constitutes an $E_{2}$-subalgebra. It is convenient to introduce the
following combinations of the generators
\begin{equation}
J_{z}=2J_{1},\quad J_{\pm }=J_{2}\pm iJ_{3},\quad P_{z}=P_{1},\qquad \text{
and\qquad }P_{\pm }=\pm P_{2}+iP_{3},
\end{equation}
such that we obtain the commutation relations 
\begin{equation}
\left[ J_{z},J_{\pm }\right] =\pm 2J_{\pm },~~\left[ J_{+},J_{-}\right]
=J_{z},~~\left[ J_{z},P_{\pm }\right] =\pm 2P_{\pm },~~\left[ J_{\pm },P_{z}
\right] =-P_{\pm },~~\left[ J_{\pm },P_{\mp }\right] =-2P_{z},  \label{E3pm}
\end{equation}
with all remaining ones vanishing. In \cite{douglas_guise} the following representation was provided for this algebra
\begin{equation}
\begin{array}{lll}
J_{z}:=x\partial _{x}-y\partial _{y},~~\quad & J_{+}:=x\partial _{y},~~ & 
J_{-}:=y\partial _{x}, \\ 
P_{z}:=-xy\partial _{z},~~ & P_{+}:=x^{2}\partial _{z},~\quad & 
P_{-}:=y^{2}\partial _{z}.
\end{array}
\label{rep}
\end{equation}
Similarly as $E_{2}$, also $E_{3}$ is left invariant with respect to various
types of $\mathcal{PT}$-symmetries
\begin{equation}
\begin{array}{llllll}
\mathcal{PT}_{1}:~~~~~ & J_{k}\rightarrow -J_{k}, & P_{k}\rightarrow -P_{k},
& i\rightarrow -i; &  &  \\ 
\mathcal{PT}_{2}: & J_{k}\rightarrow -J_{k}, & P_{k}\rightarrow P_{k}, & 
i\rightarrow -i; &  &  \\ 
\mathcal{PT}_{3}: & J_{k}\rightarrow J_{k}, & P_{1}\rightarrow P_{1}, & 
P_{2}\leftrightarrow P_{3}, & i\rightarrow -i; &  \\ 
\mathcal{PT}_{4}: & J_{1}\rightarrow -J_{1},~~~~ & J_{2/3}\rightarrow
J_{2/3},~~~~ & P_{1/3}\leftrightarrow -P_{1/3},~~~ & P_{2}\leftrightarrow
P_{2},~ & i\rightarrow -i;
\end{array}
\end{equation}
for $k=1,2,3$.

Once again we wish to find the Dyson map to map non-Hermitian Hamiltonians expressed in terms of bilinear combinations of these generators to Hermitian ones. For the $E_{3}$-algebra we take it to be of the general form
\begin{equation}
\eta =e^{\lambda _{z}J_{z}+\lambda _{+}J_{+}+\lambda _{-}J_{-}+\kappa
_{z}P_{z}+\kappa _{+}P_{+}+\kappa _{-}P_{-}},\qquad \text{\ \ \ \ \ \ \ \
for }\lambda _{z},\lambda _{\pm },\kappa _{z},\kappa _{\pm }\in \mathbb{R}.
\label{eta2}
\end{equation}
For the adjoint action of this operator on the $E_{3}$-generators we compute
\begin{equation}
\eta P_{\ell }\eta ^{-1}=\mu _{\ell z}P_{z}+\mu _{\ell +}P_{+}+\mu _{\ell
-}P_{-}\qquad \text{for }\ell =z,\pm
\end{equation}
with constant coefficients 
\begin{eqnarray*}
\mu _{zz} &=&1+2c(\omega )\lambda _{+}\lambda _{-},\quad \mu _{\pm \pm
}=1+(2\lambda _{z}^{2}+\lambda _{+}\lambda _{-})c(\omega )\pm 2s(\omega
)\lambda _{z},\quad \\
\mu _{\pm \mp } &=&c(\omega )\lambda _{\mp }^{2},\quad \mu _{\pm z}=\mp
2c(\omega )\lambda _{z}\lambda _{\mp }-2s(\omega )\lambda _{\mp },\quad \mu
_{z\pm }=\mp c(\omega )\lambda _{z}\lambda _{\pm }-s(\omega )\lambda _{\pm },
\end{eqnarray*}
and 
\begin{equation}
\eta J_{\ell }\eta ^{-1}=\nu _{\ell z}J_{z}+\nu _{\ell +}J_{+}+\nu _{\ell
-}J_{-}+\rho _{\ell z}P_{z}+\rho _{\ell +}P_{+}+\rho _{\ell -}P_{-}\qquad 
\text{for }\ell =z,\pm
\end{equation}
with constant coefficients
\begin{eqnarray*}
\nu _{zz} &=&1+2c(\omega )\lambda _{+}\lambda _{-},\quad \nu _{\pm \pm }=1+
\tilde{\omega}^{2}c(\omega )\pm 2s(\omega )\lambda _{z},\quad \nu _{\pm \mp
}=-c(\omega )\lambda _{\mp }^{2}, \\
\nu _{\pm z} &=&\mp s(\omega )\lambda _{\mp }-c(\omega )\lambda _{z}\lambda
_{\mp },\quad \nu _{z\pm }=-2c(\omega )\lambda _{z}\lambda _{\pm }\mp
2s(\omega )\lambda _{\pm },
\end{eqnarray*}
\begin{eqnarray*}
\rho _{zz} &=&4\left[ \left( \lambda _{-}\kappa _{+}-\lambda _{+}\kappa
_{-}\right) c(\omega )-\frac{\lambda _{+}\lambda _{-}}{\omega ^{2}}\mu
(c(\omega )-s(\omega ))\right] \\
\rho _{z\pm } &=&c(\omega )(\pm \lambda _{\pm }\kappa _{z}-2\lambda
_{z}\kappa _{\pm })\mp 2s(\omega )(\kappa _{\pm }+\lambda _{\pm }\kappa
_{z})\pm \frac{2c(\omega )}{\omega ^{2}}\lambda _{\pm }\nu +\frac{s(\omega )
}{\omega ^{2}}\lambda _{\pm }\left( \mu \mp 2\nu \right) \\
&&-\frac{\cosh (2\omega )}{\omega ^{2}}\mu \lambda _{\pm } \\
\rho _{\pm z} &=&c(\omega )(\lambda _{\mp }\kappa _{z}\pm 2\lambda
_{z}\kappa _{\mp })+2s(\omega )(\kappa _{\mp }-\lambda _{\mp }\kappa _{z})+
\frac{2c(\omega )}{\omega ^{2}}\lambda _{\mp }\nu \pm \frac{s(\omega )}{%
\omega ^{2}}\lambda _{\mp }\left( \mu \mp 2\nu \right) \\
&&\mp \frac{\cosh (2\omega )}{\omega ^{2}}\mu \lambda _{\mp } \\
\rho _{\pm \pm } &=&\pm c(\omega )\tilde{\mu}+s(\omega )\kappa _{z}\pm \mu 
\frac{\tilde{\omega}^{2}}{\omega ^{2}}[s(\omega )-c(\omega )]+\frac{\cosh
(2\omega )-s(\omega )}{\omega ^{2}}\lambda _{z}\mu \\
\rho _{\pm \mp } &=&-2c(\omega )\lambda _{\mp }\kappa _{\mp }\pm \frac{\mu
\lambda _{\mp }^{2}}{\omega ^{2}}[s(\omega )-c(\omega )]
\end{eqnarray*}
where we abbreviated $\omega :=\sqrt{\lambda _{z}^{2}+\lambda _{+}\lambda
_{-}}$, $\tilde{\omega}:=\sqrt{2\lambda _{z}^{2}+\lambda _{+}\lambda _{-}}$, 
$\mu :=\kappa _{z}\lambda _{z}+\kappa _{+}\lambda _{-}-\kappa _{-}\lambda
_{+}$, $\tilde{\mu}:=2\kappa _{z}\lambda _{z}+\kappa _{+}\lambda _{-}-\kappa
_{-}\lambda _{+}$, $\nu :=\kappa _{+}\lambda _{z}\lambda _{-}-\kappa
_{z}\lambda _{+}\lambda _{-}-\kappa _{-}\lambda _{z}\lambda _{+}$, $c(\omega
):=(\cosh (2\omega )-1)/(2\omega ^{2})$ and $s(\omega ):=\sinh (2\omega
)/(2\omega )$.

The construction of isospectral counterparts, if they exist, for
non-Hermitian Hamiltonians symmetric with regard to various different types
of $\mathcal{PT}$-symmetries is far more involved for this algebra.
The most generic cases are very complicated in this case as they involve 25
free parameters. One may therefore restrict the discussion to simpler
examples, such as for instance the complements of $E_{2}$ in $E_{3}$
constitutes well-defined subclasses.

For instance, we may consider a $\mathcal{PT}_{1}$-invariant Hamiltonian of  $E_{3}/E_{2}$-Lie algebraic type. Selecting $\{J_{z},P_{\pm }\}$ as the generators of the $E_{2}$-subalgebra the most general Hamiltonian of this type is 
\begin{equation}
\tilde{H}_{\mathcal{PT}_{1}}=\mu _{1}J_{+}^{2}+\mu _{2}J_{-}^{2}+\mu
_{3}P_{z}^{2}+\mu _{4}P_{z}J_{+}+\mu _{5}P_{z}J_{-}+\mu _{6}J_{+}J_{-}+i\mu
_{7}J_{+}+i\mu _{8}J_{-}+i\mu _{9}P_{z}.
\end{equation}
All the necessary tools have been provided here to find the corresponding
counterparts etc. We leave this discussion for future investigations.

\section{Discussions}
We have introduced the basic notions of the $\mathcal{PT}$-symmetric non-Hermitian and pseudo-Hermitian Hamiltonian systems and discussed advantages as well as disadvantages of the two approaches in the first two sections. In section \ref{section33}, we have constructed the exact form of the metric operator and thus the Hermitian version of the non-Hermitian systems of Euclidean Lie algebraic type Hamiltonians. We presented five different types of $\mathcal{PT}$-symmetries (\ref{PT1to5}) for the Euclidean algebra $E_{2}$ (\ref{EEE2}). Considering the most general invariant non-Hermitian Hamiltonians in terms of bilinear combinations of the generators of this algebra, we have systematically constructed isospectral counterparts from Dyson maps $\eta $ of the general form (\ref{eta}) by exploiting its adjoint action on the Lie algebraic generators. In this process some of the coupling constants involved had to be constrained. We noted that the different versions of the symmetries also lead to qualitatively quite different isospectral counterparts. For the symmetries $\mathcal{PT}_{1}$ and $\mathcal{PT}_{2}$ the required constraints rendered the original Hamiltonians $H_{\mathcal{PT}_{1/2}}$ Hermitian, such that the adjoint action of $\eta $ maps Hermitian Hamiltonians to Hermitian ones. It
should be noted that the maps are non-trivial, albeit the distinguishing features of the obtained Hamiltonians $h_{\mathcal{PT}_{1/2}}$ remain unclear. More interesting are the transformation properties of the non-Hermitian Hamiltonians invariant under the symmetries $\mathcal{PT}_{3}$, $\mathcal{PT}_{4}$ and $\mathcal{PT}_{5}$, as they lead to genuine non-Hermitian/Hermitian isospectral pairs constructed from an explicit non-perturbative Dyson map.

For the representation (\ref{rep11}) we analysed the $\mathcal{PT}_{5}$-system in further detail by solving the corresponding time-independent Schr\"{o}dinger equation. For some parameter choices we found simple transformations of the real Mathieu equation as solutions. In a subset of cases the corresponding energy spectra were identified to be entirely real, see figure \ref{fig1}. For other choices we observed spontaneously broken $\mathcal{PT}$-symmetry with region in the parameter space where the whole spectrum remained real. It is possible to consider the spectra as functions of coupling constants in such a way that its monotonic variation leads to an initial break down of the $\mathcal{PT}$-symmetry at some exceptional points which is subsequently regained, see figure \ref{fig2}. This numerically observed behaviour is completely understood from the explicit formulae for the Dyson maps, which break down at the exceptional points. We have made contact to some simple systems of optical lattices and it should be highly interesting to investigate further whether the more involved systems with richer structure we considered here may also be realized experimentally. We have verified the typical gain/loss symmetry for one of those models.

Clearly we have not exhausted the discussion for the entire parameter space
for the $\mathcal{PT}_{5}$-system and also left the analysis of time-independent Schr\"{o}dinger equation $\mathcal{PT}_{3}$ and $\mathcal{PT}_{4}$ for further investigation. An additional open problem is the analysis of alternative representations such as (\ref{rep22}) and many more not mentioned here. Also still an intriguing open challenge is the computation of the explicit Dyson map for systems of the type dealt with in section "Complex Mathieu equation". We established that they certainly require a different type of Ansatz for the Dyson map $\eta $ as the one in (\ref{eta}).

The completion of the above mentioned programme is far from being finished for the Euclidean algebra $E_{3}$. For that case we have provided the far more complicated adjoint action on the generators and left the further analysis, which can be carried out along the same lines as for $E_{2}$.


\chapter{Noncommutative Models in 3D and Minimal Volume}\label{noncommutativemodelsin3D}
In the last two chapters, we have briefly discussed the necessary tools to formulate models on a noncommutative space. In the present chapter, we look at the procedure of how to build models in this space in three dimensions. It will not be an easy task to construct models in the higher dimensions and we will face many computational difficulties, as we will see in the later part of our discussions. Certainly, the work \cite{dey_fring_gouba} has been motivated by the investigations in lower dimensions \cite{bagchi_fring,fring_gouba_scholtz,fring_gouba_bagchi_area}. 

\section{Flat Noncommutative Spaces in 3D}
Let us commence our discussion with the standard Fock space commutation relations 
\begin{equation}
\lbrack a_{i},a_{j}^{\dagger }]=\delta _{ij},\qquad \lbrack
a_{i},a_{j}]=0,\qquad \lbrack a_{i}^{\dagger },a_{j}^{\dagger }]=0,\qquad 
\text{for }i,j=1,2,3,  \label{Fock}
\end{equation}
and we assume that the relations (\ref{Fock}) are linearly related to
the standard three dimensional flat noncommutative space (\ref{nccommutator}) characterized by
the relations
\begin{equation}
\begin{array}{lll}
\lbrack x_{0},y_{0}]=i\theta _{1}, & [x_{0},z_{0}]=i\theta _{2},\qquad \quad
& [y_{0},z_{0}]=i\theta _{3}, \\ 
\lbrack x_{0},p_{x_{0}}]=i\hbar ,\qquad \quad & [y_{0},p_{y_{0}}]=i\hbar , & 
[z_{0},p_{z_{0}}]=i\hbar ,
\end{array}
~~\ \ \ \ \ \text{for }\theta _{1},\theta _{2},\theta _{3}\in \mathbb{R},
\label{cano}
\end{equation}
with all remaining commutators to be zero. $\theta _{1},\theta
_{2},\theta _{3}$ represent the noncommutative structure constants. It is worth mentioning at this point that the generators of flat noncommutative spaces will be denoted, from now and onwards, by the small letters of the corresponding observables followed by the subscript zero. The most general
linear Ansatz to relate the generators of relations (\ref{cano}) and (\ref{Fock}) reads 
\begin{equation}
\varphi _{i}=\sum\limits_{j=1}^{3}\left(\kappa _{ij}~a_{j}+\lambda
_{ij}a_{j}^{\dagger }\right),~~\ \ \ \ \ \ \text{for }\vec{\varphi}
=\{x_{0},y_{0},z_{0},p_{x_{0}},p_{y_{0}},p_{z_{0}}\},  \label{sd}
\end{equation}
where the $\kappa _{ij}$, $\lambda _{ij}$ have dimensions of length or
momentum for $i=1,2,3$ or $i=4,5,6$ respectively. The commutation relations
obeyed by the canonical variables $X$, $Y$, $Z$, $P_{x}$, $P_{y}$, $P_{z}$
associated to the deformed algebra (\ref{q2deformation}) in three dimensions are yet unknown and are subject to construction. The algebra they satisfy may be related to (\ref{q2deformation}) by similar relations as (\ref{sd}), but since the constants $\kappa_{ij}$ and $\lambda _{ij}$ are in general complex, this amounts to finding $72$ real parameters. To reduce this number to a manageable quantity one can utilize $\mathcal{PT}$-symmetry as introduced in chapter \ref{chapterPT}.

\rhead{Noncommutative Models in 3D $\&$ Minimal Volume}
\lhead{Chapter 4}
\chead{}

\subsection{The Role of $\mathcal{PT}$-Symmetry}
Whereas the momenta and coordinates in (\ref{cano}) are Hermitian operators
acting on a Hilbert space with standard inner product, this is no longer
true for the variables associated to the deformed algebra (\ref{q2deformation}) as they become in general non-Hermitian with regard to these inner products.
With reference to chapter \ref{chapterPT}, we show how to solve the difficulties in the present situation. In \cite{giri_roy}, the authors argue that the commutation relations (\ref{cano}) can not be expressed as $\mathcal{PT}$-symmetric and one is therefore forced to take the noncommutative structure constants to be complex. We reason here that this is incorrect and even the standard noncommutative space relations are in fact symmetric under many different versions of $\mathcal{PT}$-symmetry. As discussed in (\ref{PTantilinear}), all one requires to formulate a consistent quantum description is an anti-linear involutory map \cite{wigner} that leaves the relations (\ref{cano}) invariant. We identify here several possibilities:

Taking for instance $\theta _{2}=0$, the algebra (\ref{cano}) remains
invariant under the following antilinear transformations 
\begin{equation}
\begin{array}{cllll}
\mathcal{PT}_{\pm }:\quad & x_{0}\rightarrow \pm x_{0}, & y_{0}\rightarrow
\mp y_{0}, & z_{0}\rightarrow \pm z_{0}, & i\rightarrow -i, \\ 
& p_{x_{0}}\rightarrow \mp p_{x_{0}}, & p_{y_{0}}\rightarrow \pm p_{y_{0}},
& p_{z_{0}}\rightarrow \mp p_{z_{0}}. & 
\end{array}
\label{+-}
\end{equation}
We may also attempt to keep $\theta _{2}$ different from zero, in which case
we have to transform the $\theta _{2}$ as well in order to achieve the
invariance of (\ref{cano}) 
\begin{equation}
\begin{array}{cllll}
\mathcal{PT}_{\mathcal{\theta }_{\pm }}:\quad & x_{0}\rightarrow \pm x_{0},
& y_{0}\rightarrow \mp y_{0}, & z_{0}\rightarrow \pm z_{0}, & i\rightarrow
-i, \\ 
& p_{x_{0}}\rightarrow \mp p_{x_{0}}, & p_{y_{0}}\rightarrow \pm p_{y_{0}},
& p_{z_{0}}\rightarrow \mp p_{z_{0}}, & \theta _{2}\rightarrow -\theta _{2}.
\end{array}
\end{equation}
A further option, which also allows to keep $\theta _{2}$ different from
zero, would be to introduce permutations amongst the different directions
\begin{equation}
\begin{array}{cllll}
\mathcal{PT}_{xz}:\quad & x_{0}\rightarrow z_{0}, & y_{0}\rightarrow y_{0},
& z_{0}\rightarrow x_{0}, & i\rightarrow -i, \\ 
& p_{x_{0}}\rightarrow -p_{z_{0}}, & p_{y_{0}}\rightarrow -p_{y_{0}}, & 
p_{z_{0}}\rightarrow -p_{x_{0}}. & 
\end{array}
\label{xz}
\end{equation}
Clearly all of these maps are involutions $\left(\mathcal{PT}\right)^{2}=\mathbb{I}$. In
fact, there might be more options. The occurrence of various
possibilities to implement the anti-linear symmetry (\ref{PTantilinear}) is a known feature previously observed in many examples \cite{alvaredo_fring,fring_smith,assis} in dimensions larger than one. More restrictions and the explicit choice of symmetry result from the specific physical situation one wishes to describe. For instance $\mathcal{PT}_{\pm }$ might be appropriated when one deals with a problem in which one direction is singled out, $\mathcal{PT}_{\mathcal{\theta }_{\pm }}$ requires the noncommutative constant $\theta _{2}$ to appear as a parameter in the model and $\mathcal{PT}_{xz}$ suggest a symmetry across the line $x_{0}=z_{0}$. For the creation and annihilation operators this symmetry could manifest itself in different ways, for instance as $a_{i}\rightarrow \pm a_{i}$, $a_{i}^{\dagger }\rightarrow \pm a_{i}^{\dagger }$ or by the permutation of indices $a_{i}\rightarrow a_{j}^{\dagger }$, $a_{i}^{\dagger }\rightarrow a_{j}$ when they label for instance particles in different potentials, see e.g. \cite{graefe_gunther_korsch_niederle}. Once again the underlying physics will dictate which version one should select. The general reason for the occurrence of these different possibilities are just manifestations of the ambiguities in defining a metric to which the $\mathcal{PT}$-operator is directly related. What needs to be kept in mind is that we only require the symmetry under some anti-linear involution \cite{wigner} in order to obtain a meaningful quantum mechanical description.

\subsection{Oscillator Algebras of Flat Noncommutative Spaces}
Let us first see how to represent a three dimensional oscillator algebra in
terms of the canonical variables in three dimensional flat noncommutative
space. For definiteness we seek at first a description which is invariant
under $\mathcal{PT}_{\pm }$. The most generic linear Ansatz for the creation
and annihilation operators to achieve this is 
\begin{eqnarray}
a_{1} &=&\alpha _{1}x_{0}+i\alpha _{2}y_{0}+\alpha _{3}z_{0}+i\alpha
_{4}p_{x_{0}}+\alpha _{5}p_{y_{0}}+i\alpha _{6}p_{z_{0}},  \label{a1} \\
a_{2} &=&\alpha _{7}x_{0}+i\alpha _{8}y_{0}+\alpha _{9}z_{0}+i\alpha
_{10}p_{x_{0}}+\alpha _{11}p_{y_{0}}+i\alpha _{12}p_{z_{0}},  \label{a2} \\
a_{3} &=&\alpha _{13}x_{0}+i\alpha _{14}y_{0}+\alpha _{15}z_{0}+i\alpha
_{16}p_{x_{0}}+\alpha _{17}p_{y_{0}}+i\alpha _{18}p_{z_{0}},  \label{a33}
\end{eqnarray}
with dimensional real constants $\alpha _{i}$. We note that we have $\mathcal{PT}_{\pm }:a_{i}\rightarrow \pm a_{i},a_{i}^{\dagger }\rightarrow
\pm a_{i}^{\dagger }$ for $i=1,2,3$. The non-sequential ordering of the
constants in (\ref{a1})-(\ref{a33}) is chosen to perform the limit to the two
dimensional case in a convenient way. For $\alpha _{9},\ldots ,\alpha
_{18}\rightarrow 0$ we recover equation (\ref{sd}) in \cite{fring_gouba_bagchi_area}. It is useful to invoke this limit at various stages of the calculation as a consistency check. We then compute that the operators (\ref{a1})-(\ref{a33}), expressed on the three dimensional flat noncommutative space (\ref{cano}), satisfy the standard Fock space commutation relations (\ref{Fock}) provided that the following nine constraints hold:
\begin{eqnarray}
1 &=&2\sum\limits_{j=1}^{3}\left[ (2-j)\alpha _{2+k}\alpha _{j+k}\theta
_{j}-(-1)^{j}\hbar \alpha _{j+k}\alpha _{j+k+3}\right] \ \qquad \text{for }
k=0,6,12,  \label{e1} \\
0 &=&i(\alpha _{p}\alpha _{q}+\alpha _{p+2}\alpha _{q-2})\theta
_{2}+\sum\limits_{j=1}^{2}(\alpha _{j+p}\alpha _{j+q-p+2}-\alpha
_{j+p-1}\alpha _{j+q-2})\theta _{2j-1} \\
&&+\sum\limits_{j=1}^{3}(\alpha _{j+p+2}\alpha _{j+q-p-2}-\alpha
_{j+p-1}\alpha _{j+q})\hbar ~~~~\ \ \ \ \text{for }\{p,q\}=\{1,9\},\{1,15\},
\{7,15\},  \notag \\
0 &=&i(\alpha _{p}\alpha _{q}+\alpha _{p+2}\alpha _{q-2})\theta
_{2}-\sum\limits_{j=1}^{2}(-1)^{j}(\alpha _{j+p}\alpha _{j+q-p+2}+\alpha
_{j+p-1}\alpha _{j+q-2})\theta _{2j-1}  \label{e14} \\
&&-\sum\limits_{j=1}^{3}(-1)^{j}(\alpha _{j+p+2}\alpha _{j+q-p-2}+\alpha
_{j+p-1}\alpha _{j+q})\hbar ~~~\text{for }\{p,q\}=\{1,9\},\{1,15\},\{7,15\}.
\notag
\end{eqnarray}
It turns out that when keeping $\theta _{2}\neq 0$ these equations do not
admit a nontrivial solution. However, setting $\theta _{2}$ to zero we can
solve (\ref{e1})-(\ref{e14}) for instance by
\begin{eqnarray}
\alpha _{2} &=&-\frac{\alpha _{14}\left( \alpha _{7}\left( 2\hbar\alpha
_{14}\alpha _{17}-2\alpha _{13}\Delta ^{\prime }+1\right) -2\alpha
_{9}\alpha _{13}\Delta ^{\prime \prime }\right) }{2\Delta \Delta ^{\prime
\prime }}, \\
\alpha _{4} &=&\frac{\hbar\alpha _{7}\alpha _{16}\left( -2\hbar\alpha _{14}\alpha
_{17}+2\alpha _{13}\Delta ^{\prime }-1\right) +\alpha _{9}\left( 2\hbar\alpha
_{13}\alpha _{16}-1\right) \Delta ^{\prime \prime }}{2\hbar\Delta \Delta
^{\prime \prime }}, \\
\alpha _{5} &=&\frac{\alpha _{1}\Delta ^{\prime }+\alpha _{3}\Delta ^{\prime
\prime }}{\hbar\alpha _{14}}, \\
\alpha _{6} &=&\frac{2\hbar\alpha _{9}\alpha _{13}\alpha _{18}\Delta ^{\prime
\prime }+\alpha _{7}\left( 2\hbar\alpha _{13}\alpha _{18}\Delta ^{\prime
}-\alpha _{14}\left( 2\alpha _{17}\alpha _{18}\hbar^{2}+\theta _{3}\right)
\right) }{2\hbar\Delta \Delta ^{\prime \prime }}, \\
\alpha _{8} &=&\frac{\alpha _{14}\left( \alpha _{1}\left( 2\hbar\alpha
_{14}\alpha _{17}-2\alpha _{13}\Delta ^{\prime }+1\right) +2\alpha
_{3}\alpha _{13}\left( \alpha _{14}\theta _{3}-\hbar\alpha _{18}\right) \right) 
}{2\Delta \Delta ^{\prime \prime }}, \\
\alpha _{10} &=&\frac{\hbar\alpha _{1}\alpha _{16}\left( 2\hbar\alpha _{14}\alpha
_{17}-2\alpha _{13}\Delta ^{\prime }+1\right) -\alpha _{3}\left( 2\hbar\alpha
_{13}\alpha _{16}-1\right) \Delta ^{\prime \prime }}{2\hbar\Delta \Delta
^{\prime \prime }}, \\
\alpha _{11} &=&\frac{\alpha _{7}\Delta ^{\prime }+\alpha _{9}\Delta
^{\prime \prime }}{\hbar\alpha _{14}}, \\
\alpha _{12} &=&\frac{-2\hbar\alpha _{3}\alpha _{13}\alpha _{18}\Delta ^{\prime
\prime }+\alpha _{1}\left( \alpha _{14}\left( 2\alpha _{17}\alpha
_{18}\hbar^{2}+\theta _{3}\right) -2\hbar\alpha _{13}\alpha _{18}\Delta ^{\prime
}\right) }{2\hbar\Delta \Delta ^{\prime \prime }}, \\
\alpha _{15} &=&\frac{2\hbar\alpha _{14}\alpha _{17}-2\alpha \Delta ^{\prime }+1
}{2\Delta ^{\prime \prime }},
\end{eqnarray}
where we abbreviated $\Delta :=\alpha _{3}\alpha _{7}-\alpha _{1}\alpha _{9}$
, $\Delta ^{\prime }:=\hbar\alpha _{16}+\alpha _{14}\theta _{1}$, $\Delta
^{\prime \prime }:=\hbar\alpha _{18}-\alpha _{14}\theta _{3}$. Thus we still
have nine parameters left at our disposal. In other words the Ansatz (\ref
{a1})-(\ref{a33}) together with (\ref{Fock}) enforces the $\mathcal{PT}$-symmetry of the type (\ref{+-}).

Inverting the relations (\ref{a1})-(\ref{a33}) we may express the dynamical
variables in terms of the creation and annihilation operators
\begin{alignat}{1}
x_{0} &=\frac{\alpha _{9}\alpha _{17}-\alpha _{11}\alpha _{15}}{2\det M_{1}}
(a_{1}+a_{1}^{\dagger })+\frac{\alpha _{5}\alpha _{15}-\alpha _{3}\alpha
_{17}}{2\det M_{1}}(a_{2}+a_{2}^{\dagger })+\frac{\alpha _{3}\alpha
_{11}-\alpha _{5}\alpha _{9}}{2\det M_{1}}(a_{3}+a_{3}^{\dagger }),
\label{x11} \\
y_{0} &=\frac{\alpha _{10}\alpha _{18}-\alpha _{12}\alpha _{16}}{2i\det
M_{2}}(a_{1}-a_{1}^{\dagger })+\frac{\alpha _{6}\alpha _{16}-\alpha
_{4}\alpha _{18}}{2i\det M_{2}}(a_{2}-a_{2}^{\dagger })+\frac{\alpha
_{4}\alpha _{12}-\alpha _{6}\alpha _{10}}{2i\det M_{2}}(a_{3}-a_{3}^{\dagger
}),~~~~~~~ \\
z_{0} &=\frac{\alpha _{11}\alpha _{13}-\alpha _{7}\alpha _{17}}{2\det M_{1}}
(a_{1}+a_{1}^{\dagger })+\frac{\alpha _{1}\alpha _{17}-\alpha _{5}\alpha
_{13}}{2\det M_{1}}(a_{2}+a_{2}^{\dagger })+\frac{\alpha _{5}\alpha
_{7}-\alpha _{1}\alpha _{11}}{2\det M_{1}}(a_{3}+a_{3}^{\dagger }),
\end{alignat}
\begin{alignat}{1}
p_{x_{0}} &=\frac{\alpha _{12}\alpha _{14}-\alpha _{8}\alpha _{18}}{2i\det
M_{2}}(a_{1}-a_{1}^{\dagger })+\frac{\alpha _{2}\alpha _{18}-\alpha
_{6}\alpha _{14}}{2i\det M_{2}}(a_{2}-a_{2}^{\dagger })+\frac{\alpha
_{6}\alpha _{8}-\alpha _{2}\alpha _{12}}{2i\det M_{2}}(a_{3}-a_{3}^{\dagger
}),~~~~~~~ \\
p_{y_{0}} &=\frac{\alpha _{7}\alpha _{15}-\alpha _{9}\alpha _{13}}{2\det
M_{1}}(a_{1}+a_{1}^{\dagger })+\frac{\alpha _{3}\alpha _{13}-\alpha
_{1}\alpha _{15}}{2\det M_{1}}(a_{2}+a_{2}^{\dagger })+\frac{\alpha
_{1}\alpha _{9}-\alpha _{3}\alpha _{7}}{2\det M_{1}}(a_{3}+a_{3}^{\dagger }),
\\
p_{z_{0}} &=\frac{\alpha _{8}\alpha _{16}-\alpha _{10}\alpha _{14}}{2i\det
M_{2}}(a_{1}-a_{1}^{\dagger })+\frac{\alpha _{4}\alpha _{14}-\alpha
_{2}\alpha _{16}}{2i\det M_{2}}(a_{2}-a_{2}^{\dagger })+\frac{\alpha
_{2}\alpha _{10}-\alpha _{4}\alpha _{8}}{2i\det M_{2}}(a_{3}-a_{3}^{\dagger
}),  \label{x6}
\end{alignat}
where the matrices $M_{1/2}$ have entries
\begin{equation}
(M_{l})_{jk}=6j+2k+l-8\qquad \text{for\qquad }l=1,2.
\end{equation}
These expressions satisfy the commutation relations (\ref{cano}) when we
invoke the constraints (\ref{e1})-(\ref{e14}) and the standard Fock space
relation (\ref{Fock}). In that case we also have the simple relation $\det
M_{1}\det M_{2}=-1/8\hbar ^{3}$. By changing the Ansatz (\ref{a1})-(\ref{a33}) appropriately one may also obtain $\mathcal{PT}_{\mathcal{\theta }_{\pm }}$
or $\mathcal{PT}_{xz}$-invariant solutions.

\section{q-Deformed Noncommutative Spaces in 3D} \label{section4.2}

As discussed in chapter \ref{chapter2}, we start the analysis of this section by assuming a q-deformed oscillator algebra for the creation and annihilation operators $A_i^\dagger, A_i$ of the form:
\begin{equation}
A_{i}A_{j}^{\dagger }-q^{2\delta _{ij}}A_{j}^{\dagger }A_{i}=\delta
_{ij},\quad \lbrack A_{i}^{\dagger },A_{j}^{\dagger }]=0,\quad \lbrack
A_{i},A_{j}]=0,\quad \text{for }i,j=1,2,3;q\in \mathbb{R}.  \label{AAAA}
\end{equation}
In the limit $q\rightarrow 1$ we denote $A_{i}\rightarrow a_{i}$ and recover
the standard Fock space commutation relations (\ref{Fock}). Now we construct the commutation relations for the deformed noncommutative space satisfied by the deformed canonical variables $X$, $Y$, $Z$, $P_{x}$, $P_{y}$, $P_{z}$, which we express linearly in terms of the creation and annihilation
operators obeying the deformed algebra (\ref{AAAA}). Guided by the fact that
in the limit $q\rightarrow 1$ we should recover the relations (\ref{x11})-(\ref{x6}) of the previous section. We therefore make the Ansatz
\begin{eqnarray}
X &=&\hat{\kappa}_{1}(A_{1}^{\dagger }+A_{1})+\hat{\kappa}_{2}(A_{2}^{\dagger}+A_{2})+\hat{\kappa}_{3}(A_{3}^{\dagger }+A_{3}),
\label{an1} \\
Y &=&i\hat{\kappa}_{4}(A_{1}^{\dagger }-A_{1})+i\hat{\kappa}
_{5}(A_{2}^{\dagger }-A_{2})+i\hat{\kappa}_{6}(A_{3}^{\dagger }-A_{3}), \\
Z &=&\hat{\kappa}_{7}(A_{1}^{\dagger }+A_{1})+\hat{\kappa}
_{8}(A_{2}^{\dagger }+A_{2})+\hat{\kappa}_{9}(A_{3}^{\dagger }+A_{3}), \\
P_{x} &=&i\check{\kappa}_{10}(A_{1}^{\dagger }-A_{1})+i\check{\kappa}
_{11}(A_{2}^{\dagger }-A_{2})+i\check{\kappa}_{12}(A_{3}^{\dagger }-A_{3}),
\\
P_{y} &=&\check{\kappa}_{13}(A_{1}^{\dagger }+A_{1})+\check{\kappa}
_{14}(A_{2}^{\dagger }+A_{2})+\check{\kappa}_{15}(A_{3}^{\dagger }+A_{3}), \\
P_{z} &=&i\check{\kappa}_{16}(A_{1}^{\dagger }-A_{1})+i\check{\kappa}
_{17}(A_{2}^{\dagger }-A_{2})+i\check{\kappa}_{18}(A_{3}^{\dagger }-A_{3}),
\label{an6}
\end{eqnarray}
with $\hat{\kappa}_{i}=\kappa _{i}\sqrt{\hbar /(m\omega )}$ for $i=1,\ldots
,9$ having the dimension of a length and $\check{\kappa}_{i}=\kappa _{i}
\sqrt{m\omega \hbar }$ for $i=10,\ldots ,18$ possessing the dimension of a
momentum. The constants $\kappa _{i}$ for $i=1,\ldots ,18$ are therefore
dimensionless. We deliberately keep here all dimensional constants different
from $1$. With the help of the $q$-deformed oscillator algebra (\ref{AAAA})
we compute
\begin{eqnarray}
\lbrack X,Y] &=&2i\sum\nolimits_{j=1}^{3}\hat{\kappa}_{j}\hat{\kappa}_{3+j}
\left[ 1+\left( q^{2}-1\right) \right] A_{j}^{\dagger }A_{j},  \label{sdd} \\
\lbrack Y,Z] &=&-2i\sum\nolimits_{j=1}^{3}\hat{\kappa}_{3+j}\hat{\kappa}
_{6+j}\left[ 1+\left( q^{2}-1\right) \right] A_{j}^{\dagger }A_{j}, \\
\lbrack X,P_{x}] &=&2i\sum\nolimits_{j=1}^{3}\hat{\kappa}_{j}\check{\kappa}
_{9+j}\left[ 1+\left( q^{2}-1\right) \right] A_{j}^{\dagger }A_{j}, \\
\lbrack Y,P_{y}] &=&-2i\sum\nolimits_{j=1}^{3}\hat{\kappa}_{3+j}\check{\kappa
}_{12+j}\left[ 1+\left( q^{2}-1\right) \right] A_{j}^{\dagger }A_{j}, \\
\lbrack Z,P_{z}] &=&2i\sum\nolimits_{j=1}^{3}\hat{\kappa}_{6+j}\check{\kappa}
_{15+j}\left[ 1+\left( q^{2}-1\right) \right] A_{j}^{\dagger }A_{j}, \label{c5} \\
\lbrack P_{x},P_{y}] &=&-2i\sum\nolimits_{j=1}^{3}\check{\kappa}_{9+j}\check{
\kappa}_{12+j}\left[ 1+\left( q^{2}-1\right) \right] A_{j}^{\dagger }A_{j},
\label{c4} \\
\lbrack P_{y},P_{z}] &=&2i\sum\nolimits_{j=1}^{3}\check{\kappa}_{12+j}\check{
\kappa}_{15+j}\left[ 1+\left( q^{2}-1\right) \right] A_{j}^{\dagger }A_{j},
\\
\lbrack X,P_{z}] &=&2i\sum\nolimits_{j=1}^{3}\hat{\kappa}_{j}\check{\kappa}
_{15+j}\left[ 1+\left( q^{2}-1\right) \right] A_{j}^{\dagger }A_{j}, \\
\lbrack Z,P_{x}] &=&2i\sum\nolimits_{j=1}^{3}\hat{\kappa}_{6+j}\check{\kappa}
_{9+j}\left[ 1+\left( q^{2}-1\right) \right] A_{j}^{\dagger }A_{j}, \\
\lbrack X,Z] &=&[P_{x},P_{z}]=[X,P_{y}]=[Y,P_{x}]=[Y,P_{z}]=[Z,P_{y}]=0.
\label{c6}
\end{eqnarray}
Inverting now the relations (\ref{an1})-(\ref{an6}) we find that it is
indeed possible to eliminate entirely the creations and annihilation
operators from these relations. However, this leads to very lengthy
expressions, which we will not present here. Instead we report some special,
albeit still quite general, solutions obtained by setting some of the
constants to zero and imposing further constraints.

\subsection{A Particular $\mathcal{PT}_{\pm }$-Symmetric Solution}
We now make the assumption that $\hat{\kappa} _{1}=\hat{\kappa} _{4}=\hat{\kappa} _{5}=\hat{\kappa}_{8}=\check{\kappa} _{10}=\check{\kappa} _{12}=\check{\kappa} _{13}=\check{\kappa} _{14}=\check{\kappa}
_{17}=\check{\kappa} _{18}=0$. This choice still guarantees that none of the
canonical variables become mutually identical. The consistency with the
direct limit $q\rightarrow 1$ in which we want to recover (\ref{cano})
enforces the constraints
\begin{equation}
\hat{\kappa}_{2}=\frac{\hbar }{2\check{\kappa}_{11}},~\hat{\kappa}_{3}=\frac{
\theta _{1}}{2\hat{\kappa}_{6}},~\hat{\kappa}_{9}=-\frac{\theta _{3}}{2\hat{
\kappa}_{6}},~~\check{\kappa}_{15}=-\frac{\hbar }{2\hat{\kappa}_{6}},~~
\check{\kappa}_{16}=\frac{\hbar }{2\hat{\kappa}_{7}}.
\end{equation}
The only non-vanishing commutators we obtain from (\ref{sdd})-(\ref{c5}) in this case are
\begin{alignat}{1}
[X,Y] &=i \theta_1\left[1+\left(q^2-1\right)A_3^\dagger A_3\right], \label{commg1}\\
[Y,Z] &=i \theta_3\left[1+\left(q^2-1\right)A_3^\dagger A_3\right], \\
[X,P_x] &=i \hbar\left[1+\left(q^2-1\right)A_2^\dagger A_2\right], \\
[Y,P_y] &=i \hbar\left[1+\left(q^2-1\right)A_3^\dagger A_3\right], \\
[Z,P_z] &=i \hbar\left[1+\left(q^2-1\right)A_1^\dagger A_1\right]. \label{commg5}
\end{alignat}
Next we invert the relations (\ref{an1})-(\ref{an6}) to obtain
\begin{alignat}{3}
A_1 &=\frac{1}{2 \hat{\kappa}_7}Z-\frac{\theta_3}{2 \hbar \hat{\kappa}_7}P_y+\frac{i \hat{\kappa}_7}{\hbar}P_z, \quad && A_1^\dagger &&=\frac{1}{2 \hat{\kappa}_7}Z-\frac{\theta_3}{2 \hbar \hat{\kappa}_7}P_y-\frac{i \hat{\kappa}_7}{\hbar}P_z, \\
A_2 &=\frac{\check{\kappa}_{11}}{\hbar}X+\frac{i}{2 \check{\kappa}_{11}}P_x+\frac{\theta_1 \check{\kappa}_{11}}{\hbar^2}P_y, \quad && A_2^\dagger &&=\frac{\check{\kappa}_{11}}{\hbar}X-\frac{i}{2 \check{\kappa}_{11}}P_x+\frac{\theta_1 \check{\kappa}_{11}}{\hbar^2}P_y, \\
A_3 &=\frac{i}{2 \hat{\kappa}_6}Y-\frac{\hat{\kappa}_6}{\hbar}P_y, && A_3^\dagger &&=-\frac{i}{2 \hat{\kappa}_6}Y-\frac{\hat{\kappa}_6}{\hbar}P_y,
\end{alignat}
so that
\begin{alignat}{1}
A_1^\dagger A_1 &=\frac{1}{4 \hat{\kappa}_7^2}Z^2+\frac{\theta_3^2}{4 \hbar^2\hat{\kappa}_7^2}P_y^2+\frac{\hat{\kappa}_7^2}{\hbar^2}P_z^2-\frac{\theta_3}{2 \hbar\hat{\kappa}_7^2}Z P_y+\frac{i}{2 \hbar}\left[Z,P_z\right], \label{prod1}\\
A_2^\dagger A_2 &=\frac{\check{\kappa}_{11}^2}{\hbar^2}X^2+\frac{1}{4\check{\kappa}_{11}^2}P_x^2+\frac{\theta_1^2\check{\kappa}_{11}^2}{\hbar^4}P_y^2+\frac{2 \theta_1 \check{\kappa}_{11}^2}{\hbar^3}X P_y+\frac{i}{2 \hbar}\left[Y,P_y\right], \\
A_3^\dagger A_3 &=\frac{1}{4\hat{\kappa}_6^2}Y^2+\frac{\hat{\kappa}_6^2}{\hbar^2}P_y^2+\frac{i}{2 \hbar}\left[Y,P_y\right]. \label{prod3}
\end{alignat}
Substituting (\ref{prod1})-(\ref{prod3}) into the right hand sides of (\ref{commg1})-(\ref{commg5}) we obtain five equations for the five unknown commutators $[X,Y], [Y,Z], \left[X,P_x\right], \left[Y,P_y\right]$ and $\left[Z,P_z\right]$. Solving these equations, the resulting dynamical noncommutative relations become
\begin{alignat}{1}
\lbrack X,Y] &=\frac{2 i\theta _{1}}{1+q^2}+i\frac{q^{2}-1}{q^{2}+1}\frac{\theta _{1}}{
\hbar }\left( \frac{m\omega }{2\kappa _{6}^{2}}Y^{2}+\frac{2\kappa _{6}^{2}}{
m\omega }P_{y}^{2}\right) ,  \label{S11} \\
\lbrack Y,Z] &=\frac{2 i\theta _{3}}{1+q^2}+i\frac{q^{2}-1}{q^{2}+1}\frac{\theta _{3}}{
\hbar }\left( \frac{m\omega }{2\kappa _{6}^{2}}Y^{2}+\frac{2\kappa _{6}^{2}}{
m\omega }P_{y}^{2}\right) ,  \label{S12} \\
\lbrack X,P_{x}] &=\frac{2 i\hbar}{1+q^2}+i\frac{q^{2}-1}{q^{2}+1}2m\omega \left( \kappa
_{11}^{2}X^{2}+\frac{1}{4m^{2}\omega ^{2}\kappa _{11}^{2}}P_{x}^{2}+\frac{
\theta _{1}^{2}\allowbreak \kappa _{11}^{2}}{\hbar ^{2}}P_{y}^{2}+2\frac{
\theta _{1}\kappa _{11}^{2}}{\hbar }XP_{y}\right) ,~~  \label{S13} \\
\lbrack Y,P_{y}] &=\frac{2 i\hbar}{1+q^2}+i\frac{q^{2}-1}{q^{2}+1}2m\omega \left( \frac{1}{4\kappa _{6}^{2}}Y^{2}+\frac{\kappa _{6}^{2}}{m^{2}\omega ^{2}}
P_{y}^{2}\right) ,  \label{S14} \\
\lbrack Z,P_{z}] &=\frac{2 i\hbar}{1+q^2} +i\frac{q^{2}-1}{q^{2}+1}2m\omega \left( \frac{1}{4\kappa _{7}^{2}}Z^{2}+\frac{\kappa _{7}^{2}}{m^{2}\omega ^{2}}P_{z}^{2}+
\frac{\theta _{3}^{2}}{4\hbar ^{2}\kappa _{7}^{2}}P_{y}^{2}-\frac{\theta _{3}
}{2\hbar ^{2}\kappa _{7}^{2}}ZP_{y}\right) .  \label{S15}
\end{alignat}
Notice that we still have the three free parameters $\kappa _{6}$, $\kappa
_{7}$ and $\kappa _{11}$ at our disposal. It is easily verified that the
relations (\ref{S11})-(\ref{S15}) are left invariant under a $\mathcal{PT}
_{\pm }$-symmetry (\ref{cano}) in the variables $X$, $Y$, $Z$, $P_{x}$, $%
P_{y}$, $P_{z}$.

\subsubsection{Reduced Three Dimensional Solution for $q\rightarrow 1$}
The solution (\ref{S11})-(\ref{S15}) possesses a non-trivial limit leading
to an even simpler set of commutation relations. For this purpose we impose
some additional constraints by setting first $\kappa_{11}=
\kappa_{6}$, $\kappa _{7}=1/2\kappa _{6}$, $q=\exp (2\tau \kappa
_{6}^{2})$ and subsequently we take the limit $\kappa _{6}\rightarrow 0$.
The relations (\ref{S11})-(\ref{S15}) then reduce to
\begin{eqnarray}
\lbrack X,Y] &=&i\theta _{1}\left( 1+\hat{\tau}Y^{2}\right) ,\quad \lbrack
Y,Z]=i\theta _{3}\left( 1+\hat{\tau}Y^{2}\right) ,  \label{S21} \\
\lbrack X,P_{x}] &=&i\hbar \left( 1+\check{\tau}P_{x}^{2}\right) ,\quad
\lbrack Y,P_{y}]=i\hbar \left( 1+\hat{\tau}Y^{2}\right) ,\quad \lbrack
Z,P_{z}]=i\hbar \left( 1+\check{\tau}P_{z}^{2}\right),~~~ \label{S22}
\end{eqnarray}
where $\hat{\tau}=\tau m\omega /\hbar $ has the dimension of an inverse
squared length, $\check{\tau}=\tau /(m\omega \hbar )$ has the dimension of
an inverse squared momentum and $\tau ~$is dimensionless. We find a concrete
representation for this algebra in terms of the generators of the standard
three dimensional flat noncommutative space (\ref{cano})
\begin{equation}
\begin{array}{ll}
X=(1+\check{\tau}p_{_{x_{0}}}^{2})x_{0}+\frac{\theta _{1}}{\hbar }\left( 
\check{\tau}p_{_{x_{0}}}^{2}-\hat{\tau}y_{0}^{2}\right) p_{y_{0}},\qquad  & 
P_{x}=p_{x_{0}}, \\ 
Z=(1+\check{\tau}p_{z_{0}}^{2})z_{0}+\frac{\theta _{3}}{\hbar }\left( \hat{
\tau}y_{0}^{2}-\check{\tau}p_{z_{0}}^{2}\right) p_{y_{0}}, & P_{z}=p_{z_{0}},
\\ 
P_{y}=(1+\hat{\tau}y_0^{2})p_{y_{0}}, & Y=y_{0}.
\end{array}
\label{rep3d}
\end{equation}
Evidently the quantities $X$, $Z$ and $P_{y}$ are non-Hermitian in the space
in which the $x_{0}$, $y_{0}$, $z_{0}$, $p_{x_{0}}$, $p_{y_{0}}$, $p_{z_{0}}$
are Hermitian. In order to study concrete models it is very convenient to
carry out a subsequent Bopp-shift of the form $x_{0}\rightarrow x_{s}-\frac{
\theta _{1}}{\hbar }p_{y_{s}}$, $y_{0}\rightarrow y_{s}$,\ $z_{0}\rightarrow
z_{s}+\frac{\theta _{3}}{\hbar }p_{y_{s}}$, $p_{x_{0}}\rightarrow p_{x_{s}}$
, $p_{y_{0}}\rightarrow p_{y_{s}}$, $p_{z_{0}}\rightarrow p_{z_{s}}$ and
express the generators in (\ref{rep3d}) in terms of the standard canonical
variables. The same kind of shift have been utilized by many authors for their convenience, for instance one may look at \cite{curtright_fairlie_zachos,li_wang}. Since there is no explicit occurrence of $\theta _{2}$, the representation (\ref{rep3d}) is trivially invariant under $\mathcal{PT}_{\pm}$ as well as $\mathcal{PT}_{\mathcal{\theta }_{\pm }}$. Taking, however, the representation (\ref{rep3d}) and in addition $\theta _{2}\neq 0$ this evidently changes, as by direct computation one of the commutation relations is altered to $[X,Z]=i\theta _{2}\left( 1+\check{\tau}P_{x}^{2}\right)\left( 1+\check{\tau}P_{z}^{2}\right) $. Setting furthermore $\theta_{1}=\theta _{3}$ the representation in (\ref{rep3d}) is also invariant under the $\mathcal{PT}_{xz}$-symmetry stated in (\ref{xz}).

\subsubsection{Reduction into a Decoupled Two Dimensional Plus a One Dimensional Space}
The algebra (\ref{S11})-(\ref{S15}) provides a larger three dimensional
setting for a noncommutative two dimensional space decoupled from a standard
one dimensional space. This is achieved by parametrizing $q=\exp (2\tau
\kappa _{6}^{2})$, setting $\check{\kappa}_{11}=m\omega \hat{\kappa}_{6}$, $
\theta :=\theta _{1}$ and subsequently taking the limit ($\theta _{3}$, $
\kappa _{6}$)$\rightarrow 0$ reduces the algebra to a two noncommutative
dimensional space in the $X$,$Y$-direction 
\begin{equation}
\begin{array}{lll}
\lbrack X,Y]=i\theta \left( 1+\hat{\tau}Y^{2}\right) ,\qquad & 
[Y,P_{y}]=i\hbar \left( 1+\hat{\tau}Y^{2}\right) ,\qquad & [X,P_{x}]=i\hbar
\left( 1+\check{\tau}P_{x}^{2}\right) ,
\end{array}
\label{twod}
\end{equation}
decoupled from a standard one dimensional space in the $Z$-direction
\begin{equation}
\begin{array}{ll}
\lbrack Z,P_{z}]=i\hbar ,\qquad & [Y,Z]=0.
\end{array}
\end{equation}
As a representation for the algebra (\ref{twod}) in flat noncommutative
space we may simply use (\ref{rep3d}) with the appropriate limit $\theta
_{3}\rightarrow 0$. Carrying out the corresponding Bopp-shift $
x_{0}\rightarrow x_{s}-\frac{\theta }{\hbar }p_{y_{s}}$, $y_{0}\rightarrow
y_{s}$,\ $p_{x_{0}}\rightarrow p_{x_{s}}$ and $p_{y_{0}}\rightarrow
p_{y_{s}} $ yields the operators
\begin{equation}
X=x_{s}-\frac{\theta }{\hbar }p_{y_{s}}+\check{\tau}p_{x_{s}}^{2}x_{s}-\hat{
\tau}\frac{\theta }{\hbar }y_{s}^{2}p_{y_{s}},\quad Y=y_{s},\quad
P_{x}=p_{x_{s}},\quad \text{and\quad }P_{y}=p_{y_{s}}+\hat{\tau}
y_{s}^{2}p_{y_{s}},  \label{tworep}
\end{equation}
which are of course still non-Hermitian with regard to the standard inner
product.

\subsubsection{Reduction into Three Decoupled One Dimensional Spaces}
We conclude this section by noting that all three directions in the algebra (
\ref{S11})-(\ref{S15}) can be decoupled, of which one becomes a one
dimensional noncommutative space previously investigated by many authors,
e.g. \cite{kempf,bagchi_fring}. It is easy to verify that this scenario is obtained
from (\ref{S11})-(\ref{S15}) when parametrizing $q=\exp (2\tau \kappa
_{11}^{2})$ and subsequently taking the limit $(\theta _{1}$, $\theta _{3}$, 
$\kappa _{11})\rightarrow 0$. The remaining non-vanishing commutators are
then 
\begin{equation}
\lbrack X,P_{x}]=i\hbar \left( 1+\check{\tau}P_{x}^{2}\right) ,\qquad
\lbrack Y,P_{y}]=i\hbar ,\qquad \text{and\qquad }[Z,P_{z}]=i\hbar .
\label{one}
\end{equation}
Thus all three space directions are decoupled from each other. It is known
that the choices $X=(1+\check{\tau}p_{s}^{2})x_{s}$, $P_{x}=p_{s}$ or $
X^{\prime }=X^{\dagger }=x_{s}(1+\check{\tau}p_{s}^{2})$, $P_{x}^{\prime
}=p_{s}$ constitute representations for the commutation relations (\ref{one}
) in the $X$-direction.

\subsection{A Particular $\mathcal{PT}_{\mathcal{\protect\theta }_{\pm }}$-Symmetric Solution}
Instead of solving the complicated relations (\ref{sdd})-(\ref{c6}) one may
also start by making directly an Ansatz of a similar form as in (\ref{rep3d}
) without elaborating on the relation to the $q$-deformed oscillator
algebra. Proceeding in this manner with an Ansatz respecting the $\mathcal{PT
}_{\mathcal{\theta }_{\pm }}$-symmetry we find for instance the
representation
\begin{equation}
\begin{array}{ll}
X=x_{0}-\hat{\tau}\frac{\theta _{1}}{\hbar }y_{0}^{2}p_{y_{0}}-\hat{\tau}
\frac{\theta _{2}}{\hbar }y_{0}^{2}p_{z_{0}}, & P_{x}=p_{x_{0}}, \\ 
Z=z_{0}+\hat{\tau}\frac{\theta _{3}}{\hbar }y_{0}^{2}p_{y_{0}}+\hat{\tau}
\frac{\theta _{2}}{\hbar }\frac{\theta _{3}}{\theta _{1}}y_{0}^{2}p_{z_{0}},
\qquad & P_{z}=p_{z_{0}}, \\ 
P_{y}=p_{y_{0}}+\hat{\tau}y_{0}^{2}p_{y_{0}}, & Y=y_{0},
\end{array}
\label{T2}
\end{equation}
yielding the closed algebra
\begin{equation}
\begin{array}{lll}
\lbrack X,Y]=i\theta _{1}\left( 1+\hat{\tau}Y^{2}\right) ,\quad & 
[X,Z]=i\theta _{2}\left( 1+\hat{\tau}Y^{2}\right) ,\quad & [Y,Z]=i\theta
_{3}\left( 1+\hat{\tau}Y^{2}\right) , \\ 
\lbrack X,P_{x}]=i\hbar , & [Y,P_{y}]=i\hbar \left( 1+\hat{\tau}Y^{2}\right)
, & [Z,P_{z}]=i\hbar ,
\end{array}
\label{t2}
\end{equation}
with all remaining commutators vanishing. Notice that if we set $\theta
_{1}=-\theta _{3}$ the generators in (\ref{T2}) are also invariant under the 
$\mathcal{PT}_{xz}$-symmetry.

\subsection{An Additional Reduced Solution in Two Dimensions}
Here we choose the parameters $\hat{\kappa}_3=\hat{\kappa}_6=\hat{\kappa}_7=\hat{\kappa}_8=\hat{\kappa}_9
=\check{\kappa}_{12}=\check{\kappa}_{15}=\check{\kappa}_{16}=\check{\kappa}_{17}
=\check{\kappa}_{18}=0$, in such a way that the algebra reduces into the two dimensions. Thereafter employing constraints similar to those reported in \cite{fring_gouba_bagchi_area} together with the subsequent nontrivial limit $q\rightarrow 1$, we arrive at the deformed oscillator algebra
\begin{equation}
\begin{array}{lll}
\lbrack X,Y]=i\theta\left( 1+\hat{\tau}Y^{2}\right) ,\quad & 
[X,P_x]=i\hbar\left( 1+\hat{\tau}Y^{2}\right) ,\quad & [X,P_y]=0 , \\ 
\lbrack P_x,P_{y}]=i\hat{\tau}\frac{\hbar^2}{\theta}Y^2 , & [Y,P_{y}]=i\hbar \left( 1+\hat{\tau}Y^{2}\right), & [Y,P_{x}]=0 ,
\end{array}
\label{bbb}
\end{equation}
which coincides exactly to the algebra presented in \cite{fring_gouba_bagchi_area}. So far no representation for the two dimensional algebra (\ref{bbb}) was provided. Here we find that it can be represented by
\begin{equation}
X=x_{0}+\hat{\tau}y_{0}^{2}x_{0},\quad Y=y_{0},\quad P_{x}=p_{x_{0}},\quad 
\text{and\quad }P_{y}=p_{y_{0}}-\hat{\tau}\frac{\hbar }{\theta }
y_{0}^{2}x_{0}.  \label{rep2}
\end{equation}
Clearly there exist many more solutions one may construct in this systematic
manner from the Ansatz (\ref{an1})-(\ref{an6}), which will not be our
concern here. Instead we will study a concrete model, i.e. the
two-dimensional harmonic oscillator on the noncommutative space \cite{dey_fring_2DHO} described by
the algebra (\ref{bbb}).

\section{Minimal Lengths, Areas, Volumes} \label{sectionminimallengthminimalareas}
Let us now investigate the generalized uncertainty relations associated to
the algebras constructed above. As discussed in chapter \ref{chapter2}, in general, the uncertainties $\Delta A$ and $\Delta B$ resulting from a simultaneous measurement of two observables $A$ and $B$ have to obey the inequality (\ref{GUP1})
\begin{equation}
\Delta A\Delta B\geq \frac{1}{2}\left\vert \left\langle [A,B]\right\rangle
_{\rho }\right\vert .  \label{HUU}
\end{equation}
Here $\left\langle .\right\rangle _{\rho }$ denotes the inner product on a
Hilbert space with metric $\rho $ in which the operators $A$ and $B$ are
Hermitian, as discussed in more detail in \cite{bagchi_fring,fring_gouba_scholtz,fring_gouba_bagchi_area,kempf}. The minimal length $\Delta A_{\min }$, that is the precision up to which the
observable $A$ can be measured by giving up all the information on $B$ is then
computed by minimising $\Delta A\Delta B-\frac{1}{2}\left\vert \left\langle
[A,B]\right\rangle _{\rho }\right\vert $ as a function of $\Delta B$. In the
standard scenario, i.e. when $A$ and $B$ commute up to a constant, the
result is therefore usually zero. This outcome changes when the commutator $
[A,B]$ involves higher powers of $\Delta B$, in which case we encounter the
interesting scenario of non-vanishing $\Delta A_{\min }$. We now investigate
some of the solutions presented above. Depending now on the question we ask,
i.e. which quantities we attempt to measure, the minimal uncertainties for
some specific operators turn out to be different.

\subsection{A 3D Noncommutative Space Leading to Minimal Areas}
We start with our simplest three dimensional solution, that is the algebra (\ref{S21})-(\ref{S22}). If we just want to measure the position of the
particle on such a space independently of its momentum we only have to
investigate the relations (\ref{S21}). Taking $\tau >0$ and following the
logic of \cite{bagchi_fring,fring_gouba_scholtz,fring_gouba_bagchi_area}, we obtain from (\ref{S21}) for a simultaneous measurement of all space coordinates non-vanishing minimal length in two directions 
\begin{equation}
\Delta X_{\min}=\left\vert \theta _{1}\right\vert \sqrt{\hat{\tau}}\sqrt{1+
\hat{\tau}\left\langle Y\right\rangle _{\rho }^{2}},\quad \Delta Y_{\min
}=0,~ \text{and~}\Delta Z_{\min }=\left\vert \theta _{3}\right\vert 
\sqrt{\hat{\tau}}\sqrt{1+\hat{\tau}\left\langle Y\right\rangle _{\rho }^{2}}.
\label{xmin1}
\end{equation}
Thus any measurement of space will involve an unavoidable uncertainty of an
area $A$ of size $\Delta A_{0}=4\hat{\tau}\left\vert \theta _{1}\theta
_{3}\right\vert $ in the $XZ$-plane and no uncertainty in the $Y$-direction.
Changing our question and attempt to measure instead all coordinates and all
components of the momenta, we need to analyse the entire set of relations (\ref{S21})-(\ref{S22}). The analysis of the equations (\ref{S22}) alone yields
\begin{eqnarray}
\Delta X_{\min } &=&\hbar \sqrt{\check{\tau}}\sqrt{1+\hat{\tau}\left\langle
Y\right\rangle _{\rho }^{2}},~ \Delta Y_{\min }=0,~ \text{and}~
\Delta Z_{\min }=\hbar \sqrt{\check{\tau}}\sqrt{1+\hat{\tau}\left\langle
Y\right\rangle _{\rho }^{2}},~~~  \label{xmin3} \\
\Delta \left( P_{x}\right) _{\min } &=&0,\quad \ \ \ \ \  \Delta
\left( P_{y}\right) _{\min }=\hbar \sqrt{\hat{\tau}}\sqrt{1+\hat{\tau}
\left\langle Y\right\rangle _{\rho }^{2}},\quad \text{and\quad }\Delta
\left( P_{z}\right) _{\min }=0.~~  \label{xmin2}
\end{eqnarray}
Thus, depending now on whether $\left\vert \theta _{1}\right\vert $, $
\left\vert \theta _{3}\right\vert <1$ or $\left\vert \theta _{1}\right\vert $
, $\left\vert \theta _{3}\right\vert >1$ the uncertainties in (\ref{xmin1})
or (\ref{xmin3}) will be smaller, respectively. For any type of measurement
the region of uncertainty will be an area.

\subsection{A 3D Noncommutative Space Leading to Minimal Volumes}
Let us now analyse our solution (\ref{S21})-(\ref{S22}) before taking the
limit $q\rightarrow 1$. We compute the uncertainties with regard to a
measurement of all components of the coordinates and all components of the
momenta. Since now the quantities are all coupled, in the sense that we do
not have any nontrivial subalgebra, we will encounter uncertainties for all
of them and observe a different type of behaviour as indicated in the
previous subsection. Starting with a simultaneous $Y$,$P_{y}$-measurement we
compute from (\ref{HUU}) with (\ref{S14}) the uncertainties
\begin{eqnarray}
\Delta Y_{\min } &=&\left\vert \hat{\kappa}_{6}\right\vert \sqrt{\frac{1}{2}
(q^{2}-q^{-2})+(q-q^{-1})^{2}\left( \frac{1}{4\hat{\kappa}_{6}^{2}}
\left\langle Y\right\rangle _{\rho }^{2}+\frac{\hat{\kappa}_{6}^{2}}{\hbar
^{2}}\left\langle P_{y}^{2}\right\rangle _{\rho }\right) },  \label{min1} \\
\Delta \left( P_{y}\right) _{\min } &=&\frac{\hbar }{2\left\vert \hat{\kappa}
_{6}\right\vert }\sqrt{\frac{1}{2}(q^{2}-q^{-2})+(q-q^{-1})^{2}\left( \frac{1
}{4\hat{\kappa}_{6}^{2}}\left\langle Y\right\rangle _{\rho }^{2}+\frac{\hat{
\kappa}_{6}^{2}}{\hbar ^{2}}\left\langle P_{y}^{2}\right\rangle _{\rho
}\right) },  \label{min2}
\end{eqnarray}
under the assumption that $q>1$. The absolute minimal uncertainties
resulting from these expressions are therefore
\begin{equation}
\Delta Y_{0}=\frac{\left\vert \hat{\kappa}_{6}\right\vert }{\sqrt{2}}\sqrt{
q^{2}-q^{-2}},\qquad \text{and\qquad }\Delta \left( P_{y}\right) _{0}=\frac{
\hbar }{2\sqrt{2}\left\vert \hat{\kappa}_{6}\right\vert }\sqrt{q^{2}-q^{-2}}.
\label{P0}
\end{equation}
Next we carry out a simultaneous $X$,$Y$-measurement and a $Y$,$Z$-measurement by employing (\ref{S11}) and (\ref{S12}), respectively. We find the minimal lengths
\begin{alignat}{1}
\Delta X_{\min } &=\left\vert \frac{\theta _{1}}{\hat{\kappa}_{6}}
\right\vert \sqrt{\frac{1}{2}\frac{q-q^{-1}}{q+q^{-1}}+\left[ \frac{q-q^{-1}
}{q+q^{-1}}\right] ^{2}\left[ \frac{1}{4\hat{\kappa}_{6}^{2}}\left\langle
Y\right\rangle _{\rho }^{2}+\frac{\hat{\kappa}_{6}^{2}}{\hbar ^{2}}\left(
\left\langle P_{y}^{2}\right\rangle _{\rho }+\Delta\left(P_{y}\right)
_{0}^{2}\right) \right] }, \\
\Delta Z_{\min } &=\left\vert \frac{\theta _{3}}{\hat{\kappa}_{6}}
\right\vert \sqrt{\frac{1}{2}\frac{q-q^{-1}}{q+q^{-1}}+\left[ \frac{q-q^{-1}
}{q+q^{-1}}\right] ^{2}\left[ \frac{1}{4\hat{\kappa}_{6}^{2}}\left\langle
Y\right\rangle _{\rho }^{2}+\frac{\hat{\kappa}_{6}^{2}}{\hbar ^{2}}\left(
\left\langle P_{y}^{2}\right\rangle _{\rho }+\Delta \left( P_{y}\right)
_{0}^{2}\right) \right] }.
\end{alignat}
There is no minimal length in the $Y$-direction resulting from these
relations. Using the expression for $\Delta \left( P_{y}\right) _{0}$ from (
\ref{P0}), the absolute minimal values for these uncertainties are%
\begin{equation}
\Delta X_{0}=\frac{1}{2\sqrt{2}}\left\vert \frac{\theta _{1}}{\hat{\kappa}
_{6}}\right\vert \sqrt{q^{2}-q^{-2}},\qquad \text{and\qquad }\Delta Z_{0}=
\frac{1}{2\sqrt{2}}\left\vert \frac{\theta _{3}}{\hat{\kappa}_{6}}%
\right\vert \sqrt{q^{2}-q^{-2}}.
\end{equation}
Thus a measurement of the position in space will be accompanied by an
uncertainty volume $V$ of the size 
\begin{equation}
\Delta V_{0}=\frac{1}{\sqrt{2}}\left\vert \frac{\theta _{1}\theta _{3}}{\hat{
\kappa}_{6}}\right\vert \left( q^{2}-q^{-2}\right) ^{3/2}.
\end{equation}
The evaluation for the simultaneous $X$,$P_{x}$ and $Z$,$P_{z}$-measurements
are slightly more complicated due to the occurrence of the $XP_{y}$ and $
ZP_{y}$ terms in (\ref{S13}) and (\ref{S15}), respectively. We proceed
similarly as before and make also use of the well known inequalities $
\left\vert A+B\right\vert \geq \left\vert A\right\vert -\left\vert
B\right\vert $ and $\left\vert \left\langle AB\right\rangle \right\vert \leq
\Delta A\Delta B+\left\vert \left\langle A\right\rangle \left\langle
B\right\rangle \right\vert $. We report here only the final result of the
absolute minimal values
\begin{equation}
\Delta \left( P_{i}\right) _{0}=\frac{\gamma _{i}\Delta \left( P_{y}\right)
_{0}-\sqrt{\beta _{i}\left[ \alpha _{i}\gamma _{i}^{2}\Delta \left(
P_{y}\right) _{0}^{2}+\lambda _{i}(1-4\alpha _{i}\beta _{i})\right] }}{
4\alpha _{i}\beta _{i}-1}\quad ~~~\text{for }i=x,z,
\end{equation}
with 
\begin{equation}
\begin{array}{llll}
\alpha _{x}=\alpha _{2},~~ & \beta _{x}=\alpha _{11},~~ & \gamma _{x}=\frac{
2\left\vert \theta _{1}\right\vert }{\hbar }\alpha _{11},~~ & \lambda _{x}=
\frac{\hbar }{2}+\alpha _{11}\frac{\theta _{1}^{2}}{\hbar ^{2}}\Delta \left(
P_{y}\right) _{0}^{2}, \\ 
\alpha _{z}=\alpha _{7},~~ & \beta _{z}=\alpha _{16},~~ & \gamma _{z}=\frac{
\left\vert \theta _{3}\right\vert \hbar }{2}\alpha _{16},~~ & \lambda _{z}=
\frac{\hbar }{2}+\alpha _{16}\theta _{3}^{2}\Delta \left( P_{y}\right)
_{0}^{2},
\end{array}
\end{equation}
where $\alpha _{i}=$\ $\hat{\kappa}_{i}^{2}(q-q^{-1})/(q+q^{-1})\hbar $ for $
i=2,7$ and $\alpha _{i}=$\ $\check{\kappa}_{i}^{2}(q-q^{-1})/(q+q^{-1})\hbar 
$ for $i=11,16$. Further restrictions do not emerge.

By similar reasoning one finds non-vanishing $\Delta X_{\min }$, $\Delta
Z_{\min }$ and $\Delta P_{y\min }$ for the $\mathcal{PT}_{xz}$-invariant
algebra (\ref{t2}).

Note that in this section we have treated the variables $X$, $Y$, $Z$, $P_x$, $P_y$, $P_z$ as observables since they will give rise to the non-trivial uncertainty relations. In general, the set of observables is a matter of choice, especially in a non-Hermitian setting, see e.g. \cite{bender_making_sense}. In the next section, we will also make use of the variables $x_0$, $y_0$, $z_0$, $p_{x_0}$, $p_{y_0}$, $p_{z_0}$ and $x_s$, $y_s$, $z_s$, $p_{x_s}$, $p_{y_s}$, $p_{z_s}$, but simply for technical reasons we do not treat them as observables.

\section{Models on $\mathcal{PT}$-Symmetric Noncommutative Spaces}
\subsection{The 1D Harmonic Oscillator on a Noncommutative Space}\label{subsection441}
We commence with the one-dimensional harmonic oscillator on the $\mathcal{PT}
_{\pm }$-symmetric noncommutative space described by (\ref{one}). The
corresponding Hamiltonian 
\begin{equation}
H_{ncho}^{1D}=\frac{P^{2}}{2m}+\frac{m\omega ^{2}}{2}X^{2}=H_{ho}^{1D}+\frac{
m\omega ^{2}}{2}\left( \check{\tau}p_{s}^{2}x_{s}^{2}+\check{\tau}
x_{s}p_{s}^{2}x_{s}+\check{\tau}^{2}p_{s}^{2}x_{s}p_{s}^{2}x_{s}\right)
=H_{ho}^{1D}+H_{nc}^{1D},  \label{HD11}
\end{equation}
is evidently non-Hermitian with regard to the standard inner product.
However, it is $\mathcal{PT}_{\pm }$-symmetric, such that it might
constitute a well-defined self-consistent description of a physical system.
The associated Schr\"{o}dinger equation $H_{ncho}^{1D}\psi =E\psi $ is most
conveniently solved in $p$-space, i.e. with $x_{s}=i\hbar \partial _{p_{s}}$
it reads
\begin{equation}
\frac{m\omega ^{2}\hbar ^{2}}{2}(1+\check{\tau}p_{s}^{2})^{2}\psi ^{\prime
\prime }+\tau \omega \hbar p_{s}(1+\check{\tau}p_{s}^{2})\psi ^{\prime
}+\left( E-\frac{p_{s}^{2}}{2m}\right) \psi =0.  \label{diff}
\end{equation}
Using the transformation
\begin{equation}
\mu =\frac{\sqrt{1+2Em\check{\tau}}}{\tau },\qquad \nu =\frac{\sqrt{4+\tau
^{2}}}{2\tau }-\frac{1}{2}\quad \text{and\quad }z=ip_{s}\sqrt{\check{\tau}},
\label{munu}
\end{equation}
we convert (\ref{diff}) into 
\begin{equation}
(1-z^{2})\psi ^{\prime \prime }-2z\psi ^{\prime }+\left[ \nu (\nu +1)-\frac{
\mu ^{2}}{1-z^{2}}\right] \psi =0,  \label{al}
\end{equation}
which is the standard differential equation for the associated Legendre
polynomials $P_{\nu }^{\mu }(z)$ and $Q_{\nu }^{\mu }(z)$ admitting the
general solution
\begin{equation}
\psi (z)=c_{1}P_{\nu }^{\mu }(z)+c_{2}Q_{\nu }^{\mu }(z).
\end{equation}
Seeking asymptotically vanishing solutions gives rise to the quantization
condition $\mu +\nu =-n-1$ with $n\in \mathbb{N}$. With (\ref{munu}) it
follows therefore that the eigenenergies becomes
\begin{equation}
E_{n}=\omega \hbar \left( \frac{1}{2}+n\right) \sqrt{1+\frac{\tau ^{2}}{4}}
+\tau \frac{\omega \hbar }{4}(1+2n+2n^{2})\quad \quad \text{for }n\in 
\mathbb{N}_{0}.  \label{EnEnEn}
\end{equation}
The expression agrees with the one found in \cite{kempf}. The polynomial $
Q_{\nu }^{\mu }(z)$ is not defined for these values, such that $c_{2}=0$ and 
$P_{\nu }^{\mu }(z)$ reduces to 
\begin{equation}
\psi _{2n-i}(z)=c_{1}\sum_{k=i}^{2n-i}\frac{1}{k!}\left[ \prod\limits_{l=i}^{
\frac{k+i-2}{2}}2(n-l)(2n+2\nu -2l+1)\right] z^{k}\frac{(z^{2}-1)^{\frac{\nu
-2n-1+i}{2}}}{(-1)^{1+i}(1-z^{2})^{\nu }},
\end{equation}
with $i=0,1$. Clearly the $\psi _{2n-i}(z)$ vanish for $\left\vert
z\right\vert \rightarrow \infty $ if $\nu >-1$, which is always guaranteed
for $\tau m\omega >0$. The Dyson map $\eta $ which adjointly maps $
H_{ncho}^{1D}$ to a Hermitian operator was easily found \cite{kempf,bagchi_fring} to be $\eta =\left( 1+\tau P_{x}^{2}\right) ^{-1/2}$. In addition, we note
that the solutions are square integrable $\psi _{2n-i}(z)\in L^{2}(i\mathbb{R
})$ on $\left\langle \cdot \right\vert \left. \eta ^{2}\cdot \right\rangle $
and form an orthonormal basis.

An exact treatment for models in higher dimensions is more difficult,
but we may resort to perturbation theory to obtain some useful insight on
the solutions. As a quality gauge we compare here the exact solution against
perturbation theory around the standard Fock space harmonic oscillator
solution with normalized eigenstates
\begin{equation}
\left\vert n\right\rangle =\frac{(a_{x_{s}}^{\dagger })^{n}}{\sqrt{n!}}
\left\vert 0\right\rangle ,\quad a_{x_{s}}\left\vert 0\right\rangle =0,\quad
a_{x_{s}}^{\dagger }\left\vert n\right\rangle =\sqrt{n+1}\left\vert
n+1\right\rangle ,\quad a_{x_{s}}\left\vert n\right\rangle =\sqrt{n}
\left\vert n-1\right\rangle .
\end{equation}
A straightforward, albeit lengthy, computation yields the following
corrections to the harmonic oscillator energy $E_{n}^{(0)}=\omega \hbar
\left( n+\frac{1}{2}\right) $ for the eigenenergies of $H_{ncho}^{1D}$ 
\begin{equation}
E_{n}^{(p)}=E_{n}^{(0)}+E_{n}^{(1)}+E_{n}^{(2)}+\mathcal{O}(\tau ^{3})
\end{equation}
with
\begin{alignat}{1}
E_{n}^{(1)} &=\left\langle n\right\vert H_{nc}^{1D}\left\vert
n\right\rangle =\frac{\tau \omega \hbar }{4}\left( 1+2n+2n^{2}\right) +\frac{
\tau ^{2}\omega \hbar }{16}\left( 3+8n+6n^{2}+4n^{3}\right) ,~~~~ \\
E_{n}^{(2)} &=\sum_{p\neq n}\frac{\left\langle n\right\vert
H_{nc}^{1D}\left\vert p\right\rangle \left\langle p\right\vert
H_{nc}^{1D}\left\vert n\right\rangle }{E_{n}^{(0)}-E_{p}^{(0)}}=-\frac{1}{8}
\tau ^{2}\omega \hbar \left( 1+3n+3n^{2}+2n^{3}\right) +\mathcal{O}(\tau
^{3}).~~
\end{alignat}
As it should be, the expression for $E_{n}$ in (\ref{EnEnEn}) when expanded up
to order $\tau ^{3}$ coincides precisely with $E_{n}^{(p)}$. We further note
that also in a perturbative treatment the eigenenergies are strictly
positive.

The validity of these expansions is governed by the well-known sufficient
conditions for the applicability of the Rayleigh-Schr\"{o}dinger
perturbation theory to a Hamiltonian of the form $H=H_{0}+H_{1}$ around the
solutions of $H_{0}\left\vert n\right\rangle =E_{n}^{(0)}\left\vert
n\right\rangle $
\begin{equation}
\left\vert \frac{\left\langle p\right\vert H_{1}\left\vert n\right\rangle }{
E_{n}^{(0)}-E_{p}^{(0)}}\right\vert \ll 1\qquad \text{for all }p\neq n.
\end{equation}
This is guaranteed for (\ref{HD11}) when $\tau ^{2}\ll 32/(2n+13)\sqrt{
(4+n)(3+n)(2+n)(1+n)}$, such that perturbation theory will break down for
large values of $n$.

\subsection{The 2D Harmonic Oscillator on a Noncommutative Space}
Next we consider the two-dimensional harmonic oscillator on the $\mathcal{PT}
_{\pm }$-symmetric noncommutative space described by the algebra (\ref{twod}). Using the representation (\ref{tworep}) for this algebra, the
corresponding Hamiltonian reads 
\begin{alignat}{1}
H_{ncho}^{2D} &=\frac{1}{2m}(P_{x}^{2}+P_{y}^{2})+\frac{m\omega ^{2}}{2}(X^{2}+Y^{2})  \label{2dho} \\
&=H_{fncho}^{2D}+\frac{\tau \omega }{2\hbar }\left[\{p_{x_{0}}^{2}x_{0},x_{0}\}+\{y_{0}^{2}p_{y_{0}},p_{y_{0}}\}+\frac{\theta }{
\hbar }\{p_{x_{0}}^{2}p_{y_{0}},x_{0}\}-\frac{m^{2}\omega ^{2}\theta }{\hbar 
}\{y_{0}^{2}p_{y_{0}},x_{0}\}\right]   \notag \\
&~~~+\frac{\tau ^{2}\omega ^{2}m}{2\hbar ^{2}}\left[ \left\{
y_{0}^{2}p_{y_{0}},(1+\Omega )y_{0}^{2}p_{y_{0}}-\frac{\theta
p_{x_{0}}^{2}x_{0}}{\hbar }-\frac{\theta ^{2}p_{x_{0}}^{2}p_{y_{0}}}{\hbar
^{2}}\right\} +\left( \frac{p_{x_{0}}^{2}x_{0}}{m\omega }+\frac{\theta
p_{x_{0}}^{2}p_{y_{0}}}{m\omega \hbar }\right) ^{2}\right],   \notag
\end{alignat}
where we used the standard notation for the anti-commutator $\left\{
A,B\right\} =AB+BA$. The notation $H^{2D}_{fncho}$ stands for the Hamiltonian of two dimensional harmonic oscillator in flat noncommutative space. Once again this Hamiltonian is non-Hermitian with
regard to the inner product on the flat noncommutative space, but it
respects a $\mathcal{PT}_{\pm }$-symmetry. In order to be able to perturb
around the standard harmonic oscillator solution we still need to convert
flat noncommutative space into the canonical variable $x_{s}$, $y_{s}$, $
p_{x_{s}}$ and $p_{y_{s}}$. Thus when using the representation (\ref{tworep}) this Hamiltonian is converted into
\begin{eqnarray*}
H_{ncho}^{2D} &=&H_{ho}^{2D}+\frac{m\theta ^{2}\omega ^{2}}{2\hbar ^{2}}
p_{y_{s}}^{2}-\frac{m\theta \omega ^{2}}{2\hbar }\{x_{s},p_{y_{s}}\}+\frac{
\tau }{2}\left[ m\omega ^{2}\{p_{x_{s}}^{2}x_{s},x_{s}\}\right.  \\
&&+\left. \left( \frac{1}{m}+\frac{m\theta ^{2}\omega ^{2}}{\hbar ^{2}}
\right) \{y_{s}^{2}p_{y_{s}},p_{y_{s}}\}-\frac{m\theta \omega ^{2}}{\hbar }
\left( \{y_{s}^{2}p_{y_{s}},x_{s}\}+\{p_{x_{s}}^{2}x_{s},p_{y_{s}}\}\right)
\right]  \\
&&+\frac{\tau ^{2}}{2}\left[ \left( \frac{1}{m}+\frac{m\theta ^{2}\omega ^{2}
}{\hbar ^{2}}\right) \left( y_{s}^{2}p_{y_{s}}\right) ^{2}+m\omega
^{2}\left( p_{x_{s}}^{2}x_{s}\right) ^{2}-\frac{m\theta \omega ^{2}}{\hbar }
\{y_{s}^{2}p_{y_{s}},p_{x_{s}}^{2}x_{s}\}\right] , \\
&=&H_{ho}^{2D}(x_{s},y_{s},p_{x_{s}},p_{y_{s}})+H_{nc}^{2D}(x_{s},y_{s},p_{x_{s}},p_{y_{s}}).
\end{eqnarray*}
In this formulation we may now proceed as in the previous subsection and
expand perturbatively around the standard two dimensional Fock space
harmonic oscillator solution with normalized eigenstates
\begin{eqnarray}
\left\vert n_{1}n_{2}\right\rangle &=&\frac{(a_{1}^{\dagger
})^{n_{1}}(a_{2}^{\dagger })^{n_{2}}}{\sqrt{n_{1}!n_{2}!}}\left\vert
00\right\rangle ,\quad a_{i}^{\dagger }\left\vert n_{1}n_{2}\right\rangle =
\sqrt{n_{i}+1}\left\vert (n_{1}+\delta _{i1})(n_{2}+\delta
_{i2})\right\rangle ,\quad \\
a_{i}\left\vert 00\right\rangle &=&0,~~~~~\qquad \qquad \ \ \ \ \ \ \ \
a_{i}\left\vert n_{1}n_{2}\right\rangle =\sqrt{n_{i}}\left\vert
(n_{1}-\delta _{i1})(n_{2}-\delta _{i2})\right\rangle ,
\end{eqnarray}
for $i=1,2$. The energy eigenvalues for the Hamiltonian $H_{ncho}^{2D}$ then
result to 
\begin{eqnarray}
E_{nl}^{p} &=&E_{nl}^{(0)}+E_{nl}^{(1)}+E_{nl}^{(2)}+\mathcal{O}(\tau ^{2})
\\
&=&\hbar \omega (n+l+1)+\left\langle nl\right\vert H_{nc}^{2D}\left\vert
nl\right\rangle +\sum_{p,q\neq n+l=p+q}\frac{\left\langle nl\right\vert
H_{nc}^{2D}\left\vert pq\right\rangle \left\langle pq\right\vert
H_{nc}^{2D}\left\vert nl\right\rangle }{E_{nl}^{(0)}-E_{pq}^{(0)}}+\mathcal{O
}(\tau ^{2})  \notag \\
&=&E_{nl}^{(0)}+\frac{\Omega \omega \hbar }{8}\left[ (3+n+5l)-\Omega (l+
\frac{1}{2})\right] +\frac{\tau }{2}\omega \hbar \left[ 1+n+n^{2}+l+l^{2}
\right.  \notag \\
&&\left. +\frac{\Omega }{4}\left( 4+3n+n^{2}+7l+4nl+5l^{2}\right) \right] +
\mathcal{O}(\tau ^{2}),  \notag
\end{eqnarray}
where we introduced the dimensionless quantity $\Omega _{i}=m^{2}\theta
_{i}^{2}\omega ^{2}/\hbar ^{2}$. Notice that unlike as in the one
dimensional case the perturbation beyond $H_{ho}^{2D}$ also involves terms
of order $\mathcal{O}(\tau ^{0})$, such that we need to compute also $
E_{nl}^{(2)}$ to achieve a precision of first order in $\tau $. We also
notice that the energy $E_{nl}^{p}$ is only bounded from below for $\Omega
<5 $. The minus sign is an indication that we will encounter exceptional
points \cite{bender_wu} and broken $\mathcal{PT}$-symmetry in some
parameter range.

Next we construct the model which corresponds to the algebra (\ref{bbb}). Using the representation (\ref{rep2}), the corresponding Hamiltonian reads
\begin{eqnarray}
H_{ncho}^{2D} &=&\frac{1}{2m}(P_{x}^{2}+P_{y}^{2})+\frac{m\omega ^{2}}{2}
(X^{2}+Y^{2})  \label{2dho} \\
&=&H_{fncho}^{2D}+\frac{\hat{\tau}}{2}\left[ m\omega
^{2}\{y_{0}^{2}x_{0},x_{0}\}-\frac{\hbar }{m\theta }
\{y_{0}^{2}x_{0},p_{y_{0}}\}\right] +\frac{\hat{\tau}^{2}}{2}\left[ m\omega
^{2}+\frac{\hbar ^{2}}{m\theta ^{2}}\right] y_{0}^{2}x_{0}y_{0}^{2}x_{0}. 
\notag
\end{eqnarray}
Once again utilizing the Bopp shift, the Hamiltonian (\ref{2dho}) is transformed into the form
\begin{eqnarray}
H_{ncho}^{2D} &=&H_{ho}^{2D}+\frac{m\theta ^{2}\omega ^{2}}{2\hbar ^{2}}
p_{y_{s}}^{2}-\frac{m\theta \omega ^{2}}{2\hbar }\{x_{s},p_{y_{s}}\}+\frac{
\hat{\tau}}{2}\left[ m\omega ^{2}\{y_{s}^{2}x_{s},x_{s}\}-\frac{\hbar }{
m\theta }\{y_{s}^{2}x_{s},p_{y_{s}}\}\right]  \notag \\
&&+\frac{\hat{\tau}}{2}\left[ \left( \frac{1}{m}+\frac{m\theta ^{2}\omega
^{2}}{\hbar ^{2}}\right) \{y_{s}^{2}p_{y_{s}},p_{y_{s}}\}-\frac{m\theta
\omega ^{2}}{\hbar }\left(
\{y_{s}^{2}p_{y_{s}},x_{s}\}+\{y_{s}^{2}x_{s},p_{y_{s}}\}\right) \right] ~~ 
\notag \\
&&-\frac{\hat{\tau}^{2}}{2}\left[ \frac{m\theta \omega ^{2}}{\hbar }+\frac{
\hbar }{m\theta }\right] \left(
y_{s}^{2}p_{y_{s}}y_{s}^{2}x_{s}+y_{s}^{2}x_{s}y_{s}^{2}p_{y_{s}}\right) +
\frac{\hat{\tau}^{2}}{2}\left[ \frac{1}{m}+\frac{m\theta ^{2}\omega ^{2}}{
\hbar ^{2}}\right] y_{s}^{2}p_{y_{s}}y_{s}^{2}p_{y_{s}}~~  \notag \\
&&+\frac{\hat{\tau}^{2}}{2}\left[ m\omega ^{2}+\frac{\hbar ^{2}}{m\theta ^{2}
}\right] y_{s}^{2}x_{s}y_{s}^{2}x_{s}  \notag \\
&=&H_{ho}^{2D}(x_{s},y_{s},p_{x_{s}},p_{y_{s}})+H_{nc}^{2D}(x_{s},y_{s},p_{x_{s}},p_{y_{s}}).
\notag
\end{eqnarray}
The energy eigenvalues for the Hamiltonian then result to 
\begin{eqnarray}
E_{nl}^{(p)} &=&E_{nl}^{(0)}+E_{nl}^{(1)}+E_{nl}^{(2)}+\mathcal{O}(\tau ^{2})
\\
&=&E_{nl}^{(0)}+\left\langle nl\right\vert H_{nc}^{2D}\left\vert
nl\right\rangle +\sum_{p,q\neq n+l=p+q}\frac{\left\langle nl\right\vert
H_{nc}^{2D}\left\vert pq\right\rangle \left\langle pq\right\vert
H_{nc}^{2D}\left\vert nl\right\rangle }{E_{nl}^{(0)}-E_{pq}^{(0)}}+\mathcal{O
}(\tau ^{2})  \notag \\
&=&\omega \hbar \left( n+l+1\right) +\frac{1}{16}\hbar \omega \Omega \left[
2n-(2l+1)\Omega +10l+6\right]  \notag \\
&&+\frac{1}{8}\hbar \tau \omega \left[ \Omega \left(
8nl+4n+6l^{2}+10l+5\right) +10nl+5n+5l^{2}+10l+5\right] +\mathcal{O}(\tau
^{2}),  \notag
\end{eqnarray}
with $\Omega =m^{2}\theta ^{2}\omega ^{2}/\hbar ^{2}$.

\subsection{The 3D Harmonic Oscillator on a Noncommutative Space}
Let us finally consider the three-dimensional harmonic oscillator on the
noncommutative space described by the algebra (\ref{S11}) and (\ref{S12}).
Using the representation (\ref{rep3d}) together with a subsequent
Bopp-shift, the corresponding Hamiltonian can be expressed in terms of the
standard canonical coordinates
\begin{eqnarray}
H_{ncho}^{3D} &=&\frac{1}{2m}(P_{x}^{2}+P_{y}^{2}+P_{z}^{2})+\frac{m\omega
^{2}}{2}(X^{2}+Y^{2}+Z^{2})=H_{ho}^{3D}+H_{nc}^{3D} \\
&=&H_{ho}^{3D}+\frac{m\omega ^{2}}{2\hbar }\left[ \theta
_{3}\{p_{y_{s}},z_{s}\}-\theta _{1}\{x_{s},p_{y_{s}}\}+\frac{\theta
_{1}^{2}+\theta _{3}^{2}}{\hbar }p_{y_{s}}^{2}\right]   \notag \\
&&+\tau \frac{\omega }{2\hbar }\left[ \{p_{x_{s}}^{2}x_{s},x_{s}\}+
\{p_{z_{s}}^{2}z_{s},z_{s}\}+\left( 1+\Omega _{1}+\Omega _{3}\right)
\{y_{s}^{2}p_{y_{s}},p_{y_{s}}\}\right.   \notag \\
&&-\left. \frac{\theta _{1}}{\hbar }\left( m^{2}\omega
^{2}\{y_{s}^{2}p_{y_{s}},x_{s}\}+\{p_{x_{s}}^{2}x_{s},p_{y_{s}}\}\right) +
\frac{\theta _{3}}{\hbar }\left( m^{2}\omega
^{2}\{y_{s}^{2}p_{y_{s}},z_{s}\}+\{p_{z_{s}}^{2}z_{s},p_{y_{s}}\}\right) 
\right]   \notag \\
&&+\tau ^{2}\frac{1}{2m\hbar ^{2}}\left[
p_{x_{s}}^{2}x_{s}p_{x_{s}}^{2}x_{s}+p_{z_{s}}^{2}z_{s}p_{z_{s}}^{2}\text{ }
z_{s}+m^{2}\omega ^{2}\left( 1+\Omega _{1}+\Omega _{3}\right)
y_{s}^{2}p_{y_{s}}y_{s}^{2}p_{y_{s}}\right.   \notag \\
&&\left. +\frac{\theta _{3}}{\hbar }m^{2}\omega
^{2}\{p_{z_{s}}^{2}z_{s},y_{s}^{2}p_{y_{s}}\}-\frac{\theta _{1}}{\hbar }
m^{2}\omega ^{2}\{p_{x_{s}}^{2}x_{s},y_{s}^{2}p_{y_{s}}\}\right]   \notag
\end{eqnarray}
We expand now around the standard three dimensional Fock space harmonic
oscillator solution with normalized eigenstates
\begin{eqnarray}
\left\vert n_{1}n_{2}n_{3}\right\rangle  &=&\prod\nolimits_{i=1}^{3}\frac{
(a_{i}^{\dagger })^{n_{i}}}{\sqrt{n_{i}!}}\left\vert 000\right\rangle ,\quad
a_{i}^{\dagger }\left\vert n_{1}n_{2}n_{3}\right\rangle =\sqrt{n_{i}+1}
\left\vert \prod\nolimits_{j=1}^{3}(n_{j}+\delta _{ij})\right\rangle,\qquad
\\
a_{i}\left\vert 000\right\rangle  &=&0,~~~~~\qquad \qquad \ \ \ \ \ \ \ \
a_{i}\left\vert n_{1}n_{2}n_{3}\right\rangle =\sqrt{n_{i}}\left\vert
\prod\nolimits_{j=1}^{3}(n_{j}-\delta _{ij})\right\rangle .
\end{eqnarray}
for $i=1,2,3$ and compute the energy eigenvalues for $H_{ncho}^{3D}$ to
\begin{alignat}{1}
E_{nlr}^{(p)} &=E_{nlr}^{(0)}+E_{nlr}^{(1)}+E_{nlr}^{(2)}+\mathcal{O}(\tau
^{2}) \\
&=E_{nl}^{(0)}+\left\langle nlr\right\vert H_{nc}^{3D}\left\vert
nlr\right\rangle +\sum_{s,p,q\neq n+l+r=p+q+s}\frac{\left\langle
nlr\right\vert H_{nc}^{3D}\left\vert pqs\right\rangle \left\langle
pqs\right\vert H_{nc}^{3D}\left\vert nlr\right\rangle }{%
E_{nlr}^{(0)}-E_{pqr}^{(0)}}+\mathcal{O}(\tau ^{2})  \notag \\
&=\omega \hbar \left[ \frac{3}{2}+n+l+r+\frac{1}{8}\left( \Omega
_{1}+\Omega _{3}\right) \left( 3+5l\right) -\frac{1}{16}(2l+1)\left( \Omega
_{1}+\Omega _{3}\right) {}^{2}+\frac{1}{8}\left( n\Omega _{1}+r\Omega
_{3}\right) \right.   \notag \\
&+\frac{\tau }{2}\left( n^{2}+n+l^{2}+r^{2}+l+r+\frac{1}{4}\left(
n^{2}+4ln+3n+5l^{2}+7l+4\right) \Omega _{1}\right.   \notag \\
&+\left. \left. \frac{1}{4}\left( 5l^{2}+4rl+7l+r^{2}+3r+4\right) \Omega
_{3}+\frac{3}{2}\right) \right] .  \notag
\end{alignat}
As in the two dimensional case we encounter negative terms in this
expression, thus indicating that broken $\mathcal{PT}$-symmetry will be
broken in some parameter range.

\section{Discussions}
Contrary to some claims in the literature \cite{giri_roy}, we have
demonstrated that it is indeed possible to implement $\mathcal{PT}$-symmetry
on noncommutative spaces while keeping the noncommutative structure constants real. Starting from a generic Ansatz for the canonical variables obeying a
$q$-deformed oscillator algebra, we employed $\mathcal{PT}$-symmetry to restrict
the number of free parameters. The relations (\ref{sdd})-(\ref{c6})
resulting from this Ansatz turned out to be solvable. A specific $\mathcal{PT}_{\pm }$-symmetric solution was presented in (\ref{S11})-(\ref{S15}). Clearly there exist more solutions with different kinds of properties. We constructed an explicit representation for the algebra obtained in the nontrivial limit $q\rightarrow 1$ in terms of the generators of a flat
noncommutative space. With regard to the standard inner product for this space, the operators are non-Hermitian. We computed the minimal length and momenta resulting from the generalized uncertainty relations, which overall give rise to minimal areas or minimal volumes in phase space.

Despite being non-Hermitian, due to the built-in $\mathcal{PT}$-symmetry any
model formulated in terms of these variables is a candidate for a self-consistent theory with real eigenvalue spectrum. We have studied the harmonic oscillator on these spaces in one, two and three dimensions. The perturbative computation of the energy eigenvalues suggests that there exists a parameter regime for which the $\mathcal{PT}$-symmetry is broken. It would be interesting to investigate this further and determine when this transition precisely occurs. The eigenvalues will also be useful in further investigations \cite{dey_fring_squeezed} allowing for the construction of coherent states related to the algebras presented in section \ref{sectionminimallengthminimalareas}.

Obviously there are many more solutions to (\ref{sdd})-(\ref{c6}), which might be studied in their own right together with models formulated on them. Minor modifications would also allow to investigate the occurrence of upper bounds, i.e. maximal lengths and momenta \cite{nozari_etemadi}, giving rise to a second scale in special relativity, so-called doubly special relativity \cite{amelino1} constituting a possibility to explain the observation of cosmic rays with energies above the GZK-threshold \cite{bird}. However, such an upper limit might not be required as recent experiments \cite{pierreaugercollaboration1,pierreaugercollaboration2,HiRes_collaboration} seem to indicate at a consistency with the GZK-cutoff prediction, even though some inconsistencies still remain.


\chapter{A Class of Exactly Solvable Models in Noncommutative Space} \label{chapter_solvable}
In the last chapter we have revealed the possibilities of constructing various different models based on the noncommutative spaces and presented as concrete examples the harmonic oscillator in one, two and three dimensions. It is quite evident that, apart from a very few examples, in particular in the lower dimensions, it is very difficult to solve them in an exact manner and we are compelled to utilise the standard perturbation technique. However, in this chapter, we will show that there exists rather a class of solvable models that can be solved in an exact fashion.

\section{A General Construction Procedure for Solvable Non-Hermitian Potentials}
Once a Hamiltonian for a potential system is formulated on a noncommutative
space it usually ceases to be of potential type in the standard canonical variables. Our aim here is to find exact solutions for the corresponding Schr\"{o}dinger equation. Let us first explain the general method we are going to employ. It consists of four main steps: In the first we convert the system to a potential one, in the second we construct the explicit solution to that system as a function of the energy eigenvalues $E$, in the third step we employ a quantization condition by means of the choice of appropriate boundary conditions and in the final step we have to construct an appropriate metric due to the fact that the Hamiltonian might not be Hermitian.

We exploit the fact that for a large class of one dimensional models on
noncommutative spaces the Schr\"{o}dinger equation involving a Hamiltonian $
H(p)$ in momentum space acquires the general form
\begin{equation}
H(p)\psi (p)=E\psi (p)\qquad \Leftrightarrow \qquad -f(p)\psi ^{\prime
\prime }(p)+g(p)\psi ^{\prime }(p)+h(p)\psi (p)=E\psi (p),  \label{1}
\end{equation}
with $f(p)$, $g(p)$, $h(p)$ being some model specific functions and $E$
denoting the energy eigenvalue. This version of the equation may be
converted to a potential system, see for instance \cite{bhattacharjie_sudarshan,jana_roy}, 
\begin{equation}
\tilde{H}(q)\psi (q)=E\psi (q)\qquad \Leftrightarrow \qquad -\phi ^{\prime
\prime }(q)+V(q)\phi (q)=E\phi (q),  \label{pot}
\end{equation}
when transforming simultaneously the wavefunction and the momentum, 
\begin{equation}
\psi (p)=e^{\chi (p)}\phi (p),\quad \chi (p)=\int \frac{f^{\prime }(p)+2g(p)
}{4f(p)}dp,\qquad \text{and\qquad }q=\int f^{-1/2}(p)dp,  \label{psiq}
\end{equation}
respectively. In terms of the original functions $f(p)$, $g(p)$ and $h(p)$,
as defined by equation (\ref{1}), the potential is of the form%
\begin{equation}
V(q)=\left. \frac{4g^{2}+3\left( f^{\prime }\right) ^{2}+8gf^{\prime }}{16f}-
\frac{f^{\prime \prime }}{4}-\frac{g^{\prime }}{2}+h\right\vert _{q}.
\label{V}
\end{equation}

\rhead{A Class of Exactly Solvable Models in Noncommutative Space}
\lhead{Chapter 5}
\chead{}

At this stage one could simply compare with the literature on solvable
potentials in order to extract an explicit solution. However, as the
literature contains conflicting statements and ambiguous notations, we will
present here a simple and transparent construction method for the solutions
adopted from \cite{bhattacharjie_sudarshan,levai}. Furthermore, the
quantities constructed in the next step occur explicitly in the expression
for the metric. For the purpose of constructing a solvable potential we
factorize the wavefunction $\phi (q)$ in (\ref{pot}) further into
\begin{equation}
\phi (q)=v(q)F[w(q)]  \label{phisolvable}
\end{equation}
with as yet unknown functions $v(q)$, $w(q)$ and $F(w)$. This Ansatz
converts the potential equation back into a second order equation of the
type (\ref{1}), albeit for the function $F(w)$,
\begin{equation}
F^{\prime \prime }(w)+Q(w)F^{\prime }(w)+R(w)F(w)=0,  \label{2nd}
\end{equation}
where
\begin{equation}
Q(w):=\frac{2v^{\prime }}{vw^{\prime }}+\frac{w^{\prime \prime }}{\left(
w^{\prime }\right) ^{2}}\qquad \text{and\qquad }R(w):=\frac{E-V(q)}{\left(
w^{\prime }\right) ^{2}}+\frac{v^{\prime \prime }}{v\left( w^{\prime
}\right) ^{2}}.  \label{QR}
\end{equation}
Using the first relation in (\ref{QR}) we can express $v$ entirely in terms
of $w$ and $Q$
\begin{equation}
v(q)=\left( w^{\prime }\right) ^{-1/2}\exp \left[ \frac{1}{2}\int^{w(q)}Q(
\tilde{w})d\tilde{w}\right] .  \label{v}
\end{equation}
With the help of this expression we eliminate $v$ from the second relation
in (\ref{QR}) and express the difference between the energy eigenvalue and
the potential as 
\begin{equation}
E-V(q)=\frac{w^{\prime \prime \prime }}{2w^{\prime }}-\frac{3}{4}\left( 
\frac{w^{\prime \prime }}{w^{\prime }}\right) ^{2}+\left( w^{\prime }\right)
^{2}R(w)-\frac{\left( w^{\prime }\right) ^{2}Q^{\prime }(w)}{2}-\frac{\left(
w^{\prime }\right) ^{2}Q^{2}(w)}{4}.  \label{EV}
\end{equation}
Assuming now that $F$ as introduced in (\ref{phisolvable}) is a particular special
function satisfying the second order differential equation (\ref{2nd}) with
known $Q(w)$ and $R(w)$, the only unknown quantity left on the right hand
side of (\ref{EV}) is $w(q)$. In the general pursuit of constructing
solvable potentials one then selects terms on the right hand side of (\ref
{EV}) to match the constant $E$ which in turn fixes the function $w$. The
remaining terms on the right hand side must then compute to a meaningful
potential. For the case at hand it has to equal $V(q)$ as computed in (\ref
{V}). Assembling everything one has therefore obtained an explicit form for $
\phi (q)$ in (\ref{phisolvable}) and hence $\psi (p)$, as given in (\ref{psiq}),
together with the energy eigenvalues $E$.

In the next step we need to implement the appropriate boundary conditions
and quantize $\psi (p)$ to a well-defined $L^{2}(\mathbb{R})$-function $\psi
_{n}(p)$ for discrete eigenvalues $E_{n}$.

What is left is to construct an appropriate metric, since some of our
Hamiltonians are non-Hermitian, either resulting from the fact that we use a
non-Hermitian representation or from the Hamiltonian being manifestly
non-Hermitian in the first place, or a combination of both. In any of those
cases we have to re-define the metric $\rho $ on our Hilbert space to $
\left\langle \tilde{\psi}_{n}\right\vert \left. \psi _{n}\right\rangle
_{\rho }$ $:=\left\langle \tilde{\psi}_{n}\right\vert \left. \rho \psi
_{n}\right\rangle $. We could follow standard procedures as outlined in chapter \ref{chapterPT}, as for instance to solve the relation $\rho H\rho ^{-1}=H^{\dagger }$ for the operator $\rho $. However, for the scenario outlined in this section we can present a closed analytical formula. Assuming at this stage further that the function $F$, as introduced in (\ref{phisolvable}) belongs to a set of orthonormal functions, we have
\begin{equation}
\delta _{n,m}=\int \varrho (w)F_{n}\left( w\right) F_{m}\left( w\right)
^{\ast }dw=\int \varrho (p)e^{-2\text{Re}\chi (p)}\left\vert
v(p)\right\vert ^{-2}\frac{dw}{dp}\psi _{n}(p)\psi _{n}^{\ast }(p)dp,
\end{equation}
such that the metric is read off as
\begin{equation}
\rho (p)=\varrho (p)e^{-2\text{Re}\chi (p)}\left\vert v(p)\right\vert ^{-2}
\frac{dw}{dp}.  \label{rhorho}
\end{equation}
In the first integral we might need the additional metric $\varrho (w)$ in
case the special function $F\left( w\right) $ does not belong to a set of orthonormal functions. All quantities on the right hand side are explicitly known at this point of the construction allowing us to compute $\rho (p)$ directly.
We note that that the positivity of the metric in entirely governed by $dw/dp $.

\section{The Noncommutative Harmonic Oscillator in Different Representations} \label{section52}
We will focus here on a one dimensional version of a noncommutative space which results as a decoupled direction from a three dimensional version (\ref{one}) as shown in \cite{dey_fring_gouba}
\begin{equation}
\left[X,P\right]=i\hbar \left( 1+\check{\tau}P^{2}\right) .  \label{oneone}
\end{equation}
Here $\check{\tau}:=\tau /(m\omega \hbar )>0$ has the dimension of an
inverse squared momentum and $\tau ~$is therefore dimensionless. Our
intention is here to investigate different types of models for different
representations for the operators obeying these relations. We will compare
four representations, denoted as $\Pi _{(i)}$ with $i\in \{1,2,3,4\}$, for $
X $ and $P$ in relation (\ref{oneone}) expressed in terms of the standard
canonical variables $x$ and $p$ satisfying $[x,p]=i\hbar $
\begin{eqnarray}
X_{(1)} &=&(1+\check{\tau}p^{2})x, \quad P_{(1)}=p, \label{513}\\ 
X_{(2)} &=& (1+\check{\tau}p^{2})^{1/2}x(1+\check{\tau}p^{2})^{1/2},\quad P_{(2)}=p, \label{514}\\
X_{(3)} &=& x,\quad P_{(3)}=\frac{1}{\sqrt{\check{\tau}}}\tan \left(\sqrt{\check{\tau}}p\right), \label{515}\\
X_{(4)} &=& ix(1+\check{\tau}p^{2})^{1/2},\quad P_{(4)}=-ip(1+\check{\tau}p^{2})^{-1/2}. \label{516}
\end{eqnarray}
Representation $\Pi _{(1)}$ is most obvious and most commonly used, but
manifestly non-Hermitian with regard to the standard inner product. This is
adjusted in the Hermitian representation $\Pi _{(2)}$ obtained from $\Pi
_{(1)}$ by an obvious similarity transformation, i.e. $\Pi _{(2)}=(1+\check{
\tau}p^{2})^{-1/2}\Pi _{(1)}(1+\check{\tau}p^{2})^{1/2}$.

Representation $\Pi _{(3)}$ is Hermitian in the standard sense, albeit less
evident. Apart from an additional term in $X_{(3)}$ commuting with $P_{(3)}$, it appeared already in \cite{kempf_mangano_mann} where it was found to be a representation acting on the quasiposition wave function. Below we demonstrate that for some concrete models it is also related in a non-obvious way to $\Pi_{(1)}$ by the transformations to be outlined later in this section.

We have also a particular interest in representation $\Pi _{(4)}$ as it can be constructed systematically from Jordan twists \cite{castro_kullock_toppan} accompanied by an additional rotation. In \cite{castro_kullock_toppan} a closely related version of this representation, which we denote by $\Pi _{(4^{\prime })}$, occurred without the additional factors $i$ and $-i$ in $X_{(4)}$ and $P_{(4)}$, respectively. However, it is easily checked that this is incorrect and does not produce the commutation relations (\ref{oneone}), as instead this variant produces a minus sign on the right hand side in front of the $\check{\tau} P^{2}$-term. One might consider that version of a noncommutative space, which will, however, lead immediately to more severe problems such as a pole in the metric etc. We will argue here further that the construction provided in \cite{castro_kullock_toppan} results in non-physical models and requires the proposed adjustments.

We also note that representation $\Pi _{(4)}$ respects a different kind of $
\mathcal{PT}$-symmetry. Whereas the $\mathcal{PT}$-symmetry $x\rightarrow -x$, $p\rightarrow p$, $i\rightarrow -i$ of the standard canonical variables is
inherited in a one-to-one fashion by the deformed variables in representations $\Pi _{(i)}$ for $i=1,2,3$, i.e. $X_{(i)}\rightarrow-X_{(i)}$, $P_{(i)}\rightarrow P_{(i)}$, $i\rightarrow -i$, it becomes an anti-$\mathcal{PT}$-symmetry for $\Pi _{(4)}$, that is $X_{(i)}\rightarrow
X_{(i)}$, $P_{(i)}\rightarrow -P_{(i)}$, $i\rightarrow -i$. Both versions
are of course symmetries of the commutation relation (\ref{oneone}) and since
both of them are anti-linear involutions, they may equally well be employed
to ensure the reality of spectra for operators respecting the symmetry.

We expect that in concrete models the physics, such as the expectation values for observables, are independent of the representation. We will argue here that this is indeed the case.

At first we will consider the harmonic oscillator in different representations 
\begin{equation}
H_{(i)}=\frac{P_{(i)}^{2}}{2m}+\frac{m\omega ^{2}}{2}X_{(i)}^{2},\qquad 
\text{for }i=1,2,3,4,  \label{HD1}
\end{equation}
with a particular focus on $\Pi _{(4)}$, which has not been dealt with so
far. In principle the solutions for $\Pi _{(1)}$ are known, but it is
instructive to consider here briefly how they emerge in the above scheme. In
terms of the standard canonical variables the Hamiltonian reads
\begin{equation}
H_{(1)}(p)=\frac{p^{2}}{2m}+\frac{m\omega ^{2}}{2}\left( x^{2}+\check{\tau}
p^{2}x^{2}+\check{\tau}xp^{2}x+\check{\tau}^{2}p^{2}xp^{2}x\right) ,\qquad 
\text{for }p\in \mathbb{R}.  \label{H1}
\end{equation}
With $x=i\hbar \partial _{p}$, the corresponding Schr\"{o}dinger equation in
momentum space acquires the general form (\ref{1}), where we identify
\begin{equation}
f(p)=\frac{m\omega ^{2}\hbar ^{2}}{2}(1+\check{\tau}p^{2})^{2},\quad
g(p)=-\tau \hbar \omega p(1+\check{\tau}p^{2}),\quad \text{and\quad }h(p)=
\frac{p^{2}}{2m}.
\end{equation}
Then the equations (\ref{psiq}) and (\ref{V}) convert this into an equation
for a potential system with Hamiltonian $\tilde{H}_{(1)}(q)$ 
\begin{equation}
\psi (p)=\phi (p),\quad q=\sqrt{\frac{2}{\tau \omega \hbar }}\arctan \left( 
\sqrt{\check{\tau}}p\right) ,\quad \text{and}\quad V(q)=\frac{\hbar \omega }{
2\tau }\tan ^{2}\left( \sqrt{\frac{\tau \omega \hbar }{2}}q\right) .
\end{equation}
The $\tan ^{2}$-potential is well known to be solvable, which is explicitly
seen as follows. Assuming that $F(w)$ is an associated Legendre function $P_{\nu}^{\mu }(w)$ we identify from the defining differential equation for
these functions, see e.g. \cite{gradshteyn}, the coefficient functions in (\ref
{2nd}) as
\begin{equation}
Q(w)=\frac{2w}{w^{2}-1}\qquad \text{and\qquad }R(w)=\frac{\nu (\nu +1)}{
1-w^{2}}-\frac{\mu ^{2}}{\left( 1-w^{2}\right) ^{2}}.  \label{QRx}
\end{equation}
Then equation (\ref{EV}) acquires the form
\begin{equation}
E-\frac{\hbar \omega }{2\tau }\tan ^{2}\left( \sqrt{\frac{\tau \omega \hbar 
}{2}}q\right) =\left( w^{\prime }\right) ^{2}\left( \frac{\nu ^{2}+\nu +1}{
1-w^{2}}+\frac{w^{2}-\mu ^{2}}{\left( 1-w^{2}\right) ^{2}}\right) -\frac{
3\left( w^{\prime \prime }\right) ^{2}}{4\left( w^{\prime }\right) ^{2}}+
\frac{w^{\prime \prime \prime }}{2w^{\prime }}  \label{E}
\end{equation}
for the unknown function $w(q)$ and constant $E$. Assuming that the first
term on the right hand side gives rise to a constant, i.e. $\left( w^{\prime
}\right) ^{2}/(1-w^{2})=c\in \mathbb{R}^{+}$, we obtain $w(q)=\sin (\sqrt{c}
q)$ as solution of the latter equation. This function solves (\ref{E}) with
the identifications
\begin{equation}
E=\frac{\tau \omega \hbar }{8}(1+2\nu )^{2}-\frac{\hbar \omega }{2\tau }
,\quad \quad c=\frac{\tau \omega \hbar }{2},\quad \quad \text{and\quad \quad 
}\mu =\mu _{\pm }=\pm \frac{1}{\tau }\sqrt{1+\frac{\tau ^{2}}{4}}.
\label{mu}
\end{equation}
It remains to compute $v(q)$, which results from (\ref{v}), such that all
quantities assembled yield $\phi (q)$ in (\ref{phisolvable}) as
\begin{equation}
\phi (q)=\sqrt{\cos (\sqrt{\tau \omega \hbar /2}q)}P_{\nu }^{\mu _{\pm }}
\left[ \sin (\sqrt{\tau \omega \hbar /2}q)\right] .  \label{qphi}
\end{equation}
Hence with (\ref{psiq}) we obtain finally a solution to the Schr\"{o}dinger
equation in momentum space involving the Hamiltonian $H_{(1)}(x,p)$
\begin{equation}
\psi (p)=\frac{1}{(1+\check{\tau}p^{2})^{1/4}}P_{\nu }^{\mu _{\pm }}\left( 
\frac{\sqrt{\check{\tau}}p}{\sqrt{1+\check{\tau}p^{2}}}\right) .
\label{fast}
\end{equation}
At this stage the constant $\nu $ is still unspecified. Implementing now the
final step, the boundary conditions $\lim_{p\rightarrow \pm \infty }\psi (p)$
$=0$ yields the quantization condition for the energy. Using the property $
\lim_{z\rightarrow \pm 1}P_{n-m}^{m}\left( z\right) =$ $0$ for $n\in \mathbb{
N}$, $m<0$ we need to chose $\mu _{-}$ in (\ref{fast}), such that $\nu
=n+1/\tau \sqrt{1+\tau ^{2}/4}$. Therefore the asymptotically vanishing
eigenfunctions become
\begin{equation}
\psi _{n}(p)=\frac{1}{\sqrt{N_{n}}}\frac{1}{(1+\check{\tau}p^{2})^{1/4}}
P_{n-\mu _{-}}^{\mu _{-}}\left( \frac{\sqrt{\check{\tau}}p}{\sqrt{1+\check{
\tau}p^{2}}}\right) ,  \label{310}
\end{equation}
with corresponding energy eigenvalues 
\begin{equation}
E_{n}=\omega \hbar \left( \frac{1}{2}+n\right) \sqrt{1+\frac{\tau ^{2}}{4}}+
\frac{\tau \omega \hbar }{4}(1+2n+2n^{2}),  \label{EnEn}
\end{equation}
and normalization constant $N_{n}$. The expression for $E_{n}$ agrees
precisely with the one previously obtained in \cite{kempf_mangano_mann,dey_fring_gouba} by different means. The corresponding eigenfunctions $\psi _{n}(p)$ are clearly $L^{2}(\mathbb{R})$-function, but since $H_{(1)}$ is non-Hermitian we do not expect them to be orthonormal. Noting that
\begin{equation}
\delta _{n,m}=\frac{1}{\sqrt{N_{n}N_{m}^{\ast }}}\int
\nolimits_{-1}^{1}P_{n-\mu _{-}}^{\mu _{-}}\left( w\right) \left[ P_{m-\mu
_{-}}^{\mu _{-}}\left( w\right) \right] ^{\ast }dw,  \label{tPP}
\end{equation}
with normalization constant 
\begin{equation}
N_{n}:=\int\nolimits_{-1}^{1}\left\vert
P_{n-\mu _{-}}^{\mu _{-}}\left( z\right) \right\vert ^{2}dz,
\end{equation}
we use $w=\sqrt{\check{\tau}}p/\sqrt{1+\check{\tau}p^{2}}$ to compute the metric from (\ref{rhorho}). We obtain $\rho (p)=\sqrt{\check{\tau}}\left(1+\check{\tau}
p^{2}\right) ^{-1}$, which apart from an irrelevant overall factor $\sqrt{
\check{\tau}}$ is the same as the operator obtained from solving the
relations $\rho H\rho ^{-1}=H^{\dagger }$ as previously reported in \cite{bagchi_fring,dey_fring_gouba}.

Since by (\ref{514}) it follows immediately that $H_{(2)}=\rho
^{1/2}H_{(1)}\rho ^{-1/2}$, the solutions for the Hermitian Hamiltonian $
H_{(2)}$ are easily obtained from those for $H_{(1)}$ as $\rho ^{-1/2}\psi
_{n}$ with identical energy eigenvalues (\ref{EnEn}).

For the representation $\Pi _{(3)}$ we notice that the associated
Hamiltonian $H_{(3)}(p)$ is just a rescaled version of the Hamiltonian $
\tilde{H}_{(1)}(q)$, i.e. $H_{(3)}(p)=$ $\tilde{H}_{(1)}(q=p\sqrt{2/m}/\hbar
\omega )$ with $-\pi /2\sqrt{\check{\tau}}\leq p\leq \pi /2\sqrt{\check{\tau}
}$. Thus the solution for the corresponding Schr\"{o}dinger equation is
simply $\phi (q=p\sqrt{2/m}/\hbar \omega )$. The metric results to be simply
an overall constant factor $\rho (p)=\sqrt{\check{\tau}}$, which is
consistent with the fact that $\Pi _{(3)}$ is Hermitian with regard to the
standard inner product.

Leaving the aforementioned problems for $\Pi _{(4^{\prime })}$ aside, we may
still consider whether it might yield a physically meaningful Hamiltonian.
In terms of the standard canonical variables we obtain 
\begin{equation}
H_{(4^{\prime })}(p)=\frac{p^{2}}{2m(1+\check{\tau}p^{2})}+\frac{m\omega ^{2}
}{2}\left( x^{2}+\check{\tau}x^{2}p^{2}-i\hbar \check{\tau}xp\right) .
\end{equation}
In momentum space the corresponding Schr\"{o}dinger equation is of the form (
\ref{1}) with
\begin{equation}
f(p)=\frac{m\hbar ^{2}\omega ^{2}}{2}(1+\check{\tau}p^{2}),\quad g(p)=-\frac{
3}{2}\tau \hbar \omega p,\quad \text{and\quad }h(p)=\frac{p^{2}}{2m}(1+
\check{\tau}p^{2})^{-1}-\frac{\tau \hbar \omega }{2}.  \label{fgh}
\end{equation}
Then equations (\ref{psiq}) and (\ref{V}) yield
\begin{equation}
\psi (p)=(1+\check{\tau}p^{2})^{-1/2}\phi (p),~q=\sqrt{\frac{2}{\tau \omega
\hbar }}\text{sinh}^{-1}\left( \sqrt{\check{\tau}}p\right) ,~V(q)=\frac{\hbar
\omega }{2\tau }\tanh ^{2}\left( \sqrt{\frac{\tau \omega \hbar }{2}}q\right)
.  \label{psiV}
\end{equation}
With the same assumption on $F(w)$ as made previously we obtain again the
relation (\ref{E}) with the difference that the $\tan ^{2}$-potential on the
left hand side is replaced by a $\tanh ^{2}$-potential. We may produce the
latter potential by assuming $\left( w^{\prime }\right) ^{2}/(1-w^{2})=-c$,
for $c\in \mathbb{R}^{+}$, which is solved by $w(q)=i\sinh (\sqrt{c}q)$.
However, the resulting energy eigenvalues $E=\hbar \omega /2\tau -c/4(1+2\nu
)^{2}$ are not bounded from below, which renders the Hamiltonian $
H_{(4^{\prime })}$ as non-physical.

Using instead $H_{(4)}$ yields the same version of the Schr\"{o}dinger
equation, but all functions in (\ref{fgh}) are all replaced with an overall
minus sign. The corresponding quantities in (\ref{psiV}) are to be replaced
by $\psi (p)=(1+\check{\tau}p^{2})^{-1/2}P_{m-\mu _{-}}^{\mu _{-}}\left( -i
\sqrt{\check{\tau}}p\right) $ with $-i/\sqrt{\check{\tau}}\leq p\leq i/\sqrt{
\check{\tau}}$, the parameter $q$ needs to be multiplied by $-i$ and in the
potential the $\tanh ^{2}$ becomes a $\tan ^{2}$. Then the energy spectrum
becomes physically meaningful, being identical to (\ref{EnEn}). The metric
results to $\rho (p)=-i\sqrt{\check{\tau}}\left( 1+\check{\tau}p^{2}\right)
^{1/2}$ in this case.

With the explicit solutions we may now verify that the expectation values
are indeed the same for all representations. For an arbitrary function $
F\left( P_{(i)},X_{(i)}\right) $ we compute a universal expression 
\begin{equation}
\left\langle \psi _{(i)}\right\vert F\left( P_{(i)},X_{(i)}\right) \left.
\psi _{(i)}\right\rangle _{\rho _{(i)}}=\frac{1}{N}\int\nolimits_{-1}^{1}F
\left[ \frac{z}{\sqrt{\check{\tau}(1-z^{2})}},i\hbar \sqrt{\check{\tau}
(1-z^{2})}\partial _{z}\right] \left\vert P_{m-\mu _{-}}^{\mu _{-}}\left(
z\right) \right\vert ^{2}dz,  \label{exHO}
\end{equation}
for $i=1,2,3,4$. In particular we have $\left\langle \psi _{(i)}\right\vert
H_{(i)}\left. \psi _{(i)}\right\rangle _{\rho _{(i)}}=E_{n}$, $\left\langle
\psi _{(i)}\right\vert P_{(i)}\left. \psi _{(i)}\right\rangle _{\rho
_{(i)}}=0$.

\section{The Noncommutative Swanson Model in Different Representations}
Let us next consider a model which is a widely studied \cite{musumbu_geyer_heiss} solvable prototype example to investigate non-Hermitian systems, the so-called Swanson model \cite{swanson}. On a noncommutative space it reads
\begin{alignat}{1}
H_{(i)} &=\hbar \omega \left( A_{(i)}^{\dagger }A_{(i)}+\frac{1}{2}\right)
+\alpha A_{(i)}A_{(i)}+\beta A_{(i)}^{\dagger }A_{(i)}^{\dagger }\quad ~~~~~
\text{for }i=1,2,3,4, \\
&=\frac{\hbar \omega (1-\tau )-\alpha -\beta }{2m\hbar \omega }P_{(i)}^{2}+
\frac{\Omega m\omega }{2\hbar }X_{(i)}^{2}+i\left( \frac{\alpha -\beta }{
2\hbar }\right) \left( X_{(i)}P_{(i)}+P_{(i)}X_{(i)}\right),
\end{alignat}
with $A_{(j)}=\left( m\omega X_{(j)}+iP_{(j)}\right) /\sqrt{2m\hbar \omega }$
, $A_{(j)}^{\dagger }=\left( m\omega X_{(j)}-iP_{(j)}\right) /\sqrt{2m\hbar
\omega }$ and $\Omega :=\alpha +\beta +\hbar \omega $, $\alpha ,\beta \in 
\mathbb{R}$ with dimension of energy. Evidently for the standard inner
product we have in general $H_{(i)}\neq H_{(i)}^{\dagger }$ when $\alpha
\neq \beta $; even for $\tau =0$. Let us now study this model for the
different types of representations. Starting with $\Pi _{(1)}$, we obtain
the Schr\"{o}dinger equation in momentum space once again in the form of (\ref{1}), with
\begin{eqnarray}
f(p) &=&\frac{m\hbar \omega \Omega }{2}(1+\check{\tau}p^{2})^{2},\quad
g(p)=(\beta -\alpha -\tau \Omega )p(1+\check{\tau}p^{2}), \\
h(p) &=&\frac{\beta -\alpha }{2}-\frac{\tau (\alpha -\beta +\hbar \omega
)+\alpha +\beta -\hbar \omega }{2hm\omega }p^{2}.  \notag
\end{eqnarray}
Then equations (\ref{psiq}) and (\ref{V}) yield
\begin{eqnarray}
\psi (p) &=&(1+\check{\tau}p^{2})^{\frac{(\beta -\alpha )}{2\tau \Omega }
}\phi (p),\quad q=\sqrt{\frac{2}{\tau \Omega }}\arctan \left( \sqrt{\check{
\tau}}p\right) ,  \label{psiqq} \\
V_{S\tan }(q) &=&\frac{(1-\tau )\hbar ^{2}\omega ^{2}-\tau \hbar \omega
(\alpha +\beta )-4\alpha \beta }{2\tau \Omega }\tan ^{2}\left( \sqrt{\frac{
\tau \Omega }{2}}q\right) .  \label{Vq}
\end{eqnarray}
Notice that we obtain again a $\tan ^{2}$-potential, albeit with different
constants involved. Using therefore as in the previous subsection the
assumption that $F(w)$ is an associated Legendre polynomial, we compute with
(\ref{QRx}) the equation (\ref{E}) with the left hand side replaced by $
E-V_{S\tan }(q)$. With the same assumption on the function $w$, namely $
\left( w^{\prime }\right) ^{2}/(1-w^{2})=c\in \mathbb{R}^{+}$, we obtain $
w(q)=\sin (\sqrt{c}q)$ albeit now with $q$ taken from (\ref{psiqq}). The
equivalent to equation (\ref{E}) then yields
\begin{eqnarray}
E &=&\frac{\tau \Omega }{8}(1+2\nu )^{2}+\frac{4\alpha \beta +\tau \hbar
\omega (\alpha +\beta )+\hbar ^{2}(\tau -1)\omega ^{2}}{2\tau \Omega },\quad
c=\frac{\tau \Omega }{2},  \label{EE} \\
\mu _{\pm } &=&\pm \frac{\sqrt{4\left( \hbar ^{2}\omega ^{2}-4\alpha \beta
\right) +\tau \Omega (\tau \Omega -4\hbar \omega )}}{2\tau \Omega }.
\label{mumu}
\end{eqnarray}
Since $\phi (p)$ takes on the same form as in (\ref{psiq}) we obtain 
\begin{equation}
\psi _{n}(p)=\frac{1}{\sqrt{N_{n}}}(1+\check{\tau}p^{2})^{\frac{\beta
-\alpha }{2\tau \Omega }-\frac{1}{4}}P_{n-\mu _{-}}^{\mu _{-}}\left( \frac{
\sqrt{\check{\tau}}p}{\sqrt{1+\check{\tau}p^{2}}}\right) ,  \label{psi}
\end{equation}
as a solution to the Schr\"{o}dinger equation in momentum space involving
the Hamiltonian $H_{(1)}(p)$ for $p\in \mathbb{R}$. We have used the same
condition for the asymptotics of the wavefunction as stated before (\ref{310}
), such that the energy eigenvalues become%
\begin{equation}
E_{n}=\frac{1}{4}\left[ (\tau +2n\tau +2n^{2}\tau )\Omega +(2n+1)\sqrt{
4\left( \hbar ^{2}\omega ^{2}-4\alpha \beta \right) +\tau \Omega (\tau
\Omega -4\hbar \omega )}\right] .  \label{Enn}
\end{equation}
Notice that in the commutative limit $\tau \rightarrow 0$ we recover the
well-known \cite{swanson,musumbu_geyer_heiss} expression for the energy $E_{n}=(n+1/2)\sqrt{\hbar ^{2}\omega ^{2}-4\alpha \beta }$. However, we find a discrepancy with the results reported in \cite{jana_roy} when taking the parameter $\gamma $ in there to zero. The authors do not state any quantization condition, but besides that we can also not verify that the reported expression indeed satisfies the relevant Schr\"{o}dinger equation.

In figure \ref{fig51} we depict the onsets of the exceptional points as a function of
the parameters $\alpha $ and $\beta $ with the remaining parameters fixed.
We notice that for small values of $\alpha $ the domain for which the energy
is real is usually reduced, i.e. a model which still has real energy
eigenvalues on the standard space might develop complex eigenvalues on the
noncommutative space of the type (\ref{1}), e.g. for $\alpha =2$, $\beta
=0.1 $ we read off from the figure that $E_{n}(\tau =0)\in \mathbb{R}$
whereas $E_{n}(\tau =0.5)\notin \mathbb{R}$. In contrast, for larger values
of $\alpha $ complex eigenvalues might become real again once the model is
put onto the space of the type (\ref{1}), e.g. for $\alpha =15$, $\beta =0.1$
we find $E_{n}(\tau =0)\notin \mathbb{R}$ and $E_{n}(\tau =0.5)\in \mathbb{R}
$. Notice that the condition $\left( \hbar ^{2}\omega ^{2}-4\alpha \beta
\right) >\tau \Omega (\tau \Omega /4-\hbar \omega )$ which governs the
reality of the energy in (\ref{Enn}) is the same which controls the ${%
\mathcal{PT}}$-symmetry of the wavefunction $\psi (p)$, which is broken once 
$\mu _{-}\notin \mathbb{R}$.

\begin{figure}[h]
\centering   \includegraphics[width=12cm]{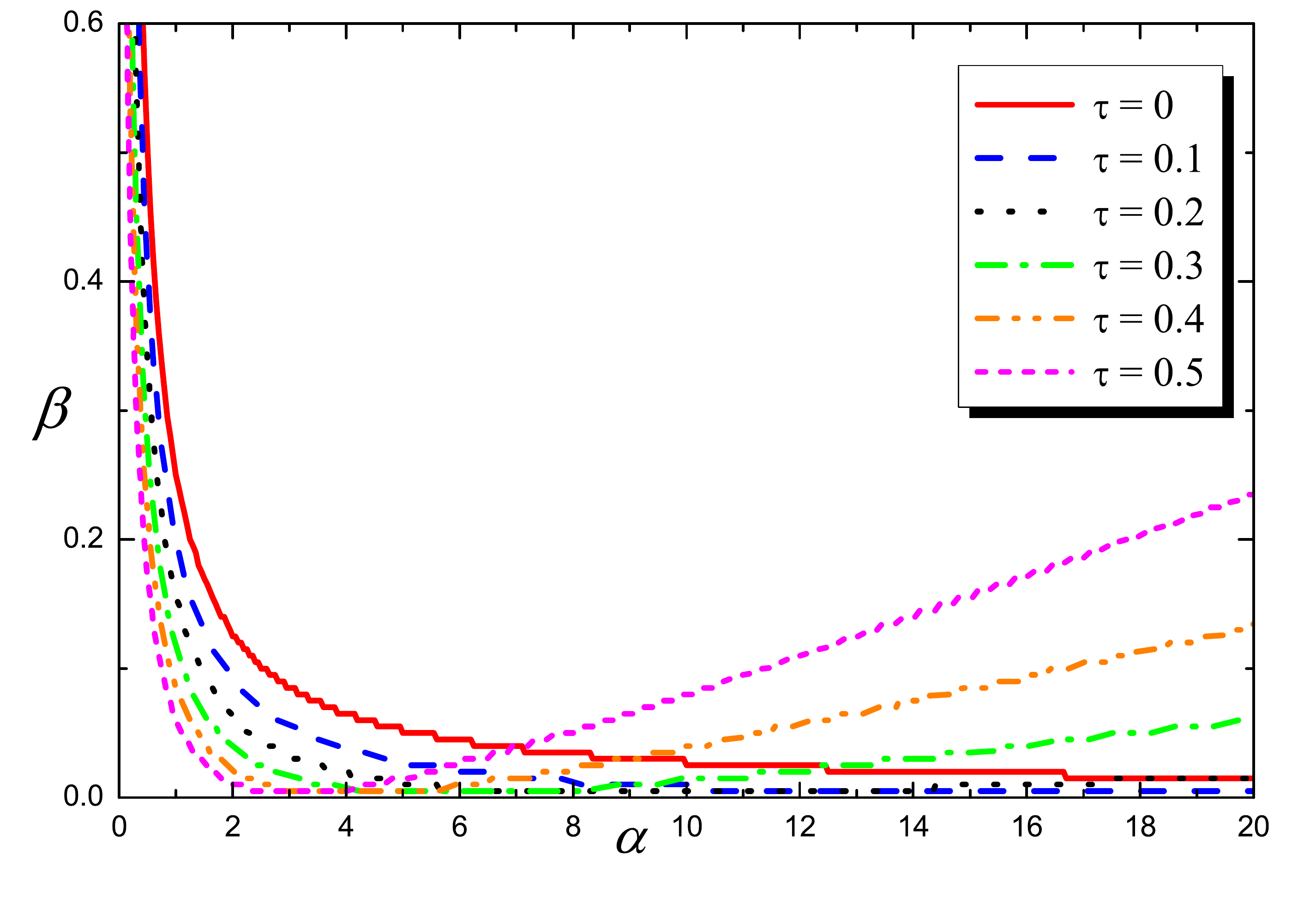}
\caption{\small{Domain of spontaneously broken (above the curve) and unbroken
(below the curve) ${\mathcal{PT}}$-symmetry for the Swanson model with $%
\protect\omega =1$, $\hbar =1$, $m=1$ and different values of $\protect\tau $%
.}}
\label{fig51}
\end{figure}
With the help of (\ref{rhorho}) the metric is now computed to $\rho (p)=\sqrt{
\check{\tau}}\left( 1+\check{\tau}p^{2}\right) ^{\frac{\alpha -\beta -\tau
\Omega }{\tau \Omega }}$.

Once again the solution for $H_{(2)}$ is $\rho ^{-1/2}\psi _{n}$ with energy
eigenvalues (\ref{Enn}) due to $H_{(2)}=\rho ^{1/2}H_{(1)}\rho ^{-1/2}$.

Next we consider the representation $\Pi _{(3)}$. The Schr\"{o}dinger
equation in momentum space acquires the general form of (\ref{1}), with
\begin{eqnarray}
f(p) &=&\frac{m\hbar \omega \Omega }{2},\quad g(p)=\frac{\beta -\alpha }{
\sqrt{\check{\tau}}}\tan \left( \sqrt{\check{\tau}}p\right) , \\
h(p) &=&\frac{\hbar \omega }{2}+\frac{\beta -\alpha -\hbar \omega }{2}\sec
^{2}\left( \sqrt{\check{\tau}}p\right) +\frac{\hbar \omega -\alpha -\beta }{
2\tau }\tan ^{2}\left( \sqrt{\check{\tau}}p\right) .  \notag
\end{eqnarray}
In this case the equations (\ref{psiq}) and (\ref{V}) yield
\begin{equation}
\psi (p)=\left[ \cos \left( \sqrt{\check{\tau}}p\right) \right] ^{\frac{
\alpha -\beta }{\tau \Omega }}\phi (p),\quad q=\sqrt{\frac{2}{m\hbar \omega
\Omega }}p,\quad V(q)=V_{S\tan }(q).
\end{equation}
Notice that in the $q$-variables the potential obtained is exactly the same
as the one previously computed (\ref{Vq}) for representation $\Pi _{(3)}$.
Thus we obtain the same equation (\ref{EE}) and (\ref{mumu}) for the energy
and the parameter $\mu $, respectively. However, the corresponding
wavefunctions differ, resulting in this case to 
\begin{equation}
\psi _{n}(p)=\frac{1}{\sqrt{N_{n}}}\left[ \cos \left( \sqrt{\check{\tau}}
p\right) \right] ^{\frac{\alpha -\beta }{\tau \Omega }+\frac{1}{2}}P_{n-\mu
_{-}}^{\mu _{-}}\left[ \sin \left( \sqrt{\check{\tau}}p\right) \right] ,
\end{equation}
for $-\pi /2\sqrt{\check{\tau}}\leq p\leq \pi /2\sqrt{\check{\tau}}$. We
compute $\rho (p)=\sqrt{\check{\tau}}\left[ \cos \left( \sqrt{\check{\tau}}
p\right) \right] ^{\frac{2(\beta -\alpha )}{\tau \Omega }}$ from (\ref{rhorho})
as relevant metric. Notice that $\rho (p)$ reduces to the standard metric
for $\alpha =\beta $ reflecting the fact that $H_{(3)}$ is Hermitian for
these values.

Since representation $\Pi _{(4^{\prime })}$ was identified as being
non-physical in the previous subsection, it is clear that this will also be
the case for the Swanson model and we will therefore not treat it any
further here.

For $\Pi _{(4)}$ the Schr\"{o}dinger equation in momentum space is also of
the form (\ref{1}), with
\begin{alignat}{1}
f(p) &=\frac{m\hbar \omega \Omega }{2}(1+\check{\tau}p^{2}),\quad
g(p)=p\left( \beta -\alpha +\frac{3}{2}\tau \Omega \right) , \\
h(p) &=\frac{1}{2(1+\check{\tau}p^{2})}\left\{ (\beta -\alpha +\tau \Omega
)+\frac{p^{2}}{m\hbar \omega }\left[ \alpha +\beta -\hbar \omega +\tau
(2\beta -2\alpha +\hbar \omega )+\tau ^{2}\Omega \right] \right\} .~~~~ 
\notag
\end{alignat}
In this case the equations (\ref{psiq}) and (\ref{V}) yield
\begin{equation}
\psi (p)=(1+\check{\tau}p^{2})^{\frac{\alpha -\beta }{2\tau \Omega }-\frac{1
}{2}}\phi (p),\quad q=-i\sqrt{\frac{2}{\tau \Omega }}\text{sinh}^{-1}\left( 
\sqrt{\check{\tau}}p\right) ,\quad V(q)=V_{S\tan }(q).
\end{equation}
Notice that in the $q$-variables the potential obtained is exactly the same
as the one previously computed (\ref{Vq}) for representation $\Pi _{(3)}$.
Thus we obtain the same equations (\ref{EE}) and (\ref{mumu}) for the energy
eigenvalues and the parameter $\mu $, respectively. However, the final
wavefunction differs, resulting, after imposing the boundary conditions, to 
\begin{equation}
\psi _{n}(p)=\frac{1}{\sqrt{N_{n}}}(1+\check{\tau}p^{2})^{\frac{\alpha
-\beta }{2\tau \Omega }-\frac{1}{4}}P_{n-\mu _{-}}^{\mu _{-}}\left( -i\sqrt{
\check{\tau}}p\right) ,
\end{equation}
with $-i/\sqrt{\check{\tau}}\leq p\leq i/\sqrt{\check{\tau}}$. Now we
evaluate $\rho (p)=-i\sqrt{\check{\tau}}\left( 1+\check{\tau}p^{2}\right) ^{
\frac{\beta -\alpha }{\tau \Omega }+\frac{1}{2}}$as metric from our general
formula (\ref{rhorho}).

We compute again the expectation values for some arbitrary function $F\left(
P_{(i)},X_{(i)}\right) $ in all four representations 
\begin{equation}
\left\langle \psi _{(i)}\right\vert F\left( P_{(i)},X_{(i)}\right) \left.
\psi _{(i)}\right\rangle _{\rho _{(i)}}=\frac{1}{N}\int\nolimits_{-1}^{1}F
\left[ \frac{z}{\sqrt{\check{\tau}(1-z^{2})}},i\hbar \sqrt{\check{\tau}
(1-z^{2})}\partial _{z}\right] \left\vert P_{m-\mu _{-}}^{\mu _{-}}\left(
z\right) \right\vert ^{2}dz,
\end{equation}
which looks formally exactly the same as (\ref{exHO}) with the difference
that $\mu _{-}$ is given by the expression in (\ref{mumu}).

\section{A Noncommutative P\"{o}schl-Teller Potential in Disguise}
In the previous sections we observed that simple models on a noncommutative
space may lead to more unexpected solvable potential systems when expressed
in terms of the standard canonical variables and a subsequent
transformation. We may also reverse the question and explore which type of
model on a noncommutative space one obtains when we start from a well-known
solvable potential in the standard canonical variables. For instance, we
wish to construct the widely studied P\"{o}schl-Teller potential \cite
{poschl_teller}. Since the transformations are difficult to invert, we use
trial and error and find that this indeed achieved when starting with the
Hamiltonian 
\begin{equation}
H_{(i)}=\frac{\beta }{2m}P_{(i)}^{2}+\frac{\hbar \omega \alpha }{2\check{\tau
}}P_{(i)}^{-2}+\frac{m\omega ^{2}}{2}X_{(i)}^{2}+\frac{\hbar \omega \alpha }{
2}+\frac{\beta }{2m\check{\tau}}\quad \text{for }i=1,2,3,4;\alpha ,\beta \in 
\mathbb{R}.
\end{equation}
We note that this Hamiltonian can not be viewed as a deformation of a model
on a standard commutative space as it is intrinsically noncommutative, in
the sense that it does not possess a trivial commutative limit $\tau
\rightarrow 0$. Proceeding as in the previous subsections we find for the
representation $\Pi _{(1)}$ that the Schr\"{o}dinger equation in momentum
space is once more of the general form of (\ref{1}), with
\begin{eqnarray}
f(p) &=& \frac{m\hbar ^{2}\omega ^{2}}{2}(1+\check{\tau}p^{2})^{2},\qquad
g(p)=-m\hbar ^{2}\omega ^{2}\check{\tau}p(1+\check{\tau}p^{2}), \notag \\ 
h(p) &=& \frac{(1+\check{\tau}p^{2})(\alpha m\hbar \omega +\beta p^{2})}{2m\check{\tau}p^{2}}.
\end{eqnarray}
From equation (\ref{psiq}) we obtain now
\begin{equation}
\psi (p)=\phi (p),\quad q=\sqrt{\frac{2}{\tau \hbar \omega }}\arctan \left( 
\sqrt{\check{\tau}}p\right) ,\quad
\end{equation}
and as anticipated we compute a P\"{o}schl-Teller potential with the help of
equation (\ref{V}) 
\begin{equation}
V_{PT}(q)=\frac{\hbar \omega \alpha }{2}\csc ^{2}\left( q\sqrt{\frac{\hbar
\omega \tau }{2}}\right) +\frac{\beta }{2m\check{\tau}}\sec ^{2}\left( q
\sqrt{\frac{\hbar \omega \tau }{2}}\right) .  \label{VPT}
\end{equation}
Assuming now that the special function $F(w)$ in (\ref{2nd}) is a Jacobi
polynomial $P_{n}^{(a,b)}(w)$, with $n\in \mathbb{N}_{0}$, $a,b\in \mathbb{R}
$, we identify from its defining differential equation, see e.g. \cite{gradshteyn}, the coefficient functions in (\ref{2nd}) as
\begin{equation}
Q(w)=\frac{b-a-(2+a+b)w}{1-w^{2}}\qquad \text{and\qquad }R(w)=\frac{
n(n+1+a+b)}{1-w^{2}}.
\end{equation}
Then equation (\ref{EV}) is evaluated to
\begin{eqnarray}
E-V_{PT}(q) &=&\frac{n\left( w^{\prime }\right) ^{2}(a+b+n+1)}{1-w^{2}}-\frac{\left( w^{\prime }\right) ^{2}\left[ b-a-w(a+b+2)\right] ^{2}}{
4\left( 1-w^{2}\right) ^{2}} \\
&&+\frac{\left( w^{\prime }\right) ^{2}\left[ w^{2}(a+b+2)+2w(a-b)+a+b+2\right] 
}{2\left( 1-w^{2}\right) ^{2}}-\frac{3\left( w^{\prime \prime }\right) ^{2}}{
4\left( w^{\prime }\right) ^{2}}+\frac{w^{\prime \prime \prime }}{2w^{\prime}}, \notag
\end{eqnarray}
with as yet unknown function $w(q)$ and constant $E$. As in the previous
section we\ assume again that the first term on the right hand side gives
rise to a constant, i.e. $\left( w^{\prime }\right) ^{2}/(1-w^{2})=c\in 
\mathbb{R}^{+}$, but this time we choose the solution $w(q)=\cos (\sqrt{c}q)$
, which solves (\ref{E}) with the identifications
\begin{eqnarray}
E_{n} &=& \frac{\hbar \omega \tau }{2}(1+2n+a+b)^{2},\qquad c=2\tau \omega \hbar
, \notag \\
a_{\pm } &=& \pm \frac{1}{2}\sqrt{1+\frac{4\alpha }{\tau },}\qquad b_{\pm
}=\pm \frac{1}{2}\sqrt{1+\frac{4\beta }{\tau ^{2}}}.  \label{EPT}
\end{eqnarray}
Computing $v(q)$ by means of (\ref{v}) we assemble everything into the
solution of the Schr\"{o}dinger equation involving $H_{(1)}(x,p)$ 
\begin{equation}
\psi _{n}(p)=\frac{1}{\sqrt{N_{n}}}p^{1/2+a_{+}}(1+\check{\tau}
p^{2})^{-(1+a_{+}+b_{+})/2}P_{n}^{(a_{+},b_{+})}\left( \frac{1-\check{\tau}
p^{2}}{1+\check{\tau}p^{2}}\right) .  \label{ps}
\end{equation}
We have selected here $a_{+}$ and $b_{+}$ in order to implement the
appropriate boundary conditions $\lim\nolimits_{p\rightarrow \pm \infty
}\psi _{n}(p)=0$ together with $\psi _{n}(0)=0$. We note that the energy
eigenvalues are real and bounded from below as long as $\alpha >-\tau /4$
and $\beta >-\tau ^{2}/4$. The occurrence of exceptional points is due to
the ${\mathcal{PT}}$-symmetry breaking of the wavefunction $\psi _{n}(p)$
when $a_{+}$, $b_{+}\notin \mathbb{R}$.

Following the same procedure as in the previous subsections we find for the
remaining representations 
\begin{equation}
\tilde{H}_{(1)}(q)=\tilde{H}_{(2)}(q)=\tilde{H}_{(3)}(q)=\tilde{H}_{(4)}(q),
\end{equation}
where $q$ is related to $p$ differently in each case. Converting between the
different variables and computing the relevant pre-factors as in the
previous subsection we then find 
\begin{alignat}{1}
\psi _{(2)}(p) &=\rho _{(1)}^{-1/2}\psi _{(1)}(p), \\
\psi _{(3)}(p) &=\frac{1}{\sqrt{N_{n}}}\frac{\left[ 1-\cos (2p\sqrt{\check{
\tau}}))\right] ^{\frac{1+a_{+}}{2}}\left[ \cos (2p\sqrt{\check{\tau}})+1)
\right] ^{\frac{1+b_{+}}{2}}}{\sqrt{\sin (2p\sqrt{\check{\tau}})}}
P_{n}^{(a_{+},b_{+})}\left[ \cos \left( 2p\sqrt{\check{\tau}}\right) \right]
,~~~~ \\
\psi _{(4)}(p) &=\frac{1}{\sqrt{N_{n}}}p^{a_{+}+1/2}(1+\check{\tau}p^{2})^{
\frac{2b_{+}-1}{4}}P_{n}^{(a_{+},b_{+})}\left( 1+2\check{\tau}p^{2}\right) ,
\end{alignat}
for the energy eigenvalue (\ref{EPT}) where $p>0$ for $\Pi _{(2)}$, $-\pi /2
\sqrt{\check{\tau}}\leq p\leq \pi /2\sqrt{\check{\tau}}$ for $\Pi _{(3)}$
and $-i/\sqrt{\check{\tau}}\leq p\leq i/\sqrt{\check{\tau}}$ for $\Pi _{(4)}$
. Using the orthogonality relation for the Jacobi polynomial,
\begin{equation*}
\int_{-1}^{1}(1-x)^{a}(1+x)^{b}P_{n}^{(a,b)}\left( x\right)
P_{m}^{(a,b)}\left( x\right) dx=\delta _{n,m}N_{n}~\text{\ \ \ \ \ for }
\text{Re}~a,\text{Re}~b>-1,
\end{equation*}
with $N_{n}=\frac{2^{a+b+1}\Gamma \left( a+n+1\right) \Gamma \left(
b+n+1\right) }{n!\Gamma \left( a+b+n+1\right) \Gamma \left( a+b+2n+1\right) }
$. we compute the metric from (\ref{rhorho}) to 
\begin{alignat}{1}
\rho _{(1)}(p) &=-2\sqrt{\check{\tau}}(1+\check{\tau}p^{2})^{-1}, \qquad \rho _{(2)}(p)=1, \notag \\ 
\rho_{(3)}(p) &=-2\sqrt{\check{\tau}} \quad \text{and}\quad \ \rho _{(4)}(p)=2i\sqrt{\check{\tau}}
(1+\check{\tau}p^{2})^{1/2}.
\end{alignat}

We also note that for representation $\Pi _{(4^{\prime })}$ we obtain the
same potential (\ref{VPT}) with $\csc ^{2}\rightarrow \text{csch}^{2}$, $
\sec ^{2}\rightarrow -\text{sech}^{2}$ plus an overall constant, which is
once again non-physical in the sense of leading to an unbounded spectrum from
below.

Finally we compute the expectation values for some arbitrary function $
F\left( P_{(i)},X_{(i)}\right) $ 
\begin{eqnarray}
&&\left\langle \psi _{(i)}\right\vert F\left( P_{(i)},X_{(i)}\right) \left.
\psi _{(i)}\right\rangle _{\rho _{(i)}} \\
&&=\frac{1}{N}\int\nolimits_{-1}^{1}F
\left[ \frac{z}{\sqrt{\check{\tau}(1-z^{2})}},i\hbar \sqrt{\check{\tau}
(1-z^{2})}\partial _{z}\right] \left\vert P_{n}^{(a_{+},b_{+})}\left(
z\right) \right\vert ^{2}dz \notag,
\end{eqnarray}
which is again the same for all four representations.

\section{Discussions}
We have shown how different representations for the operators $X$ and $P$
obeying a generalized version of Heisenberg's uncertainty relation are related to each other by the transformations outlined in section \ref{section52}. We have demonstrated their equivalence within the setting of three characteristically different types of solvable models, a Hermitian one, a non-Hermitian one and an intrinsically noncommutative one. In all cases we showed that an appropriate metric can be found such that expectation values result to be representation independent. We provided an explicit formula for this metric, involving the quantities computed in the first two steps of the general procedure. The computations were carried out in momentum space, but naturally the method works equally well in standard $x$-space. In both cases the order of the differential equation imposes a limitation on the type of models which may be considered.

For representation $\Pi _{(4^{\prime })}$ proposed in \cite{castro_kullock_toppan} we found that it does not lead to the uncertainty relations (\ref{oneone}) and moreover that for the models investigated it always gives rise to non-physical spectra which are not bounded from below. This suggests that the general procedure of Jordan twists requires a mild modification as outlined in the manuscript.

Clearly it would be interesting to extend this analysis to different types
of full three dimensional algebras for noncommutative spaces and investigate
alternative representations, such as for instance for those already reported
in \cite{dey_fring_gouba}.


\chapter{Noncommutative Coherent States} \label{chapter_coherent}
So far, we have been discussing different possibilities of constructing models based on the noncommutative space-time structure and most of the attention has been paid to the analysis of the algebras which are albeit more mathematical. In this chapter, we focus more on the physical nature of the systems, which is actually where one may experience many exciting situations. We introduce the coherent states of the noncommutative models respecting the modified uncertainty relation, which have been explored fairly little in the literature.

Coherent states are the specific quantum states of a system whose dynamics most closely resemble the oscillating behaviour of the corresponding classical systems. The first example was derived by Erwin Schr{\"o}dinger in 1926 \cite{schrodinger}, while searching for the solutions of the Schr{\"o}dinger equation that satisfy the Bohr's correspondence principle. In his discovery, he constructed an interesting type of wave packet, consisting of a large number of harmonic oscillator wave functions, which does not spread out with time and the behaviour of which is quite similar to a solitary wave. In modern notations, this can be written down as 
\begin{equation} \label{wavepacket}
\langle x \vert \alpha \rangle=\pi^{-1/4} \text{exp}\left(-\frac{1}{2}x^2+\sqrt{2}x \alpha-\frac{\alpha^2}{2}-\frac{\vert \alpha \vert ^2}{2}\right).
\end{equation}
Much later, in 1951, the same wave packet was derived from the other way around for the first time by Iwata \cite{iwata}, who considered first, the coherent states to be the eigenstates of the non-Hermitian annihilation operator
\begin{equation} \label{eigen}
\hat{a}\vert \alpha \rangle=\alpha \vert \alpha \rangle,
\end{equation}
and then successfully derived the wavepacket (\ref{wavepacket}). After that, many authors arrived at the same equation (\ref{wavepacket}) from many different arguments and finally it was Glauber \cite{glauber}, who actually carried out a more systematic analysis to present them in a more compact formula in terms of the Fock basis and entitled them as the "coherent states" for the first time in the literature. We briefly go through the construction procedure here. Operating with integer powers of the creation operator $a^\dagger$ on the vacuum state $\vert 0\rangle$, one builds the excited states of a harmonic oscillator, which when normalised, acquire the form
\begin{equation}
\vert n\rangle=\frac{\left(a^\dagger\right)^n}{\sqrt{n !}}\vert0\rangle, \qquad (n=0,1,2 ....).
\end{equation}
The coherent states $\vert \alpha \rangle$ can be expressed in terms of the orthonormal set $\vert n \rangle$, satisfying the completeness relation 
\begin{equation}\label{alphaalpha}
\vert \alpha \rangle :=\displaystyle\sum_{n}\vert n \rangle \langle n \vert \alpha \rangle,
\end{equation}
where the inner product $\langle n \vert \alpha \rangle=\frac{\alpha^n}{\sqrt{n !}}\langle 0 \vert \alpha \rangle$ could easily be calculated from the assumption that the coherent states are the eigenstates of the annihilation operator, so that the states (\ref{alphaalpha}) can be written as

\rhead{Noncommutative Coherent States}
\lhead{Chapter 6}
\chead{}

\begin{equation}\label{glauberunnormalised}
\vert \alpha \rangle=\langle 0 \vert \alpha \rangle\displaystyle\sum_{n=0}^\infty \frac{\alpha^n}{\sqrt{n!}}\vert n \rangle.
\end{equation}
With the requirement that the states are normalised $\langle\alpha\vert\alpha\rangle=1$, the normalisation constant is computed to be $\langle 0\vert\alpha\rangle=e^{-\vert \alpha \vert^2/2}$, such that the coherent states of the ordinary harmonic oscillator (\ref{glauberunnormalised}) obtain the compact form in terms of the standard Fock basis
\begin{equation} \label{GlauberCoherent}
\vert \alpha \rangle=e^{-\frac{\vert \alpha \vert^2}{2}}\displaystyle\sum_{n=0}^\infty\frac{\alpha^n}{\sqrt{n!}}\vert n \rangle, \qquad \forall\quad \alpha \in \mathbb{C}.
\end{equation}
Note that the coherent states (\ref{GlauberCoherent}) can also be constructed from the vacuum with the application of the unitary displacement operator $\hat{D}$ acting on it
\begin{equation}
\vert \alpha \rangle=\hat{D}(\alpha)\vert 0 \rangle, \qquad \text{where} \qquad \hat{D} (\alpha)=e^{\alpha \hat{a}^\dagger-\alpha^\ast\hat{a}},
\end{equation}
which was used by Feynman \cite{feynman} and Glauber \cite{glauber1951} back in 1951. Notice that the states (\ref{GlauberCoherent}) have very interesting mathematical features that are very different from the usual Fock states. For instance, two different coherent states are not orthogonal to each other due to the property that the annihilation operator is not a self adjoint operator by definition.

Several generalisations of different type of coherent states have been proposed afterwords and their properties have been analysed, and now the concept is not limited to the case of harmonic oscillator model only. That is the reason why the coherent states corresponding to the quantum harmonic oscillator are referred to as the "canonical coherent states" or standard coherent states or the Gaussian states. Sometimes they are also called the minimum uncertainty wavepacket, due to the property $\Delta x \Delta p=\hbar/2$, with $\Delta x=\Delta p=\sqrt{\hbar/2}$. For a concrete review on this subject up to 2001, one may follow up the reference \cite{dodonov}. In our discussion, however, we will focus on the generalised version of the Glauber coherent states, which was proposed and successfully tested for the hydrogen atom and other systems by Klauder \cite{klauder1995,klauder1996}. Let us quickly discuss the general properties of those kind of states for our reference purpose.

\section{Klauder Coherent States}
We first look at the properties of the canonical coherent states. First of all, they are continuous in $\alpha$, namely 
\begin{equation}
\Vert~\vert\alpha+\delta \alpha \rangle-\vert\alpha\rangle~\Vert^2 \rightarrow 0, \qquad \text{as} \qquad \delta \alpha \rightarrow 0.
\end{equation}
Secondly, by definition, they are normalised and most importantly the time evolution of any such coherent state remains within the family of coherent states
\begin{equation}
e^{-i H t}\vert \alpha \rangle=e^{-\frac{\vert \alpha\vert^2}{2}}\displaystyle\sum_{n=0}^\infty\frac{\alpha^n e^{-i \omega n t}}{\sqrt{n !}}\vert n\rangle=\left| e^{-i \omega t}\alpha \right\rangle.
\end{equation}
Looking up at the canonical coherent states carefully, a direct generalisation is possible \cite{klauder1995,klauder1996,gazeau_klauder,gazeau_monceau} by first considering the kets $\vert n \rangle$ to be the eigenstates of the Hamiltonian $H$ other than the harmonic oscillator corresponding to the eigenvalues $E_n=\omega e_n$ and then replacing the square root of $n !$ by the generalised factorial $\left[e_n\right]!=e_1 e_2.....e_n$, in analogy to the pioneering work of Jackson \cite{jackson}. For a review on this topic, one may look for instance at \cite{ali_antoine_gazeau}. The virtue of the generalised coherent states is that they can be applied to any generalised setting rather than that to a particular case of harmonic oscillator.

For a Hermitian Hamiltonian $h$ with discrete bounded below and nondegenerate eigenspectrum $E_n=\omega e_n$, and orthonormal eigenstates $\vert \phi_n\rangle$, the Klauder coherent states are defined as a two parameter set of $J$ and a dimensionless $\gamma$, with $\gamma=\omega t, \omega > 0$
\begin{equation} \label{klaudercoherent}
\vert J, \gamma\rangle=\frac{1}{\mathcal{N}(J)}\displaystyle\sum_{n=0}^\infty \frac{J^{n/2}e^{-i \gamma e_n}}{\sqrt{\rho_n}}\vert \phi_n\rangle, \qquad J \in \mathbb{R}_0^+, \gamma \in \mathbb{R}.
\end{equation}
The probability distribution and the normalisation constant can be computed from
\begin{equation}\label{probdensity}
\rho_n=\displaystyle\prod_{k=1}^n e_k \qquad \text{and} \qquad \mathcal{N}^2(J)=\displaystyle\sum_{k=0}^\infty \frac{J^k}{\rho_k},
\end{equation}
with $\rho_0=1$, where the later result is from the requirement $\langle J,\gamma\vert J,\gamma\rangle=1$. It is clear that one could obtain the coherent states for any generalised settings with known eigenvalues $E_n$ and eigenvectors $\vert \phi_n\rangle$ by plugging them into the equations (\ref{klaudercoherent}) and (\ref{probdensity}). These states are called coherent states for several reasons. Firstly they are the eigenvectors of the operator $a(\gamma)$ by construction,
\begin{equation}
a(\gamma)\vert J,\gamma\rangle=J\vert J,\gamma\rangle, \quad \text{with}\quad a(\gamma)\equiv e^{-i \gamma H/\omega}a e^{i \gamma H/\omega}.
\end{equation}
Secondly, they are continuous in time and $J$ and finally, by definition, they are normalised and temporarily stable
\begin{equation}
e^{-i H t}\vert J,\gamma \rangle=\vert J,\gamma+\omega t \rangle,
\end{equation}
which are same as that of the canonical coherent states. An additional property could be achieved, when one introduces a sequence of dimensionless real numbers $e_i$ and impose the constraint
\begin{equation}
0=e_0<e_1<e_2<.....\qquad \text{with} \qquad \rho_0=\left[e_0\right]!=1,
\end{equation}
further, such that,
\begin{alignat}{1}\label{actionangle}
\langle J,\gamma\vert H \vert J,\gamma\rangle & = \frac{1}{\mathcal{N}^2(J)}\displaystyle\sum_{n,m=0}^\infty \frac{J^{\frac{n+m}{2}}}{\sqrt{\rho_m \rho_n}}\omega e_n\langle\phi_m\vert\phi_n\rangle =\frac{1}{\mathcal{N}^2(J)}\displaystyle\sum_{n=0}^\infty \frac{J^n}{\rho_n}\omega e_n \nonumber \\
& =\frac{\omega}{\mathcal{N}^2(J)} \displaystyle\sum_{n=1}^\infty \frac{J^n}{\rho_{n}} e_n =\omega \frac{\displaystyle\sum_{n=1}^\infty J^n/\rho_{n-1}}{\displaystyle\sum_{n=0}^\infty J^n/\rho_{n}} =\omega \frac{\displaystyle\sum_{n=0}^\infty J^{n+1}/\rho_{n}}{\displaystyle\sum_{n=0}^\infty J^n/\rho_{n}}=\omega J,
\end{alignat}
Notice that, the property (\ref{actionangle}) is very similar to the classical action-angle identity, where $J$ and $\gamma$ can be compared with the classical action-angle variables. The Klauder coherent states with this additional property (\ref{actionangle}) is often called the Gazeau-Klauder (GK) coherent states \cite{gazeau_klauder}. In their investigation \cite{gazeau_klauder}, the authors also explored similar kind of coherent states for the case of the Hamiltonians with a nondegenerate continuous spectrum. The degenerate case has been discussed in detail in \cite{ali_bagarello}, which one may look at for their general interests.

\section{GK-Coherent States for Non-Hermitian Hamiltonians}
In this section, we would like to develop the method and extend the ideas of coherent states to the case of non-Hermitian Hamiltonians $H$, because of the fact that the noncommutative spaces often produce Hamiltonians which are of non-Hermitian nature. In these situations, the coherent states are easily adoptable when we assume the Hamiltonian $H$ is pseudo/quasi Hermitian i.e. the non-Hermitian Hamiltonian $H$ and the Hermitian Hamiltonian $h$ are related by a similarity transformation $h=\eta H \eta^{-1}$, with $\eta^\dagger \eta$ being a positive definite operator playing the role of metric (\ref{hermiticity}). When computing expectation values for operators associated to the non-Hermitian system, we need to change the metric \cite{bender_boettcher,bender_making_sense,scholtz_geyer_hahne,mostafazadeh1,mostafazadeh5}, because the observables are expected to be Hermitian. The same reasoning has to be adopted for the evaluation of expectation values with regard to the coherent states. Therefore the expectation value
for a non-Hermitian operator $\mathcal{O}$ related to a Hermitian operator $%
o $ by a similarity transformation $o=\eta \mathcal{O}\eta ^{-1}$ is
computed as
\begin{equation}
\left\langle J,\gamma ,\Phi \right\vert \mathcal{O}\left\vert J,\gamma ,\Phi
\right\rangle _{\eta }:=\left\langle J,\gamma ,\Phi \right\vert \eta
^{\dagger }\eta \mathcal{O}\left\vert J,\gamma ,\Phi \right\rangle
=\left\langle J,\gamma ,\phi \right\vert o\left\vert J,\gamma ,\phi
\right\rangle ,  \label{O}
\end{equation}
if $\left\vert J,\gamma ,\phi\right\rangle$ belongs to the domain of $\eta$. Our notation is to be understood in the sense that in the state $\left\vert
J,\gamma ,\phi \right\rangle $ and $\left\vert J,\gamma ,\Phi \right\rangle $
we sum over the eigenstates of the Hermitian Hamiltonian $h$ and
non-Hermitian Hamiltonian $H$, respectively. These states are continuous in
the two variables $(J,\gamma )$, provide a resolution of the identity, are
temporarily stable, in the sense that they remain coherent states under time
evolution, and satisfy the action identity
\begin{equation}
\left\langle J,\gamma ,\Phi \right\vert H\left\vert J,\gamma ,\Phi
\right\rangle _{\eta }=\left\langle J,\gamma ,\phi \right\vert h\left\vert
J,\gamma ,\phi \right\rangle =\hbar \omega J.  \label{AI}
\end{equation}
This identity ensures that $(J,\gamma )$ are action angle variables \cite{gazeau_klauder,antoine_gazeau_monceau_klauder_penson}.

The main purpose is here to consider a model on a noncommutative space with
nontrivial commutation relations for their canonical variables giving rise
to minimal uncertainties. It is then interesting to investigate how close
the GK-states approach the minimum uncertainty product and eventually might
even become squeezed states. Thus for a simultaneous measurement of two
observables $A$ and $B$ in this system we need to evaluate the left and
right hand side of the generalized version of Heisenberg's uncertainty
relation 
\begin{equation}
\Delta A\Delta B\geq \frac{1}{2}\left\vert \left\langle J,\gamma ,\Phi
\right\vert [A,B]\left\vert J,\gamma ,\Phi \right\rangle _{\eta }\right\vert.  \label{HU}
\end{equation}
The uncertainties are computed as $\left(\Delta A\right)^2=\left\langle J,\gamma ,\Phi
\right\vert A^{2}\left\vert J,\gamma ,\Phi \right\rangle _{\eta
}-\left\langle J,\gamma ,\Phi \right\vert A\left\vert J,\gamma ,\Phi
\right\rangle _{\eta }^{2}$ and analogously for $\Delta B$. In order to test
the quality of the coherent states, i.e. to see how closely they resemble
to the classical mechanics, we may also test Ehrenfest's theorem for an operator $A$
\begin{equation}
i\hbar \frac{d}{dt}\left\langle J,\gamma +t\omega ,\Phi \right\vert
A\left\vert J,\gamma +t\omega ,\Phi \right\rangle _{\eta }=\left\langle
J,\gamma +t\omega ,\Phi \right\vert [A,H]\left\vert J,\gamma +t\omega ,\Phi
\right\rangle _{\eta }.  \label{Ehren}
\end{equation}
We used in (\ref{Ehren}) the fact that the time evolution for the states $
\left\vert J,\gamma ,\Phi \right\rangle $ is simply implemented as $\exp
(-iHt/\hbar )\left\vert J,\gamma ,\Phi \right\rangle =\left\vert J,\gamma
+t\omega ,\Phi \right\rangle $, see \cite{gazeau_klauder,antoine_gazeau_monceau_klauder_penson}. Specifying the
operators $A$ and $B$ we will also test below the correspondence principle.

\section{Squeezed Coherent States for the Noncommutative Perturbative Harmonic Oscillator}\label{section63}
Here we present a perturbative treatment of the above considerations around $
h_{0}$ for a Hamiltonian decomposable as $h=h_{0}+h_{1}$, with $h_{0}\left\vert n\right\rangle =e_{n}^{(0)}\left\vert n\right\rangle $.
According to standard Rayleigh-Schr\"{o}dinger perturbation theory the first
order expansions of the eigenenergies and the eigenstates are
\begin{equation}
e_{n}=e_{n}^{(0)}+\left\langle n\left\vert h_{1}\right\vert n\right\rangle +
\mathcal{O}(\tau ^{2})\quad \text{and}\quad\left\vert \phi
_{n}\right\rangle =\left\vert n\right\rangle +\displaystyle\sum_{k\neq n}\frac{
\left\langle k\right\vert h_{1}\left\vert n\right\rangle }{
e_{n}^{(0)}-e_{k}^{(0)}}\left\vert k\right\rangle +\mathcal{O}(\tau ^{2}),
\label{ersteO}
\end{equation}
respectively, where $\tau$ is the perturbative parameter. Wherever appropriate, we then simply use these expressions in (\ref{klaudercoherent}) for our computations.

We will now construct the GK-coherent states and various expectation values
for the one dimensional harmonic oscillator \cite{bagchi_fring,kempf,dey_fring_gouba}
\begin{equation}
H=\frac{P^{2}}{2m}+\frac{m\omega ^{2}}{2}X^{2}-\hbar \omega \left( \frac{1}{2
}+\frac{\tau }{4}\right)   \label{HD1}
\end{equation}
defined on the noncommutative space satisfying 
\begin{equation}
\lbrack X,P]=i\hbar \left( 1+\check{\tau}P^{2}\right) ,\qquad X=(1+\check{
\tau}p^{2})x,\qquad P=p.  \label{oneoneone}
\end{equation}
Here $\check{\tau}:=\tau /(m\omega \hbar )$ has the dimension of an inverse
squared momentum with $\tau ~$being dimensionless. We have also provided in (\ref{oneoneone}) a representation for the noncommutative variables in terms of the standard canonical variables $x$, $p$ satisfying $[x,p]=i\hbar $. The ground
state energy is conveniently shifted to allow for a factorization of the
energy. The Hamiltonian in (\ref{HD1}) in terms of $x$, $p$ differs from the
one treated recently in \cite{ghosh_roy} as we take a different
representation for $X$ and $P$, which we believe to be incorrect in \cite
{ghosh_roy} even up to $\mathcal{O}(\tau )$. The so-called Dyson map $\eta $,
whose adjoint action relates the non-Hermitian Hamiltonian in (\ref{HD1}) to
its isospectral Hermitian counterpart $h$, is easily found to be $\eta =(1+%
\check{\tau}p^{2})^{-1/2}$. With the help of this expression we evaluate
\begin{equation}
h=\eta H\eta ^{-1}=\frac{p^{2}}{2m}+\frac{m\omega ^{2}}{2}x^{2}+\frac{\omega
\tau }{4\hbar }(p^{2}x^{2}+x^{2}p^{2}+2xp^{2}x)-\hbar \omega \left( \frac{1}{
2}+\frac{\tau }{4}\right) +\mathcal{O}(\tau ^{2}).  \label{h}
\end{equation}
Taking now $h_{0}$ to be the standard harmonic oscillator and
\begin{equation}
h_1=\frac{\omega\tau }{4\hbar }(p^{2}x^{2}+x^{2}p^{2}+2xp^{2}x)-\hbar \omega \left( \frac{1}{2}+\frac{\tau }{4}\right),
\end{equation}
the energy eigenvalues for $H$ and $h$ were computed to lowest order in perturbation theory \cite{kempf,dey_fring_gouba} to 
\begin{equation}
E_{n}=\hbar \omega e_{n}=\hbar \omega n\left[ 1+\frac{\tau }{2}(1+n)\right] +
\mathcal{O}(\tau ^{2}).  \label{En}
\end{equation}
According to (\ref{ersteO}) we now calculate the first order expression for
the wavefunctions to
\begin{equation}
\left\vert \phi _{n}\right\rangle =\left\vert n\right\rangle -\frac{\tau }{16
}\sqrt{(n-3)_{4}}\left\vert n-4\right\rangle +\frac{\tau }{16}\sqrt{(n+1)_{4}
}\left\vert n+4\right\rangle +\mathcal{O}(\tau ^{2}).  \label{phi}
\end{equation}
where $(x)_{n}:=\Gamma (x+n)/\Gamma (x)$ denotes the Pochhammer symbol. We
stress that it is vital to include the second and third term in $\left\vert
\phi _{n}\right\rangle $ in order to achieve an accuracy of order $\tau $
for expectation values. In \cite{ghosh_roy}, where a similar computation was
attempted, these terms were incorrectly ignored. The expression for $E_{n}$
coincides with the one found in \cite{ghosh_roy} for $\tau \rightarrow
2\lambda $, as their computation only involves $\left\langle k\right\vert
h_{1}\left\vert n\right\rangle $. Given $e_{n}$ as defined by the
relation (\ref{En}), we compute the probability density (\ref{probdensity}) and the expansions of its inverse
\begin{equation}
\rho _{n}=\frac{1}{2^{n}}\tau ^{n}n!\left( 2+\frac{2}{\tau }\right)
_{n}+\mathcal{O}(\tau ^{2})\qquad \text{and\qquad }\frac{1}{\rho _{n}}=\frac{1}{n!}-\frac{3+n}{
4(n-1)!}\tau +\mathcal{O}(\tau ^{2}),
\end{equation}
We use the latter expression to evaluate the normalization constant in (\ref{probdensity}) 
\begin{equation}\label{normsqueezed}
\mathcal{N}^{2}(J)=e^{J}\left( 1-\tau J-\frac{\tau }{4}J^{2}\right) +
\mathcal{O}(\tau ^{2}).
\end{equation}
We have now assembled all the necessary quantities to define the GK-coherent
states $\left\vert J,\gamma ,\phi \right\rangle $ in (\ref{klaudercoherent}) and are in the position to verify the validity of some of the crucial requirements on them, test their behaviour and compute expectation values.

As is well known \cite{bender_boettcher,bender_making_sense,scholtz_geyer_hahne,mostafazadeh1,mostafazadeh5}, in a non-Hermitian setting the observables are not dictated by the Hamiltonian
and it becomes a matter of choice to select them or equivalently the metric 
\cite{scholtz_geyer_hahne}. In fact, this is also true for a Hermitian system, where, however, the choice of the standard metric seems to be the most natural one. Here we are mainly interested in the Hamiltonian $H$ of (\ref{HD1}) with $X$ and $P$ as observables, but it will also be instructive to consider first
the Hermitian system described by $h$ with $x$ and $p$ being the observables
of choice.

\subsection{Observables in the Hermitian System}
At first we consider the Hamiltonian $h$ in (\ref{h}) as fundamental and
treat the variables $x$ and $p$ as observables in that system. Expectation values are then most easily computed by taking the states $\left\vert n\right\rangle $ to be the normalized standard Fock space eigenstates of the harmonic oscillator with usual properties $a^{\dagger }\left\vert n\right\rangle =\sqrt{n+1}\left\vert n+1\right\rangle $ and $a\left\vert n\right\rangle =\sqrt{n}\left\vert n-1\right\rangle $. To first order in $\tau $, we then compute the expectation values of the creation and annihilation operators which are explicitly shown below. 

Using the expression (\ref{klaudercoherent}), we first calculate
\begin{equation}
\left\langle J,\gamma ,\phi \right\vert a\left\vert J,\gamma ,\phi
\right\rangle =\frac{1}{\mathcal{N}^{2}}\sum\limits_{n,m=0}^{\infty }\frac{
J^{(m+n)/2}\exp \left[ i\gamma (e_{m}-e_{n})\right] }{\sqrt{\rho _{m}\rho
_{n}}}\left\langle \phi _{m}\right\vert a\left\vert \phi _{n}\right\rangle .
\end{equation}
With the expansion of $\left\vert \phi _{n}\right\rangle $ (\ref{phi}) to first order in $\tau $ we obtain
\begin{eqnarray}
\left\langle \phi _{m}\right\vert a\left\vert \phi _{n}\right\rangle &=&
\sqrt{n}\delta _{m,n-1}+\frac{\tau }{16}\left( \sqrt{(n+1)_{4}}\sqrt{n}-
\sqrt{(n-3)_{4}}\sqrt{n-4}\right) \delta _{m,n-5} \notag\\
&&~~~~~~~~~~~~~\ +\frac{\tau }{16}\left( \sqrt{(n+1)_{4}}\sqrt{n+4}-\sqrt{
(n-3)_{4}}\sqrt{n}\right) \delta _{m,n+3},  \notag \\
&=&\sqrt{n}\delta _{m,n-1}+\frac{\tau }{4}\sqrt{(n+1)(n+2)(n+3)}\delta
_{m,n+3},  
\end{eqnarray}
such that
\begin{equation}
\left\langle J,\gamma ,\phi \right\vert a\left\vert J,\gamma ,\phi
\right\rangle =\sum\limits_{n=1}^{\infty }\frac{\sqrt{n}J^{n-1/2}e^{i\gamma
(e_{n-1}-e_{n})}}{\mathcal{N}^{2}\sqrt{\rho _{n-1}\rho _{n}}}+\tau
\sum\limits_{n=0}^{\infty }\frac{J^{n+3/2}\sqrt{(n+1)_{3}}e^{i\gamma
(e_{n+3}-e_{n})}}{4\mathcal{N}^{2}\sqrt{\rho _{n+3}\rho _{n}}}+\mathcal{O}
(\tau ^{2}).  \label{a3}
\end{equation}
The last sum has been ignored in \cite{ghosh_roy}, but is an important
contribution to order $\tau $. Using $e_{n-1}-e_{n}=-1-n\tau $ and $\rho
_{n}=\rho _{n-1}e_{n}$ the first sum in (\ref{a3}) is evaluated as
\begin{eqnarray}
&&\frac{e^{-i\gamma }}{\mathcal{N}^{2}}\sum\limits_{n=1}^{\infty} \frac{\sqrt{n}J^{n-1/2}e^{-i \gamma n \tau}}{\rho_{n-1}\sqrt{e_n}} =\frac{e^{-i\gamma }}{\mathcal{N}^{2}\sqrt{J}}\sum\limits_{n=1}^{\infty }\frac{
J^n e^{-i\gamma n\tau }}{\rho _{n-1}\sqrt{1+\frac{\tau }{2}(1+n)}} \notag \\
&=&\frac{e^{-i\gamma }}{\mathcal{N}^{2}\sqrt{J}}\sum\limits_{n=1}^{\infty }\frac{J^n}{\rho _{n-1}}\left[ 1-\frac{\tau }{4}(1+n+4i\gamma n)\right] +
\mathcal{O}(\tau ^{2}),~~~~~  \notag \\
&=&\frac{e^{-i\gamma }}{\sqrt{J}}\left[ \left( 1-\frac{\tau }{4}\right) 
\frac{\sum\limits_{n=1}^{\infty }\frac{J^{n}}{\rho _{n-1}}}{
\sum\limits_{n=0}^{\infty }\frac{J^{n}}{\rho _{n}}}-\frac{\tau }{4}
(1+4i\gamma )\frac{\sum\limits_{n=1}^{\infty }\frac{nJ^{n}}{\rho _{n-1}}}{
\sum\limits_{n=0}^{\infty }\frac{J^{n}}{\rho _{n}}}\right] +\mathcal{O}(\tau
^{2}), \notag \\
&=&\sqrt{J}e^{-i\gamma }\left[ 1-\frac{\tau }{4}\left( 2+J+4i\gamma
(1+J)\right) \right] +\mathcal{O}(\tau ^{2}).  \label{a6}
\end{eqnarray}
For the second sum in (\ref{a3}) we use $e_{n+3}-e_{n}=3+3\tau (2+n)$ and $
\rho _{n+3}=\rho _{n}e_{n+3}e_{n+2}e_{n+1}$, such that it becomes
\begin{equation}
\tau \sum\limits_{n=0}^{\infty }\frac{J^{n+3/2\sqrt{(n+1)_{3}}}e^{i\gamma
(3+3\tau (2+n))}}{4\mathcal{N}^{2}\rho _{n}\sqrt{e_{n+3}e_{n+2}e_{n+1}}}+
\mathcal{O}(\tau ^{2})=\frac{\tau J^{3/2e^{3i\gamma }}}{4\mathcal{N}^{2}}
\sum\limits_{n=0}^{\infty }\frac{J^{n}}{\rho _{n}}+\mathcal{O}(\tau ^{2})=
\frac{\tau J^{3/2e^{3i\gamma }}}{4}+\mathcal{O}(\tau ^{2}).  \label{a7}
\end{equation}
Collecting (\ref{a6}) and (\ref{a7}), we obtain 
\begin{equation}
\left\langle J,\gamma ,\phi \right\vert a\left\vert J,\gamma ,\phi
\right\rangle =\sqrt{J}e^{-i\gamma }\left[ 1-\frac{\tau }{4}\left(
2+J+4i\gamma (1+J)\right) \right] +\frac{\tau }{4}J^{3/2}e^{3i\gamma }+
\mathcal{O}(\tau ^{2}).~~  \label{a}
\end{equation}
Following the same procedure, we can also compute the expectation value of the creation operator as 
\begin{equation}
\left\langle J,\gamma ,\phi \right\vert a^{\dagger }\left\vert J,\gamma
,\phi \right\rangle =\sqrt{J}e^{i\gamma }\left[ 1-\frac{\tau }{4}\left(
2+J-4i\gamma (1+J)\right) \right] +\frac{\tau }{4}J^{3/2}e^{-3i\gamma }+
\mathcal{O}(\tau ^{2}).  \label{ad}
\end{equation}
In what follows we will often drop the explicit mentioning of the order in $
\tau $, understanding that all our computations are carried out to first
order. Using the fact that $x=\sqrt{\hbar /(2m\omega )}(a^{\dagger }+a)$ and 
$p=i\sqrt{\hbar m\omega /2}(a^{\dagger }-a)$, the expectation values
\begin{alignat}{1}
\left\langle J,\gamma ,\phi \right\vert x\left\vert J,\gamma ,\phi
\right\rangle  &=\sqrt{\frac{2J\hbar }{m\omega }}\left[ \cos \gamma -\tau
\left[ \gamma \sin \gamma +\frac{\cos \gamma }{2}+J\sin \gamma \left( \gamma
+\frac{\sin 2\gamma }{2}\right) \right] \right], \label{xp} \\
\left\langle J,\gamma ,\phi \right\vert p\left\vert J,\gamma ,\phi
\right\rangle  &=-\sqrt{2Jm\omega \hbar }\left[ \sin \gamma +\tau \left[
\gamma \cos \gamma -\frac{\sin \gamma }{2}+J\cos \gamma \left( \gamma -\frac{
\sin 2\gamma }{2}\right) \right] \right] ,  \notag
\end{alignat}
then follow trivially from (\ref{a}) and (\ref{ad}). Expanding $x^{2}$ and $
p^{2}$ in terms of $a^{\dagger }$ and $a$, a similar, albeit more lengthy,
computation yields
\begin{eqnarray}
\left\langle J,\gamma ,\phi \right\vert x^{2}\left\vert J,\gamma ,\phi
\right\rangle  &=&\frac{\hbar }{2m\omega }\left[ 1+4J\cos ^{2}\gamma -\tau
J\left( 6\gamma \sin 2\gamma +\cos 2\gamma +2\right) \right.   \label{xx2} \\
&&\left. \qquad \quad -\tau J^{2}(4\gamma \sin 2\gamma -\cos 4\gamma +1)
\right] ,  \notag \\
\left\langle J,\gamma ,\phi \right\vert p^{2}\left\vert J,\gamma ,\phi
\right\rangle  &=&\frac{\hbar m\omega }{2}\left[ 1+4J\sin ^{2}\gamma +\tau
J\left( 6\gamma \sin 2\gamma +\cos 2\gamma -2\right) \right.   \label{p2} \\
&&\left. \qquad \quad +\tau J^{2}(4\gamma \sin 2\gamma +\cos 4\gamma -1)
\right] .  \notag
\end{eqnarray}
These two expressions may be used to compute the expectation value of $h$,
as defined in (\ref{h}), with regard to the GK-coherent states. The
remaining term in $h$ only needs to be computed to zeroth order to achieve
an overall accuracy of order $\tau $. We therefore calculate
\begin{equation}
\left\langle J,\gamma ,\phi \right\vert
p^{2}x^{2}+x^{2}p^{2}+2xp^{2}x\left\vert J,\gamma ,\phi \right\rangle =\hbar
^{2}(1+4J+2J^{2}-2J^{2}\cos 4\gamma )+\mathcal{O}(\tau ).
\end{equation}
Summing the contributions from (\ref{xp}), (\ref{xx2}) and (\ref{p2}),
together with the required pre-factors to make up the Hamiltonian $h$,
yields the action identity (\ref{AI}) as expected. We remark that this
crucial identity was violated in \cite{ghosh_roy}.

Employing the above quantities we can also investigate how close the
coherent states approach the minimum uncertainty product. Assembling the
required expectation values we then obtain
\begin{eqnarray}
\Delta x^{2} &=&\left\langle J,\gamma ,\phi \right\vert x^{2}\left\vert
J,\gamma ,\phi \right\rangle -\left\langle J,\gamma ,\phi \right\vert
x\left\vert J,\gamma ,\phi \right\rangle ^{2} \notag\\
&=&\frac{\hbar }{2m\omega }\left[
1+\tau J(\cos 2\gamma -2\gamma \sin 2\gamma )\right] ,~~~~~~ \\
\Delta p^{2} &=&\left\langle J,\gamma ,\phi \right\vert p^{2}\left\vert
J,\gamma ,\phi \right\rangle -\left\langle J,\gamma ,\phi \right\vert
p\left\vert J,\gamma ,\phi \right\rangle ^{2} \notag\\
&=&\frac{\hbar m\omega }{2}\left[
1-\tau J(\cos 2\gamma -2\gamma \sin 2\gamma )\right] ,
\end{eqnarray}
and therefore 
\begin{equation}
\Delta x\Delta p=\frac{\hbar }{2}+\mathcal{O}(\tau ^{2}).
\end{equation}
Thus the states $\left\vert J,\gamma ,\phi \right\rangle $ saturate the
minimal uncertainty in a simultaneous measurement of $x$ and $p$ and
therefore constitute squeezed states for all values of $J$ and $\gamma $ up
to first order in perturbation theory.

Using (\ref{xp}) we also verify Ehrenfest's theorem (\ref{Ehren}) for the
operators $x$ 
\begin{eqnarray}
&&i\hbar \frac{d}{dt}\left\langle J,\gamma +t\omega ,\phi \right\vert
x\left\vert J,\gamma +t\omega ,\phi \right\rangle =\left\langle J,\gamma
+t\omega ,\phi \right\vert [x,h]\left\vert J,\gamma +t\omega ,\phi
\right\rangle , \\
&=&\left\langle J,\gamma +t\omega ,\phi \right\vert \frac{i\hbar }{m}p+\frac{
i\tau \omega }{2}(px^{2}+x^{2}p+2xpx)\left\vert J,\gamma +t\omega ,\phi
\right\rangle  \notag \\
&=&-i\hbar ^{3/2}\sqrt{\frac{2J\omega }{m}}\left[ \sin \hat{\gamma}+\tau 
\left[ \frac{1}{2}\sin \hat{\gamma}+\cos \hat{\gamma}\left( (J+1)\hat{\gamma}
+\frac{3}{2}J\sin 2\hat{\gamma}\right) \right] \right]  \notag
\end{eqnarray}
and $p$
\begin{eqnarray}
&&i\hbar \frac{d}{dt}\left\langle J,\gamma +t\omega ,\phi \right\vert
p\left\vert J,\gamma +t\omega ,\phi \right\rangle =\left\langle J,\gamma
+t\omega ,\phi \right\vert [p,h]\left\vert J,\gamma +t\omega ,\phi
\right\rangle , \\
&=&\left\langle J,\gamma +t\omega ,\phi \right\vert -i\hbar m\omega
^{2}x-i\tau \omega (px^{2}+x^{2}p)\left\vert J,\gamma +t\omega ,\phi
\right\rangle  \notag \\
&=&-i\sqrt{2Jm}\hbar ^{3/2}\omega ^{3/2}\left[ \cos \hat{\gamma}+\frac{\tau 
}{4}\left[ (3J+2)\cos \hat{\gamma}-4(J+1)\hat{\gamma}\sin \hat{\gamma}
-3J\cos 3\hat{\gamma}\right] \right] .  \notag
\end{eqnarray}
For convenience we abbreviated here $\hat{\gamma}:=\gamma +t\omega $.

\subsection{Observables in the Non-Hermitian System}
As stated, the system we actually wish to investigate is described by the
non-Hermitian Hamiltonian (\ref{HD1}) with a non-trivial commutation
relation (\ref{oneoneone}) for its associated observables $X$ and $P$. In order to
test the inequality (\ref{HU}) we need to compute 
\begin{eqnarray}
\left\langle J,\gamma ,\Phi \right\vert X\left\vert J,\gamma ,\Phi
\right\rangle _{\eta } &=&\sqrt{\frac{2J\hbar }{m\omega }}\left[ \cos \gamma
+\frac{\tau }{2}\sin \gamma (J\sin 2\gamma -2\gamma (1+J))\right] ,
\label{X} \\
\left\langle J,\gamma ,\Phi \right\vert X^{2}\left\vert J,\gamma ,\Phi
\right\rangle _{\eta } &=&\frac{\hbar }{2m\omega }\left[ 1+4J\cos ^{2}\gamma
+\tau \lbrack 1+J(2-2\cos 2\gamma -6\gamma \sin 2\gamma )\right. \notag \\
&&+\left. 2J^{2}\sin 2\gamma (\sin 2\gamma -2\gamma )]\right] \label{XX}.
\end{eqnarray}
We note here that the actual computation has been carried out by translating
first all quantities to a Hermitian setting and then following the same
reasoning as in the previous subsection. Combining (\ref{X}) and (\ref{XX})
then yields 
\begin{eqnarray}
\Delta X^{2} &=&\left\langle J,\gamma ,\Phi \right\vert X^{2}\left\vert
J,\gamma ,\Phi \right\rangle _{\eta }-\left\langle J,\gamma ,\Phi
\right\vert X\left\vert J,\gamma ,\Phi \right\rangle _{\eta }^{2}
\label{DXX} \\
&=&\frac{\hbar }{2m\omega }\left[ 1+\tau \left( 1+J(2-2\gamma \sin 2\gamma
-\cos 2\gamma )\right) \right] .  \notag
\end{eqnarray}
The computation for the expectation values of $P$ is simpler, since the
metric commutes with $p$, such that
\begin{eqnarray}
\left\langle J,\gamma ,\Phi \right\vert P\left\vert J,\gamma ,\Phi
\right\rangle _{\eta } &=&\left\langle J,\gamma ,\phi \right\vert p\left\vert
J,\gamma ,\phi \right\rangle , \notag \\
\text{and}\qquad\left\langle J,\gamma ,\Phi\right\vert P^{2}\left\vert J,\gamma ,\Phi \right\rangle _{\eta} &=&\left\langle J,\gamma ,\phi \right\vert p^{2}\left\vert J,\gamma ,\phi
\right\rangle ,
\end{eqnarray}
and therefore
\begin{equation}
~~\Delta P^{2}=\Delta p^{2}.  \label{DP}
\end{equation}
Expanding finally (\ref{DXX}) and (\ref{DP}), we obtain
\begin{equation}
\Delta X\Delta P=\frac{\hbar }{2}\left[ 1+\frac{\tau }{2}\left( 1+4J\sin
^{2}\gamma \right) \right] =\frac{\hbar }{2}\left( 1+\hat{\tau}\left\langle
J,\gamma ,\Phi \right\vert P^{2}\left\vert J,\gamma ,\Phi \right\rangle
\right) .
\end{equation}
This means that also in the non-Hermitian setting the minimal uncertainty product for the observables $X$ and $P$, commuting as specified in (\ref{oneoneone}), is saturated. Thus to first order in perturbation theory also the GK-coherent states $\left\vert J,\gamma ,\Phi \right\rangle $ are squeezed states, remarkably this holds irrespective of the values for $J$ and $\gamma$.

Apparently this result was also obtained in \cite{ghosh_roy}, but our disagreement with the results presented in there is at least fourfold. Firstly, the authors used the incorrect representation for the canonical variables $X$ and $P$ as mentioned earlier. Secondly the authors computed conceptually the wrong expectations values even when using their representation. Thirdly, the authors only take the first order in (\ref{phi}) into account and therefore miss out various terms contributing to the first order calculation in $\tau $. Finally, even ignoring the previous three points and adopting all the wrong concepts used in \cite{ghosh_roy}, we disagree on a purely computational level with many of the expressions presented in there.

Next we also verify Ehrenfest's theorem (\ref{Ehren}) for the operators $X$ 
\begin{eqnarray}
&&i\hbar \frac{d}{dt}\left\langle J,\gamma +t\omega ,\Phi \right\vert
X\left\vert J,\gamma +t\omega ,\Phi \right\rangle _{\eta }=\left\langle
J,\gamma +t\omega ,\Phi \right\vert [X,H]\left\vert J,\gamma +t\omega ,\Phi
\right\rangle _{\eta },\notag \\
&=&\left\langle J,\gamma +t\omega ,\Phi \right\vert \frac{i\hbar }{m}(P+
\check{\tau}P^{3})\left\vert J,\gamma +t\omega ,\Phi \right\rangle _{\eta } 
 \label{EE1} \\
&=&-i\hbar ^{3/2}\sqrt{\frac{2J\omega }{m}}\left[ \sin \hat{\gamma}+\tau
\left[ (J+1)\hat{\gamma}\cos \hat{\gamma}+\frac{1}{2}\sin \hat{\gamma}
(2+J-3J\cos 2\hat{\gamma})\right] \right]  \notag
\end{eqnarray}
and the operator $P$
\begin{eqnarray}
&&i\hbar \frac{d}{dt}\left\langle J,\gamma +t\omega ,\Phi \right\vert
P\left\vert J,\gamma +t\omega ,\Phi \right\rangle _{\eta }=\left\langle
J,\gamma +t\omega ,\Phi \right\vert [P,H]\left\vert J,\gamma +t\omega ,\Phi
\right\rangle _{\eta },  \notag \\
&=&\left\langle J,\gamma +t\omega ,\Phi \right\vert -im\hbar \omega
^{2}\left( X+\frac{\check{\tau}}{2}XP^{2}+\frac{\check{\tau}}{2}
P^{2}X\right) \left\vert J,\gamma +t\omega ,\Phi \right\rangle _{\eta } 
\label{EE2} \\
&=&-i\sqrt{2Jm}\hbar ^{3/2}\omega ^{3/2}\left[ \cos \hat{\gamma}+\frac{\tau 
}{4}\left[ (3J+2)\cos \hat{\gamma}-4(J+1)\hat{\gamma}\sin \hat{\gamma}
-3J\cos 3\hat{\gamma}\right] \right] .  \notag
\end{eqnarray}
Taking now for simplicity $\gamma =0$, differentiating (\ref{EE1}) once again
and combining it with (\ref{EE2}) we obtain the identity corresponding  to
Newton's equation of motion
\begin{equation}
\left\langle J,t\omega ,\Phi \right\vert \ddot{X}\left\vert J,t\omega ,\Phi
\right\rangle _{\eta }=-\omega ^{2}\left\langle J,t\omega ,\Phi \right\vert
X+\frac{\check{\tau}}{2}(3XP^{2}+3P^{2}X+2PXP)\left\vert J,t\omega ,\Phi
\right\rangle _{\eta },  \label{Newton}
\end{equation}
The relations (\ref{EE1}) and (\ref{EE2}) were not recovered in \cite{ghosh_roy}, where the comparison between the left and right hand sides mismatched. Instead of (\ref{Newton}) the authors proposed a "correspondence principle with twist". According to our argumentation this is incorrect and there is in fact no reason to assume the Newton's equation is simply the same as the one for the standard harmonic oscillator. The reason for the discrepancy are the aforementioned conceptual and computational mistakes in \cite{ghosh_roy}.

\section{Coherent States for the Noncommutative Non-Perturbative Harmonic Oscillator} \label{section64}
In this case, we set out by considering an operator $a$ and its adjoint $a^\dagger$, acting in a Hilbert space with basis $\vert n \rangle, n=0,1,2...$, such that
\begin{alignat}{2}
a^\dagger\vert n \rangle &= \sqrt{n+1}~\vert n+1\rangle, \qquad~~ \vert n \rangle &&=\frac{\left(a^\dagger\right)^n}{\sqrt{n!}}~\vert 0\rangle \nonumber \\
a\vert n \rangle &= \sqrt{n}~\vert n-1\rangle, \qquad \qquad a\vert 0\rangle &&= 0, \qquad \text{and} \qquad [a,a^\dagger]=1. \label{FockFock}
\end{alignat}
Now, following refs. \cite{sun_fu,macfarlane,biedenharn,kulish_damaskinsky,arik_coon}, we define a $q$-deformed version of the Fock space (\ref{FockFock}) with basis $\vert n \rangle_q$ involving $q$-deformed integers $[n]_q$, such that
\begin{alignat}{2} \label{qdefalgebra}
A^\dagger\vert n \rangle_q &= \sqrt{[n+1]_q}~\vert n+1\rangle_q, \qquad~~ \vert n \rangle_q &&=\frac{\left(A^\dagger\right)^n}{\sqrt{[n]_q!}}~\vert 0\rangle \nonumber \\
A\vert n \rangle_q &= \sqrt{[n]_q}~\vert n-1\rangle_q, \qquad \qquad A\vert 0\rangle &&= 0, \qquad \text{and} \qquad A A^\dagger-q^2 A^\dagger A=1,
\end{alignat}
with $q \leq 1$ and $[n]_q!=[1]_q[2]_q....[n]_q$. Therefore, we seek an explicit value of $[n]_q$, which can be defined as
\begin{equation}\label{qint}
[n]_q := \frac{1-q^{2n}}{1-q^2}.
\end{equation}
Using the expression of $[n]_q$ from (\ref{qint}), one can easily verify that the Fock space defined in  (\ref{qdefalgebra}) satisfies the $q$-deformed oscillator algebra (\ref{q2deformation}, \ref{AAAA}) $A A^\dagger-q^2 A^\dagger A=1$, which essentially produces the $q$-deformed noncommutative spaces, as explored in section \ref{section22} and \ref{section4.2}. Furthermore, from (\ref{qdefalgebra}) and (\ref{qint}), it can also be deduced that the states $\vert n \rangle_q$ form an orthonormal basis, i.e.,$~_{q}\!\left\langle n\right\vert \!\left. m\right\rangle _{q}=\delta_{n,m}$. As was first argued in \cite{arik_coon}, the $q$-deformed Hilbert space $\mathcal{H}_{q}$ is then spanned by the vectors 
$\left\vert \psi \right\rangle :=\sum_{n=0}^\infty c_{n}\left\vert n\right\rangle _q$ with $c_{n}\in \mathbb{C}$, such that $\left\langle \psi \right\vert \!\left. \psi\right\rangle =\sum_{n=0}^{\infty }\left\vert c_{n}\right\vert^{2}<\infty $.

Let us now work out a concrete example that follows from the deformed algebra (\ref{qdefalgebra}). In section \ref{section22}, we started with the q-deformed oscillator algebra generators (\ref{qXP}) $X=\alpha\left(A^\dagger+A\right)$ and $P=i\beta\left(A^\dagger-A\right)$ and constructed the commutation relation (\ref{XPcompartial})
\begin{equation}\label{XPcompartial1}
\left[X,P\right]=\frac{4i\alpha\beta}{1+q^2}\left[1+\frac{q^2-1}{4}\left(\frac{X^2}{\alpha^2}+\frac{P^2}{\beta^2}\right)\right].
\end{equation}
Here instead of taking the nontrivial limit to reduce the expression (\ref{XPcompartial1}) into the simpler form (\ref{ncsimplecommutator}), we choose the parameters $\alpha$ and $\beta$ to be of the form
\begin{equation}\label{alphabeta}
\alpha=\frac{1}{2}\sqrt{1+q^2}\sqrt{\frac{\hbar}{m \omega}} \qquad \text{and} \qquad \beta=\frac{1}{2}\sqrt{1+q^2}\sqrt{\hbar m \omega}
\end{equation}
so that we end up with the commutator
\begin{equation}\label{XPcomcomplete}
\left[X,P\right]=i\hbar+i\frac{q^2-1}{q^2+1}\left(m \omega X^2+\frac{1}{m \omega}P^2\right).
\end{equation}
The interesting feature about this version of a noncommutative space-time is that it leads to a minimal length as well as a minimal momentum as discussed in chapter \ref{chapter2} and \ref{noncommutativemodelsin3D}. Later, we will analyse the generalised uncertainty relation in the context of coherent states and focus on the comparison between the classical and quantum description of the system as done in the last section. However, before doing that, we seek a concrete noncommutative model based on the algebra (\ref{XPcompartial1}), on which we build the coherent states up. 

Unlike the previous case where we picked up a perturbative noncommutative model, here we choose the Harmonic oscillator Hamiltonian in the noncommutative space
\begin{equation}\label{nchoexact}
H=\hbar \omega\left(A^\dagger A+1\right),
\end{equation}
where $A$ and $A^\dagger$ satisfy the $q$-deformed algebra (\ref{qdefalgebra}). Therefore the eigenvalue and the eigenstate are written as
\begin{equation} \label{nchoeigen}
E_n=\hbar \omega e_n=\hbar \omega [n]_q \qquad \text{and} \qquad \left\vert\phi_n\right\rangle=\left\vert n\right\rangle_q
\end{equation}
respectively. By replacing the $q$-deformed creation and annihilation operators
\begin{equation}
A^\dagger=\frac{1}{2}\left(\frac{X}{\alpha}-\frac{i P}{\beta}\right) \qquad \text{and} \qquad A=\frac{1}{2}\left(\frac{X}{\alpha}+\frac{i P}{\beta}\right)
\end{equation} 
together with the values of $\alpha$ and $\beta$ (\ref{alphabeta}) and utilizing the generalised uncertainty relation (\ref{XPcomcomplete}), we obtain the explicit form of the noncommutative Harmonic oscillator in terms of the position and momentum observables in noncommutative space as
\begin{equation}
H=\frac{2}{\left(1+q^2\right)^2}\left[m \omega^2 X^2+\frac{1}{m}P^2+\frac{\hbar \omega}{8}\left(q^4-2q^2-3\right)\right].
\end{equation}
We are now ready to construct the coherent states for the non-perturbative Harmonic oscillator model. However, before proceeding to the construction of the coherent states it would be interesting to produce the Hermitian version of the representation of the non-Hermitian observables $X$ and $P$, so that we can argue that the model is more physical in spite of the fact that the Hamiltonian is non-Hermitian. Notice that when we assume that the conjugation of $A$ and $A^{\dagger }$ yield $A^{\dagger }$ and $A$, respectively, the operators $X$ and $P$ can be seen as Hermitian. In that case the metric $\eta $ is taken to be the
standard one, possibly with some change to ensure proper self-adjointness
and the convergence of the inner products. Indeed, in \cite{macfarlane,atakishiyev_nagiyev} such a representation on a unit circle acting on Rogers-Sz\"{e}go polynomials \cite{andrews} was derived,
\begin{equation}
A=\frac{i}{\sqrt{1-q^{2}}}\left( e^{-i\check{x}}-e^{-i\check{x}/2}e^{2\tau 
\check{p}}\right) ,\qquad \text{and\qquad }A^{\dagger }=\frac{-i}{\sqrt{
1-q^{2}}}\left( e^{i\check{x}}-e^{2\tau \check{p}}e^{i\check{x}/2}\right) .
\label{rep1}
\end{equation}
Here we used the dimensionless variables $\check{x}=x\sqrt{m\omega /\hbar 
\text{ }}$ and $\check{p}=p/\sqrt{m\omega \hbar \text{ }}$ with $x$, $p$
being the standard canonical coordinates satisfying $\left[ x,p\right]
=i\hbar $. Evidently $A^{\dagger }$ is the conjugate of $A$ for $q<1$ and
consequently it follows that also the operators $X$ and $P$ satisfying (\ref{XPcomcomplete}) are Hermitian in this representation, i.e. $X^{\dagger }=X$, $P^{\dagger }=P$. We notice further that for the representation (\ref{rep1}) the $\mathcal{PT}$-symmetry of the standard canonical variables $\mathcal{PT}$: $x\rightarrow -x$, $p\rightarrow p$, $i\rightarrow -i$ is inherited by canonical variables on the noncommutative space $\mathcal{PT}$: $X\rightarrow -X$, $P\rightarrow P$, $i\rightarrow -i$.

There exist also alternative representations \cite{burban_klimyk} 
\begin{equation}
A=\frac{1}{1-q^{2}}D_{q},\qquad \text{and\qquad }A^{\dagger
}=(1-x)-x(1-q^{2})D_{q},  \label{Dq}
\end{equation}
in terms of Jackson derivatives $D_{q}f(x):=[f(x)-f(q^{2}x)]/[x(1-q^{2})]$
introduced in \cite{jackson1}. The operators in (\ref{Dq}) satisfy the fifth relation of (\ref{qdefalgebra}) when acting on eigenvectors constructed from normalized Rogers-Sz\"{e}go polynomials. It is less obvious to see whether this representation can be made Hermitian. For our purposes it is important that at least one such representation exists and we may compute expectation values on the $q$-deformed Fock space with the standard metric.

Having discussed the physical implications of the representation, we now construct the coherent states. Using the harmonic oscillator model (\ref{nchoexact}) that we have constructed above, together with the eigenvalues and eigenvectors defined in (\ref{nchoeigen}), we obtain the probability distribution (\ref{probdensity}) $\rho_n=[n]_q!$. We use the standard convention $\rho_0=[0]_q!=1$, so that the Gazeau-Klauder axiom (\ref{actionangle}) is satisfied. Furthermore, the normalisation condition $\langle J,\gamma\vert J,\gamma\rangle=1$ yields the $q$-deformed exponential $E_q(J)$ as the normalisation constant
\begin{equation}
E_{q}(J):=\displaystyle\sum_{n=0}^{\infty }\frac{J^{n}}{[n]_{q}!}=\mathcal{N}
^{2}(J).
\end{equation}
Thus our normalized coherent state 
\begin{equation}
\left\vert J,\gamma \right\rangle _{q}:=\frac{1}{\sqrt{E_{q}(J)}}
\displaystyle\sum_{n=0}^{\infty }\frac{J^{n/2}\exp (-i\gamma e_{n})}{\sqrt{[n]_{q}!
}}\left\vert n\right\rangle _{q},  \label{coho}
\end{equation}
coincides with the coherent state $\left\vert z\right\rangle $, as defined
already in \cite{arik_coon}, for the specific choice $\left\vert
z^{2},0\right\rangle _{q}$, that is for $t=0$. Let us now investigate some
properties of these states and in particular investigate to which kind of
expectation values they lead for observables and compare with the results
for the nontrivial $q\rightarrow 1$ limit studied in the previous section \cite{dey_fring_squeezed}. In the latter case these states were found to be squeezed states up to first order in perturbation theory in $\tau $ when parametrizing the deformation parameter as $q^{\tau }$. Most importantly we wish to investigate whether these states respect the generalized uncertainty relations.

\subsection{Generalised Heisenberg's Uncertainty Relations}
Let us first analyse the generalized version of Heisenberg's uncertainty relation for a simultaneous measurement of the two observables $X$ and $P$ projected onto the normalized coherent states $\left\vert J,\gamma \right\rangle _{q}$ as defined in equation (\ref{coho}) 
\begin{equation}
\left. \Delta X\Delta P\right\vert _{\left\vert J,\gamma \right\rangle
_{q}}\geq \frac{1}{2}\left\vert \left( _{q}\!\left\langle J,\gamma
\right\vert [X,P]\left\vert J,\gamma \right\rangle _{q}\right) _{\eta
}\right\vert .  \label{GHU}
\end{equation}
The uncertainty for $X$ is computed as $\Delta X^{2}=\left(
_{q}\!\left\langle J,\gamma \right\vert X^{2}\left\vert J,\gamma
\right\rangle _{q}\right) _{\eta }-\left( _{q}\!\left\langle J,\gamma
\right\vert X\left\vert J,\gamma \right\rangle _{q}\right) _{\eta }^{2}$ and
analogously for $P$ with $X\rightarrow P$. The $\eta $ indicates that we
might have to change to a nontrivial metric when $X$ and/or $P$ are
non-Hermitian following the prescriptions provided in the recent literature
on non-Hermitian systems \cite{bender_boettcher,bender_making_sense,scholtz_geyer_hahne,mostafazadeh1,mostafazadeh5} or more specifically for this particular setting in \cite{dey_fring_squeezed}.

In order to verify the inequality (\ref{GHU}) for the states (\ref{coho}) we
compute first the expectation values for the creation and annihilation
operators
\begin{equation}
_{q}\!\left\langle J,\gamma \right\vert A\left\vert J,\gamma \right\rangle
_{q}=\frac{1}{E_q(J)}\displaystyle\sum_{m,n=0}^\infty\frac{J^{(m+n)/2}e^{i\gamma\left(e_m-e_n\right)}\sqrt{[n]_q}}{\sqrt{[n]_q![m]_q!}}\delta_{m,n-1} \nonumber
\end{equation}
Using $e_{n-1}-e_n=[n-1]_q-[n]_q=-q^{2n-2}$, we can write
\begin{equation}
_{q}\!\left\langle J,\gamma \right\vert A\left\vert J,\gamma \right\rangle
_{q}=\frac{1}{E_q(J)}\displaystyle\sum_{n=1}^\infty \frac{J^{n-1/2}e^{-i\gamma q^{2n}}}{[n-1]_q!}=\frac{1}{E_q(J)}\displaystyle\sum_{n=0}^\infty \frac{J^{n+1/2}}{[n]_q!}e^{-i\gamma q^{2n}} \nonumber
\end{equation}
Introducing the function
\begin{equation}\label{Fq}
F_{q}(J,\gamma ):=\displaystyle\sum_{n=0}^{\infty }\frac{J^{n}e^{i\gamma q^{2n}}}{
[n]_{q}!}=\displaystyle\sum_{n=0}^{\infty }\frac{i^{n}}{n!}E_{q}(q^{2n}J)\gamma
^{n},
\end{equation}
we can write down the expectation values of the creation and annihilation operators as
\begin{equation}
_{q}\!\left\langle J,\gamma \right\vert A\left\vert J,\gamma \right\rangle
_{q}=J^{1/2}\frac{F_{q}(J,-\gamma )}{E_{q}(J)},\qquad \text{and\qquad }
_{q}\!\left\langle J,\gamma \right\vert A^{\dagger }\left\vert J,\gamma
\right\rangle _{q}=J^{1/2}\frac{F_{q}(J,\gamma )}{E_{q}(J)}~. \label{AA}
\end{equation}
Notice that the function (\ref{Fq}) reduces to the $q$-deformed exponential $F_{q}(J,0)=E_{q}(J)$ and also the duality in the derivatives with respect to
the two parameters. The standard derivative with respect to $\gamma $
corresponds to a $q$-deformation in the parameter $J$
\begin{equation}
-i\frac{d}{d\gamma }F_{q}(J,\gamma )=F_{q}(q^{2}J,\gamma )  \label{ID1}
\end{equation}
and in turn the Jackson derivative acting on $J$ is identical to a deformation in the second parameter
\begin{equation}
D_{q}F_{q}(J,\gamma )=\frac{F_{q}(J,\gamma )-F_{q}(q^{2}J,\gamma )}{
J(1-q^{2})}=F_{q}(J,q^{2}\gamma ).  \label{ID2}
\end{equation}
These identities are easily derived from the defining relations for $F_{q}$
and will be made use of below. Using the representations for $X=\alpha \left(A^\dagger+A\right)$ and $P=i\beta \left(A^\dagger-A\right)$ in terms of the creation and annihilation operators, it follows directly with the help of (\ref{AA}) that
\begin{eqnarray}
_{q}\!\left\langle J,\gamma \right\vert X\left\vert J,\gamma \right\rangle
_{q} &=&\frac{\alpha J^{1/2}}{E_{q}(J)}\left[ F_{q}(J,\gamma
)+F_{q}(J,-\gamma )\right] , \\
_{q}\!\left\langle J,\gamma \right\vert P\left\vert J,\gamma \right\rangle
_{q} &=&\frac{i\beta J^{1/2}}{E_{q}(J)}\left[ F_{q}(J,\gamma
)-F_{q}(J,-\gamma )\right] .
\end{eqnarray}
To compute the expectation values for $X^{2}$ and $P^{2}$, we use once again the representation of $X$ and $P$ to express them in terms of the $A^{\dagger }$ and $A$. Thus we evaluate 
\begin{eqnarray}
_{q}\!\left\langle J,\gamma \right\vert A^{\dagger }A^{\dagger }\left\vert
J,\gamma \right\rangle _{q} &=&J\frac{F_{q}(J,\gamma (1+q^{2}))}{E_{q}(J)},
\label{AA1} \\
_{q}\!\left\langle J,\gamma \right\vert AA\left\vert J,\gamma \right\rangle
_{q} &=&J\frac{F_{q}(J,-\gamma (1+q^{2}))}{E_{q}(J)}, \\
_{q}\!\left\langle J,\gamma \right\vert A^{\dagger }A\left\vert J,\gamma
\right\rangle _{q} &=&J, \\
_{q}\!\left\langle J,\gamma \right\vert AA^{\dagger }\left\vert J,\gamma
\right\rangle _{q} &=&1+q^{2}J,  \label{AA4}
\end{eqnarray}
and with $X^{2}=\alpha ^{2}(A^{\dagger }A^{\dagger }+A^{\dagger
}A+AA^{\dagger }+AA)$ and $P^{2}=-\beta ^{2}(A^{\dagger }A^{\dagger
}-A^{\dagger }A-AA^{\dagger }+AA)$ we assemble this to 
\begin{alignat}{1}
_{q}\!\left\langle J,\gamma \right\vert X^{2}\left\vert J,\gamma
\right\rangle _{q} &=\alpha ^{2}\left[ J\frac{F_{q}(J,\gamma
(1+q^{2}))+F_{q}(J,-\gamma (1+q^{2}))}{E_{q}(J)}+1+J+q^{2}J\right] , \\
_{q}\!\left\langle J,\gamma \right\vert P^{2}\left\vert J,\gamma
\right\rangle _{q} &=-\beta ^{2}\left[ J\frac{F_{q}(J,\gamma
(1+q^{2}))+F_{q}(J,-\gamma (1+q^{2}))}{E_{q}(J)}-1-J-q^{2}J\right] .
\end{alignat}
From these expressions we find that the right hand side of the generalized
Heisenberg's inequality (\ref{GHU}) is always a constant value independent
of $\gamma $, i.e. time, 
\begin{equation}
\frac{1}{2}\left\vert _{q}\!\left\langle J,\gamma \right\vert \hbar +\frac{
q^{2}-1}{q^{2}+1}\left( m\omega X^{2}+\frac{1}{m\omega }P^{2}\right)
\left\vert J,\gamma \right\rangle _{q}\right\vert =\frac{\hbar }{4}(1+q^{2})
\left[ 1+(q^{2}-1)J\right] .  \label{RHS}
\end{equation}
The square of the left hand side of (\ref{GHU}) \ can be written as 
\begin{equation}
\left. \Delta X^{2}\Delta P^{2}\right\vert _{\left\vert J,0\right\rangle
_{q}}=\alpha ^{2}\beta ^{2}\left[ 1+(1+q^{2})J+G_{q}-G_{c}^{2}(\gamma )
\right] \left[ 1+(1+q^{2})J-G_{q}-G_{s}^{2}(\gamma )\right] ,  \label{LHS}
\end{equation}
where we introduced the functions
\begin{equation}
G_{c}(\gamma ):=\frac{2\sqrt{J}}{E_{q}(J)}\displaystyle\sum_{n=0}^{\infty }\frac{
J^{n}}{[n]_{q}!}\cos (\gamma q^{2n}),~\qquad ~\text{~}G_{s}(\gamma ):=\frac{
2i\sqrt{J}}{E_{q}(J)}\displaystyle\sum_{n=0}^{\infty }\frac{J^{n}}{[n]_{q}!}\sin
(\gamma q^{2n}),  \label{G}
\end{equation}
and $G_{q}:=\sqrt{J}G_{c}(\gamma +\gamma q^{2})$. Noting that $\lim_{\gamma \rightarrow 0}G_{q}=2J$, $\lim_{\gamma\rightarrow 0}G_{c}(\gamma )=2\sqrt{J}$ and $\lim_{\gamma\rightarrow 0}G_{s}(\gamma )=0$, it is easy to see that for $\gamma =0$ the
expression (\ref{LHS}) becomes the square of (\ref{RHS}), such that the
minimal uncertainty product for the observables $X$ and $P$ is saturated.
From the expressions in (\ref{G}) we deduce that the range for these
functions is $-2J\leq G_{q}\leq 2J$, $0\leq G_{c}^{2}(\gamma )\leq 4J$ and $
-4J\leq G_{s}^{2}(\gamma )\leq 0$. Recognizing next that the inequality
holds when each of the brackets in (\ref{LHS}) is greater than $1+(q^{2}-1)J$
, this requires that $2J\geq G_{c}^{2}(\gamma )-G_{q}$ and at the same time $
2J\geq G_{s}^{2}(\gamma )+G_{q}$. This means $4J\geq G_{c}^{2}(\gamma
)+G_{s}^{2}(\gamma )$, which by the previous estimates is indeed the case.
Overall this implies that for $\gamma \neq 0$ the uncertainty relation (\ref
{GHU}) is always respected.

Next we verify Ehrenfest's theorem. For the time evolution of the operator $X$ we compute directly
\begin{equation}
i\hbar \frac{d}{dt}~_{q}\!\left\langle J,\omega t\right\vert X\left\vert
J,\omega t\right\rangle _{q}=-\frac{\omega \hbar \alpha J^{1/2}}{E_{q}(J)}
\left[ F_{q}(q^{2}J,\omega t)-F_{q}(q^{2}J,-\omega t)\right] ,  \label{E1}
\end{equation}
and compare it to
\begin{equation}
_{q}\!\left\langle J,\omega t\right\vert \left[ X,H\right] \left\vert
J,\omega t\right\rangle _{q}=-\frac{\omega \hbar \alpha J^{1/2}}{E_{q}(J)}
\displaystyle\sum_{s=\pm \omega t}\frac{s}{\omega t}F_{q}(J,s)+\frac{s}{\omega t}
J(q^{2}-1)F_{q}(J,q^{2}s),  \label{E2}
\end{equation}
which is easily computed from the expectation values
\begin{eqnarray}
_{q}\!\left\langle J,\gamma \right\vert A^{\dagger }A^{\dagger }A\left\vert
J,\gamma \right\rangle _{q} &=&J^{3/2}\frac{F_{q}(J,q^{2}\gamma )}{E_{q}(J)},
\label{AAA1} \\
_{q}\!\left\langle J,\gamma \right\vert A^{\dagger }AA^{\dagger }\left\vert
J,\gamma \right\rangle _{q} &=&J^{1/2}\frac{F_{q}(J,\gamma )}{E_{q}(J)}
+q^{2}J^{3/2}\frac{F_{q}(J,q^{2}\gamma )}{E_{q}(J)}, \\
_{q}\!\left\langle J,\gamma \right\vert A^{\dagger }AA\left\vert J,\gamma
\right\rangle _{q} &=&J^{3/2}\frac{F_{q}(J,-q^{2}\gamma )}{E_{q}(J)}, \\
_{q}\!\left\langle J,\gamma \right\vert AA^{\dagger }A\left\vert J,\gamma
\right\rangle _{q} &=&J^{1/2}\frac{F_{q}(J,-\gamma )}{E_{q}(J)}+q^{2}J^{3/2}
\frac{F_{q}(J,-q^{2}\gamma )}{E_{q}(J)}\text{.}  \label{AAA4}
\end{eqnarray}
The equality of (\ref{E1}) and (\ref{E2}) follows from the identities (\ref
{ID1}) and (\ref{ID2}). Similarly we verified the validity of Ehrenfest's
theorem also for the operator $P$.

\section{Fractional Revival Structure} \label{section65}
In the last two sections, we have analysed the mathematical properties of coherent states for both the case of perturbative and non-perturbative noncommutative harmonic oscillator and observed the fact that the GK-coherent states for the perturbative noncommutative harmonic oscillator indeed yield a squeezed coherent state for both the Hermitian and non-Hermitian settings up to the first order. On the other hand the non-perturbative case produces a squeezed state for a specific value of $\gamma$. We have also illustrated the fact that both of the models satisfy the classical action angle identity and other necessary relations for the comparison between the classical and quantum description of the systems. In this section, we explore further physical insights into the comparison from the analysis of the revival structure of the wave packets.

It was at the dawn of the quantum mechanics, when people raised the problem of comparing the quantum and classical dynamics of the systems. The idea was immensely enhanced thereafter by many people and was studied in connection with the quasi-classical quantization of highly excited multi-dimensional quantum systems \cite{littlejohn}. In the classical limit, $\hbar \rightarrow 0$ (i.e. at the energy range which corresponds to the large quantum numbers $n$), the energy spectrum of systems has a quasi-equidistant characteristics and the energy difference between two adjacent levels is proportional to the classical frequency ($\omega_{\text{cl}}$) of the system, so that the classical period becomes
\begin{equation}
T_{\text{cl}}=\frac{2 \pi \hbar}{\Delta E_n}.
\end{equation}
The transition to the classical description also requires the consideration of the wave packet, i.e. the number of states constituting the wave packet must be large enough. It is because of the fact that the wave packet consisting of small number of waves manifest nonclassical behaviour even for large quantum numbers. 

However, a wave packet consisting of sufficiently large number of waves, might even spread out in the situation when one operates in much smaller time interval than the classical period. But, quite interestingly, the spreading does not sustain forever, rather it regains its initial shape completely after the classical period, due to the equidistant character of the spectrum of the states it is formed by. In this sense, the wave packet dynamics might be interpreted by means of a quantum beats among the large number of states \cite{averbukh_kovarsky_perelman} and the correspondence between the quantum and classical dynamics is retained for infinitely long time.

At this point one should note that the picture explained above holds only for strict equality of the energy level spacing, for instance for the case of simple harmonic oscillator. This is no longer true in the general scenario, rather, the wave packet is affected in this case (even in the domain of high quantum numbers), by the slight anharmonicity produced by the unequal energy spacing and therefore the resulting quantum dephasing will inevitably lead to the destruction of the wave packet and will restrict the duration of its classical like evolution. 

However, in several papers devoted to the numerical investigations of the long term evolution of Rydberg wave packets \cite{parker_stroud,alber_ritsch_zoller} and to the evolution of some nonlinear systems \cite{milburn,yurke_stoler,mecozzi_tombesi}, it was discovered that this dephasing does not correspond to the spreading out of the wave packet, rather it is also possible to regain its shape completely after a time $t=T_\text{rev}$, known as the "revival time" in the literature, which was calculated later in an analytical fashion in \cite{averbukh_perelman}. 

Consider a wave packet consisting of large number of waves from the highly excited discrete states of a quantum system
\begin{equation}\label{wavepacketstrong}
\left\vert\psi(t)\right\rangle =\displaystyle\sum_{n\geq 0} c_n e^{-i E_n t/\hbar}\vert \phi_n\rangle \qquad \text{with} \qquad \displaystyle\sum_{n=0}^\infty \left\vert c_n\right\vert^2=1.
\end{equation}
To understand the localisation of the wavepacket, it is crucial to provide some general precision of the statistical nature of the weighting function $\left\vert c_n\right\vert^2$ and that is why it is worthwhile to compare the weighting function $\left\vert c_n\right\vert^2$ with the Poisson distribution function
\begin{equation}
\frac{\langle n \rangle^n e^{-\langle n\rangle}}{n!}.
\end{equation}
The deviation from the Poissonian case can be estimated by the quantitative measurement of the Mandel parameter $Q$ \cite{mandel,mandel1,short_mandel}, which is defined as follows
\begin{equation}\label{mandel}
Q=\frac{\left(\Delta n\right)^2}{\langle n \rangle}-1.
\end{equation}
Therefore, in the Poissonian case, the value of the Mandel parameter $Q=0$. In the case $Q<0$, we say that the weighting distribution is sub-Poissonian and in the case of $Q>0$, we say the distribution to be super-Poissonian. We also refer the case $Q\approx 0$ to the quasi-Poissonian case. Note that the super-Poissonian case corresponds to the spreading of the wave packet, whereas the Poissonian, quasi-Poissonian and the sub-Poissonian case imply to the well localised wave packet, which is actually what we need in our analysis. 

Now considering the fact that the wave packet (\ref{wavepacketstrong}) is strongly weighted around a mean value $\langle n \rangle$ for the number operator $N$ ($N\vert n\rangle=n\vert n \rangle$) :
\begin{equation}
\langle \psi \vert N\vert\psi\rangle=\displaystyle\sum_{n=0}^\infty n\vert c_n\vert^2\equiv\langle n \rangle, 
\end{equation}
and the spreading $\sigma \approx\Delta n \equiv \sqrt{\left\langle n^2\right\rangle-\langle n\rangle^2}$ is small compared with $\langle n \rangle$, we can expand the energy eigenvalue $E_n$ in a Taylor series in $n$ around $\bar{n}$
\begin{equation}\label{energyTaylor}
E_n\simeq E_{\bar{n}}+E_{\bar{n}}'\left(n-\bar{n}\right)+\frac{1}{2!}E_{\bar{n}}''\left(n-\bar{n}\right)^2+\frac{1}{3!}E_{\bar{n}}'''\left(n-\bar{n}\right)^3+....~~,
\end{equation}
where $\bar{n} \in N$ be the integer closest to $\langle n\rangle$, so that the classical period, revival time, super revival time etc. are defined to be
\begin{equation}\label{clrev}
T_{\text{cl}}=\frac{2 \pi \hbar}{\left\vert E_{\bar{n}}'\right\vert},\qquad T_{\text{rev}}=\frac{2 \pi \hbar\times 2!}{\left\vert E_{\bar{n}}''\right\vert} \qquad T_{\text{suprev}}=\frac{2 \pi \hbar\times 3!}{\left\vert E_{\bar{n}}'''\right\vert}.
\end{equation}
In order to visualize all the above analysis into a pictorial presentation, an efficient method is to calculate the autocorrelation function \cite{bluhm_kostelecky_porter} of the wave packet (\ref{wavepacketstrong})
\begin{equation}\label{autocorrelation}
A(t)=\langle \psi(0)\vert \psi(t)\rangle=\displaystyle\sum_{n=0}^\infty \vert c_n \vert ^2e^{-iE_n t/\hbar}~~.
\end{equation}
Numerically $\left\vert A(t)\right\vert^2$ varies between $0$ and $1$. The maximum $\left\vert A(t)\right\vert^2=1$ is reached when $\psi(t)$ exactly matches the initial wave packet $\psi(0)$, and the minimum corresponds to the case where $\psi(t)$ is far from the initial state. On the other hand, fractional revivals and fractional super revivals appear as periodic peaks in $\left\vert A(t)\right\vert^2$ with periods that are rational fractions of the revival time $T_{\text{rev}}$ and super-revival time $T_{\text{suprev}}$.

\subsection{Fractional Revival Structure from Perturbative Noncommutative Harmonic Oscillator}
Let us now see how we obtain the effects explained above from some specific examples. First we go back to the case of perturbative harmonic oscillator case (section \ref{section63}), where we defined the coherent states as
\begin{equation}\label{infsum}
\vert J,\gamma,\phi\rangle=\displaystyle\sum_{n=0}^\infty c_n(J)e^{-i E_n t/\hbar}\vert\phi_n\rangle \qquad \text{with} \qquad c_n(J)=\frac{J^{n/2}}{\mathcal{N}(J)\sqrt{\rho_n}}~~.
\end{equation}
Note that in our considerations, the situation does not correspond to the simple harmonic oscillator case with equidistant energy spectrum in the domain of high quantum numbers, rather the case correlates with the unequal energy spacing which will inevitably lead to the fractional revival structure as illustrated above. The average values of the number operator $N$ and its square $N^2$, can be computed as
\begin{equation} \label{NNA}
\langle n \rangle=\frac{J}{\mathcal{N}^2(J)}\frac{d}{dJ}\mathcal{N}^2(J), \qquad \langle n^2 \rangle=\frac{J}{\mathcal{N}^2(J)}\frac{d}{dJ}J\frac{d}{dJ}\mathcal{N}^2(J)~~.
\end{equation}
We can calculate the precise expressions (\ref{NNA}) with the substitution of the value of the normalisation constant $\mathcal{N}^2(J)$, which has been calculated  for this case in (\ref{normsqueezed})
\begin{equation}
\left\langle n\right\rangle =J-\tau \left( J+\frac{J^{2}}{2}\right) +
\mathcal{O}(\tau ^{2}),\quad \text{and\quad }\left\langle n^{2}\right\rangle
=J+J^{2}-\tau \left( J+3J^{2}+J^{3}\right) +\mathcal{O}(\tau ^{2}),
\label{nb}
\end{equation}
such that
\begin{equation}
\Delta n^{2}=\left\langle n^{2}\right\rangle -\left\langle n\right\rangle
^{2}=J-\tau \left( J+J^{2}\right) +\mathcal{O}(\tau ^{2}).
\end{equation}
Consequently the Mandel parameter (\ref{mandel}) turns out to be negative
\begin{equation}
Q:=\frac{\Delta n^{2}}{\left\langle n\right\rangle }-1=-\frac{J\tau }{2}+
\mathcal{O}(\tau ^{2})<0,
\end{equation}
suggesting a sub-Poissonian statistics. This implies that we have a strong
localization around $\bar{n}$, which is essential to obtain the revival structure of the classical-like sub-wave packet. Therefore we can expand the energy eigenvalue (\ref{En}) in a Taylor series around $\bar{n}\approx n$ as stated in (\ref{energyTaylor}), to obtain the classical period and the revival time (\ref{clrev})
\begin{equation}
T_{\text{cl}}=\frac{2\pi }{\omega }-\frac{\tau }{\omega }(1+2J)\pi ,\quad
\quad \text{and\quad \quad }T_{\text{rev}}=\frac{4\pi }{\omega \tau }.
\label{TT}
\end{equation}
Observe that there is no super-revival time here, because the energy is a quadratic function of $n$. We now use all the above quantities to analyse the behaviour of the autocorrelation function (\ref{autocorrelation})
\begin{equation}
A(t):=\left\vert \left\langle J,\gamma ,\phi \right. \left\vert J,\gamma
+t\omega ,\phi \right\rangle \right\vert ^{2}=\left\vert \left\langle
J,\gamma ,\Phi \right. \left\vert J,\gamma +t\omega ,\Phi \right\rangle
_{\eta }\right\vert ^{2}.
\end{equation}
In order to find a set of meaningful values for our free parameters $J$, $
\tau $ and also to find an appropriate upper limit cutoff in the sum (\ref
{infsum}), let us first investigate the weighting function $c_{n}(J)$.
\begin{figure}[h]
\centering   \includegraphics[width=7.5cm,height=6.0cm]{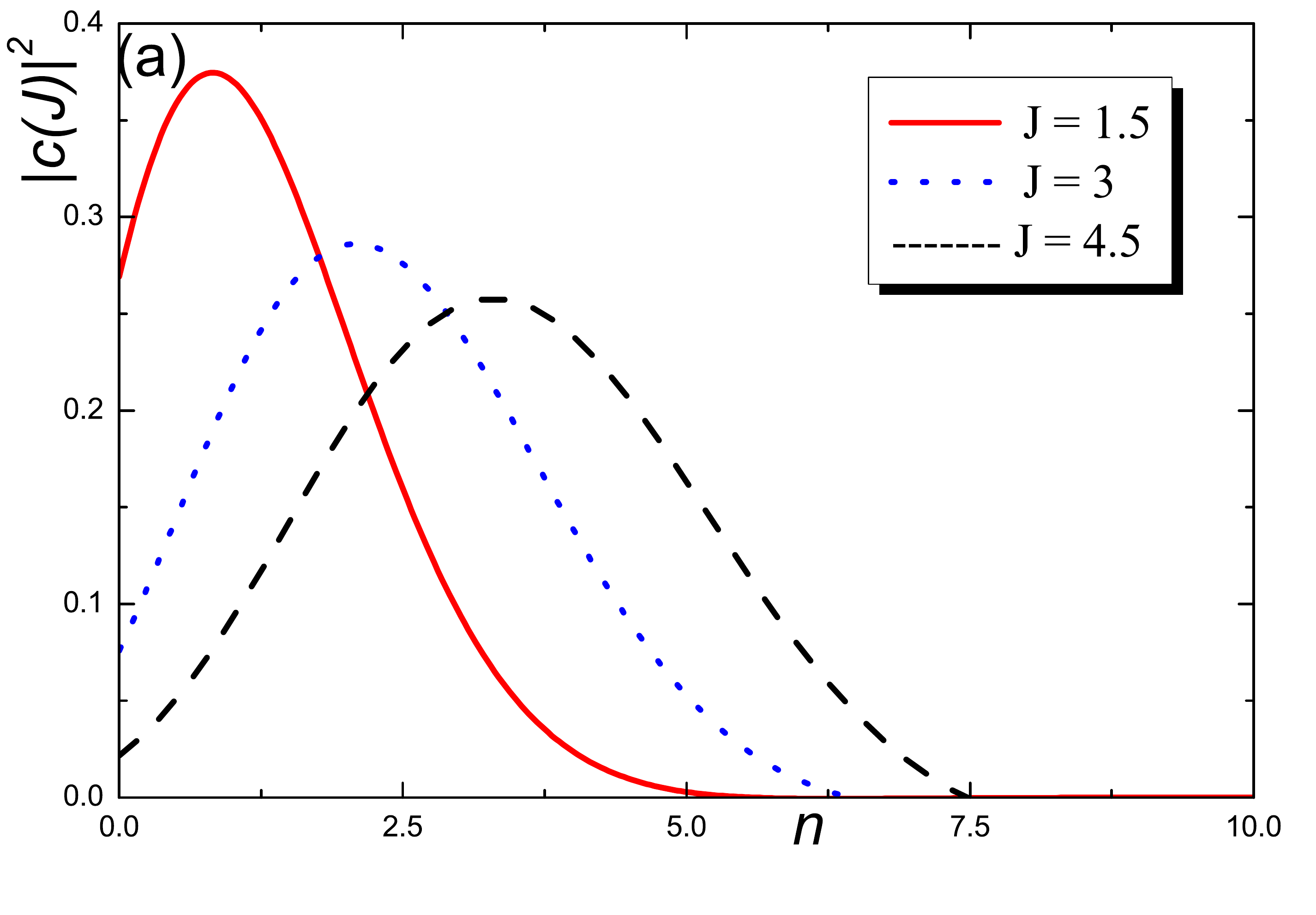} %
\includegraphics[width=7.5cm,height=6.0cm]{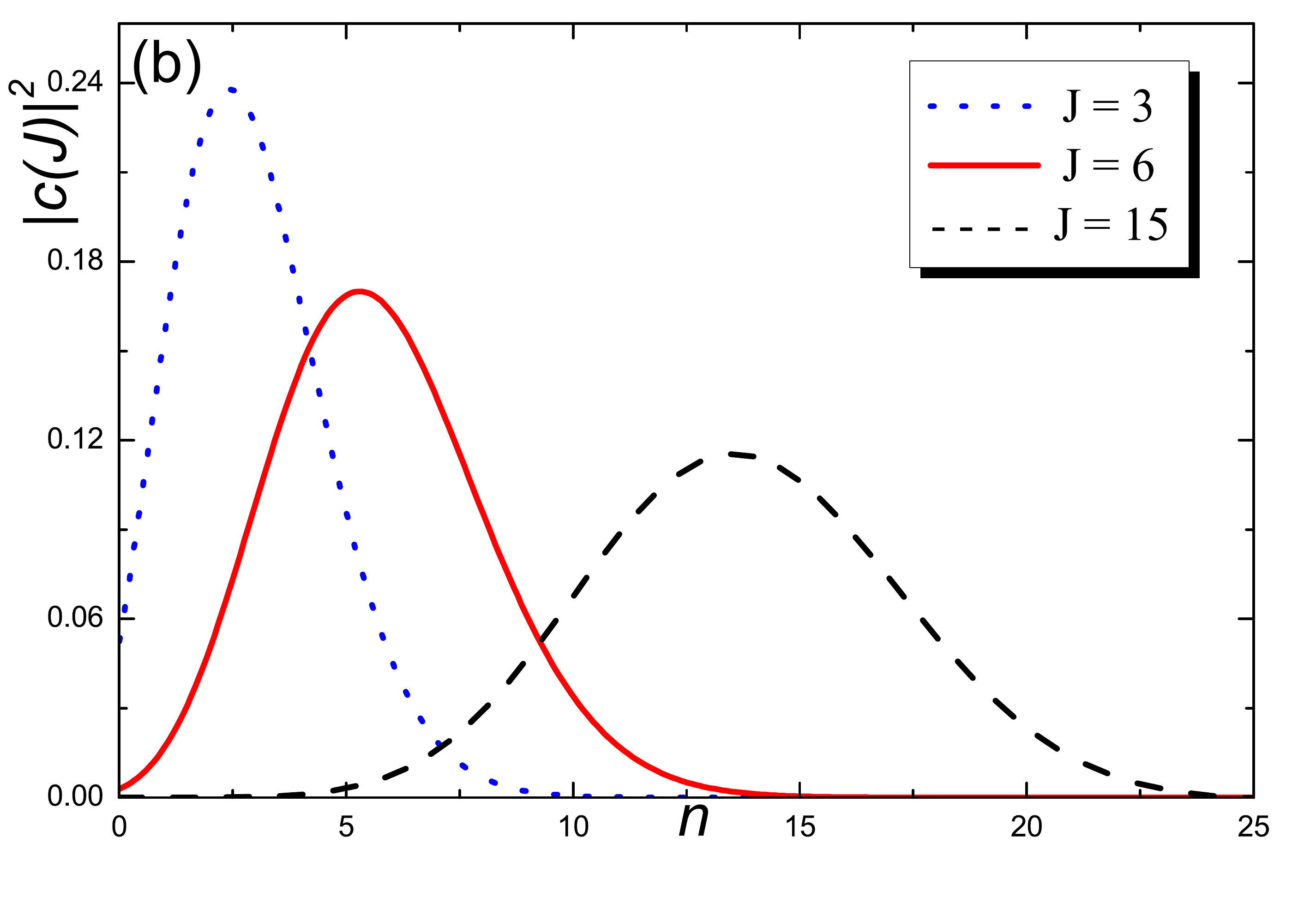}
\caption{\small{(a) Weighting function for $\protect\tau =0.1$ with $\left\langle n\right\rangle =1.24,2.25,3.04$ for $J=1.5,3,4.5$, respectively and (b) $\protect\tau =0.01$ with $\left\langle n\right\rangle =2.93,5.76,13.72$ for $J=3,6,15$, respectively.}}
\label{F2}
\end{figure}

For the chosen values we observe in figure \ref{F2} that the wave packets
are well localized around $\bar{n}$ resulting from (\ref{nb}), such that the
prerequisite for the validity of the analysis in \cite{averbukh_perelman} is given. Increasing the values of $J$ for fixed $\tau $ we observe negative values
for $|c(J)|^{2}$ for large values of $n$, which clearly indicates that our
perturbative expressions are no longer valid in that regime. We also note
that $n\approx 50$ will be a sufficiently good value to terminate the sum in
the expression for the autocorrelation function (\ref{infsum}) analysed in
figure \ref{F1}.
\begin{figure}[h!]
\centering   \includegraphics[width=7.5cm,height=6.0cm]{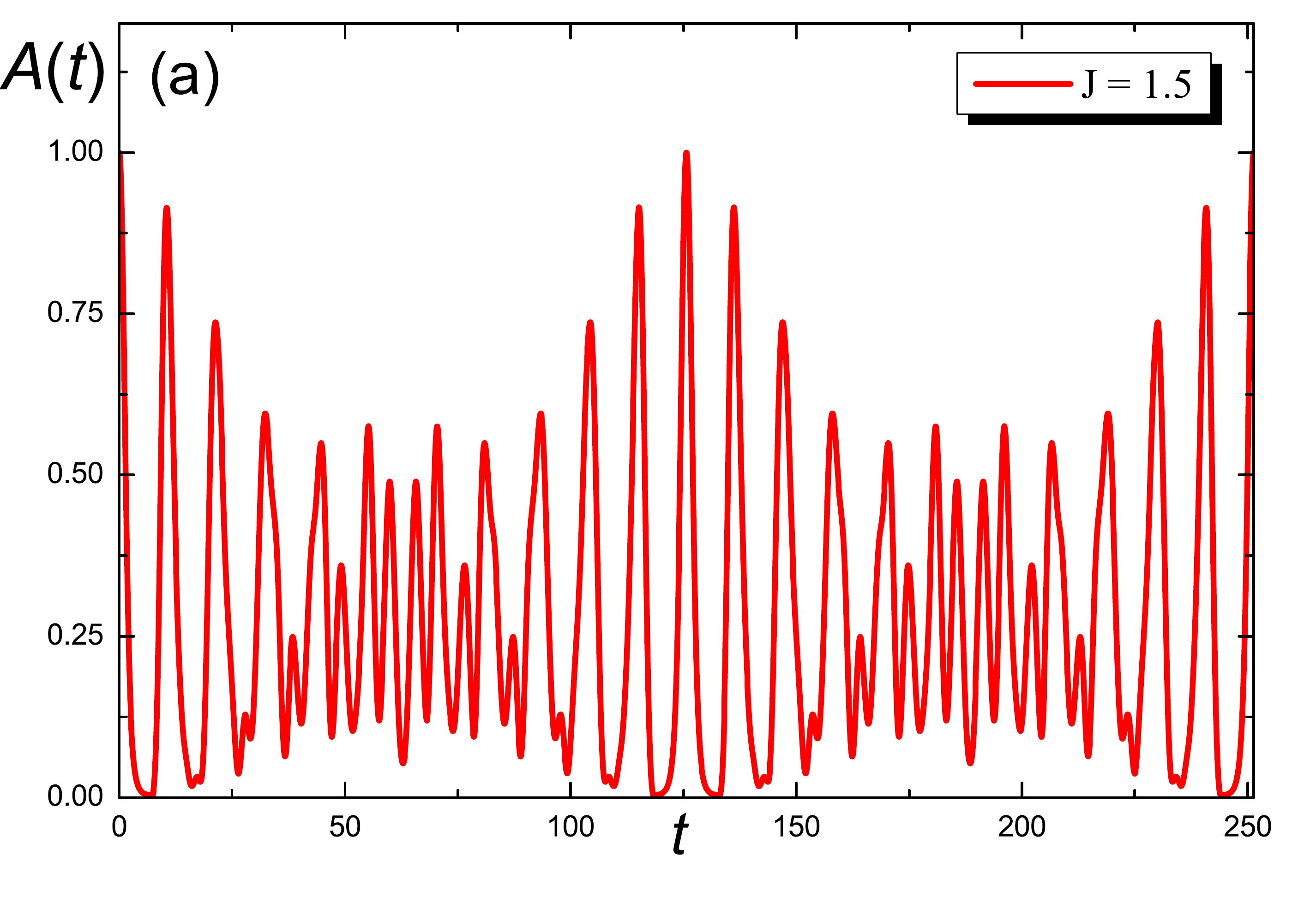} %
\includegraphics[width=7.5cm,height=6.0cm]{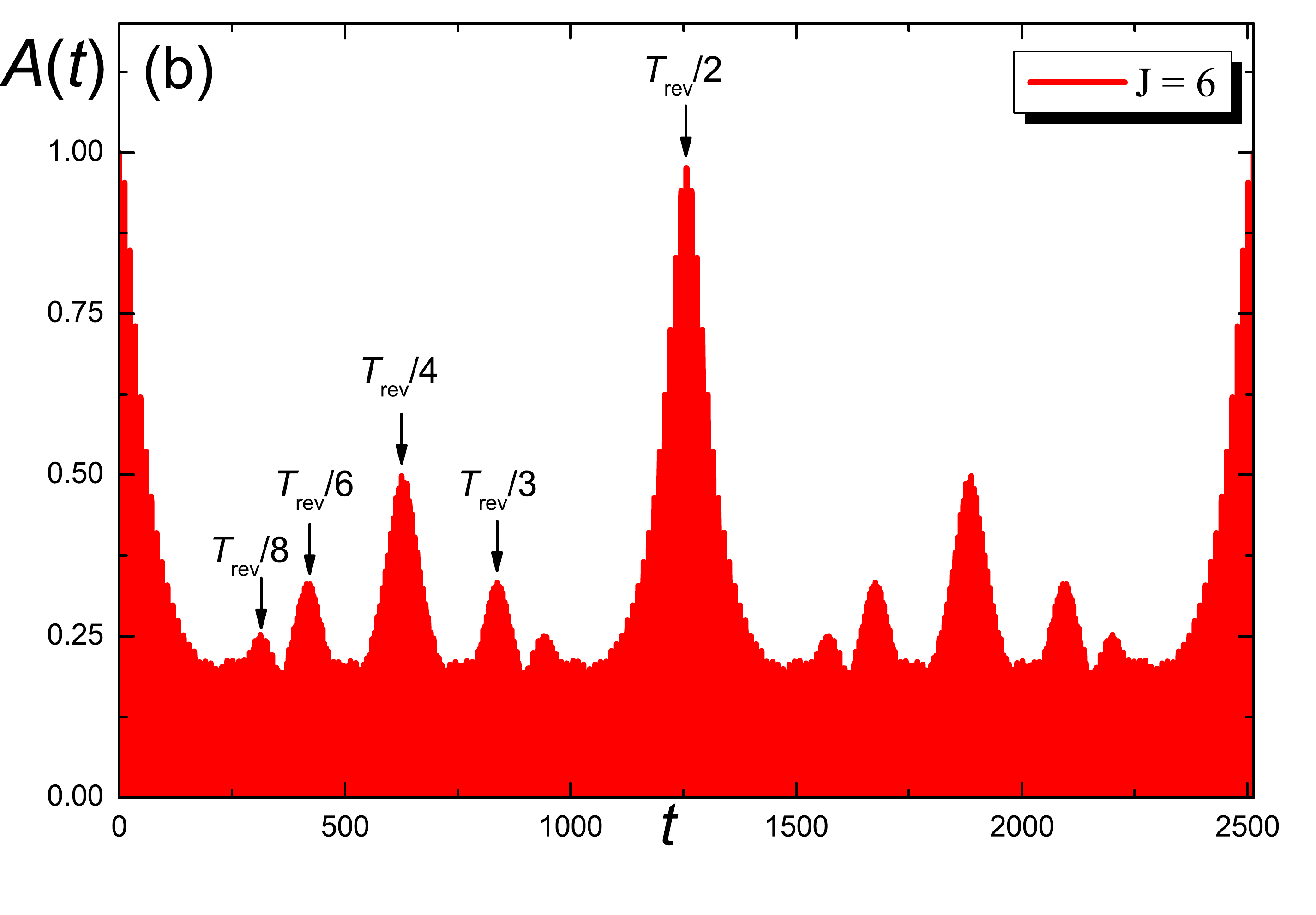}
\caption{\small{(a) Autocorrelation function as a function of time for $J=1.5$, $\protect\tau =0.1$, $\protect\omega =0.5$, $\hbar =1$, $\protect\gamma =0$, $
T_{\text{cl}}=10.05$ and $T_{\text{rev}}=251.32$; (b) Autocorrelation
function as a function of time for $J=6$, $\protect\tau =0.01$, $\protect%
\omega =0.5$, $\hbar =1$, $\protect\gamma =0$, $T_{\text{cl}}=11.74$ and $T_{
\text{rev}}=2513.27$.}}
\label{F1}
\end{figure}
In panel (a) of figure \ref{F1} we clearly observe local maxima at multiples
of the classical period $T_{\text{cl}}$. As explained in \cite{averbukh_perelman} the first full reconstruction of the original wave packet is obtained at $T_{\text{rev}}/2$ which is clearly visible in panel (a). The fractional revivals are better observed for smaller values of $\tau $ as depicted in panel (b). In that scenario the classical periods are so small as compared to the revival time that they are no longer resolved. We clearly observe a number of fractional revivals, such as for instance $T_{\text{rev}}/4$
corresponding to the superposition of two classical-like sub-wave packets
and others as indicated in the figure.

Notice that our expressions and our analysis presented here differ once
again from the one in \cite{ghosh_roy}, where for instance the mandatory
revival at $T_{\text{rev}}/2$ was not observed. It should be noted that in the considered case the wave packet revival time (\ref{TT}) depends explicitly on the deformation parameter $\tau $, such that a possible measurement could distinguish between a noncommutative and a standard commutative space. For instance, in the order of femtoseconds half and quarter revivals have been observed experimentally \cite{vrakking_villeneuve_stolow} for molecular wave packets described by anharmonic oscillator potentials with eigenenergies similar to (\ref{En}). Up to this point we have only analysed the case to first order perturbation theory in $\tau $, let us now extend it to the exact case.

\subsection{Fractional Super-Revival Structure from Non-Perturbative Noncommutative Harmonic Oscillator}
As we have already seen that the revival structure is directly linked to the dependence of the energy eigenvalues $E_{n}$ on the quantum number $n$, i.e. the existence of the $k$-th derivative $d^{k}E_{\bar{n}}/d\bar{n}^{k}$ with respect to some average value $\bar{n}$ at which the wave packet $\psi =\sum c_{n}\phi _{n}$ is well localized. For the case at hand with the energy eigenvalue (\ref{nchoeigen}) being $E_n=\hbar \omega [n]_q$, with $[n]_q=\frac{1-q^{2n}}{1-q^2}$ (\ref{qint}), we would expect that these derivatives exist to all orders, such that we expect infinitely many revival times  (\ref{clrev}) to exist.

At the smallest scale one obtains the classical period $T_{\text{cl}}=2\pi\hbar/\vert E_{\bar{n}}' \vert$, thereafter at large scale the fractional revivals for the revival time $T_{\text{rev}}=4\pi\hbar/\vert E_{\bar{n}}'' \vert$, then the super-revival structure for super-revival time $T_{\text{suprev}}=12\pi\hbar/\vert E_{\bar{n}}'''$ and so on. 

The peak of the wave packet is computed in this case to $\bar{n}:=\left\langle n\right\rangle =Jd\ln \mathcal{N}^{2}(J)/dJ$. Noting that $d^{k}E_{n}/dn^{k}=\hbar \omega 2^{k}q^{2n}\ln ^{k}q/(q^{2}-1)$ we obtain the times 
\begin{equation}
T_{\text{cl}}=\frac{\pi }{\omega }\left\vert \frac{q^{2}-1}{q^{2\bar{n}}\ln q
}\right\vert ,\quad \quad T_{\text{rev}}=\frac{\pi }{\omega }\left\vert 
\frac{q^{2}-1}{q^{2\bar{n}}\ln ^{2}q}\right\vert ,\qquad \text{and\quad
\quad }T_{\text{suprev}}=\frac{3\pi }{2\omega }\left\vert \frac{q^{2}-1}{q^{2
\bar{n}}\ln ^{3}q}\right\vert .  \label{TT}
\end{equation}

\begin{figure}[h]
\centering   \includegraphics[width=7.5cm,height=6.0cm]{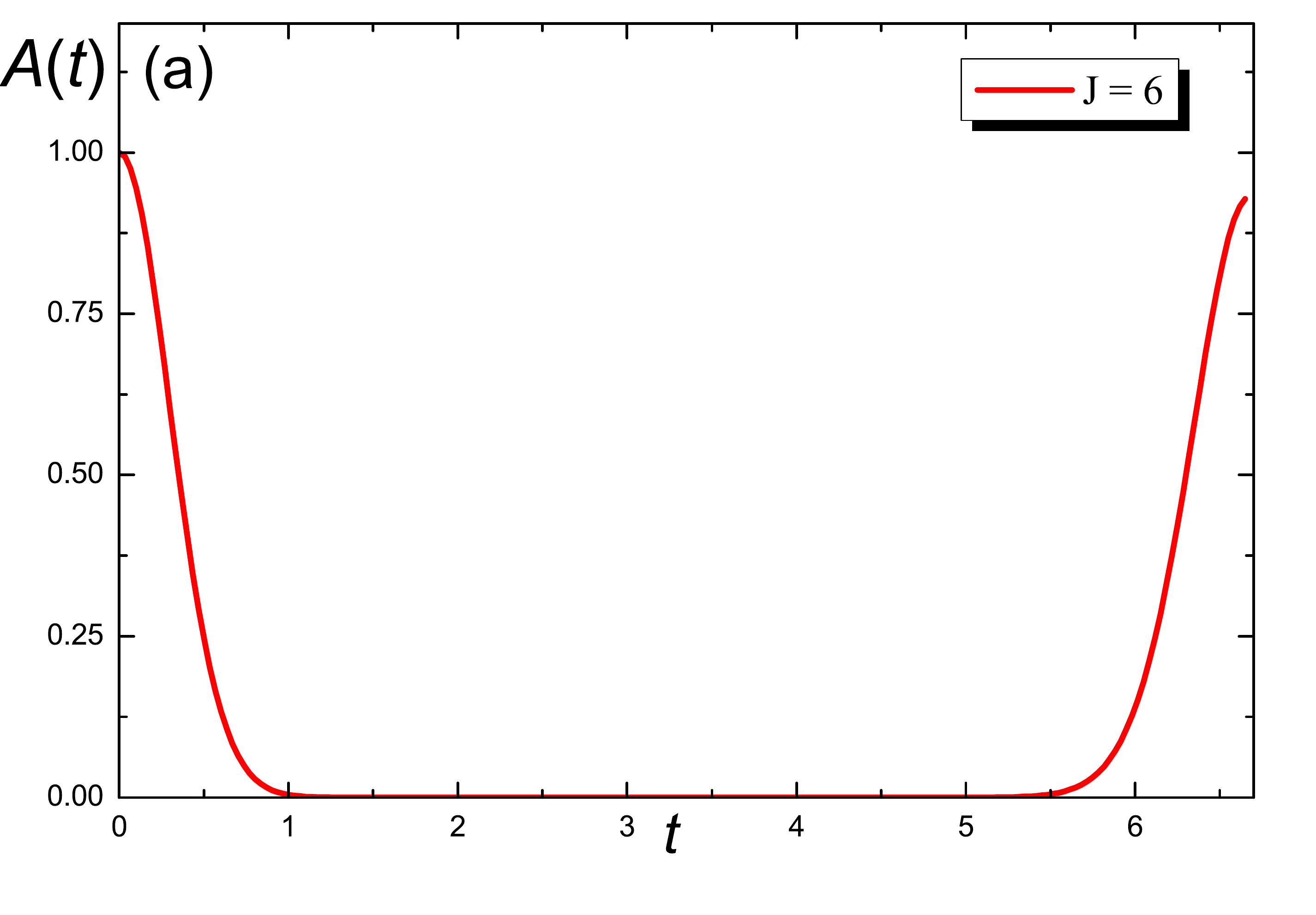}
\caption{\small{Autocorrelation function as a function of time at different scales for $\hbar=1$, $\protect\omega= 1$, $q=e^{-0.005}$, $J=6, \bar{n}
=6.1875 $ and classical period at $T_{\text{cl}}=6.65$. }}
\label{FF1}
\end{figure}

In figure \ref{FF1} and \ref{FF2}, we present the autocorrelation function as a function of time at different scales. In Figure \ref{FF1}, the revival after the classical period is clearly visible.

\begin{figure}[h]
\centering   \includegraphics[width=7.5cm,height=6.0cm]{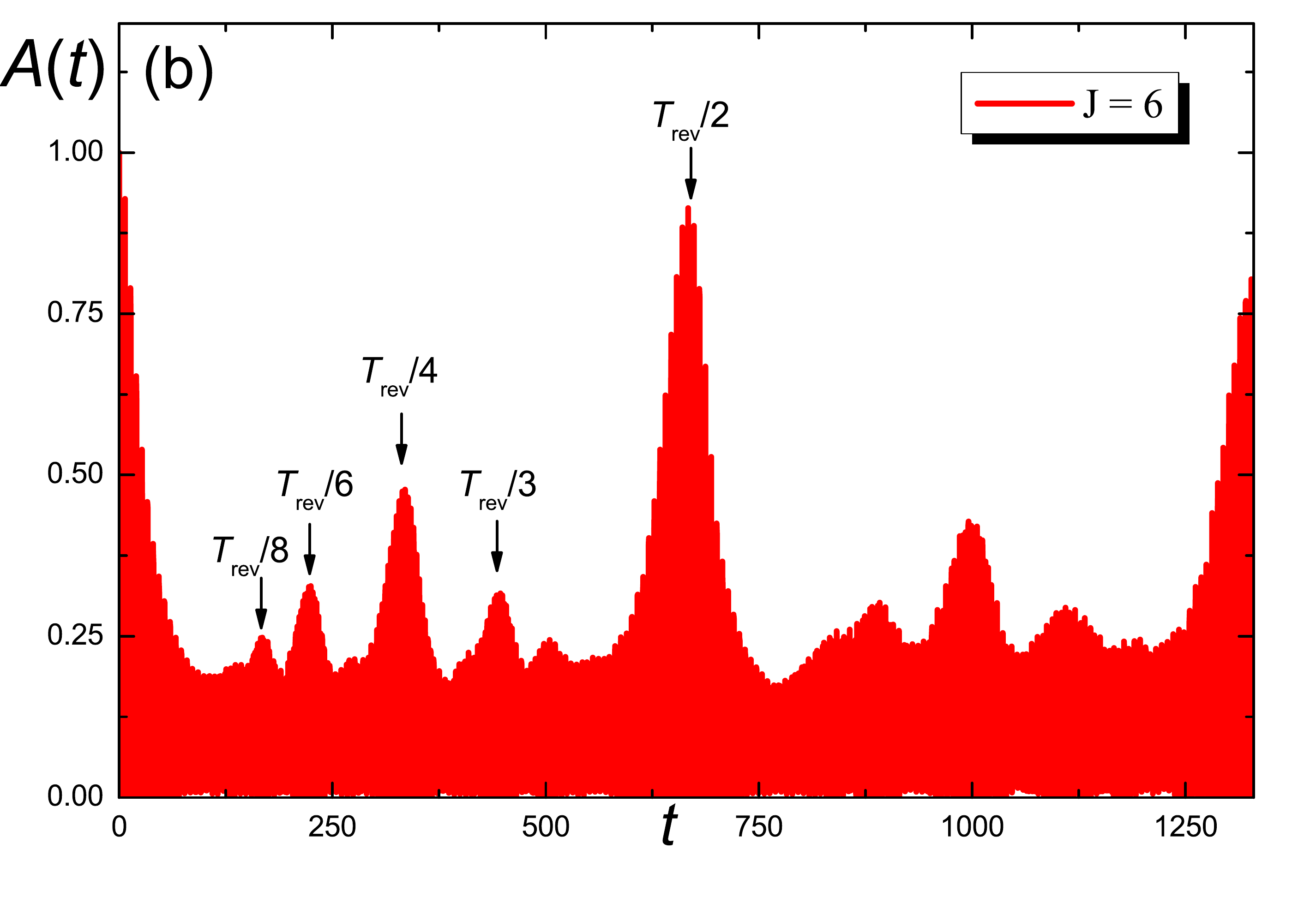}
\includegraphics[width=7.5cm,height=6.0cm]{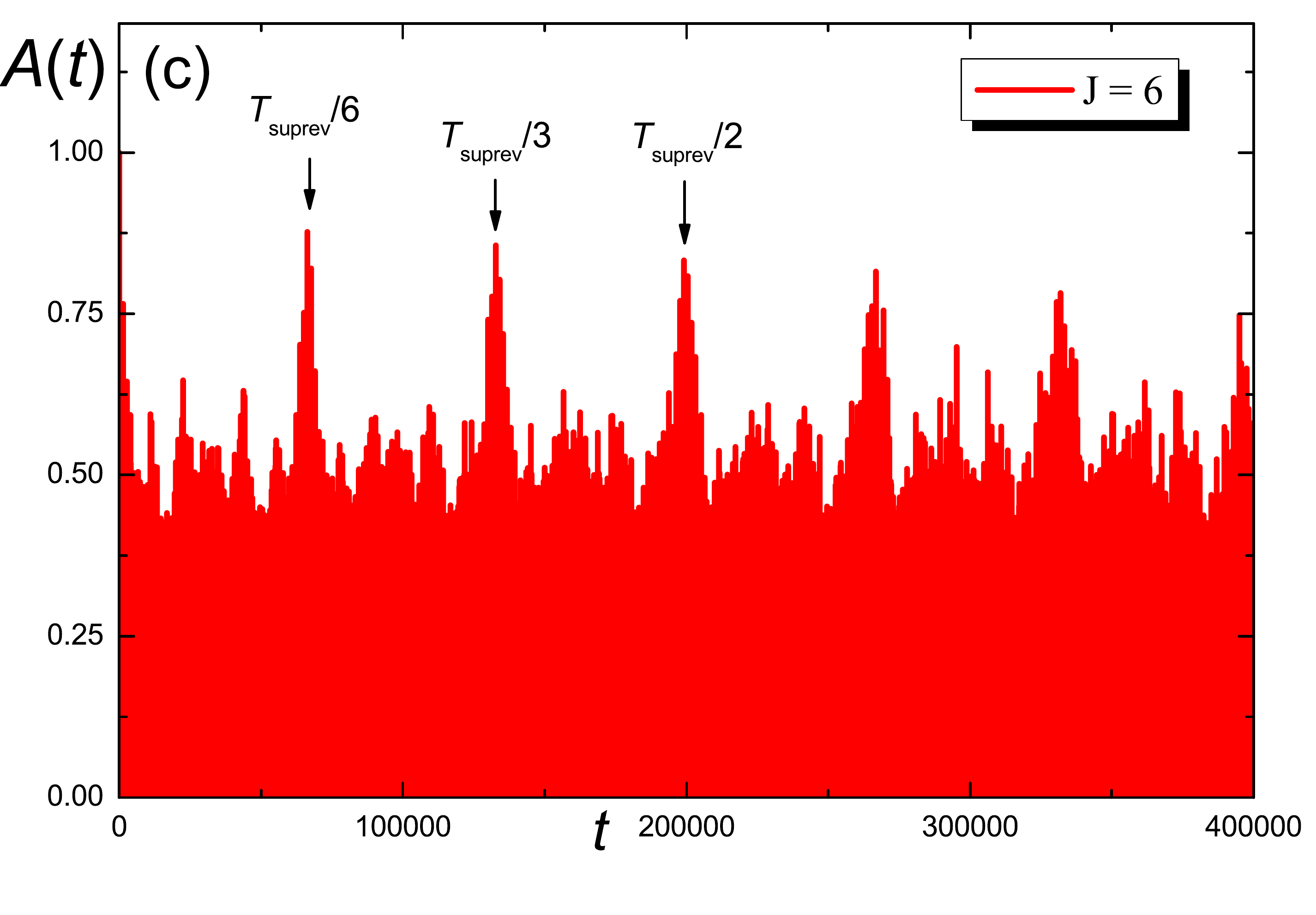}
\caption{\small{Autocorrelation function as a function of time at different scales
for $\hbar=1$, $\protect\omega= 1$, $q=e^{-0.005}$, $J=6$ and $\bar{n}
=6.1875 $. (b) fractional
revival times for $T_{\text{rev}}=1330.19$ and (c) fractional super-revival
times for $T_{\text{suprev}}=3999056$.}}
\label{FF2}
\end{figure}

The parameters have been chosen in a way that $T_{\text{rev}}/T_{\text{cl}}\approx 200$, such that at the revival time scale the revivals due to the classical periods have died out and only the revival due to $T_{\text{rev}}$ are exhibited as clearly visible in the computation presented in panel (a) of figure \ref{FF2}. With $T_{\text{suprev}}/T_{\text{rev}}\approx 300$ this type of behaviour is repeated at the super-revival time scale as seen in panel (b) of figure \ref{FF2}. Due to the aforementioned dependence of the energy eigenvalues on  $n$, we conjecture here that this behaviour is repeated order by order. However, the verification of this feature posses a more challenging numerical problem which we do not address here.

\section{Discussions}
Our central results is the construction of explicit expressions for the GK-coherent states for non-Hermitian systems on a noncommutative space leading to a generalized version of Heisenberg's uncertainty relation. We built up coherent states for harmonic oscillator models for two different cases, initially for the perturbative one and finally for the exact case. We showed that these states are squeezed, in the former case, for all values of $J$ and $\gamma$, whereas in the later case, for a specific value of $\gamma$, as they saturate the minimal uncertainty. In both cases, we established the fact that two of the nontrivial GK-axioms are satisfied. First of all the states are shown to be temporarily stable, i.e. they remain coherent under time evolution, and secondly the states satisfy the action identity (\ref{actionangle}) allowing for a close relation to a classical description in terms of action angle variables. We also demonstrated that when using appropriate metric, Ehrenfest's theorem is satisfied for the observables $X$ and $P$.

The desired resemblance of the coherent states with a classical description was further underpinned by an analysis of the revival structure exhibiting the typical quasi-classical evolution of the original wave packet. The revival structure being linked to the dependence of the energy eigenvalues $E_n$ on the quantum number $n$, the perturbative case produces only a revival structure, because the energy eigenvalues were computed up to the the first order (\ref{En}) in perturbation theory in this case. Instead, for the exact case they are not limited to a single revival structure only, but there exist infinitely many revival times. As an example, the existence of super-revival time were shown explicitly at a much larger time scale in figure \ref{FF2}. Although it is a computational challenge to explore higher order revival structures, in principle they exist.

There are various open problems left for future investigations, such as the study of different types of models on the type of noncommutative spaces investigated here. Especially an extension to higher dimensional models would be very interesting. A systematic comparison with different types of coherent states would be insightful, especially with rare construction related to non-Hermitian Hamiltonians \cite{graefe_schubert}. More computational power should also allow us to confirm our conjecture about the existence of revival time structure at much larger time scales for the case of nonperturbative harmonic oscillator, such as super-super-revival time structure.


\chapter{Quality Assessment of Coherent States} \label{chapterBohmian}
In the last chapter, we have illustrated the notions of coherent states in detail and we are convinced enough that the Klauder coherent states produce qualitatively very good coherent states even for the noncommutative structure of the space-time. The analysis of the fractional revival structure revealed the underlying truth, which is more relevant than the analysis of mathematical properties for the understanding of the physical behaviour of the systems. However, from the experimental point of view, these are probably not enough and require further investigations. In this chapter, we present a simple but efficient procedure based on the trajectory interpretation and endeavour the qualitative comparison between the dynamics of the coherent states with that of the classical particle, which is clearly more convincing than that of the previous method explored in the last chapter.  

The usual interpretation of the quantum theory is self consistent, but it involves an assumption that cannot be tested experimentally , i.e. the most complete possible specification of an individual system is in terms of a wave function that determines only probable results of actual measurement processes. This assumption has been the object of severe criticism, notably on the part of Einstein, who has always believed that, even at the quantum level, there must exist precisely definable elements or dynamical variables determining (as in Classical Physics) the actual behaviour of each individual system and not merely its probable behaviour.

Many physicists have felt that objections such as those raised by Einstein are not relevant, firstly, because the present form of the quantum theory with its usual probability interpretation is in excellent agreement with an extremely wide range of experiments, and, secondly, because no consistent alternative interpretations have been suggested.

However, in 1952 David Bohm \cite{bohm1,bohm2} intended to provide a radically new approach to quantum theory in terms of "hidden" variables. In contrast to the usual interpretation, this alternative interpretation permits us to conceive of each individual system as being in a precisely definable state, whose changes with time are determined by definite laws, analogous to (but not identical with) the classical equation of motion.

\section{Bohmian Mechanics}
Bohmian mechanics was originally proposed sixty years ago \cite{bohm1}
to address some of the difficulties present in the standard formulation of
quantum mechanics based on the Copenhagen interpretation and its aim was to
provide an alternative ontological view. Its central purpose is to avoid the
need for the collapse of the wavefunction and instead provide a trajectory-based scheme allowing for a causal interpretation. While this metaphysical
discussion is still ongoing and is in parts very controversial \cite{holland,hiley,bohm_hiley_book}, it needs to be stressed that Bohmian mechanics leads to the same predictions of measurable quantities as the orthodox framework. Here we will leave the interpretational issues aside and build on the fact that the Bohmian formulation of quantum mechanics has undoubtedly proven to be a successful technical tool for the study of some concrete physical scenarios. For instance, it has been applied successfully to study photodissociation problems \cite{mayor_askar_rabitz}, scattering problems \cite{chou_wyatt_scattering}, tunneling processes\cite{lopreore_wyatt}, atom diffraction by surfaces \cite{sanz_borondo_miret,wang_darling_holloway,guantes_sanz} and high harmonic generation \cite{wu_augstein_faria}. Whereas these applications are mainly based on an analysis of real valued quantum trajectories, more recently there has also been the suggestion for a formulation of Bohmian mechanics based on complex trajectories \cite{leacock_padgett,leacock_padgett1,goldfarb_degani_tannor,chou_wyatt_complex}. We will discuss here both versions, but it is this latter formulation on which we will place our main focus and which will be the main subject of our investigations.

\rhead{Quality Assessment of Coherent States}
\lhead{Chapter 7}
\chead{}

The starting point for the construction of the Bohmian quantum trajectories is usually a solution of the time-dependent Schr\"{o}dinger equation involving a
potential $V(x)$ 
\begin{equation}
i\hbar \frac{\partial \psi (x,t)}{\partial t}=-\frac{\hbar ^{2}}{2m}\frac{
\partial ^{2}\psi (x,t)}{\partial x^{2}}+V(x)\psi (x,t)\text{.}  \label{SE}
\end{equation}
The two variants leading either to real or complex trajectories are
distinguished by different parametrizations of the wavefunctions.

\subsection{Real Bohmian Mechanics}
The standard formulation of real Bohmian mechanics starts by writing the wave function in terms of the real amplitude $R$ and the real action function $S$ as
\begin{equation}
\psi (x,t)=R(x,t)e^{i S(x,t)/\hbar},\text{\qquad with \ \ }R(x,t),S(x,t)\in 
\mathbb{R}\text{.}  \label{phi1}
\end{equation}
and substituting this wave function (\ref{phi1}) into the time dependent Schr{\"o}dinger equation (\ref{SE}), yielding a system of two coupled partial differential equations, which after separation of real and imaginary parts read as
\begin{equation}
S_{t}+\frac{(S_{x})^{2}}{2m}+V(x)-\frac{\hbar ^{2}}{2m}\frac{R_{xx}}{R}
=0,\qquad \text{and\qquad }mR_{t}+R_{x}S_{x}+\frac{1}{2}RS_{xx}=0,
\label{r1}
\end{equation}
usually referred to as the quantum Hamilton-Jacobi equation (QHJE) and continuity equation respectively. When considering these equations from a
classical point of view, the second term in the first equation of (\ref{r1})
is interpreted as the kinetic energy, such that the real velocity $v(t)$ results to
\begin{equation}
mv(x,t)=S_{x}=\frac{\hbar }{2i}\left[ \frac{\psi ^{\ast }\psi _{x}-\psi \psi
_{x}^{\ast }}{\psi ^{\ast }\psi }\right]. \label{realvel}
\end{equation}
The QHJE (\ref{r1}) differs from the classical Hamilton-Jacobi equation by the addition of the term $-\frac{\hbar^2}{2 m}\frac{R_{xx}}{R}$, known as "quantum potential", which can be expressed as
\begin{equation} \label{realqp}
Q(x,t)=-\frac{\hbar ^{2}}{2m}\frac{R_{xx}}{R}=\frac{\hbar ^{2}}{4m}\left[ \frac{\left( \psi ^{\ast }\psi\right) _{x}^{2}}{2\left( \psi ^{\ast }\psi \right) ^{2}}-\frac{\left( \psi^{\ast }\psi \right) _{xx}}{\psi ^{\ast }\psi }\right] .
\end{equation}
Therefore, instead of using the Hamilton-Jacobi formalism (where one must use the Jacobi's theorem to obtain trajectories), one might compute the quantum trajectories by solving the velocity equation (\ref{realvel}). The corresponding time-dependent effective potential is therefore $V_{\text{eff}}(x,t)=V(x)+Q(x,t)$. Then one has two options to compute quantum trajectories. One can either solve directly the equation (\ref{realvel}) for $x(t)$ or employ the effective potential $V_{\text{eff}}$ solving $m\ddot{x}=-\partial V_{\text{eff}}/\partial x$ instead. Due to the different order of the differential equations to be solved, we have then either one or two free parameter available. Thus for the two possibilities to coincide the initial momentum is usually not free of choice, but the initial position $x(t=0)=x_{0}$ is the only further input. The connection to the standard quantum mechanical description is then achieved by computing expectation values from an ensemble of $n$ trajectories, e.g. $\left\langle x(t)\right\rangle_{n}=1/n\sum\nolimits_{i=1}^{n}x_{i}(t)$.

\subsection{Complex Bohmian Mechanics}\label{subcomplexbohmian}
In contrast, the complex version of the Bohmian mechanics is obtained by expressing the wave function in terms of the complex action function $\tilde{S}$ as
\begin{equation}
\psi (x,t)=e^{i\tilde{S}(x,t)/\hbar},\text{\qquad with}\qquad \tilde{S}(x,t)\in 
\mathbb{C},  \label{phi2}
\end{equation}
and substituting into the time-dependent Schr{\"{o}}dinger equation, yielding the single complex valued QHJE as
\begin{equation}
\tilde{S}_{t}+\frac{(\tilde{S}_{x})^{2}}{2m}+V(x)-\frac{i\hbar }{2m}\tilde{S}
_{xx}=0.  \label{c1}
\end{equation}
Interpreting this equation in a similar way as in the previous subsection,
but now as a complex quantum Hamilton-Jacobi equation, the second term in (\ref{c1}) yields a complex velocity and the last term becomes a complex
quantum potential 
\begin{equation}
m\tilde{v}(x,t)=\hat{S}_{x}=\frac{\hbar }{i}\frac{\psi _{x}}{\psi },~\ ~
\tilde{Q}(x,t)=-\frac{i\hbar }{2m}\tilde{S}_{xx}=-\frac{\hbar ^{2}}{2m}\left[
\frac{\psi _{xx}}{\psi }-\frac{\psi _{x}^{2}}{\psi ^{2}}\right] .
\label{complex}
\end{equation}
The corresponding time-dependent effective potential is now $\tilde{V}_{
\text{eff}}(x,t)=V(x)+\tilde{Q}(x,t)$. Once again one has two options to
compute quantum trajectories, either solving the first equation in (\ref
{complex}) for $x(t)$, which is, however, now a complex variable.
Alternatively, we may also view the effective Hamiltonian $H_{\text{eff}
}=p^{2}/2m+\tilde{V}_{\text{eff}}(x,t)=H_{r}+iH_{i}$ in its own right and
simply compute the canonical equations of motion directly from 
\begin{alignat}{2}
\dot{x}_{r} &= \frac{1}{2}\left( \frac{\partial H_{r}}{\partial p_{r}}+\frac{
\partial H_{i}}{\partial p_{i}}\right) ,\qquad~~ \dot{x}_{i} &&= \frac{1}{2}\left( \frac{\partial H_{i}}{\partial p_{r}}-\frac{\partial H_{r}}{\partial p_{i}}\right),  \label{H1} \\
\dot{p}_{r} &= -\frac{1}{2}\left( \frac{\partial H_{r}}{\partial x_{r}}+
\frac{\partial H_{i}}{\partial x_{i}}\right) ,~~~~~~\dot{p}_{i} &&= \frac{1}{2}
\left( \frac{\partial H_{r}}{\partial x_{i}}-\frac{\partial H_{i}}{\partial
x_{r}}\right) ,  \label{H2}
\end{alignat}
where we use the notations $x=x_{r}+ix_{i}$ and $p=p_{r}+ip_{i}$ with $x_{r}$, $x_{i}$, $p_{r}$, $p_{i}\in \mathbb{R}$. For the complex case the relation
to the conventional quantum mechanical picture is less well established
although some versions have been suggested to extract real expectation
values, e.g. based on taking time-averaged mean values \cite{cdyang}, seeking for isochrones \cite{goldfarb_degani_tannor,chou_wyatt_complex} or using imaginary part of the velocity field of particles on the real axis \cite{mvjohn}. 

At this point we should clearly mention the pathway of what we are going to explore in this chapter. We will compute the dynamics of the coherent states using the Bohmian mechanics as illustrated above and the dynamics of the classical particle using the standard process independently and compare them together to look at the qualitative difference that they possess. Once again we use the Klauder coherent states (\ref{klaudercoherent}) as a generalised coherent states, as clarified in the last chapter. We evaluate the expressions for the velocity, the quantum potential and the resulting trajectories for different solvable potentials commencing with different choices of solutions $\psi $.

\section{The Harmonic Oscillator}
The harmonic oscillator 
\begin{equation}
\mathcal{H}_{\text{ho}}=\frac{p^{2}}{2m}+\frac{1}{2}m\omega ^{2}x^{2},
\label{VHO}
\end{equation}
constitutes a very instructive example on which many of the basic features
can be understood. We will therefore take it as a starting point. Many results may already be found in the literature, but for completeness we also report them here together with some new findings.

\subsection{Real Case}
As reported for instance by Holland \cite{holland}, using the formulas (\ref{realvel}) and (\ref{realqp}) it is easy to see that for any stationary state $\psi_{n}(x,t)=\phi _{n}(x)e^{-iE_{n}t/\hbar }$, with $\phi _{n}(x)$ being a solution of the stationary Schr\"{o}dinger, the velocity (\ref{realvel})
results to $v(t)=0$. This is compatible with the values obtained from the
use of the quantum potential $Q(x)=E_{n}-V(x)$, because this corresponds to
a classical motion in a constant effective potential $V_{\text{eff}}(x,t)=E_{n}$. This is of course qualitatively very far removed from our
original potential (\ref{VHO}), such that classical trajectories obtained
from $\mathcal{H}_{\text{ho}}$ and its effective version are fundamentally
of qualitatively different nature.

Instead we would expect, that when starting from coherent states we end up
with a behaviour much closer to the classical behaviour resulting from the
original Hamiltonian. By direct computation shown in \cite{holland}, using
the standard Gaussian wavepackets of the form
\begin{equation}
\psi _{c}(x,t)=\left( \frac{m\omega }{\hbar \pi }\right) ^{1/4}e^{-\frac{
m\omega }{2\hbar }(x-a\cos \omega t)^{2}-\frac{i}{2}\left[ \omega t+\frac{
m\omega }{\hbar }\left( 2xa\sin \omega t-\frac{1}{2}a^{2}\sin 2\omega
t\right) \right] },  \label{fc}
\end{equation}
to compute the above quantities with $a$ being the centre of the wavepacket
at $t=0$, one obtains from equations (\ref{realvel}) and (\ref{realqp}) for the velocity and the quantum potential as
\begin{equation}
v(x,t)=-a\omega \sin \omega t,\qquad \text{and\qquad }~Q(x,t)=\frac{\hbar
\omega }{2}-\frac{1}{2}m\omega ^{2}(x-a\cos \omega t)^{2}.  \label{vQ}
\end{equation}
Solving now the first equation in (\ref{vQ}) with $dx/dt=v(x,t)$ for $
x(t)=a(\cos \omega t-1)+x_{0}$ with initial condition $x(0)=x_{0}$, we may
construct the corresponding potential from $m\ddot{x}=-\partial V/\partial x$
. The result is compatible with the effective potential obtained from $
Q(x,t)+V(x)$ when replacing the explicit time dependent terms with
expressions in $x(t)$. Alternatively, from the effective potential 
\begin{equation}
V_{\text{eff}}[x(t)]=\frac{1}{2}m\omega ^{2}(x(t)-x_{0}-a)^{2}+\frac{\hbar
\omega }{2},
\end{equation}
Newton's equation will give the above solution for $x(t)$. Thus for the
states $\psi _{c}(x,t)$ the Bohmian trajectories for the harmonic oscillator
potential $\mathcal{H}_{\text{ho}}$ are indeed the same as those resulting
from the motion in a classical harmonic oscillator potential.

What has not been analysed so far is the use of the more general Klauder
coherent states (\ref{klaudercoherent}) as input into the evaluation of the Bohmian trajectories and corresponding quantum potential $Q(x,t)$. Since we have $e_{n}=n$ for the case at hand, the probability distribution and normalization constant (\ref{probdensity}) are computed to $\rho _{n}=n!$ and $\mathcal{N}(J)=e^{J/2}$, respectively. The solution for the stationary Schr\"{o}dinger equation is well known to be the normalized wavefunction 
\begin{equation}
\phi _{n}(x)=(\frac{m\omega }{\pi \hbar })^{1/4}\exp \left( -\frac{mx^{2}\omega }{2\hbar }\right) H_{n}\left(x/\sqrt{\frac{\hbar }{m\omega }}\right) /\sqrt{2^{n}n!}~~,
\end{equation}
with $H_{n}\left( x\right)$ denoting Hermite polynomials. The Mandel parameter $Q$ in (\ref{mandel}) always equals zero independently of $J$, such that we are always dealing with a Poissonian distribution and the localisation of wave packet is ensured.

Due to the fact that $\psi _{J}(x,t)$ involves an infinite sum, it is
complicated to compute analytic expressions for the quantities in (\ref{vQ}). However, since we expect a close resemblance to the expressions obtained
from standard coherent states $\psi _{c}(x,t)$, we suggest here that the
corresponding Bohmian trajectories and quantum potential are given by 
\begin{equation}
x(t)=x_{\text{max}}^{J}(\cos \omega t-1)+x_{0}\quad \text{and\quad }
~Q(x,t)=\frac{\hbar \omega }{2}-\frac{1}{2}m\omega ^{2}(x-x_{\text{max}
}^{J}\cos \omega t)^{2},  \label{func}
\end{equation}
respectively. Our conjecture is guided by the analogy to the previous case, $
x_{\text{max}}^{J}$ is taken here to be the value where the wavepacket has its maximum, i.e., $x_{\text{max}}^{J}=$ $\max \left\vert \psi _{J}(x,0)\right\vert $. In this case we compute the quantities of interest numerically.  We take $\omega =1$, $\hbar =1$ and $m=1$ in all numerical computations throughout the work. We observe stability for  $\psi _{J}(x,t)$ computed from (\ref{klaudercoherent}) up to six digits, when terminating the sum at $n=150$. Our results for the trajectories and quantum potential are depicted in figure \ref{figbohm1}(a) and \ref{figbohm1}(b), respectively.

\begin{figure}[h]
\centering   \includegraphics[width=7.5cm,height=6.0cm]{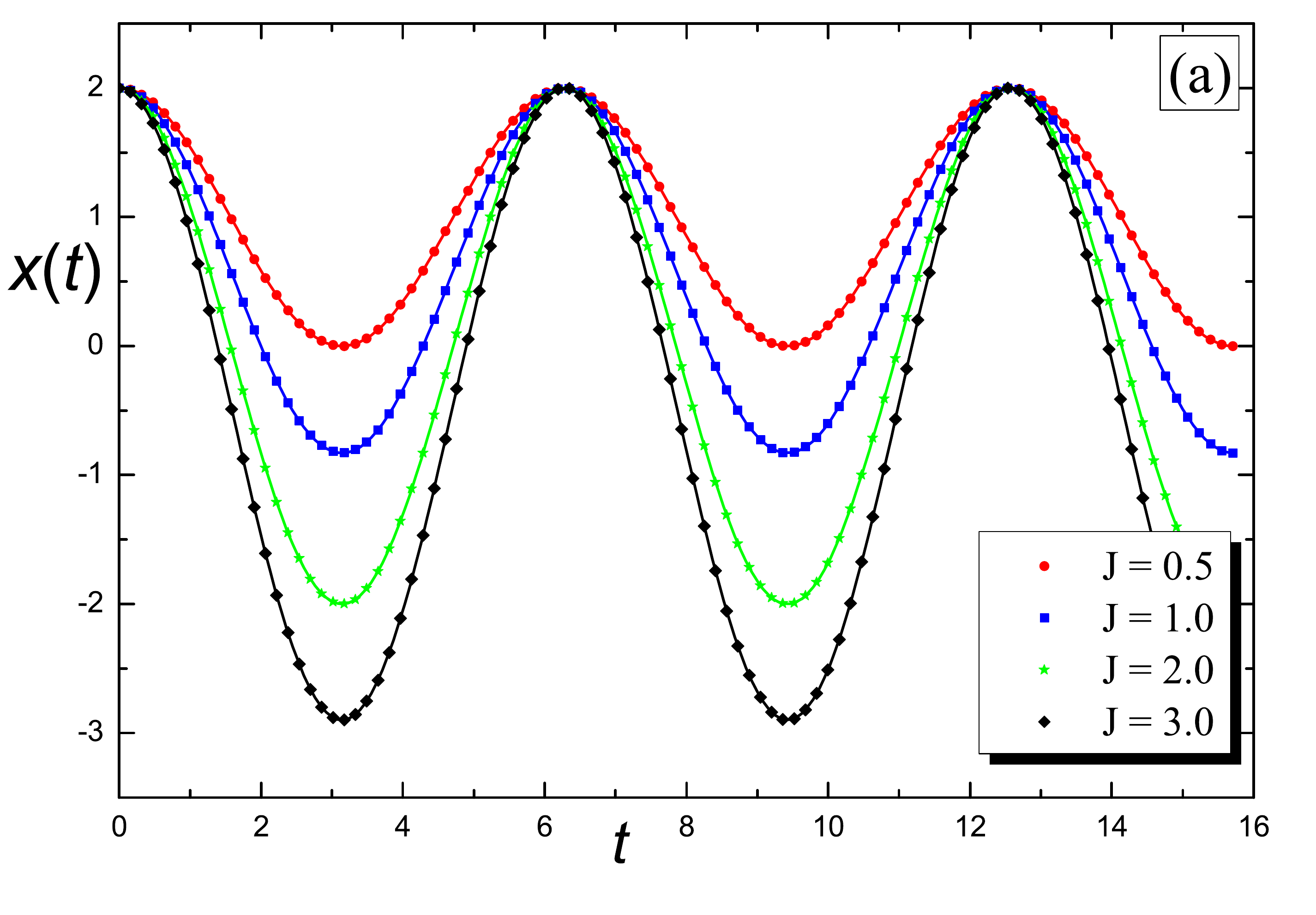}
\includegraphics[width=7.5cm,height=6.0cm]{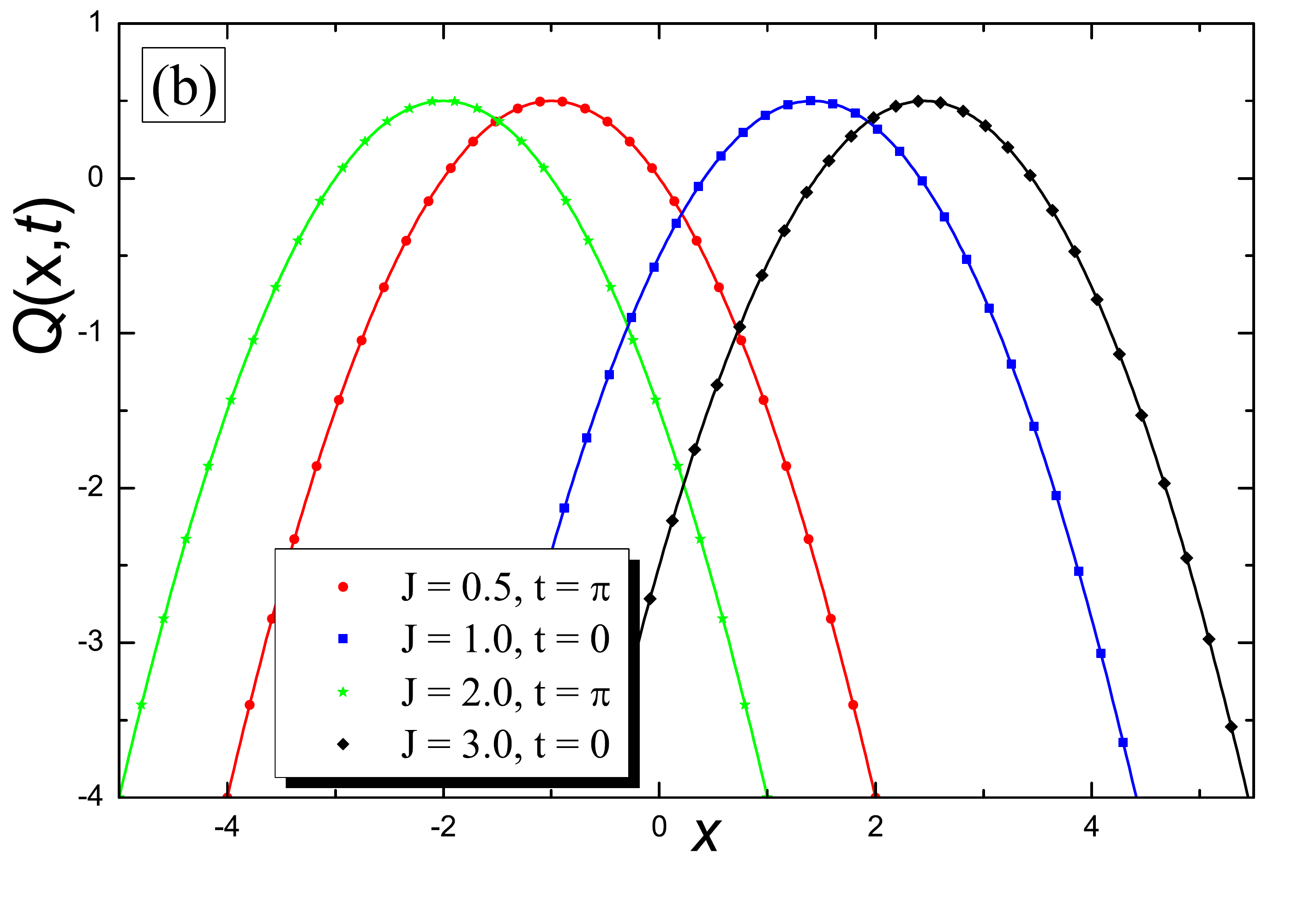}
\caption{\small{(a) Real Bohmian quantum trajectories as functions of time from
Klauder coherent states (dotted) versus classical trajectories
corresponding to (\protect\ref{func}) (solid lines). (b) Quantum potential
from Klauder coherent states (dotted) versus conjectured formula (\protect
\ref{func}) (solid lines). We have taken $x_{0}=2$ and computed the maxima
to $x_{\text{max}}^{0.5}=1$, $x_{\text{max}}^{1}=1.4142$, $x_{\text{max}
}^{2}=2$, $x_{\text{max}}^{3}=2.4495$.}}
\label{figbohm1}
\end{figure}

We observe perfect agreement between the numerical computation of $x(t)$
from solving the equation (\ref{realvel}) using the expression (\ref{klaudercoherent}) for the Klauder coherent states for various values of $J$ and the conjectured analytical expression (\ref{func}) for $x(t)$ in which we only compute the value for $x_{\text{max}}^{J}$ numerically. We find a similar agreement for the computation of the quantum potential $Q(x,t)$, either
numerically using the expression (\ref{klaudercoherent}) in the second equation in (\ref{realqp}) or from the conjectured analytical expression in (\ref{func}). We also find agreement between the two computations solving either directly the equation (\ref{realvel}) for $x(t)$ or employing the effective potential $V_{\text{eff}}$ to solve $m\ddot{x}=-\partial V_{\text{eff}}/\partial x$ instead. What remains is the interesting challenge to compute the infinite sums together with the subsequent expressions explicitly in an analytical manner.

\subsection{Complex Case}
As in the previous subsection we start again with stationary states $\psi
_{n}(x,t)=\phi _{n}(x)e^{-iE_{n}t/\hbar }$ as basic input into our
computation as outlined in subsection \ref{subcomplexbohmian}. A fundamental difference to the real case is that now we do not obtain a universal answer for all models. From (\ref{complex}) we compute
\begin{eqnarray}
\tilde{v}_{0}(x,t) &=&i\omega x,\qquad \qquad ~~\quad ~~~\tilde{Q}_{0}(x,t)=
\frac{\hbar \omega }{2},  \label{vq1} \\
\tilde{v}_{1}(x,t) &=&i\omega x-\frac{i\hbar }{mx},~~~~~~\quad ~~\tilde{Q}
_{1}(x,t)=\frac{\hbar \omega }{2}+\frac{\hbar ^{2}}{2mx^{2}}.  \label{vq2}
\end{eqnarray}
As discussed in \cite{mvjohn1}, for $n=1,2$ the explicit analytical solutions
may be found in these cases. By direct integration of the first equations in
(\ref{vq1}), (\ref{vq2}) or from $m\ddot{x}=-\partial V_{\text{eff}
}/\partial x$ we compute 
\begin{equation}
x_{0}(t)=x_{0}e^{i\omega t},\quad \text{and\quad }x_{1}(t)=\pm \sqrt{\frac{
\hbar }{m\omega }+e^{2it\omega }\left( x_{0}^{2}-\frac{\hbar }{m\omega }
\right) }.
\end{equation}
For larger values of $n$ we obtain more complicated equations for the
velocities and quantum potentials, which may be solved numerically for $x(t)$, see also \cite{mvjohn1,cdyang1}. For instance, we obtain from (\ref{complex})
\begin{alignat}{1}
\tilde{v}_{5}(x,t) &=ix\omega -\frac{5i\hbar }{mx}+\frac{60i\hbar
^{3}-40i\hbar ^{2}mx^{2}\omega }{15\hbar ^{2}mx-20\hbar m^{2}x^{3}\omega
+4m^{3}x^{5}\omega ^{2}}, \notag\\
\tilde{Q}_{5}(x,t) &=\frac{\hbar \left( 225\hbar ^{5}+225\hbar
^{4}mx^{2}\omega +200\hbar ^{2}m^{3}x^{6}\omega ^{3}-80\hbar
m^{4}x^{8}\omega ^{4}+16m^{5}x^{10}\omega ^{5}\right) }{2m\left( 15\hbar
^{2}x-20\hbar mx^{3}\omega +4m^{2}x^{5}\omega ^{2}\right) ^{2}}~~\notag
\end{alignat}
The solutions for $x_{1}(t)$ and $x_{5}(t)$ are depicted in figure \ref{figbohm2}.

\begin{figure}[h]
\centering   \includegraphics[width=7.5cm,height=6.0cm]{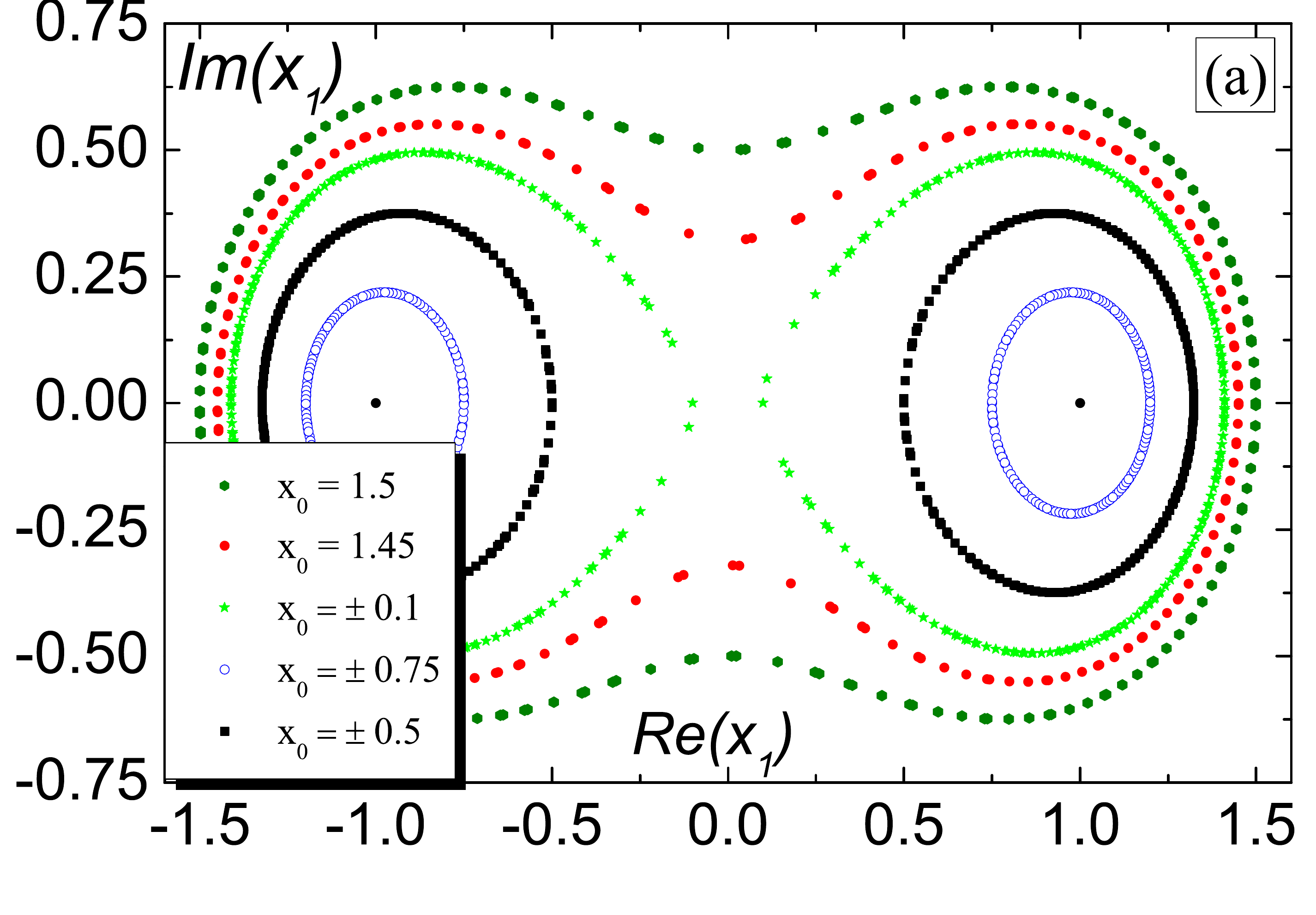}
\includegraphics[width=7.5cm,height=6.0cm]{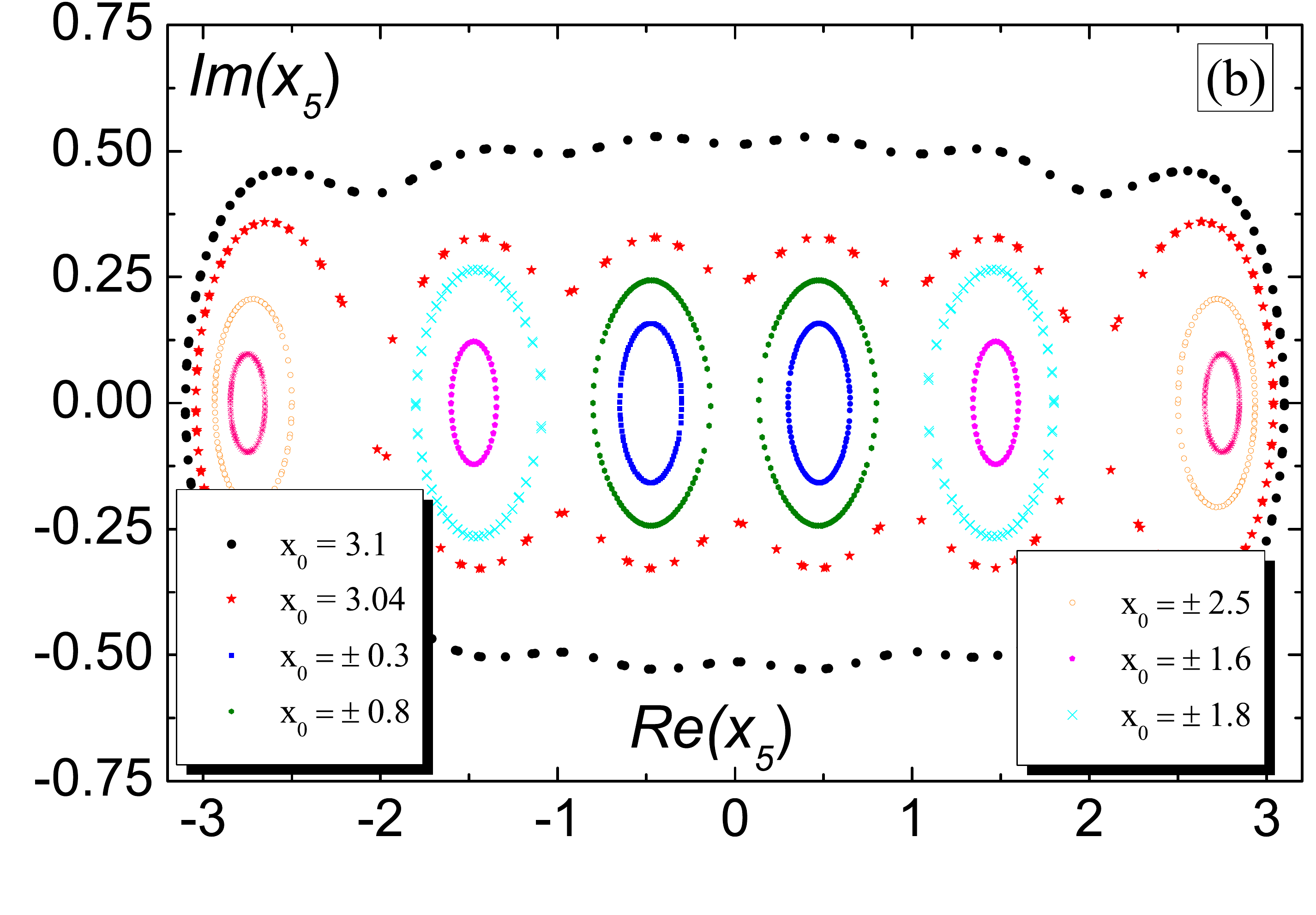}
\caption{\small{Complex Bohmian quantum trajectories as functions of time for
different initial values $x_{0}$ resulting from stationary states $\protect
\psi _{1}(x,t)$ and $\protect\psi _{5}(x,t)$ in panel (a) and (b),
repectively.}}
\label{figbohm2}
\end{figure}

In both cases we observe that the fixed points, at $\pm 1$ for $\tilde{v}
_{1}(x,t)$ and at $\pm 0.476251$, $\pm 1.47524$, $\pm 2.75624$ for $\tilde{v}
_{5}(x,t)$, are centres surrounded by closed limit cycles. For large enough
initial values we also observe bounded motion surrounding all fixed points.

Next we use once more the Gaussian wavepackets (\ref{fc}) as input to
evaluate the velocity and the quantum potential from (\ref{complex})
\begin{equation}
\tilde{v}(x,t)=-\omega (x_{i}+a\sin \omega t)+i\omega (x_{r}-a\cos \omega
t),\qquad \text{and\qquad }~\tilde{Q}(x,t)=\frac{\hbar \omega }{2},
\label{vv}
\end{equation}
The value for the constant quantum potential was also found in \cite{chou_wyatt_complex}. Solving now the equation of motion with $\tilde{v}(t)$ for $x_{r}$ and $ x_{i} $ we obtain a complex trajectory
\begin{equation}
x(t)=\left( \frac{a}{2}+c_{1}\right) \cos \omega t-c_{2}\sin \omega t+i\left[
c_{2}\cos \omega t+\left( c_{1}-\frac{a}{2}\right) \sin \omega t\right] ,
\label{x1}
\end{equation}
with integration constants $c_{1}$ and $c_{2}$. We compare this with the
classical result computed from the complex effective Hamiltonian
\begin{equation}
\mathcal{H}_{\text{eff}}=\frac{1}{2m}(p_{r}^{2}-p_{i}^{2})+\frac{m\omega ^{2}
}{2}(x_{r}^{2}-x_{i}^{2})+i\left( \frac{1}{m}p_{r}p_{i}+m\omega
^{2}x_{r}x_{i}\right) +\frac{\hbar \omega }{2}.  \label{Heff}
\end{equation}
We may think of this Hamiltonian as being $\mathcal{PT}$-symmetric, where
the symmetry is induced by the complexification and realized as $\mathcal{PT}
$: $x_{r}\rightarrow -x_{r}$, $x_{i}\rightarrow x_{i}$, $p_{r}\rightarrow
p_{r}$, $p_{i}\rightarrow -p_{i}$, $i\rightarrow -i$. The equations of
motion are then computed according to (\ref{H1}) and (\ref{H2}) to
\begin{equation}
\dot{x}_{r}=\frac{p_{r}}{m},\quad \dot{x}_{i}=\frac{p_{i}}{m},\quad \dot{p}
_{r}=-m\omega ^{2}x_{r},\quad \text{and\quad }\dot{p}_{i}=-m\omega ^{2}x_{i}.
\end{equation}
As these equations decouple, they are easily solved. We find
\begin{equation}
x(t)=x_{r}(0)\cos \omega t+\frac{p_{r}(0)}{m\omega }\sin \omega t+i\left[
x_{i}(0)\cos \omega t+\frac{p_{i}(0)}{m\omega }\sin \omega t\right] .
\label{x2}
\end{equation}

\begin{figure}[h]
\centering   \includegraphics[width=7.5cm,height=6.0cm]{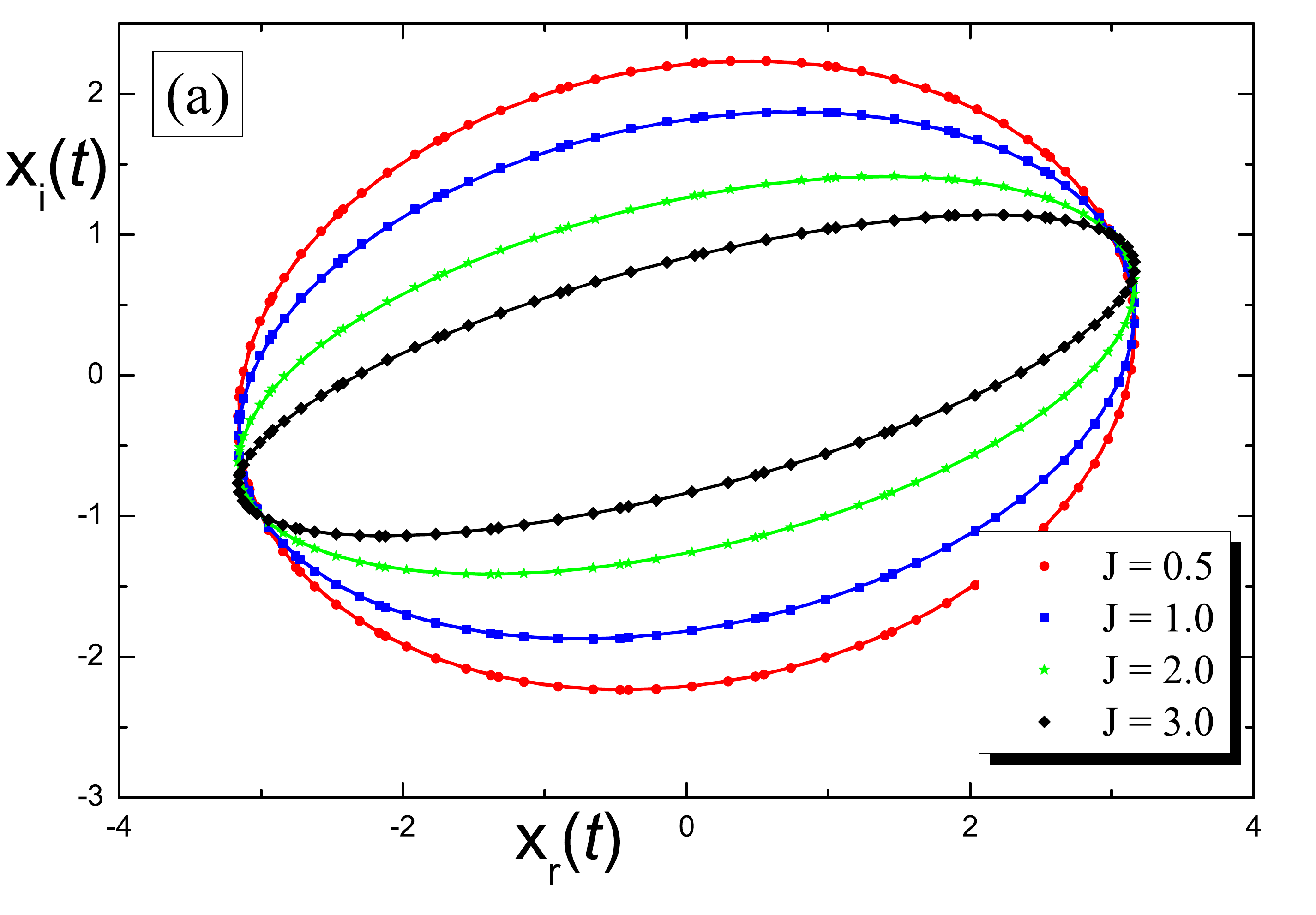}
\includegraphics[width=7.5cm,height=6.0cm]{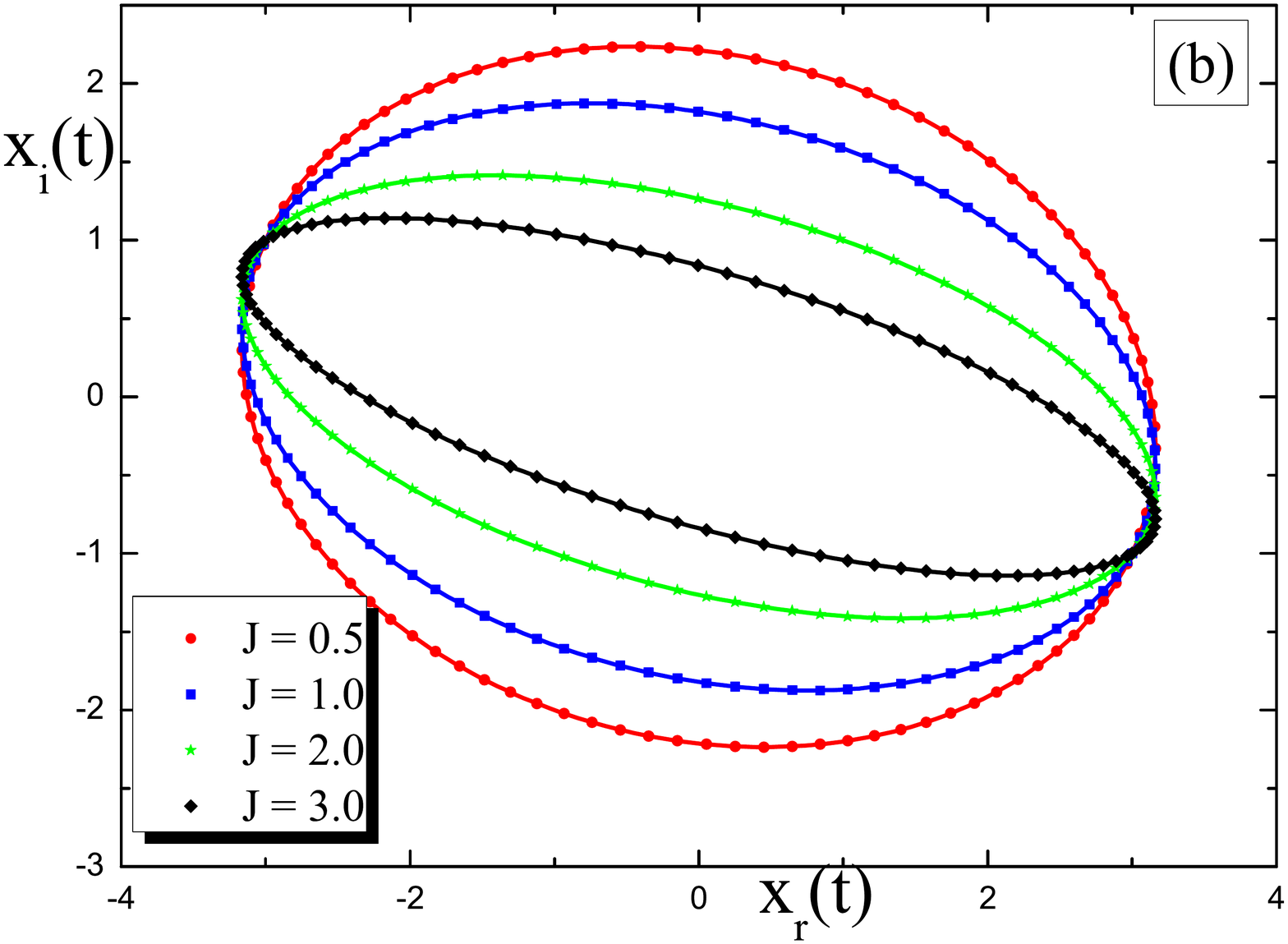}
\caption{\small{Complex Bohmian trajectories resulting from Klauder coherent states (dotted) compared to the purely classical computation (\protect\ref{x2})
(lined) for different values of $J$, with initial value (a) $x_{0}=3+i$ and
(b) $x_{0}=3-i$ with maximal values $x_{\text{max}}^{0.5}=1$, $x_{\text{max}
}^{1}=1.4142$, $x_{\text{max}}^{2}=2$, $x_{\text{max}}^{3}=2.4495$.}}
\label{figbohm3}
\end{figure}

In order to compare this now with the outcome from taking general Klauder
coherent states (\ref{klaudercoherent}) in the evaluation of $\tilde{v}(t)$ and $\tilde{Q}(x,t)$ and the corresponding trajectories we require the four initial
values. In contrast, solving the first order differential equation for the
velocity (\ref{complex}) we only require the two initial values for the
complex position. To compute the initial values for the momentum we can take
again the results for the Gaussian wavepackets as a guide and compare (\ref{x1}) and (\ref{x2}). The compatibility between the two then requires
\begin{equation}
x_{r}(0)=\frac{p_{i}(0)}{m\omega }+x_{\text{max}}^{J},\qquad \text{and\qquad 
}x_{i}(0)=-\frac{p_{r}(0)}{m\omega },  \label{con}
\end{equation}
where we have replaced $a$ by $x_{\text{max}}^{J}$. We can now either simply
solve this for the initial values for the momentum (\ref{con}) or
alternatively use directly the same initial values obtained from the
solution of (\ref{complex}). Comparing the direct parametric plot of (\ref
{x2}) for the stated initial conditions with the numerical computation of
the complex Bohmian trajectories resulting from Klauder coherent states, we
find perfect agreement as depicted in figure \ref{figbohm3}.

Thus under these constraints for the initial conditions the trajectories
resulting from a classical analysis of the effective Hamiltonian (\ref{Heff}) and the integration of the complex Bohmian trajectories resulting from Klauder coherent states are identical. Notice that the quantum nature of $\mathcal{H}_{\text{eff}}$ is only visible in form of the overall constant $\hbar \omega /2$, which does, however, not play any role in the computation of the equations of motion.

\section{The P{\"o}schl-Teller Potential}
Next we discuss the Bohmian trajectories associated with the P\"{o}schl-Teller Hamiltonian \cite{poschl_teller} of the form 
\begin{equation}
\mathcal{H}_{\text{PT}}=\frac{p^{2}}{2m}+\frac{V_{0}}{2}\left[ \frac{\lambda
(\lambda -1)}{\cos ^{2}(x/2a)}+\frac{\kappa (\kappa -1)}{\sin ^{2}(x/2a)}
\right] -\frac{V_{0}}{2}(\lambda +\kappa )^{2}~~~\text{for }0\leq x\leq a\pi,  \label{HPT}
\end{equation}
with $V_{0}=\hbar ^{2}/(4ma^{2})$. This model has been widely discussed in
the mathematical physics literature, e.g. \cite{antoine_gazeau_monceau_klauder_penson,kleinert_mustapic}, since it has the virtue of being exactly solvable, classically as well as quantum mechanically. For a given energy $E$ a classical solution is known to be
\begin{equation}
x(t)=a\arccos \left[ \frac{\alpha -\beta }{2}+\sqrt{\gamma }\cos \left( 
\sqrt{\frac{2E}{m}}\frac{t}{a}\right) \right] ,  \label{PTcl}
\end{equation}
with $\alpha =\lambda (\lambda -1)V_{0}/E$, $\beta =\kappa (\kappa
-1)V_{0}/E $ and $\gamma =\alpha ^{2}/4+\beta ^{2}/4-\alpha \beta /2-\alpha
-\beta +1$. The time dependent Schr\"{o}dinger equation is solved by discrete eigenfunctions 
\begin{equation}
\psi _{n}(x,t)=\frac{1}{\sqrt{N_{n}}}\cos ^{\lambda }\left( \frac{x}{2a}
\right) \sin ^{\kappa }\left( \frac{x}{2a}\right) ~_{2}F_{1}\left[
-n,n+\kappa +\lambda ;k+\frac{1}{2};\sin ^{2}\left( \frac{x}{2a}\right)
\right] e^{-iE_{n}t/\hbar }  \label{PTST}
\end{equation}
$\allowbreak $with $_{2}F_{1}$ denoting the Gauss hypergeometric function.
The corresponding energy eigenvalues and the normalization factor are given
by 
\begin{eqnarray}
E_{n}&=&\frac{\hbar ^{2}}{2ma^{2}}n(n+\kappa +\lambda ),  \\
N_{n}&=&a2^{n}n!\frac{\Gamma (\kappa +1/2)\Gamma (n+\lambda +1/2)}{\Gamma
(2n+1+\lambda +\kappa )}\prod\limits_{l=1}^{n}\frac{n-1+l+\kappa +\lambda }{
2l-1+2\kappa },\notag
\end{eqnarray}
respectively. We will use these solutions in what follows.

\subsection{Real Case} \label{section731}
As in the previous case we start with the construction of the trajectories
from stationary states (\ref{PTST}). Once again for the real case the
computation is unspectacular in this case since the velocity computed from (\ref{realvel}) is $v(t)=0$ and the corresponding quantum potential results
again simply to $Q(x)=E_{n}-V_{\text{PT}}(x)$, such that classical
trajectories correspond to a motion in a constant effective potential $V_{\text{eff}}(x,t)=E_{n}$.

More interesting, and qualitatively very close to the classical behaviour,
are the trajectories resulting from the Klauder coherent states given by the
general expression (\ref{klaudercoherent}). In this case the probability distribution (\ref{probdensity}) is computed with $e_{n}=n(n+\kappa +\lambda )$ to $\rho _{n}=n!(n+\kappa+\lambda )_{n}$, where $(x)_{n}:=\Gamma (x+n)/\Gamma (x)$ denotes the Pochhammer symbol. With these expressions the normalization constant reduces to a confluent hypergeometric function $\mathcal{N}^{2}(J)=~_{0}F_{1}\left(1+\kappa +\lambda;J\right) $, from which we compute the Mandel parameter (\ref{mandel}) to be
\begin{equation}
Q(J,\kappa +\lambda )=\frac{J}{2+\kappa +\lambda }~\frac{_{0}F_{1}\left(
3+\kappa +\lambda ;J\right) }{_{0}F_{1}\left( 2+\kappa +\lambda ;J\right) }-
\frac{J}{1+\kappa +\lambda }~\frac{_{0}F_{1}\left( 2+\kappa +\lambda
;J\right) }{_{0}F_{1}\left( 1+\kappa +\lambda ;J\right) }~.
\end{equation}
Using the relation between the confluent hypergeometric function and the
modified Bessel function this is easily converted into the expression found
in \cite{antoine_gazeau_monceau_klauder_penson}. We agree with the finding therein that $Q$ is always negative, but disagree with the statement that $Q$ tends to zero for large $J $ for fixed $\kappa $, $\lambda $. Instead we argue that for fixed coupling constants the Mandel parameter $Q$ is a monotonically decreasing function of $J$ with $Q(0,\kappa +\lambda )=0$. Assuming that the coherent states closely resemble a classical behaviour, we conjecture here in analogy to the classical solution (\ref{PTcl}) that the quantum trajectories acquire the general form
\begin{equation}
x(t)=a\arccos \left[ \frac{X_{+}}{2}+\frac{X_{-}}{2}\cos \left( 2\pi \frac{t
}{T}\right) \right] ,  \label{PTx}
\end{equation}
with $X_{\pm }=\cos (x_{0}/a)\pm \cos (x_{m}/a)$, $T$ denoting the period
and $x_{m}=x(T/2)=\max [x(t)]$. Our conjecture is based on an extrapolation
of the analysis of the relations between $\alpha $ and $\beta $ and functions of $x(0)$ and $x(T/2)$. The effective potential computed from (\ref{PTx}) is then of P\"{o}schl-Teller type 
\begin{equation}
V_{\text{eff}}=\frac{2ma^{2}\pi ^{2}}{T^{2}}\left[ \frac{\cos
^{2}(x_{0}/2a)\cos ^{2}(x_{m}/2a)}{\cos ^{2}(x/2a)}+\frac{\sin
^{2}(x_{0}/2a)\sin ^{2}(x_{m}/2a)}{\sin ^{2}(x/2a)}\right] .
\end{equation}
As in the previous case we will compute the quantum trajectories numerically. We take $a=2$ in all numerical computations in this section. Our results
from solving (\ref{realvel}) are depicted in figure \ref{figbohm4}.

\begin{figure}[H]
\centering   \includegraphics[width=7.5cm,height=6.0cm]{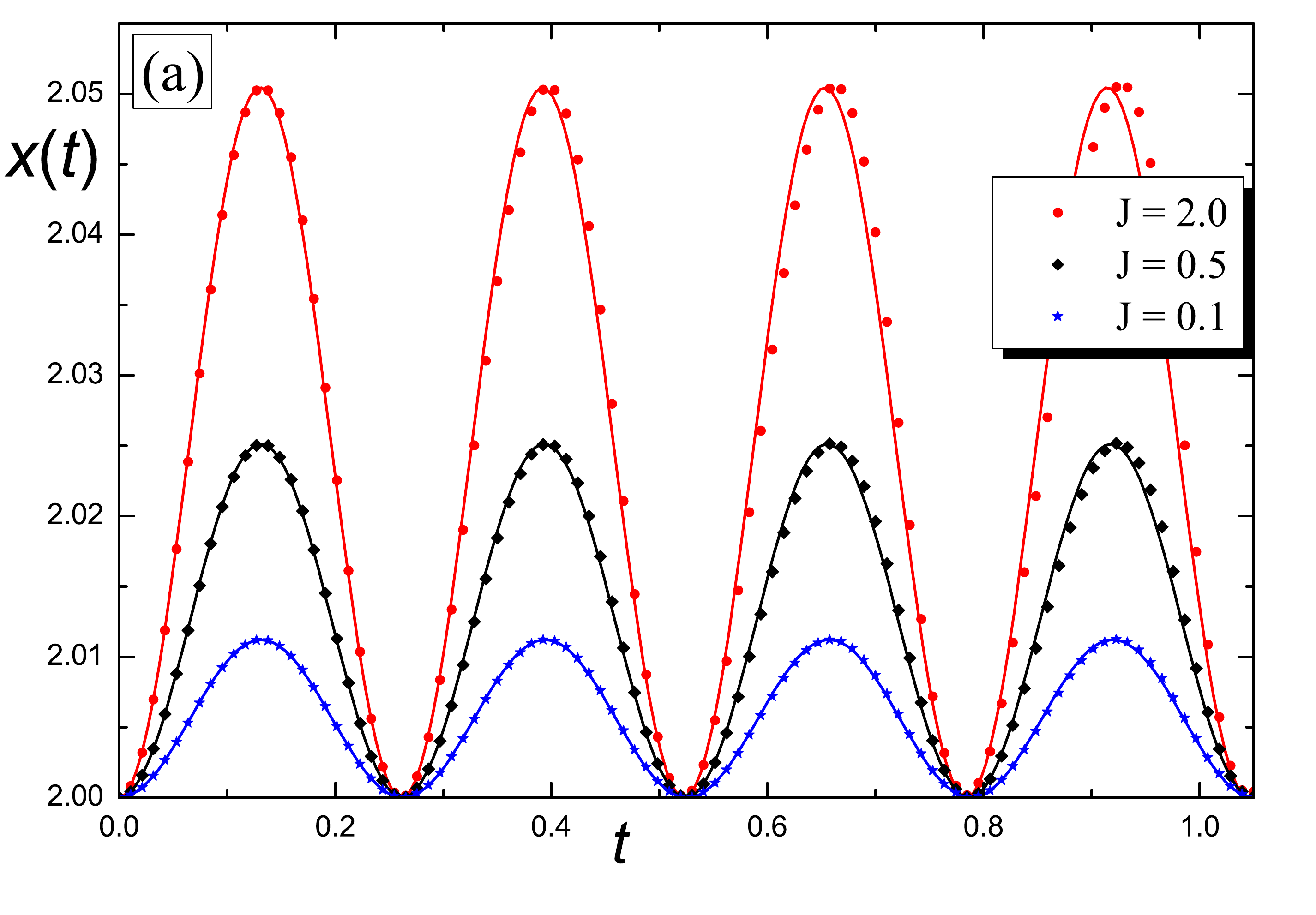}
\includegraphics[width=7.5cm,height=6.0cm]{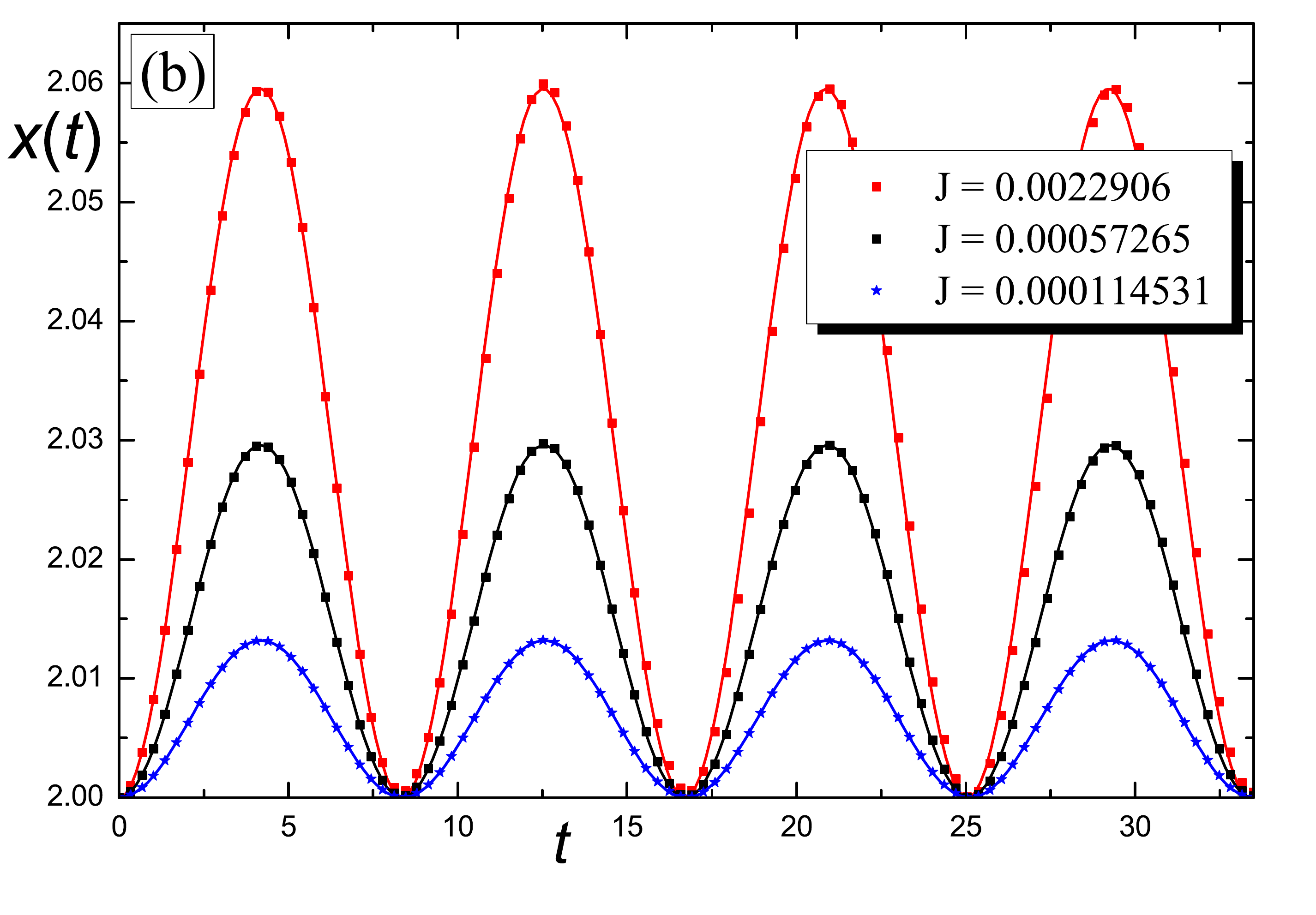} \centering
\includegraphics[width=7.5cm,height=6.0cm]{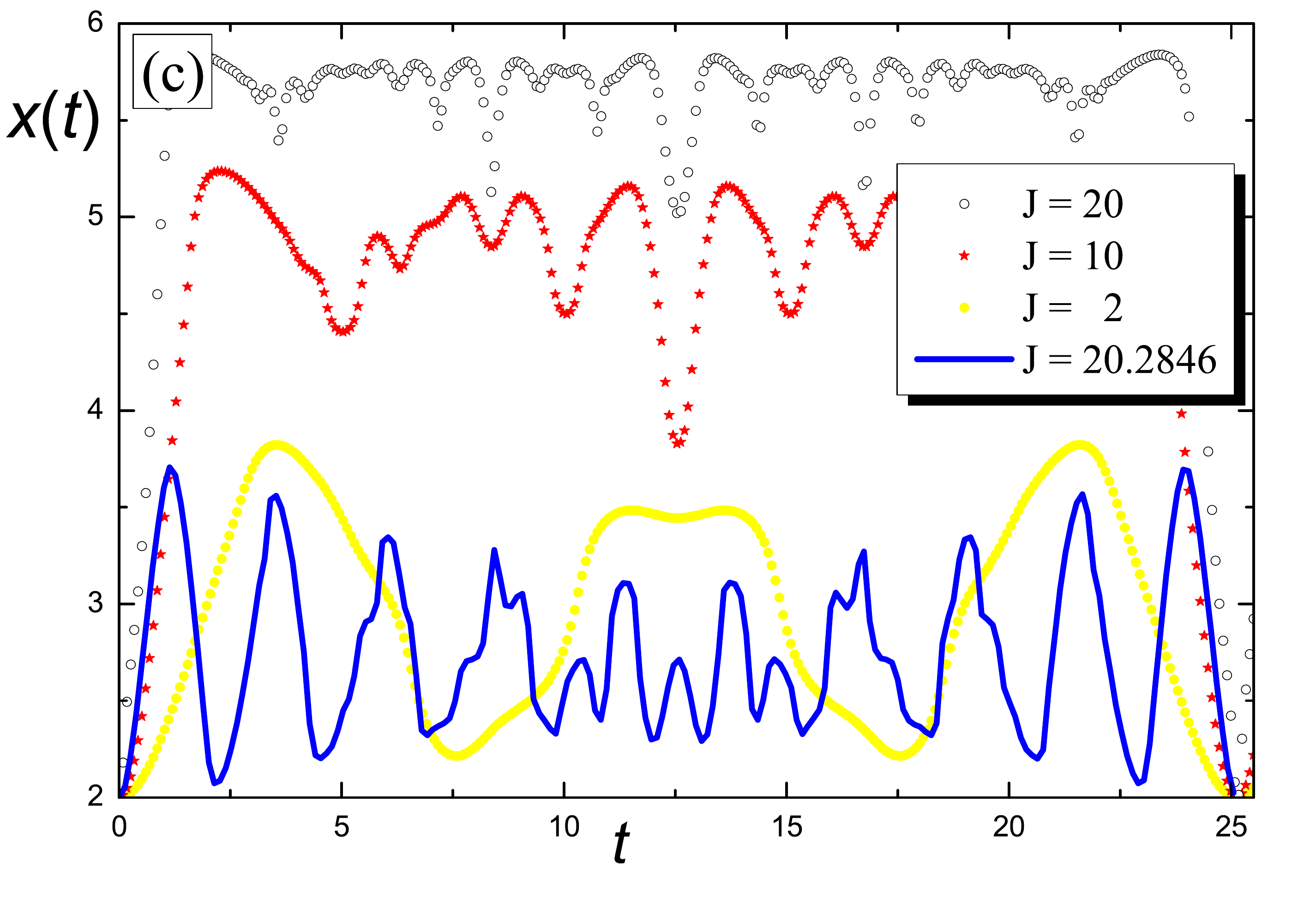}
\includegraphics[width=7.5cm,height=6.0cm]{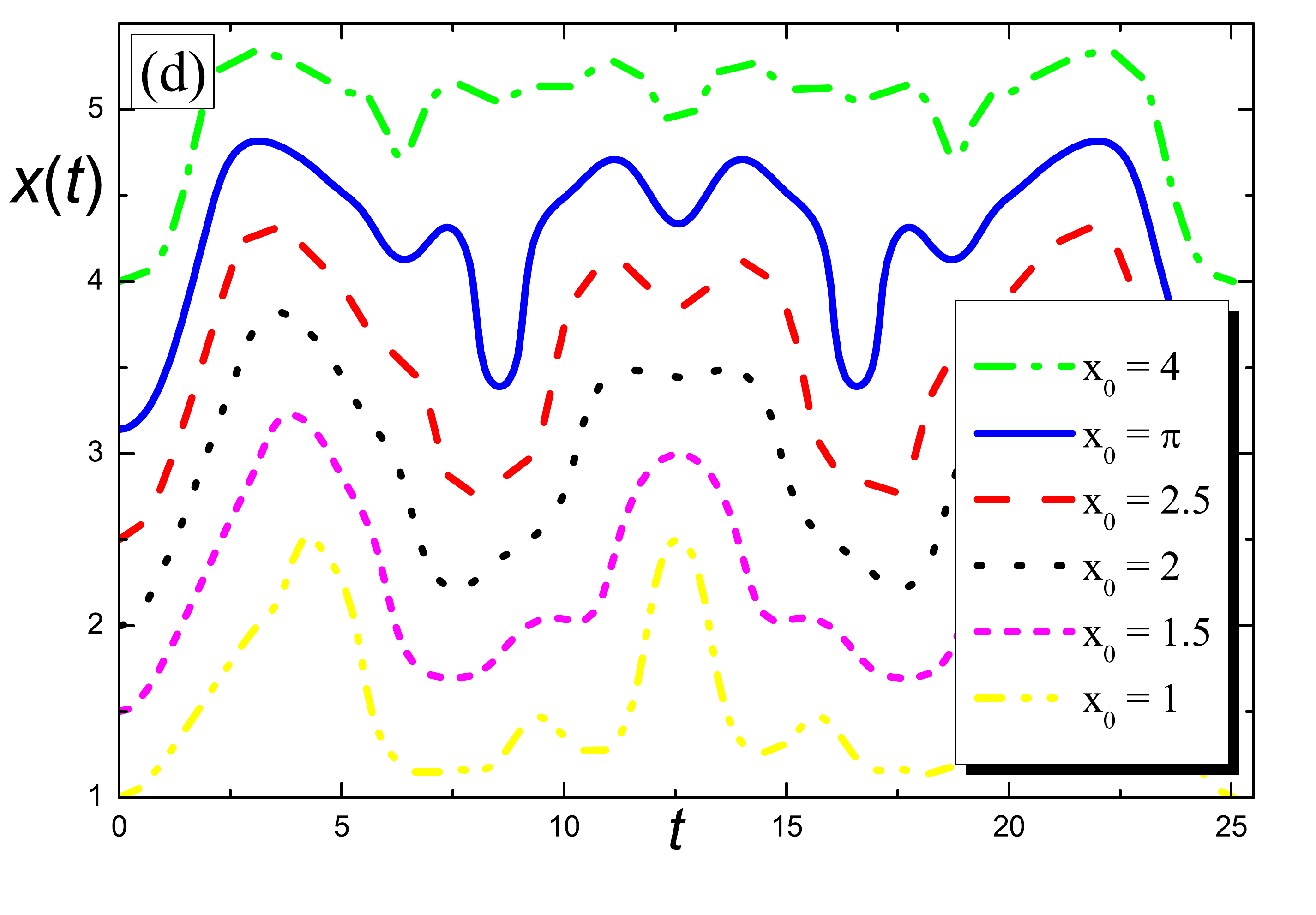}
\caption{\small{Real Bohmian trajectories as functions of time from Klauder
coherent states (dotted) versus classical trajectories (solid lines)
corresponding to (\protect\ref{PTx}). (a) Quasi-Poissonian distribution $(Q\approx 0)$ with initial value $x_{0}=2$, coupling constants $\protect\kappa =90$, $\protect\lambda =100$, maxima $x_{\text{m}}^{2}=2.0504447$, $x_{\text{m}
}^{0.5}=2.0251224$, $x_{\text{m}}^{0.1}=2.0112100$ and periods $
T^{2}=0.2612875$, $T^{0.5}=0.2622200$, $T^{0.1}=0.2627495$ for different
values of $J$. (b) Quasi-Poissonian distribution $(Q\approx 0)$ with initial value $x_{0}=2$, coupling constants $\protect\kappa =2$, $\protect\lambda =3$, maxima $x_{\text{m}}^{0.0022906}=2.059522$, $x_{\text{m}}^{0.00057265}=2.0295876$, $x_{
\text{m}}^{0.000114531}=2.0131884$ and periods $T^{0.0022906}=8.34795$, $
T^{0.00057265}=8.36305$, $T^{0.000114531}=8.37129$ for different values of $
J $. (c) Sub-Poissonian distribution with initial value $x_{0}=2$, coupling
constants $\protect\kappa =2$, $\protect\lambda =3$ for $J=2,10,20 $
(dotted) and $\protect\kappa =9$, $\protect\lambda =10$ for $J=20.2846$
(solid). (d) Sub-Poissonian distribution for various initial values with
coupling constants $\protect\kappa =2$, $\protect\lambda =3$ for $J=2$.}}
\label{figbohm4}
\end{figure}

Most importantly we observe that the behaviour of the trajectories is
entirely controlled by the values of the Mandel parameter $Q$. Panel (a) and
(b) of figure \ref{figbohm4} show trajectories for different values of $J$ with pairwise identical values of the Mandel parameter, that is $Q(2,190)=Q(0.0022906,5)=-0.000054529 $, $Q(0.5,190)=Q(0.00057265,5)=-0.000013634$ and $Q(0.1,190)=Q(0.000114531,5)=-2.72691\times 10^{-6}$. We notice that the
overall qualitative behaviour is simply rescaled in time. We further observe
a small deviation from the periodicity growing with increasing time. As a
consequence the matching between the quantum trajectories obtained from
solving (\ref{realvel}) and our conjectured analytical expression (\ref{PTx})
is good for small values of time, but worsens as time increases. The
agreement improves the closer the Mandel parameter approaches the Poissonian
distribution function, i.e. $Q=0$. Once the Mandel parameter becomes very negative
the correlation between the classical motion and the Bohmian trajectories is
entirely lost as shown in panel (c) of figure \ref{figbohm4} for $Q(2,5)=-0.0425545 $, $Q(10,5)=-0.149523$ and $Q(20,5)=-0.218944$. We also
notice from panel (c) that the qualitative similarity observed for equal
values of the Mandel parameter seen in panels (a) and (b) is lost once the
states do not resemble a classical behaviour. This is seen by comparing the yellow dotted line and the solid blue line corresponding to the same
values $Q(2,5)=Q(20.2846,19)=-0.0425545$. Panel (d) shows the sensitivity
with regard to the initial values $x_{0}$. Whereas for the trajectories
resembling the classical motion (\ref{PTx}) this change does not affect the
overall qualitative behaviour, it produces a more significant variation in
the non-classical regime.

\begin{figure}[H]
\centering   \includegraphics[width=7.5cm,height=6.0cm]{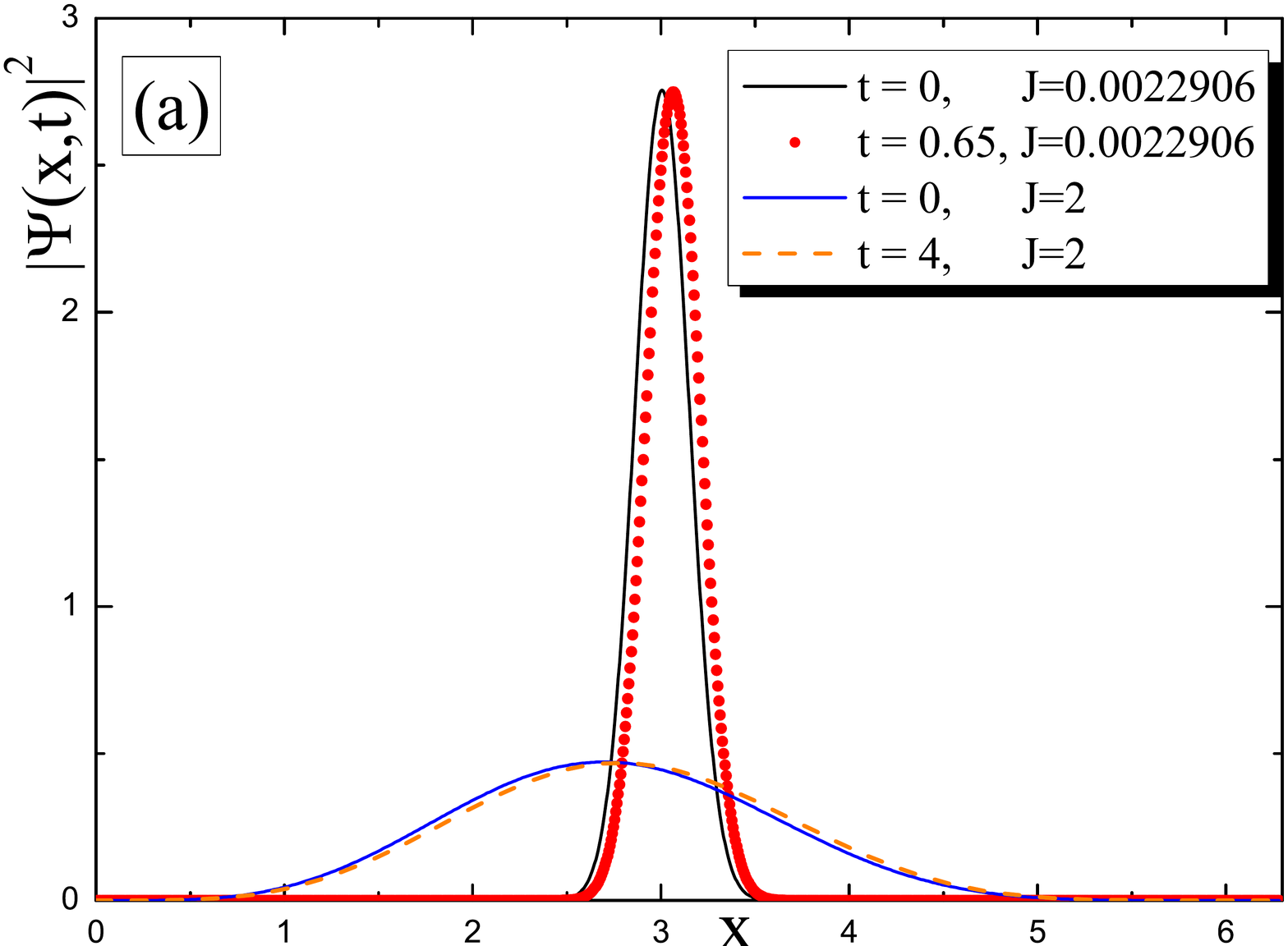}
\includegraphics[width=7.5cm,height=6.0cm]{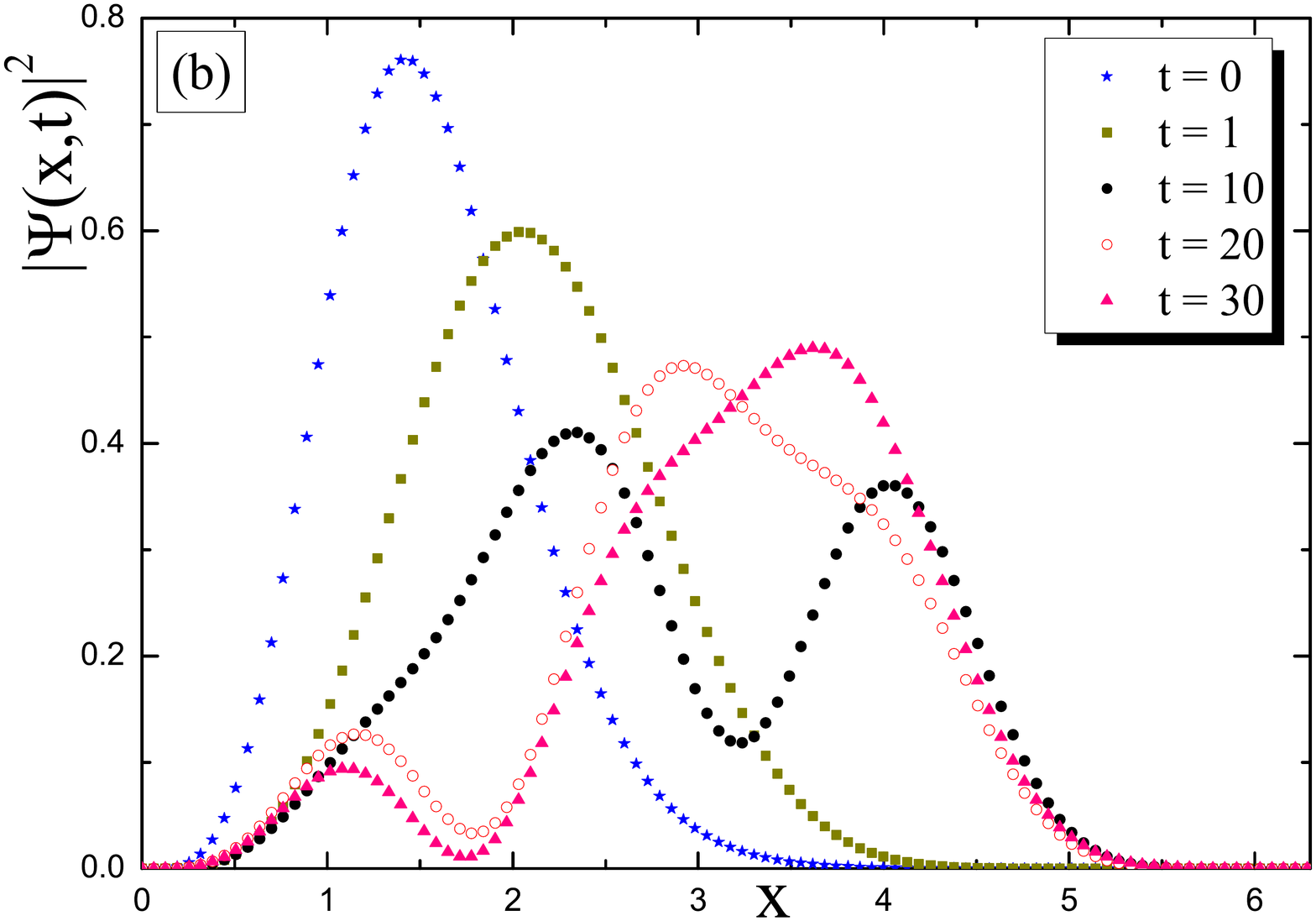}
\caption{\small{(a) Periodic soliton like motion in the quasi-Poissonian regime for $\protect\kappa =90$, $\protect\lambda =100$ (thin) and $\protect\kappa =2$, 
$\protect\lambda =3$ (broad) with $Q=-0.000054529$ identical in both cases.
(b) Spreading wave in the sub-Poissonian regime with $Q=-0.0917752$ for $J=5$
and $\protect\kappa =2$, $\protect\lambda =3$.}}
\label{figbohm5}
\end{figure}

The observation for this behaviour is that in the quasi-Poissonian regime
the coherent states evolve as soliton like structures keeping their shape
carrying out a periodic motion in time. In contrast, in the sub-Poissonian
regime the motion is no longer periodic and the initial Gaussian shape of
the wave is dramatically changed under the evolution of time. These features
are demonstrated in figure \ref{figbohm5}.

\begin{figure}
\centering   \includegraphics[width=7.5cm,]{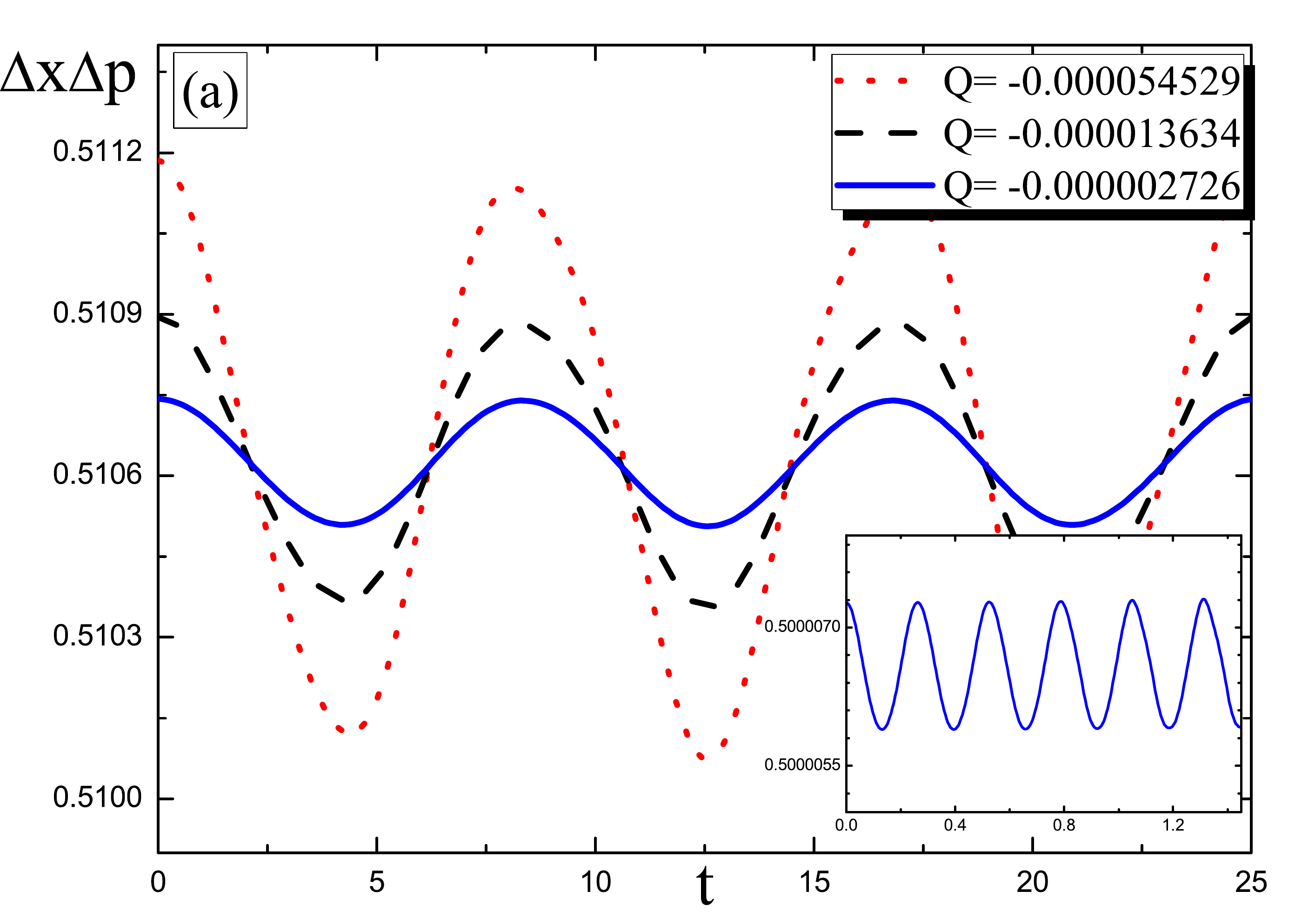}
\includegraphics[width=7.5cm]{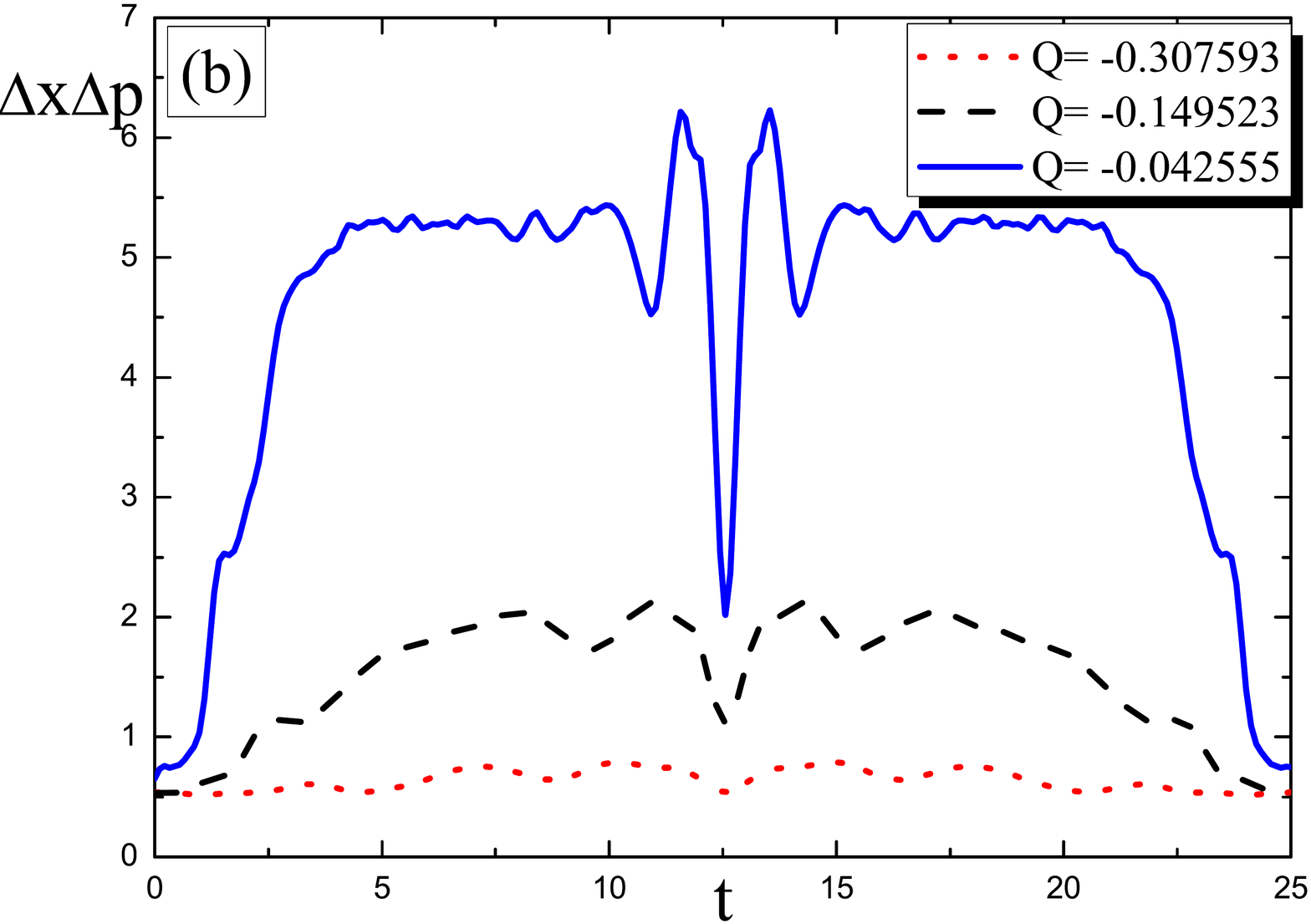}
\caption{\small{Product of the position and momentum uncertainty as functions of
time for different values of the Mandel parameter $Q$. (a) The coupling
constants are $\protect\kappa =2,\protect\lambda =3$, with $J=0.0022906$
(red dotted), $J=0.00057265$ (black dashed), $J=0.000114531$ (blue solid)
and $J=0.1$ for the subpanel with $\protect\kappa =90,\protect\lambda =100$.
(b) The coupling constants are $\protect\kappa =2,\protect\lambda =3$ with $
J=2$ (red dotted), $J=10$ (black dashed) and $J=50$ (blue solid).}}
\label{figbohm6}
\end{figure}

In figure \ref{figbohm6} we plot the uncertainty relations for comparison. In panel (a) of figure \ref{figbohm6}, we observe that in the quasi-Poissonian regime the saturation level is almost reached with $\Delta x \Delta p$ being very close to $1/ 2$, oscillating around $0.5106$ with a deviation of $\pm 0.0006$ and in the subpanel oscillating around $0.5000065$ with a deviation of $\pm 0.0000008$. This is of course compatible with the very narrow soliton like
structure observed in figure \ref{figbohm5} leading to a classical type of behaviour. However, in the sub-Poissonian regime the uncertainty becomes larger, as seen in panel (b), corresponding to a spread out wave behaving very non-classical.

\subsection{Complex Case}
Let us now consider the complex Bohmian trajectories starting once again
with the construction from stationary states $\psi _{n}(x,t)=\phi_{n}(x)e^{-iE_{n}t/\hbar }$. For the lowest states we may compute analytical expressions from (\ref{complex}) for the velocities 
\begin{eqnarray}
\tilde{v}_{0}(x,t) &=&\frac{\hbar \left[ (\kappa +\lambda )\cos \left( \frac{
x}{a}\right) +\kappa -\lambda \right] }{i2am\sin \left( \frac{x}{a}\right) }
,~~ \\
\tilde{v}_{1}(x,t) &=&\frac{\hbar \left[ (2\kappa ^{2}+\kappa )\cot \left( 
\frac{x}{2a}\right) +(2\lambda ^{2}+\lambda )\tan \left( \frac{x}{2a}\right)
-(\kappa +\lambda +1)(\kappa +\lambda +2)\sin \left( \frac{x}{a}\right)
\right] }{i2am\left[ (\kappa +\lambda +1)\cos \left( \frac{x}{a}\right)
+\kappa -\lambda \right] },  \notag
\end{eqnarray}
and the quantum potentials
\begin{eqnarray}
\tilde{Q}_{0}(x,t) &=&V_{0}\frac{\left[ (\kappa -\lambda )\cos \left( \frac{x
}{a}\right) +\kappa +\lambda \right] }{\sin ^{2}\left( \frac{x}{a}\right) },
\label{Q1} \\
\tilde{Q}_{1}(x,t) &=&\frac{V_{0}}{2}\left[ \frac{4(\kappa +\lambda
+1)\left( (\kappa -\lambda )\cos \left( \frac{x}{a}\right) +\kappa +\lambda
+1\right) }{\left[ (\kappa +\lambda +1)\cos \left( \frac{x}{a}\right)
+\kappa -\lambda \right] ^{2}}+\frac{\kappa }{\sin ^{2}\left( \frac{x}{2a}
\right) }+\frac{\lambda }{\cos ^{2}\left( \frac{x}{2a}\right) }\right] . 
\notag  \label{Q2}
\end{eqnarray}
We note that the quantum potential $\tilde{Q}_{1}(x,t)$ resembles a P\"{o}schl-Teller potential apart from its first term. For $n=0$ we solve (\ref{complex}) analytically for the trajectories 
\begin{equation}
x_{0}(t)=\pm a\arccos \left\{ \frac{\left[ (\kappa +\lambda )\cos \left( 
\frac{x_{0}}{a}\right) +\kappa -\lambda \right] e^{\frac{iht(\kappa +\lambda
)}{2a^{2}m}}+\lambda -\kappa }{\kappa +\lambda }\right\} .
\end{equation}
For excited states one may easily solve these equations numerically as
depicted in figure \ref{figbohm7}.

\begin{figure}
\centering   \includegraphics[width=7.5cm,height=6.0cm]{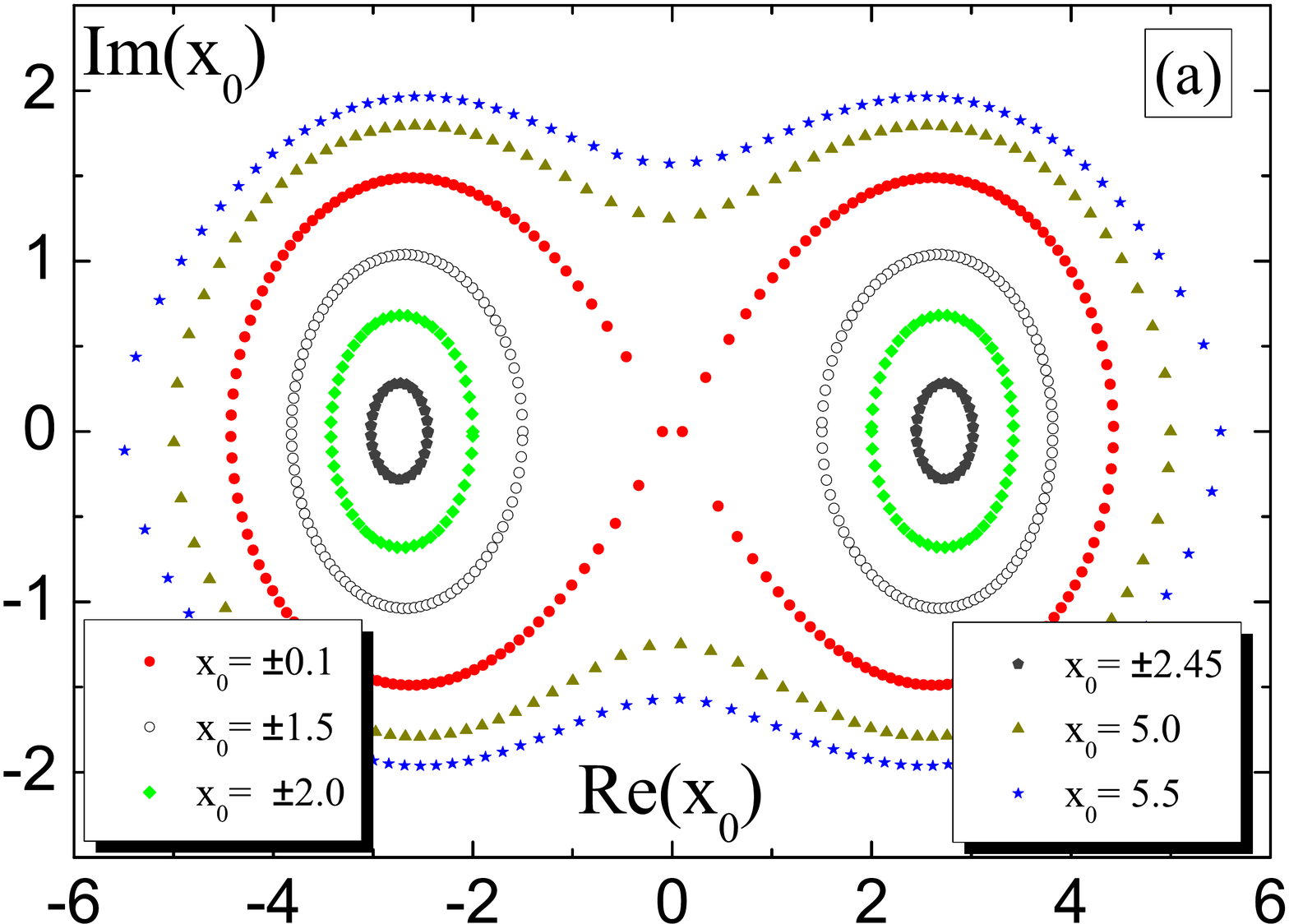}
\includegraphics[width=7.5cm,height=6.0cm]{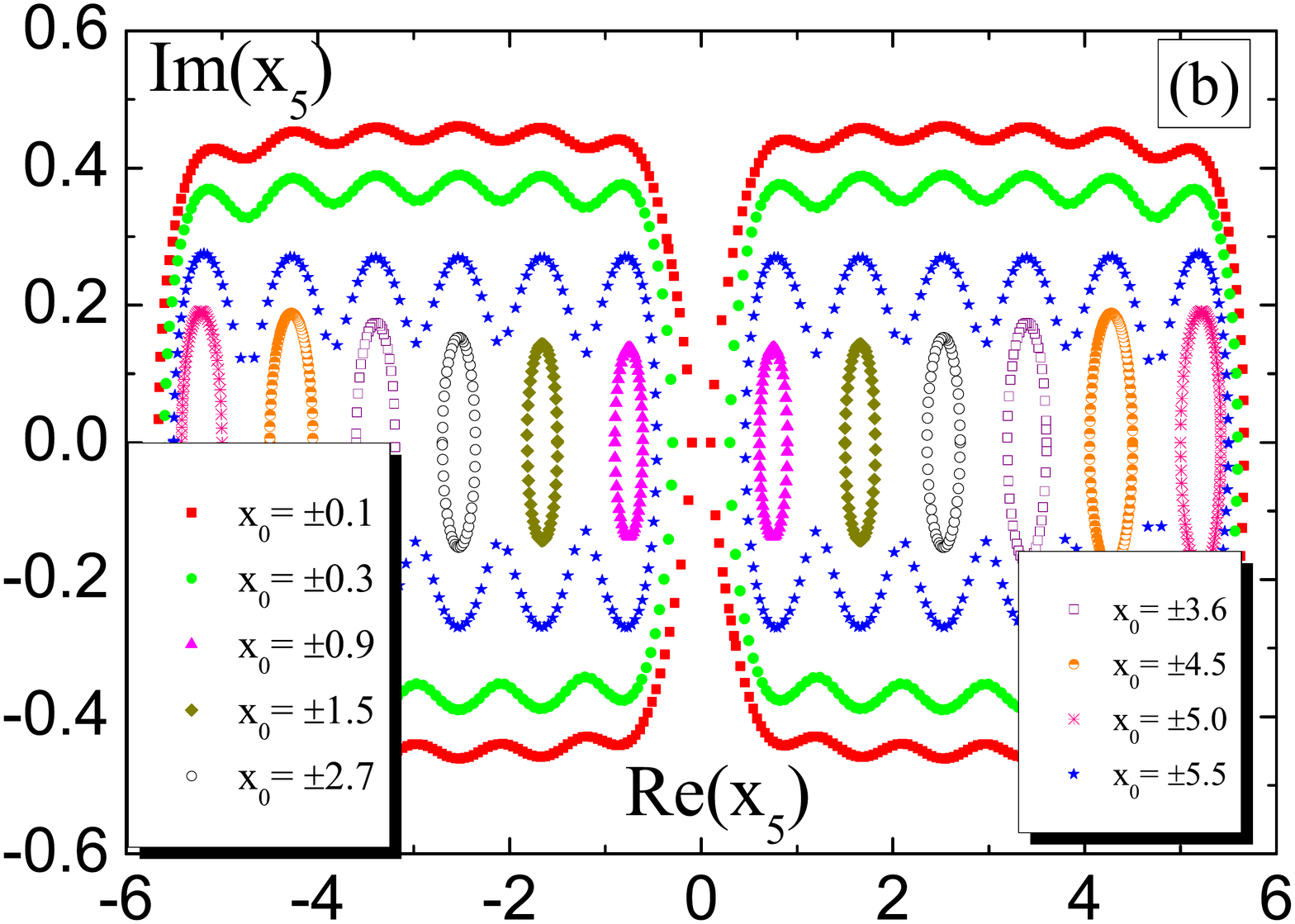}
\caption{\small{Complex Bohmian trajectories as functions of time for different
initial values $x_{0}$ resulting from stationary states $\Psi _{0}(x,t)$ and 
$\Psi _{5}(x,t)$ in panel (a) and (b), respectively.}}
\label{figbohm7}
\end{figure}

We observe the usual appearance of the only possible types of fixed points
in a Hamiltonian system, that is centres and saddle points. For the ground
state we observe a close resemblance of the qualitative behaviour with the
solution of the first excited state obtained for the harmonic oscillator as
shown in figure \ref{figbohm2}.

Unlike the trajectories resulting from coherent states those obtained from
stationary states are not expected to have a similar behaviour to the purely
classical ones obtained from solving directly the equations of motion (\ref{H1}) and (\ref{H2}). Complexifying the variables as specified after (\ref{H1}) and (\ref{H2}), we may split the Hamiltonian into its real and
imaginary part $\mathcal{H}_{\text{PT}}=H_{r}+iH_{i}$ with
\begin{eqnarray}
H_{r} &=&\frac{p_{r}^{2}-p_{i}^{2}}{2m}-\frac{V_{0}}{2}(\lambda +\kappa )^{2} \notag \\
&&+V_{0}\left[ \frac{(\lambda
^{2}-\lambda )\left[ \cosh \left( \frac{x_{i}}{a}\right) \cos \left( \frac{
x_{r}}{a}\right) +1\right] }{\left[ \cosh \left( \frac{x_{i}}{a}\right)
+\cos \left( \frac{x_{r}}{a}\right) \right] {}^{2}}-\frac{(\kappa
^{2}-\kappa )\left[ \cosh \left( \frac{x_{i}}{a}\right) \cos \left( \frac{
x_{r}}{a}\right) -1\right] }{\left[ \cos \left( \frac{x_{r}}{a}\right)
-\cosh \left( \frac{x_{i}}{a}\right) \right] {}^{2}}\right],  \notag \\
H_{i} &=&\frac{p_{i}p_{r}}{m}+V_{0}\left[ \frac{(\lambda ^{2}-\lambda )\sinh
\left( \frac{x_{i}}{a}\right) \sin \left( \frac{x_{r}}{a}\right) }{\left[
\cosh \left( \frac{x_{i}}{a}\right) +\cos \left( \frac{x_{r}}{a}\right)
\right] ^{2}}-\frac{(\kappa ^{2}-\kappa )\sinh \left( \frac{x_{i}}{a}\right)
\sin \left( \frac{x_{r}}{a}\right) }{\left[ \cos \left( \frac{x_{r}}{a}
\right) -\cosh \left( \frac{x_{i}}{a}\right) \right] ^{2}}\right] .
\label{PTcomplex}
\end{eqnarray}
This Hamiltonian also respects the aforementioned $\mathcal{PT}$-symmetry $\mathcal{PT}$: $x_{r}\rightarrow -x_{r}$, $x_{i}\rightarrow x_{i}$, $p_{r}\rightarrow p_{r}$, $p_{i}\rightarrow -p_{i}$, $i\rightarrow -i$.
Contourplots of the potential are shown in figures \ref{figbohm9} and \ref{figbohm10} with the colourcode convention being associated to the spectrum of light decreasing from red to violet. The corresponding equations of motion are easily computed from (\ref{H1}) and (\ref{H2}), albeit not reported here as
they are very lengthy, and solved numerically as shown for some parameter
choices in the figures \ref{figbohm8}, \ref{figbohm9} and \ref{figbohm10} as solid lines.

Let us now compare them with the complex Bohmian trajectories computed from
the Klauder coherent states (\ref{klaudercoherent}). A previous initial attempt to compute these trajectories has been made in \cite{john_mathew}, however,
the preliminary computations presented there do not agree with our findings.
We start by depicting a case for the quasi-Poissonian distribution in figure
\ref{figbohm8}.

Remarkably, in that case we find a perfect match between these two entirely
different computations. We observe that unlike as for the real trajectories,
for which we required an effective potential to achieve agreement, these
computations are carried out in both cases for exactly the same coupling
constants $\kappa $ and $\lambda $ with no adjustments made. Thus, just as
for the harmonic oscillator, this suggests that the complex quantum
potential is simply a constant such that the effective potential essentially
coincides with the original one in (\ref{HPT}). From the trajectories with
larger radii in panel (a) of figure \ref{figbohm8} we observe that the trajectories do not close and are not perfect ellipses. Prolonging the time beyond the cut-off time in the panel (a) scenario we find inwardly spiralling trajectories. The coincidence between the purely classical calculation and the quantum trajectories still persists for larger values of $J$ and more asymmetrical initial conditions closer to the boundary of the potential as shown for an example in panel (b). For larger values of time we encounter numerical problems due to the poor convergence of the series for large values of $J$.

Our initial values in \ref{figbohm8}(a) are chosen to lie on the isochrones, that is the set of all points which when evolved in time will arrive all
simultaneously, say at $t_{f}$, on the real axis. The isochrone is indicated
in figure \ref{figbohm8}(a) by a blue line and an additional arrow attached to it
pointing in the direction in which the real axis is reached. In our example
the arrival time is chosen to be $t=0.04$. As discussed for instance in \cite{goldfarb_degani_tannor,chou_wyatt_complex} the wavefunction defined on the isoclines can be thought of as leading to physical information as their corresponding complex quantum trajectories acquire real values. Moreover, as shown in \cite{chou_wyatt_complex}, one may even reconstruct the precise form of the entire wavefunction from the knowledge of the isochrone and the information transported by the action. We will follow up this line of enquiry elsewhere.

\begin{figure}[h]
\centering   \includegraphics[width=7.5cm,height=6.0cm]{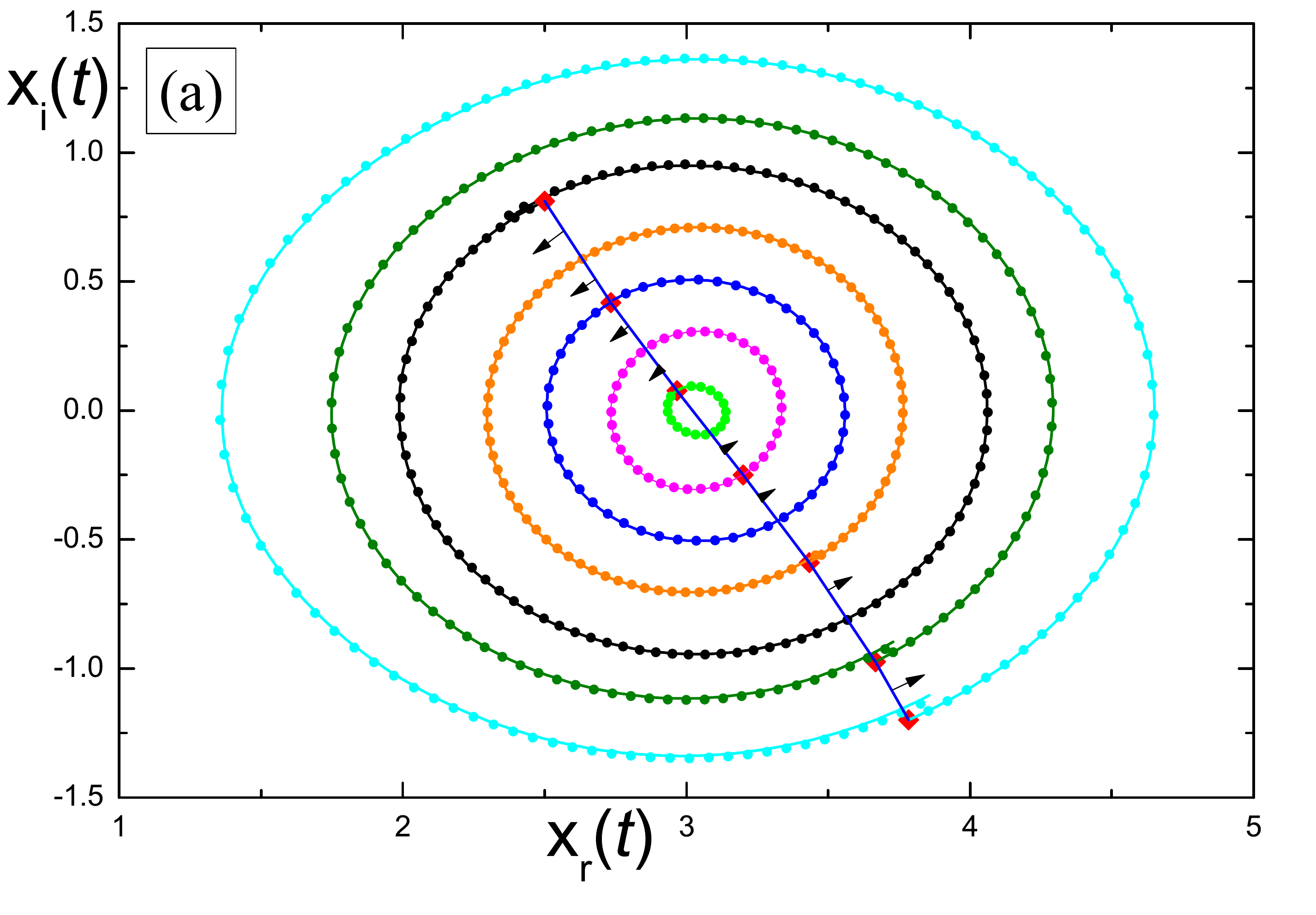}
\includegraphics[width=7.5cm,height=6.0cm]{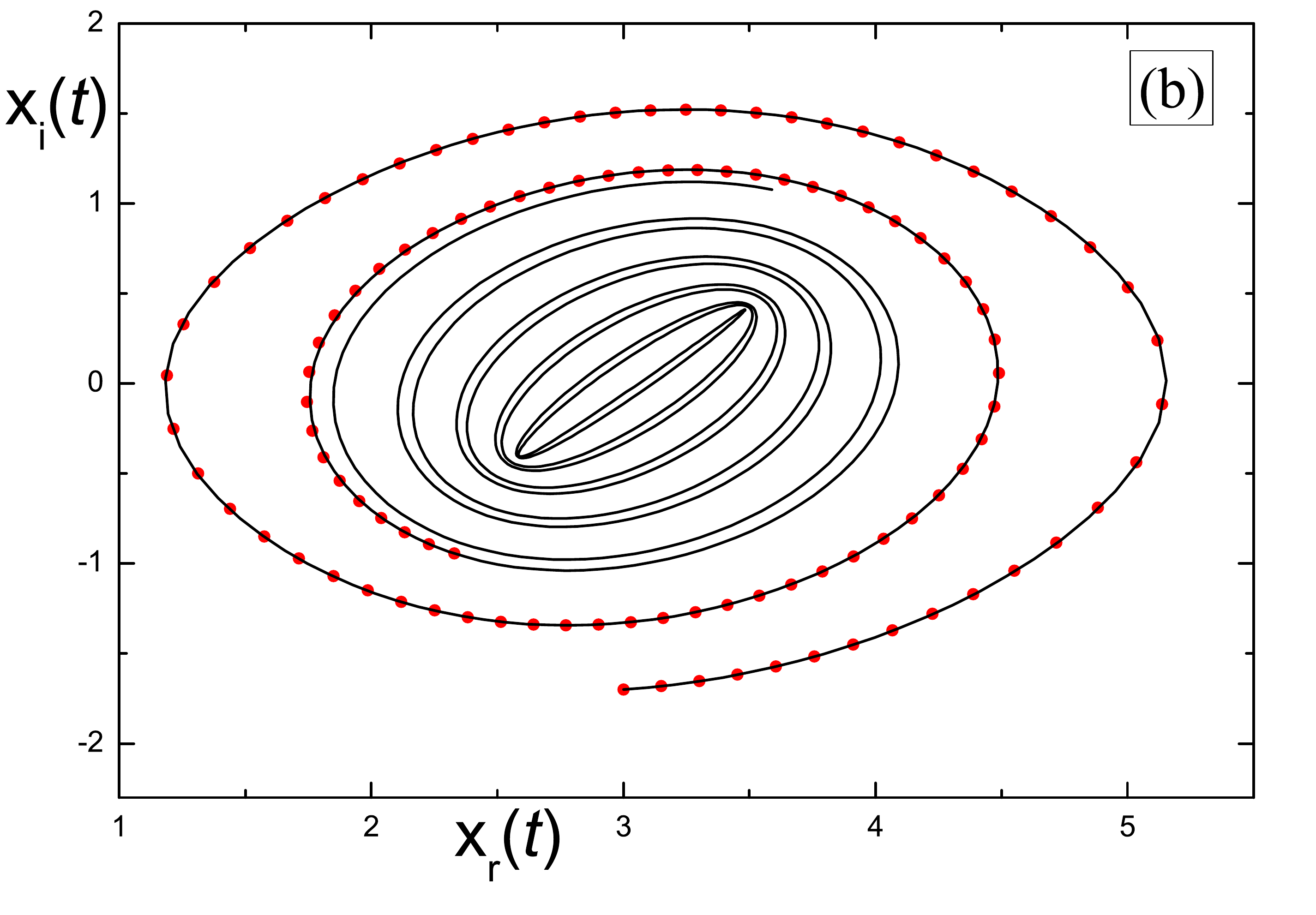}
\caption{\small{Complex Bohmian trajectories as functions of time from Klauder
coherent states (dotted) versus classical trajectories corresponding to
solutions of (\protect\ref{H1}) and (\protect\ref{H2}) for the complex P\"{o}
schl-Teller Hamiltonian $\mathcal{H}_{\text{PT}}$ (\protect\ref{PTcomplex})
(solid lines) for quasi-Poissonian distribution with $\protect\kappa =90$, $
\protect\lambda =100$. (a) The evolution is shown from $t=0$ to $t=0.27$
with initial values on the isochrone with $t_{f}=0.04$ for $J=0.5$ and in
(b) from $t=0$ to $t=0.5$ (quantum) and $t=0$ to $t=3.0$ (classical) with
initial value $x_{0}=3-1.7i$ and $p_{0}=32.907+3.16416i$ for $J=20$. The
energy for this trajectory is therefore $E=\mathcal{H}_{\text{PT}
}(x_{0},p_{0})=-8.44045+102.134i$.}}
\label{figbohm8}
\end{figure}

The qualitative behaviour of the classical trajectories can be understood by
considering the motion in the complex potential. We consider first a trajectory of a particle with real energy depicted in figure \ref{figbohm9} as ellipse.

\begin{figure}
\centering   \includegraphics[width=7.5cm,height=6.5cm]{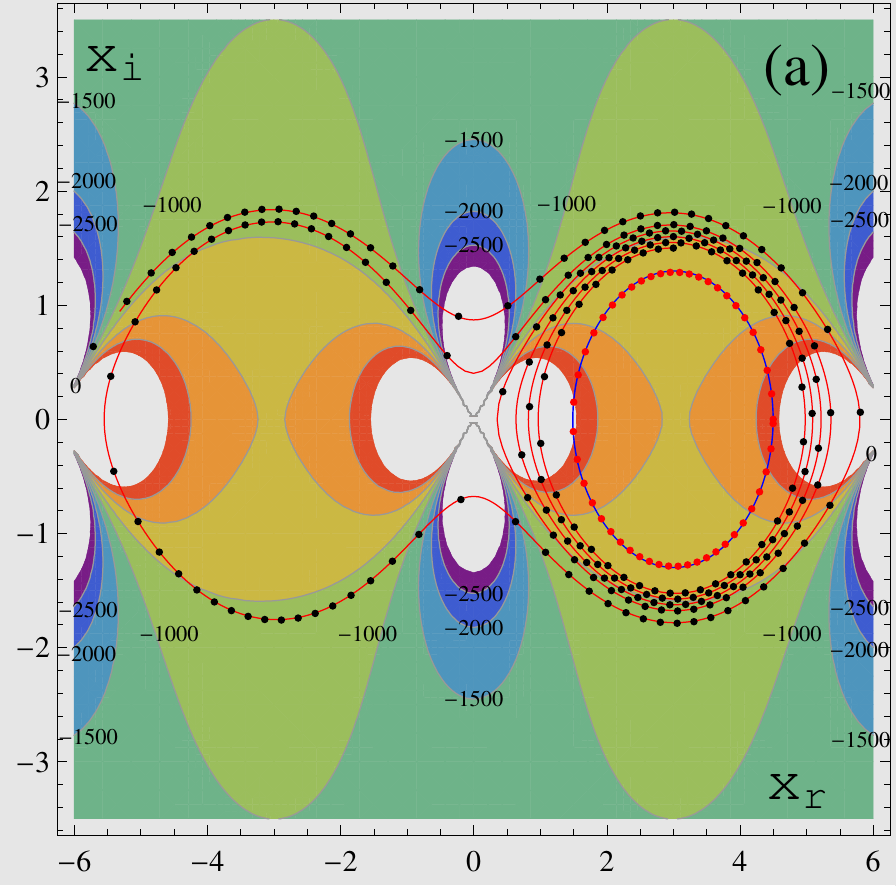}
\includegraphics[width=7.5cm,height=6.5cm]{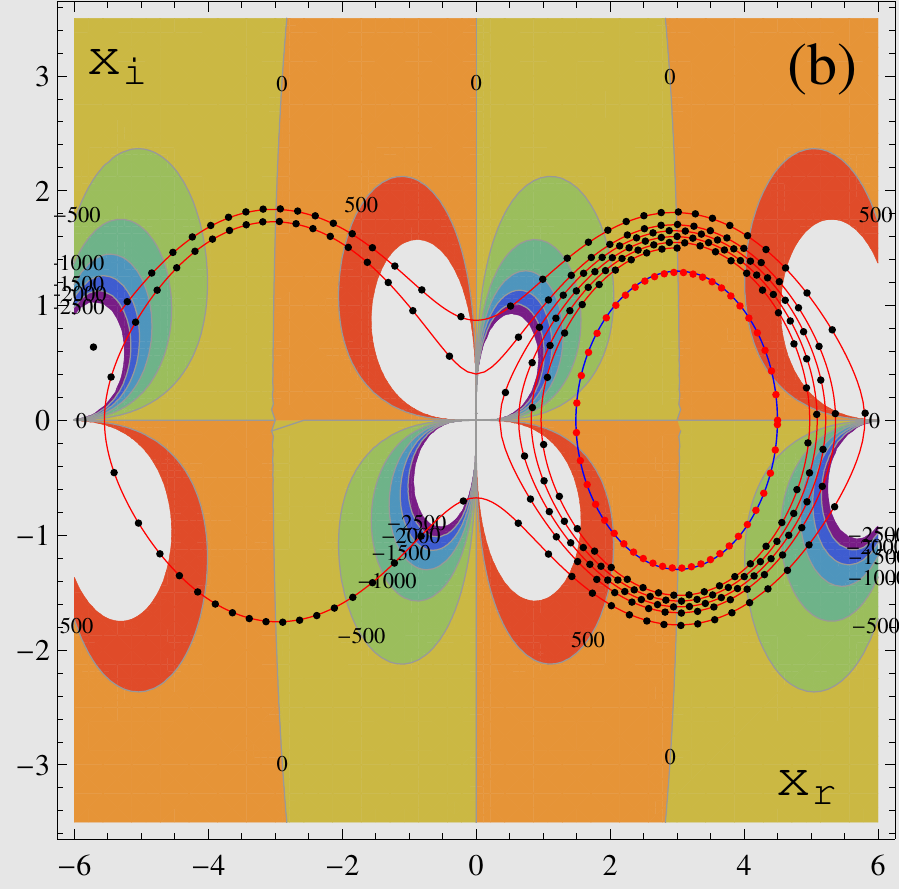}
\caption{\small{Complex Bohmian quantum trajectories as functions of time from $
J=0.5$ Klauder coherent states (dotted) versus classical trajectories
(solid lines) corresponding to solutions of (\protect\ref{H1}) and (\protect
\ref{H2}) for the complex P\"{o}schl-Teller potential $V_{\text{PT}}$ (a)
real part, (b) imaginary part for quasi-Poissonian and localized
distribution from $t=0$ to $t=1.78$ with $\protect\kappa =90$, $\protect
\lambda =100$. The initial values are $x_{0}=3+1.5i$, $p_{0}=-30.1922+0.385121i$ such that $E=-6.55991-13.5182i$ (red solid, black scattered) and $x_{0}=4.5$, $p_{0}=41.8376 i$ with real energy $E=-31.7564$
(blue solid, red dotted).}}
\label{figbohm9}
\end{figure}

The initial position is taken to be on the real axis with the particle
getting a kick parallel to the imaginary axis. Within the real part of the
potential the particle starts on a higher potential level and would simply
roll down further up into the upper half plane towards the imaginary axis
due to the curvature of the potential. However, the particle is also
subjected to the influence of the imaginary part of the potential and at
roughly $x_{r}=3$ this effect is felt when the particle reaches a turning
point, pulling it back to the real axis which is reached at the point when
reflecting $x_{i}$ at the turning point. At that point it has reached a
higher potential level from which it rolls down back to the initial position
through the lower half plane in a motion similar to the one performed in the
upper half plane.

Trajectories with complex values for the initial values have in general also
complex energies. As can be seen in figure \ref{figbohm9} for a specific case, we obtain at first a qualitatively similar motion to the real case, with the
difference that the particle spirals outwards. In the real part of the
potential this has the effect that after a few turns the particle is
eventually attracted by the sink on top of the origin. The momentum it gains
through this effect propels it into the region with negative real part. Thus
the particle has bypassed the infinite potential barrier at the origin on
the real axis, tunnelling to the next potential minimum, i.e. to the
forbidden region in the real scenario. Similar effects have been observed in
the purely classical treatment of a complex elliptic potential in \cite{bender_hook_kooner}. The continuation of this trajectory and scenarios for other parameter choices can be understood in a similar manner. For instance in
figure \ref{figbohm10} we depict a trajectory which does not spiral at first, but the particle has instead already enough momentum that allows it to tunnel
directly into the negative region.

\begin{figure}
\centering   \includegraphics[width=7.5cm,height=6.5cm]{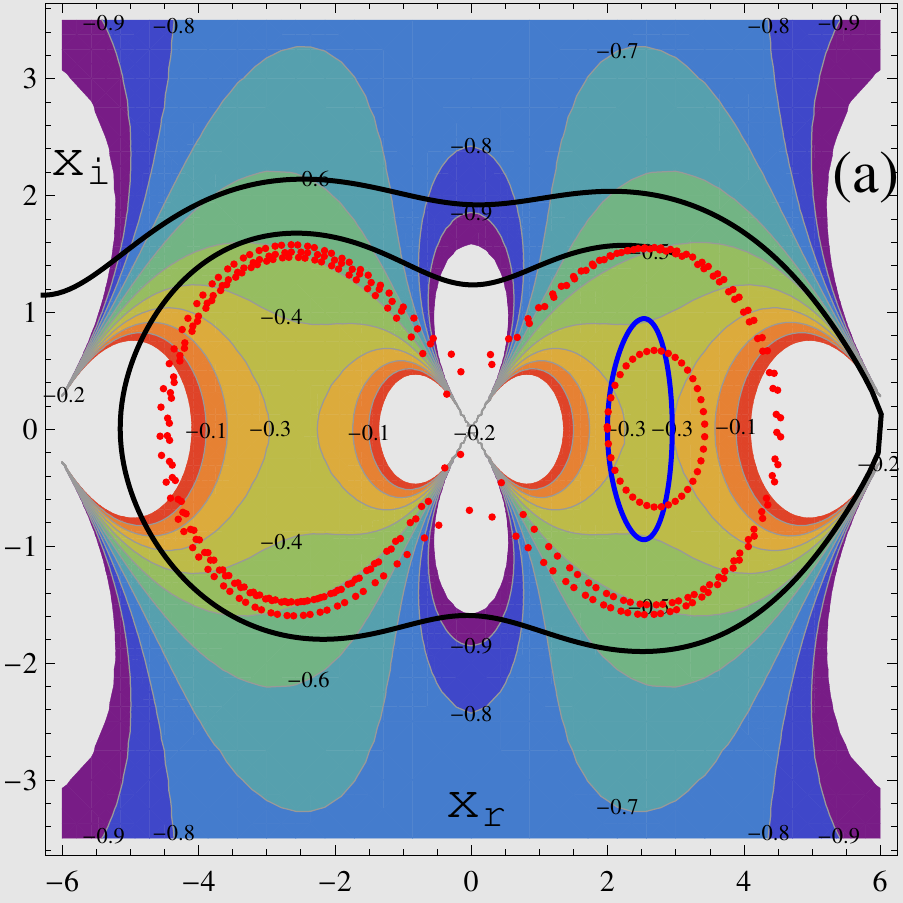}
\includegraphics[width=7.5cm,height=6.5cm]{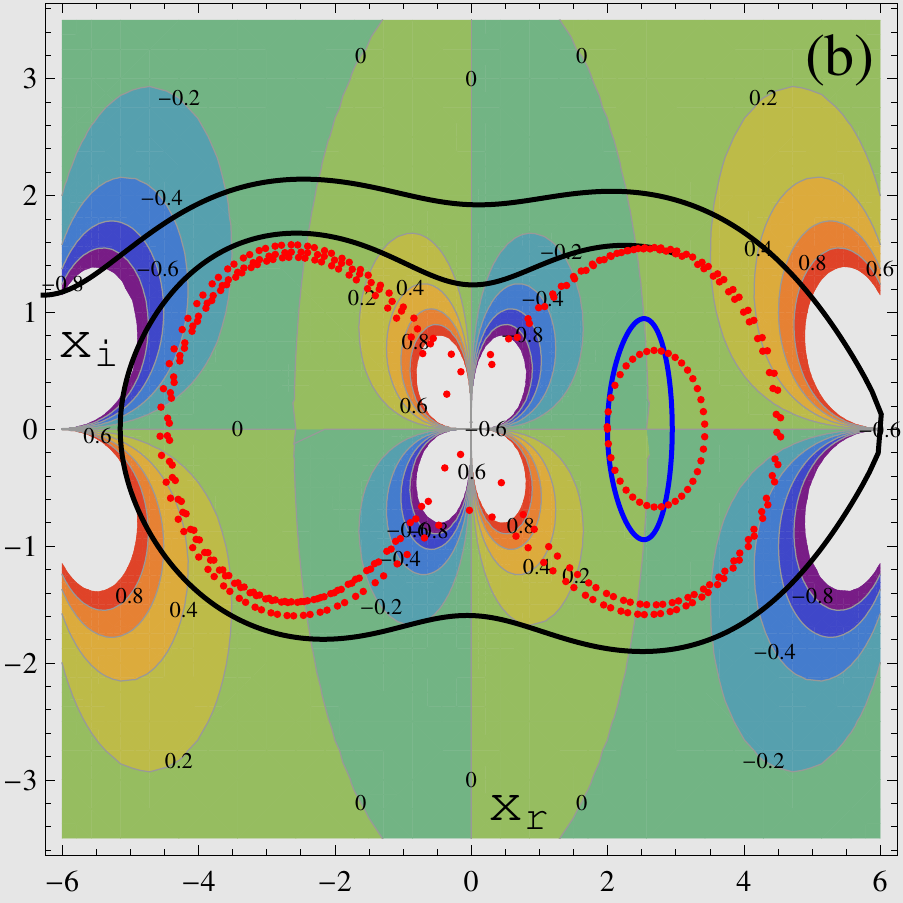}
\caption{\small{Complex Bohmian trajectories as functions of time from $
J=0.00057265$ Klauder coherent states (dotted) versus classical (solid
lines) corresponding to solutions of (\protect\ref{H1}) and (\protect\ref{H2}
) for the complex P\"{o}schl-Teller potential $V_{\text{PT}}$ (a) real part,
(b) imaginary part for the quasi-Poissonian regime and spread distribution
from $t=0$ to $t=100$ (quantum) and $t=0$ to $t=32$ (classical) with $
\protect\kappa =2$, $\protect\lambda =3$. The initial values are $
x_{0}=3+1.5i$ and $p_{0}=-0.788329+0.157336i$ such that $
E=-0.187539-0.0275087i$ (black solid, red scattered) and $x_{0}=2$, $
p_{0}=-0.49446 i$ with real energy $E=-0.277833$ (blue solid, red dotted).}}
\label{figbohm10}
\end{figure}

As in the real scenario the Mandel parameter controls the overall
qualitative behaviour, although in the complex case this worsens for
non-real initial values that is complex energies. In quasi-Poissonian regime
pictured in figure \ref{figbohm8} and \ref{figbohm9} we observe a complete match
between the purely classical and the quantum computation. However, this agreement ceases to exist in figure \ref{figbohm10}, despite the fact that it is
showing a quasi-Poissonian case with the same value for $Q$. The difference
is that in the latter case the wavefunction is less well localized as we saw
in figure \ref{figbohm5}. As can be seen in figure \ref{figbohm10}, for real energies we still have the same qualitative behaviour, but for complex energies the less localized wave is spread across a wide range of potential levels such that it can no longer mimic the same classical motion. However, qualitatively we
can see that in principle it is still compatible with the motion in a
complex classical P\"{o}schl-Teller potential.

\begin{figure}[H]
\centering   \includegraphics[width=7.5cm,height=6.5cm]{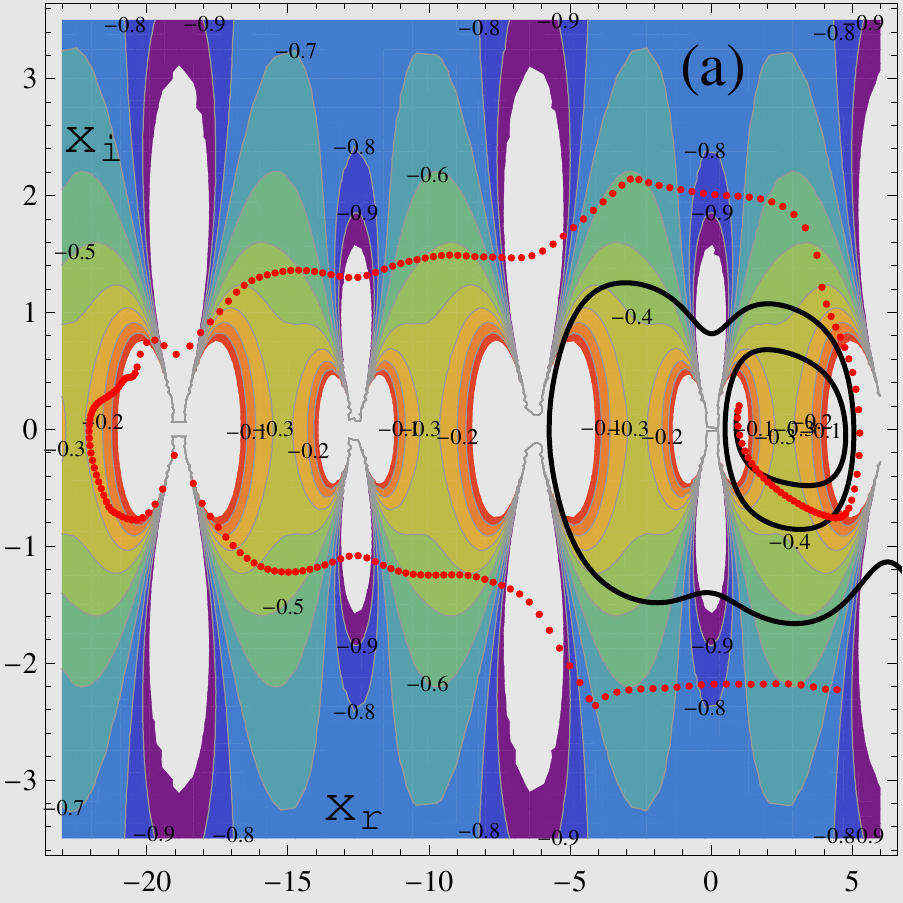}
\includegraphics[width=7.5cm,height=6.5cm]{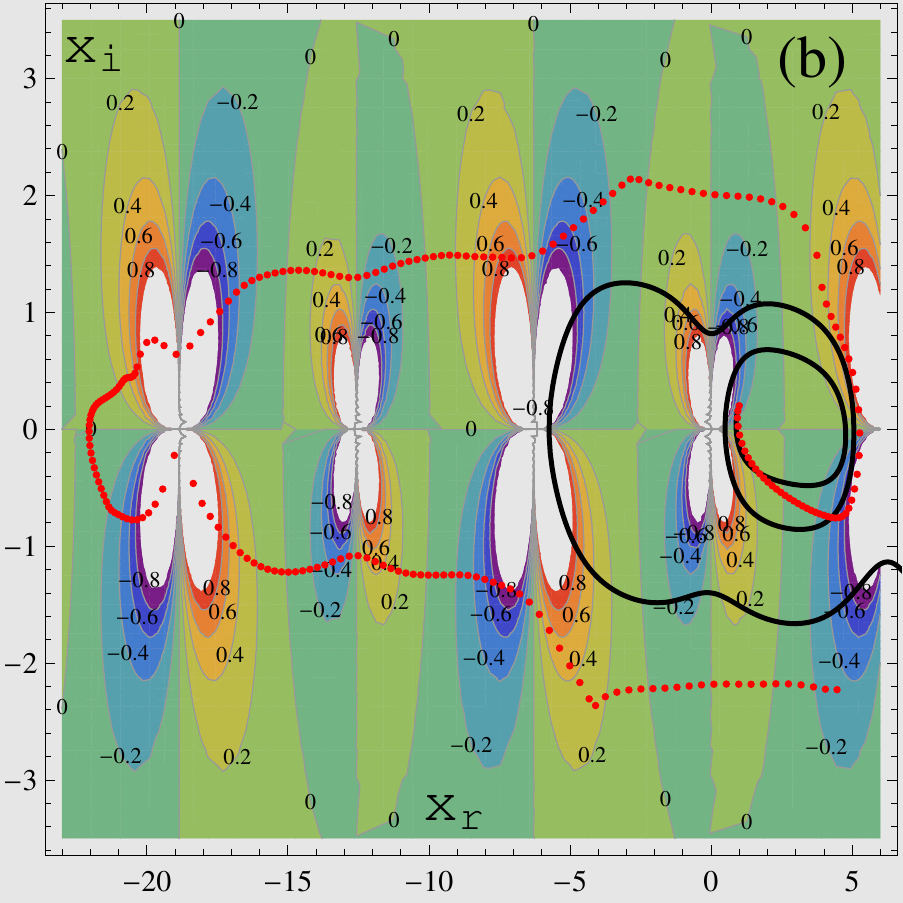}
\caption{\small{Complex Bohmian trajectories as functions of time from $J=10$ Klauder coherent states (dotted) versus classical (solid lines)
corresponding to solutions of (\protect\ref{H1}) and (\protect\ref{H2}) for
the complex P\"{o}schl-Teller potential $V_{\text{PT}}$ (a) real part, (b)
imaginary part for the sub-Poissonian regime from $t=0$ to $t=21$ (quantum)
and $t=0$ to $t=32$ (classical) with $\protect\kappa =2$, $\protect\lambda
=3 $. The initial values are $x_{0}=2+0.2i$ and $p_{0}=-0.665052-0.406733i$
such that $E=-0.0941366-0.364635i$ (black solid, red dotted).}}
\label{figbohm11}
\end{figure}

As can be seen in figure \ref{figbohm11} this resemblances ceases to exist when we enter the sub-Poissonian regime.

We observe that the correlation between the two behaviours is now entirely
lost. Notably, the quantum trajectories enter into regions not accessible to
the classical ones. However, in a very coarse sense we can still explain the
overall behaviour of the quantum trajectories by appealing to the complex
potential.

\section{A Calogero Type Potential}
Now we consider the solvable Hamiltonian system \cite{haar} of the form
\begin{equation}\label{Hamsingularity}
H=\frac{p^2}{2m}+V_0\left(\frac{a}{x}-\frac{x}{a}\right)^2~.
\end{equation}
The potential becomes a Calogero type potential and the harmonic oscillator for large and small values of a, respectively. The singularity of the potential function at the origin gives rise to the interesting possibility of tunnelling. Demanding $\psi_n(x=0,t)$ to be finite, the discrete eigenfunctions of the time dependent Schr{\"o}dinger equation \cite{haar} are given by
\begin{equation}
\psi_n(x,t)=C_n x^{\nu } e^{-\sqrt{\frac{m V_0}{2\hbar^2 a^2}}x^2}F_1\left[-n,\nu+\frac{1}{2},\sqrt{\frac{2mV_0}{\hbar^2 a^2}}x^2\right]e^{-iE_nt/\hbar},
\end{equation}
with
\begin{equation}
\nu=\frac{1}{2} \left(\sqrt{\frac{8 a^2 m V_0}{\hbar^2}+1}+1\right),
\end{equation}
where, $F_1$ denotes Kummer's confluent hypergeometric function. The corresponding eigenvalues were found to be
\begin{equation}
E_n=\sqrt{\frac{8 V_0 \hbar^2}{ma^2}}\left[n+\frac{1}{2}+\frac{1}{4}\left(\sqrt{\frac{8mV_0a^2}{\hbar^2}+1}-\sqrt{\frac{8mV_0a^2}{\hbar^2}}\right)\right]~.
\end{equation}

\subsection{Real Case}
Let us start our analysis with the Hamiltonian to be real. As we wish to look at the dynamical behaviour of the particle emerging out of the coherent states and compare with the classical behaviour, we first solve the Canonical equations of motion
\begin{equation}
\dot{x}=\frac{p}{m}\qquad \text{and}\qquad\dot{p}=\frac{2V_0\left(a^4-x^4\right)}{a^2x^3}~.
\end{equation}
It is easy to solve the above equations numerically to obtain the trajectories of the classical particle as depicted in figure \ref{figbohm12} (a). The more challenging and complicated part is to compute the trajectories of the quantum particle resulting from the coherent states, for which one requires to calculate the sum (\ref{klaudercoherent}) of infinite number of states and subsequently solve the velocity (\ref{realvel}) of the real version of Bohmian mechanics. However we find the series to be converging fast, which makes it easier to draw the dynamics of the coherent states numerically. Before doing that, again one must ensure the fact that the wave packet is well localised so that it behaves as a soliton like particle as discussed in section \ref{section65}. In our present problem the Mandel parameter depends on the values of $V_0$ and $a$, which can be adjusted to suitable set of numbers to obtain the classical behaviour of the coherent states. Apart from the scaling factor, we found a precise qualitative matching of the trajectories as shown in figure \ref{figbohm12}. It is possible to adjust the scaling factor also, by simply conjecturing the Hamiltonian as indicated in the case of P{\"o}schl-Teller model in section \ref{section731}, which is absolutely not necessary in our present problem to examine the qualitative behaviour of the particle. We have chosen the initial value of position $x_0=3$ for both the cases of quantum and classical trajectories. The initial value of momentum is computed from the initial velocity from (\ref{realvel}), which is required to acquire the solution of the canonical equations.   

\begin{figure}
\begin{minipage}[t]{0.5\textwidth}
\includegraphics[scale=0.26]{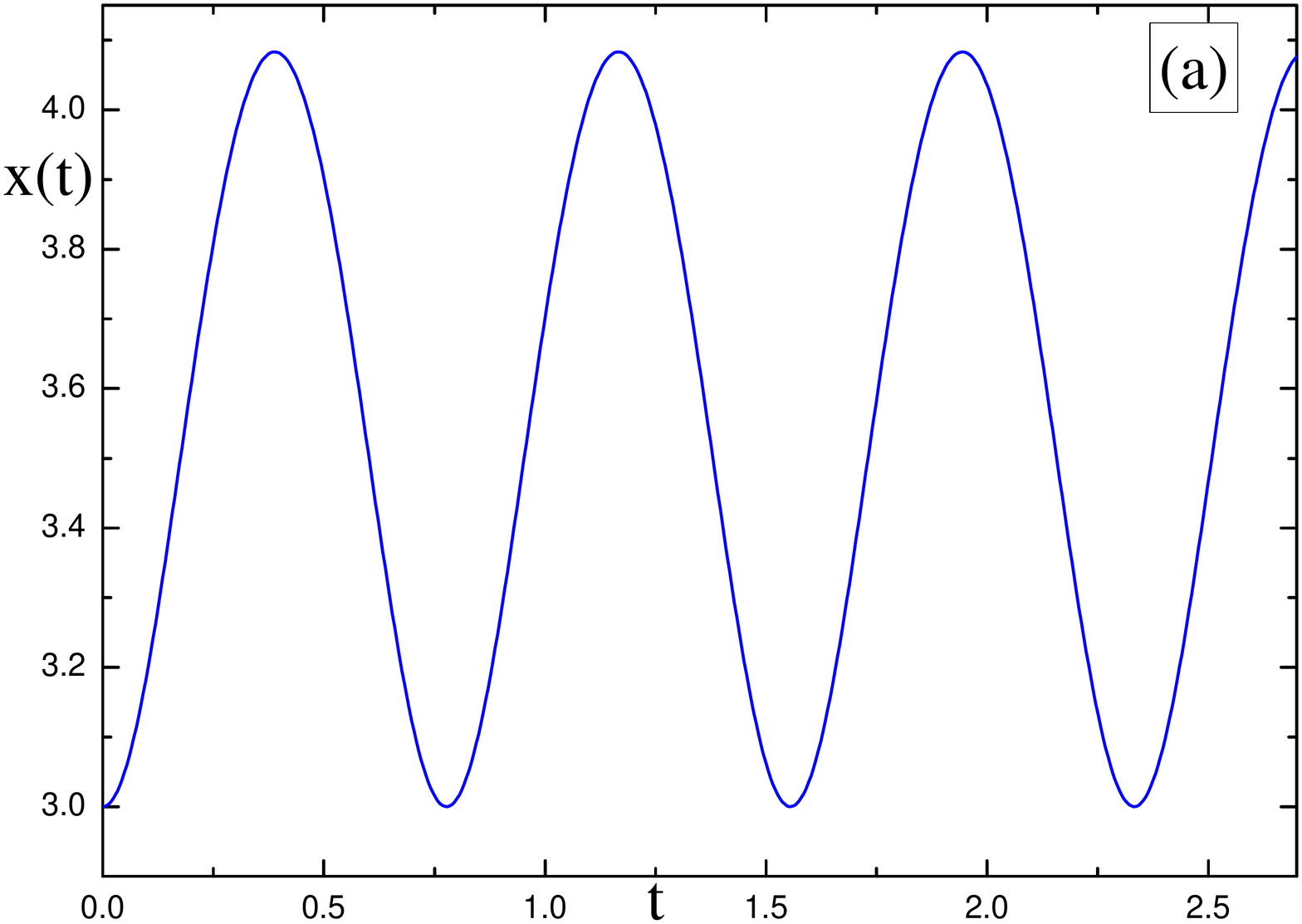}
\end{minipage}
\begin{minipage}[t]{0.5\textwidth}
\includegraphics[scale=0.26]{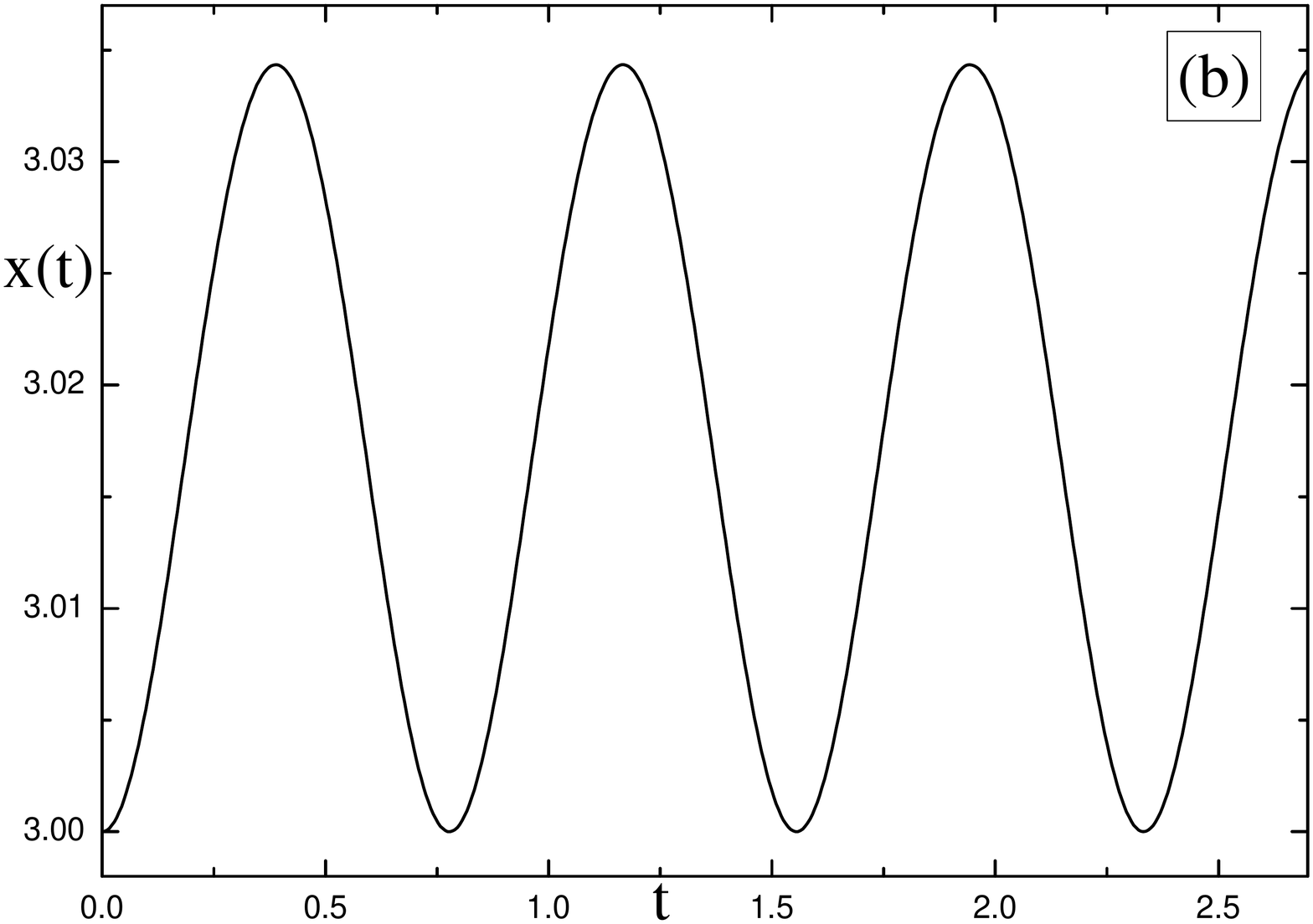}
\end{minipage}
\caption{\small{(a) Classical trajectories (b) Real Bohmian quantum trajectories from Klauder coherent states as a function of time. We have chosen $J=1, a=3.5, V_0=100$, so that $Q=0.0158626$}}
\label{figbohm12}
\end{figure}

\subsection{Complex Case}
Let us now consider the more interesting case, namely the complex Bohmian quantum
trajectories compared to the complex classical trajectories. For that scenario we complexify the momentum, position and coupling constant of the model, i.e. we take $x=x_r+ix_i$, $p=p_r+ip_i$ and $a=a_r+ia_i$ in the Hamiltonian (\ref{Hamsingularity}) with $x_r,x_i,p_r,p_i,a_r,a_i \in \mathbb{R}$. Separating the Hamiltonian into its real and imaginary part $H=H_r+iH_i$ with
\begin{eqnarray}
H_r &=& \frac{p_r^2-p_i^2}{2m}-V_0\frac{\left[\left(a_r^2-a_i^2\right)\left(x_i^2-x_r^2\right)-4a_ra_ix_rx_i\right]\left(\vert a\vert^4+\vert x\vert^4\right)}{\vert a\vert^4\vert x\vert^4}-2V_0 \nonumber \\
H_i &=& \frac{p_ip_r}{m}+\frac{2V_0(a_ix_r-a_rx_i)(a_ix_i+a_rx_r)\left(\vert a\vert^4-\vert x\vert^4\right)}{\vert a\vert^4\vert x\vert^4}~,
\end{eqnarray}
we observe that for $a_r=0$ or $a_i=0$ this Hamiltonian respects the $\mathcal{PT}$ symmetries, $\mathcal{PT} : x_r\rightarrow \pm x_r,~x_i \rightarrow \mp x_i,~p_r\rightarrow\pm p_r,~p_i\rightarrow\mp p_i,~i\rightarrow-i$. Following the ideas and techniques recently developed \cite{bender_boettcher,bender_making_sense,mostafazadeh5} this means that potentially even the non-Hermitian version of this model constitutes a well defined self-consistent quantum mechanical system. The classical canonical equations of motion are obtained from (\ref{H1}) and (\ref{H2})
\begin{eqnarray}
\dot{x}_r &=& \frac{1}{2}\left(\frac{\partial H_r}{\partial p_r}+\frac{\partial H_i}{\partial p_i}\right)=\frac{p_r}{m}, \qquad \dot{x}_i=\frac{1}{2}\left(\frac{\partial H_i}{\partial p_r}-\frac{\partial H_r}{\partial p_i}\right)=\frac{p_i}{m}, \label{canobohm1}\\
\dot{p}_r &=& -\frac{1}{2}\left(\frac{\partial H_r}{\partial x_r}+\frac{\partial H_i}{\partial x_i}\right) \label{canobohm2}\\
&=& 2V_0\left(\frac{(a_i^2-a_r^2)x_r-2a_ia_rx_i}{\vert a\vert^4}+\frac{2a_ia_rx_i(3x_r^2-x_i^2)+(a_r^2-a_i^2)(x_r^3-3x_i^2x_r)}{\vert x\vert^6}\right),\nonumber \\
\dot{p}_i &=& \frac{1}{2}\left(\frac{\partial H_r}{\partial x_i}-\frac{\partial H_i}{\partial x_r}\right) \label{canobohm3}\\
&=& 2V_0\left(\frac{x_i(a_i^2-a_r^2)+2a_ia_rx_r}{\vert a\vert^4}+\frac{(3x_ix_r^2-x_i^3)(a_i^2-a_r^2)+2a_ia_rx_r(x_r^2-3x_i^2)}{\vert x\vert^6}\right) \nonumber.
\end{eqnarray}
They are now fairly straight-forward to solve numerically, as depicted in figures \ref{figbohm13}, \ref{figbohm14}, \ref{figbohm15}, \ref{figbohm16} and \ref{figbohm17}, together with the trajectories resulting from coherent states and the contourplots of the potential. The depth and the height of the contourplots are presented with the colour convention being associated with the spectrum of light decreasing from red to violet as mentioned earlier. It is worthwhile to attach the classical and quantum trajectories together with the contour plots by which the dynamics of the particle can be explained when one looks at the motion of the particle through the  potential hills and wells.

The quantum trajectories resulting from coherent states are computed utilising the complex version of the Bohmian mechanics according to equation (\ref{complex}). We have chosen the values of the parameters $a$ and $V_0$ once again to a suitable number, such that the Mandel parameter stays close to zero as discussed in the real case.

\begin{figure}[H]
\centering   \includegraphics[width=7.5cm,height=6.5cm]{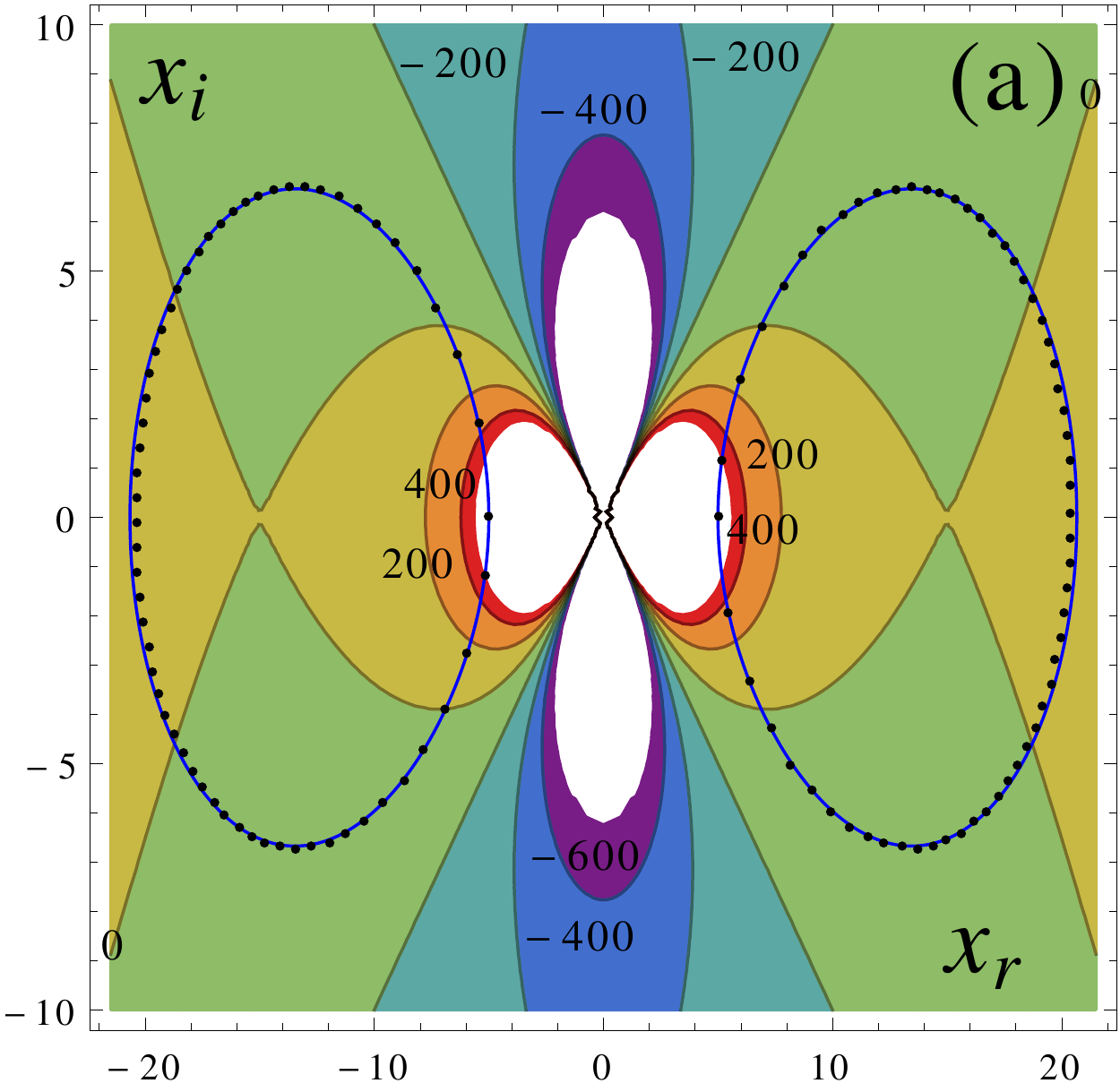}
\includegraphics[width=7.5cm,height=6.5cm]{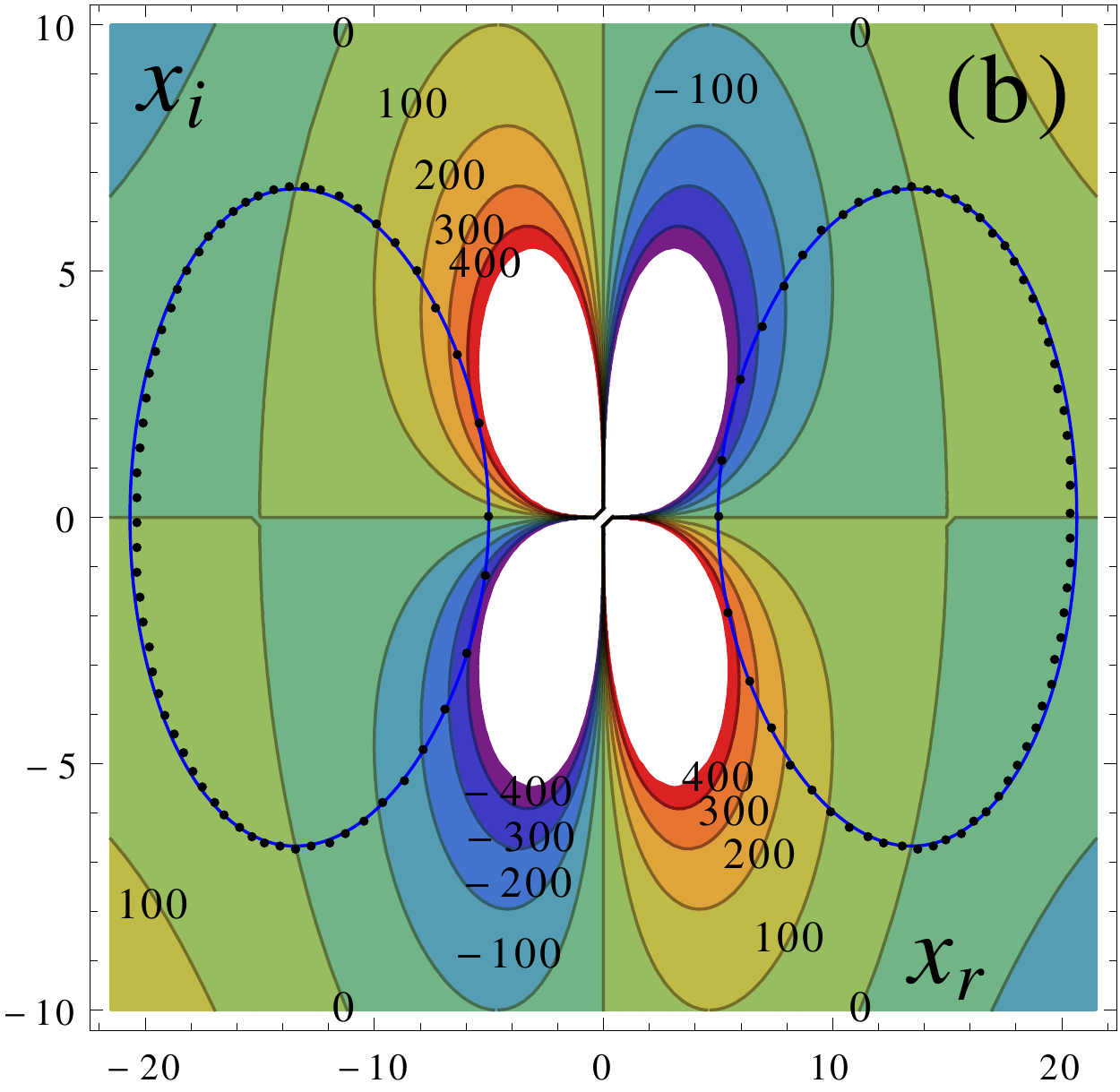}
\caption{\small{Closed complex Bohmian trajectories from coherent states(black scattered) versus classical trajectories (blue solid) for the values $J=0.1, a=15, V_0=100$, with initial values $x_0=\pm 5, p_0=\pm 37.8033i$, real energy $E=-3.43552$ and $Q=0.00695196$. (a) Real part (b) Imaginary part of the potential.}}
\label{figbohm13}
\end{figure}

Quite remarkably the precise agreement between the classical trajectories and the trajectories of the coherent states has been found in all cases as depicted in figures \ref{figbohm13}-\ref{figbohm17} for different sets of initial values and other parameters.

In figure \ref{figbohm13} the initial value of position is taken to be a purely real number such that the initial value of momentum is extracted from $\left(\text{velocity}\right)_{x=x_r+ix_i}$ at time $t=0$ to be a purely imaginary number directing along the negative imaginary axis. As a result, the particle is kicked downward at $x_r=3$ and then follows the potential hills and wells to complete the trajectory along a deformed ellipse.

\begin{figure}[H]
\centering   \includegraphics[width=7.5cm,height=6.5cm]{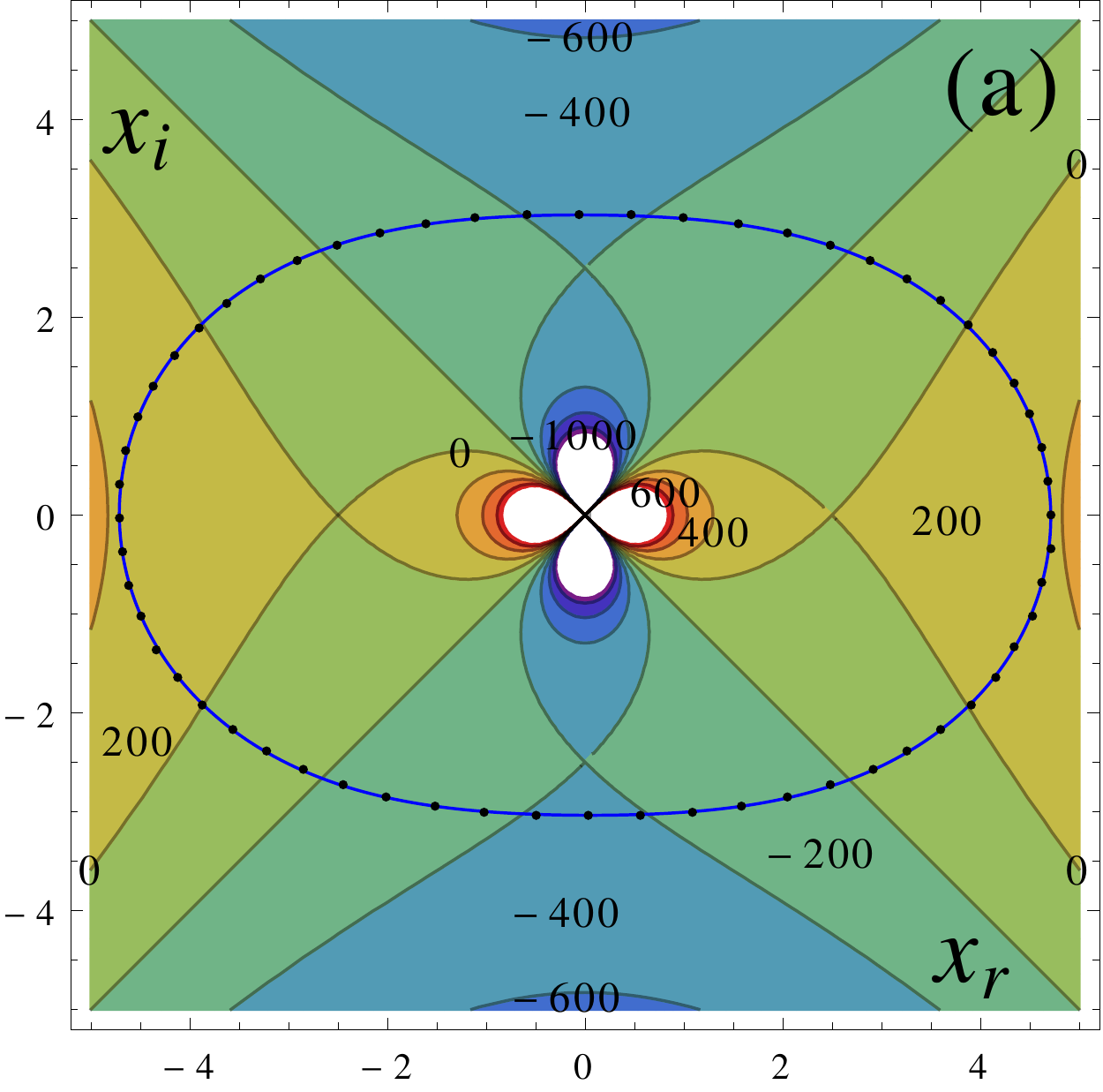}
\includegraphics[width=7.5cm,height=6.5cm]{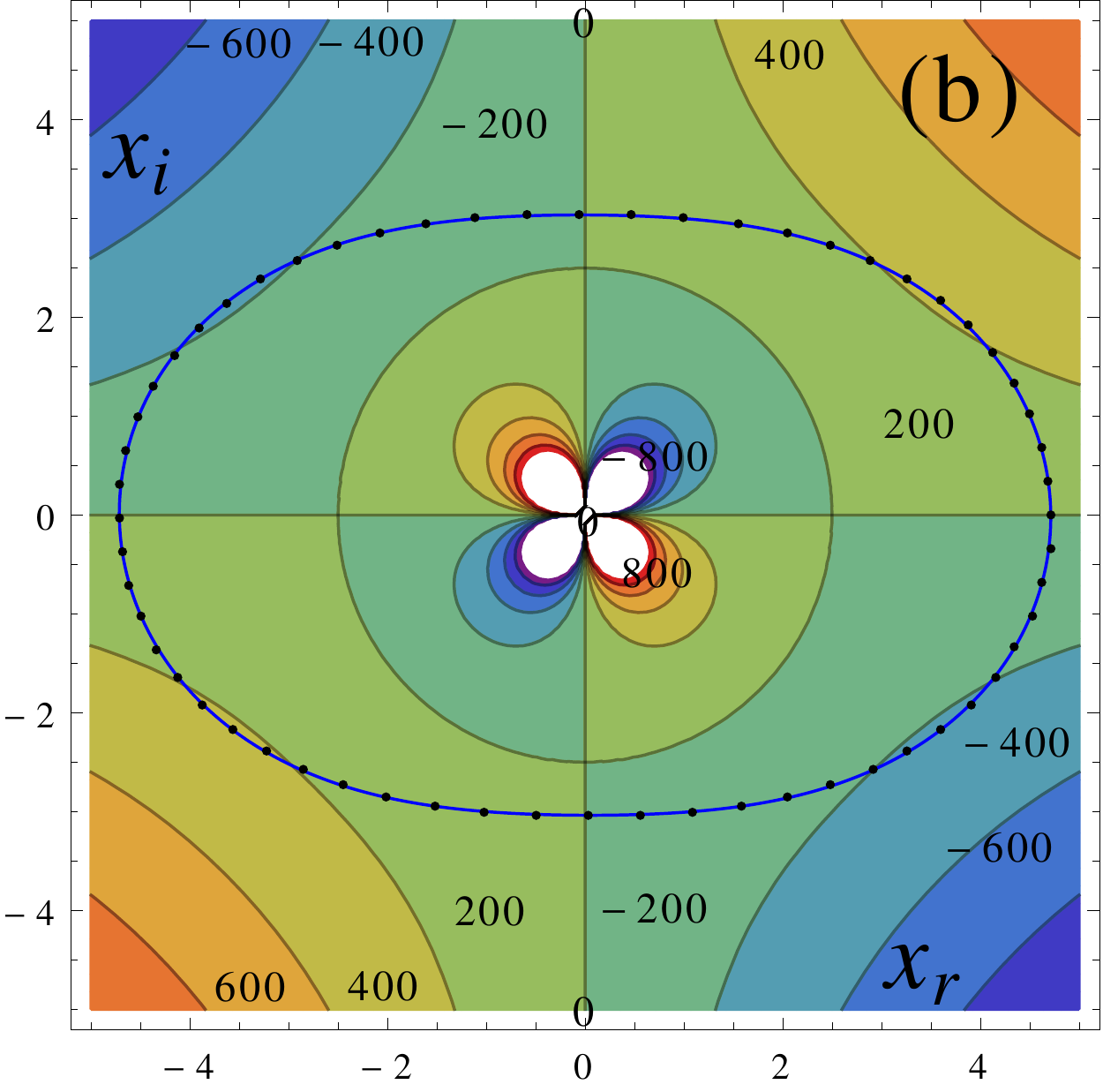}
\caption{\small{Closed complex Bohmian trajectories from coherent states(black scattered) versus classical trajectories (blue solid) for the values $J=0.5, a=2.5, V_0=100$, with initial values $x_0=1+3i, p_0=-27.92+2.15i$, complex energy $E=9.42-1.60i$ and $Q=0.0058182$. (a) Real part (b) Imaginary part of the potential.}}
\label{figbohm14}
\end{figure}

Before further discussion on the trajectories obtained, let us quickly mention few important investigations in the context of complex classical mechanics \cite{nanayakkara,bender_holm_hook,bender_holm_hook_2,bender_brody_hook, arpornthip_bender,bender_feinberg_hook_weir,bender_hook_kooner, anderson_bender_morone,cavaglia_fring_bagchi}. Complex classical mechanics is a rich and largely unexplored area of mathematical physics. In past few years the behaviour of classical particles were examined in many analytical complex potentials. It was found that the trajectories of classical particles in complex potentials having real energy are always closed and periodic, whereas the trajectories are generally open and chaotic when the energy is complex \cite{bender_holm_hook,bender_holm_hook_2}, which was examined later numerically in \cite{bender_brody_hook}. In those papers it was emphasized that this behaviour is consistent with the Bohr-Sommerfeld quantization condition
\begin{equation}
\oint p~dx=\left(n+\frac{1}{2}\right)\pi,
\end{equation}
which can only be applied if the classical orbits are closed. Therefore they claimed that there exists an association between real energies and the existence of closed classical trajectories and thus an association of closed trajectories with the $\mathcal{PT}$-symmetry of the Hamiltonians.

\begin{figure}[H]
\centering   \includegraphics[width=7.5cm,height=6.5cm]{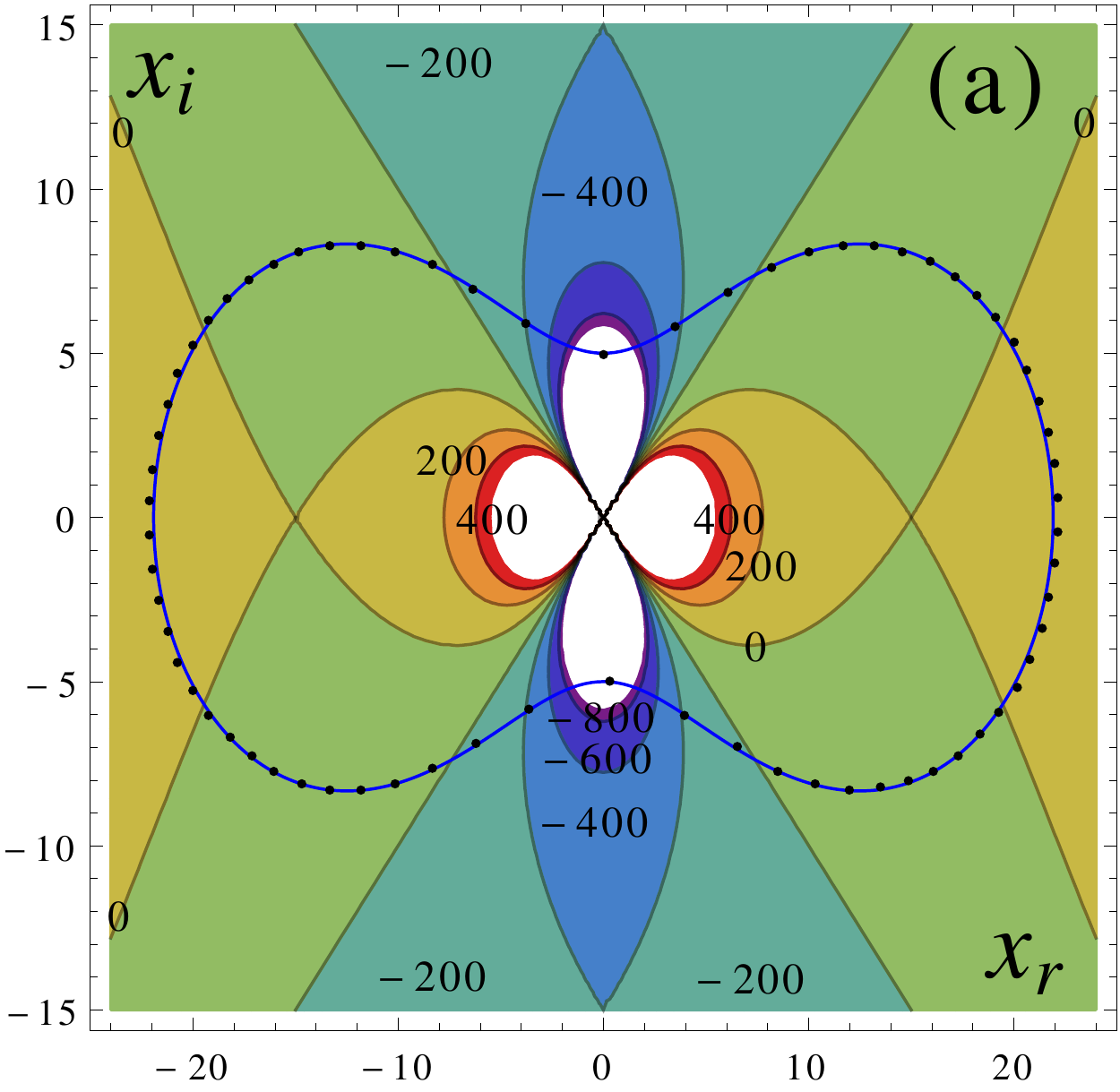}
\includegraphics[width=7.5cm,height=6.5cm]{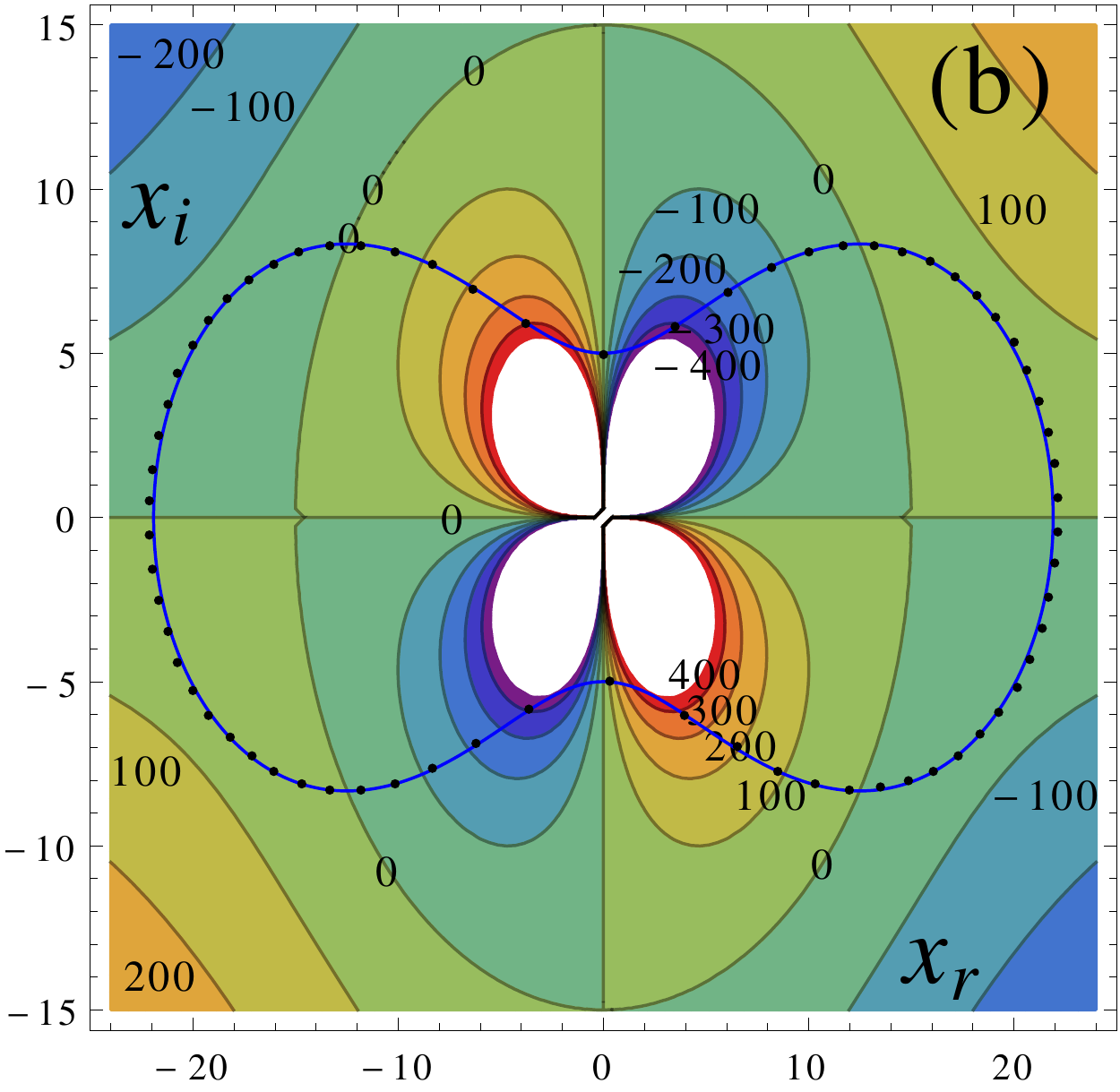}
\caption{\small{Closed complex Bohmian trajectories from coherent states(black dotted) versus classical trajectories (blue solid) for the values $J=1, a=15.0, V_0=100$, with initial values $x_0=5i, p_0=-47.2789$, real energy $E=6.5379$ and $Q=0.0590412$. (a) Real part (b) Imaginary part of the potential.}}
\label{figbohm15}
\end{figure}

\begin{figure}[H]
\centering   \includegraphics[width=7.5cm,height=6.5cm]{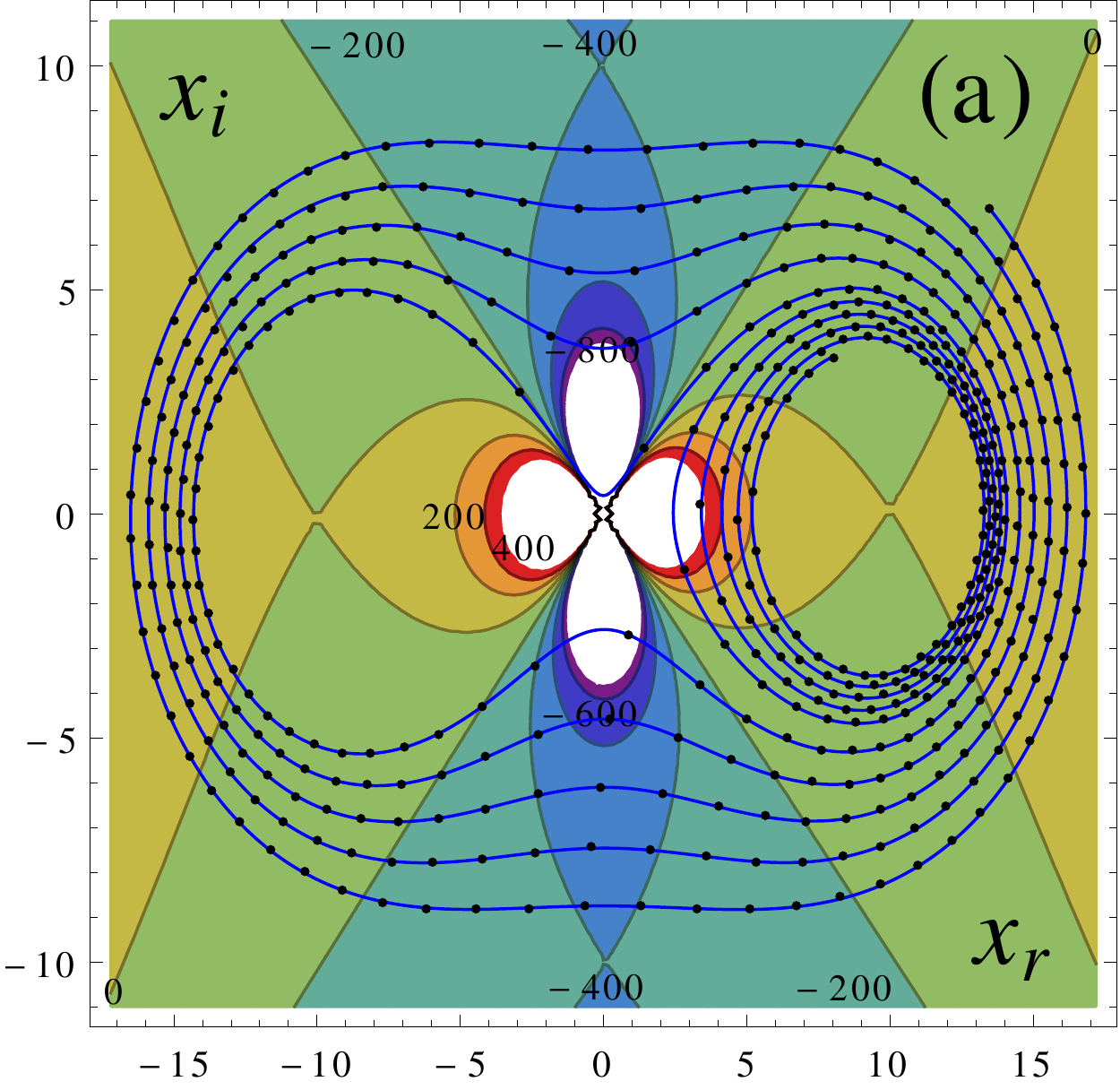}
\includegraphics[width=7.5cm,height=6.5cm]{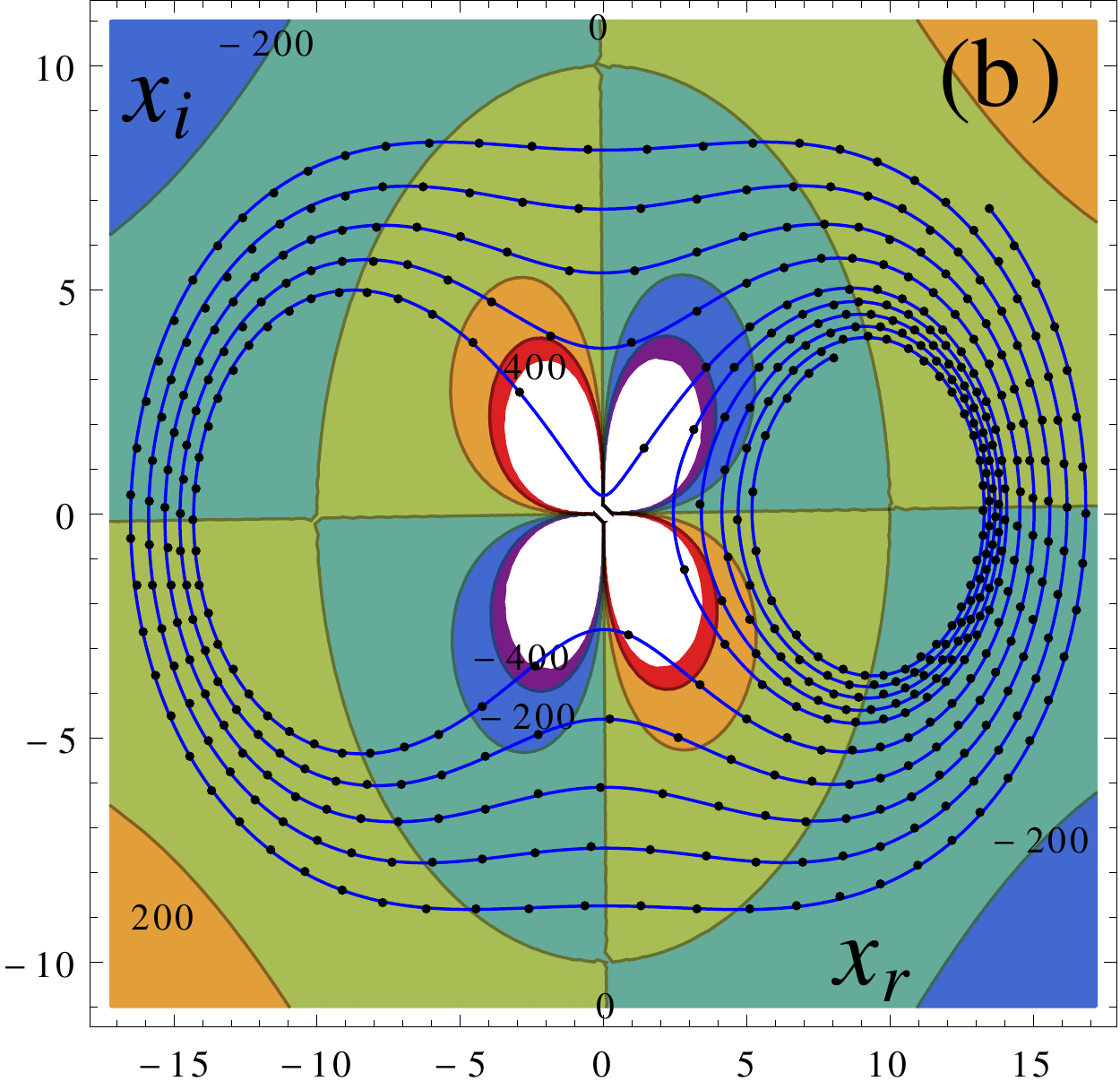}
\caption{\small{Open complex Bohmian trajectories from coherent states(black dotted) versus classical trajectories (blue solid) for the values $J=1, a=10+0.1i, V_0=100$, with initial values $x_0=8+5i, p_0=-14.8377-1.35193i$, complex energy $E=1.00459-0.746714i$ and $Q=0.0418894 + 0.000367417 i$. (a) Real part (b) Imaginary part of the potential.}}
\label{figbohm16}
\end{figure}

However, in our computation \cite{dey_fring_bohmian_2} we found that this statement is not true in general. In figures \ref{figbohm14}, we observe that the trajectory is closed and periodic inspite of the energy being complex and in figure \ref{figbohm17} the trajectory is open, though the energy is real. In fact we have presented all four possibilities, namely, close and periodic orbit for real energy (figure \ref{figbohm13}) and complex energy (figure \ref{figbohm14}); and open orbit for real energy (figure \ref{figbohm17}) and complex energy (figure \ref{figbohm16}). Albeit these possibilities were also indicated in \cite{cavaglia_fring_bagchi}.

\begin{figure}[H]
\centering   \includegraphics[width=7.5cm,height=6.5cm]{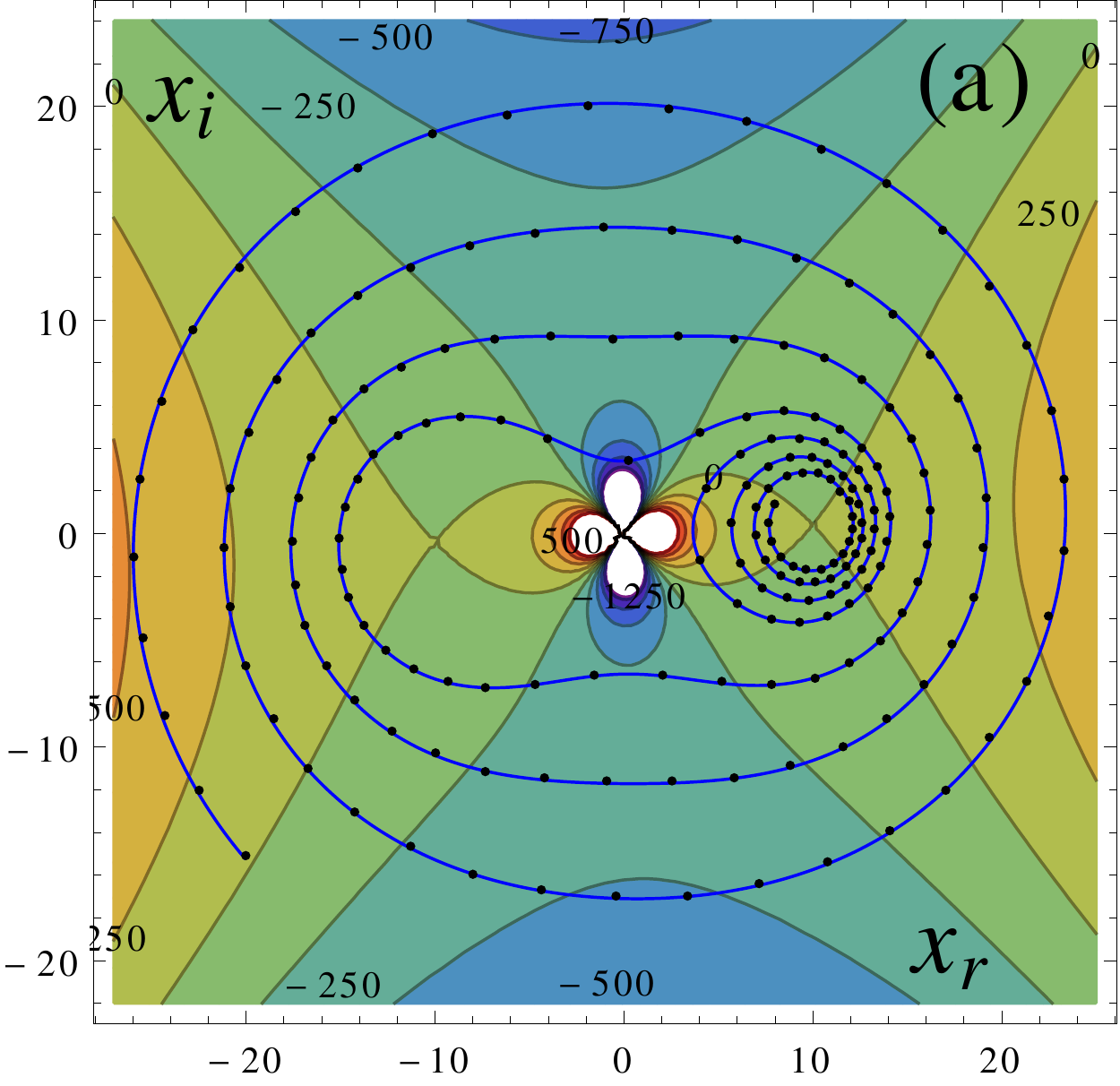}
\includegraphics[width=7.5cm,height=6.5cm]{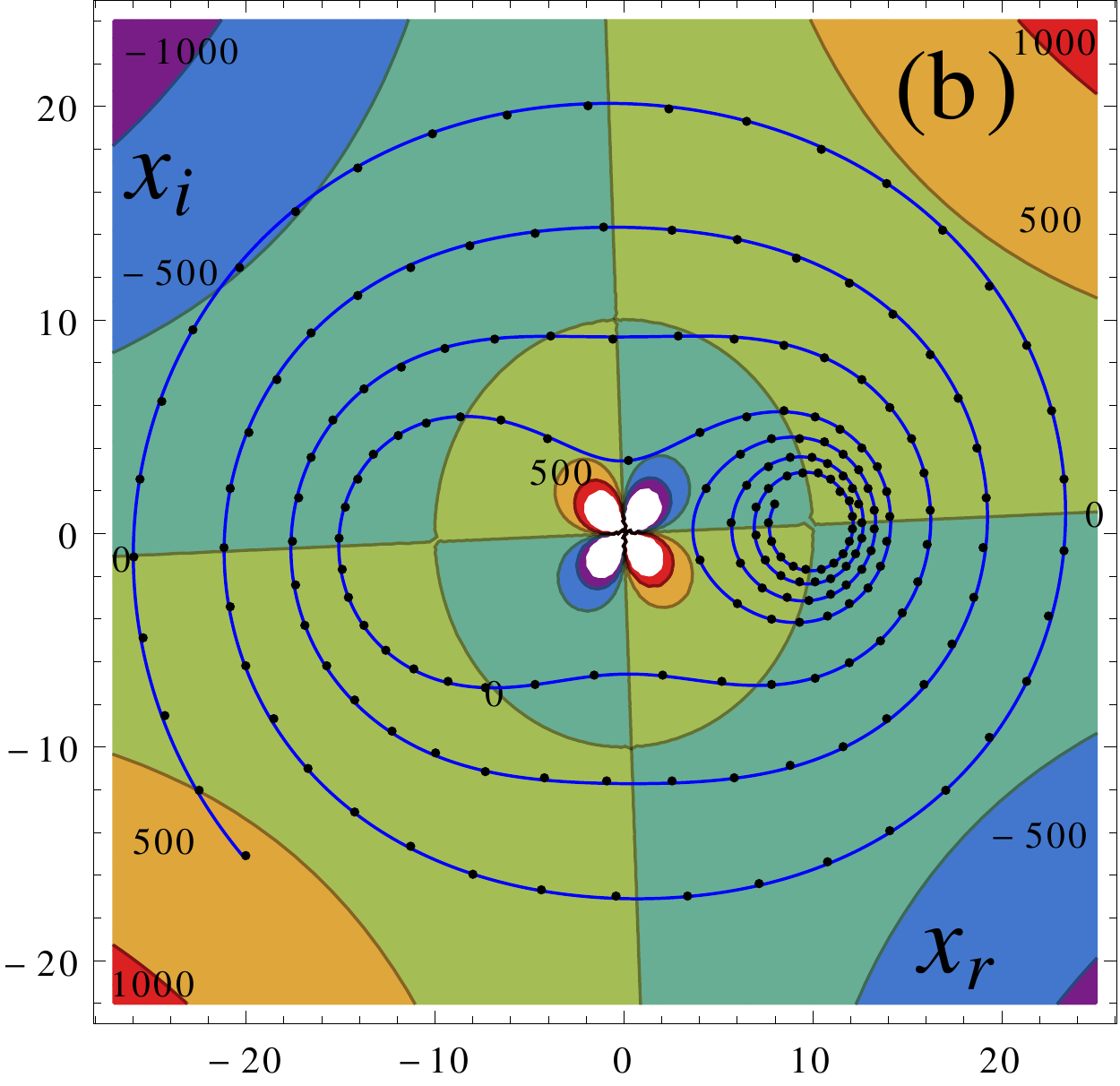}
\caption{\small{Open complex Bohmian trajectories from coherent states(black dotted) versus classical trajectories (blue solid) for the values $J=1, a=10+0.4i, V_0=100$, with initial values $x_0=8+1.35314894i, p_0=-3.69416-5.90461i$, real energy $E=0.210195$ and $Q=0.0418969 + 0.00146966 i$. (a) Real part (b) Imaginary part of the potential.}}
\label{figbohm17}
\end{figure}

In figure \ref{figbohm15}, we take initial value of position $x_0$ to be purely imaginary so that the initial momentum $p_0$ becomes purely real negative number and that is why the particle is forced to move along the negative real axis. The initial momentum being high, it does not feel the infinite sink in the origin and does not fall into the well, instead the particle is tunnelled to the negative region of the potential. Because of the suitable circumstances and having sufficient momentum, the tunnelling behaviour was found in other cases too, for example for the cases of figures \ref{figbohm14}-\ref{figbohm17}. Notice that this behaviour is certainly very surprising, because the particle is not supposed to exist in the negative region of the potential classically. However as the particle is being kicked strongly along the negative real axis, the particle feels the width of the potential sink in both the real and complex to be more thin compared to the case of figure \ref{figbohm13} and that is why it is tunnelled to the negative part of the potential which was also found in \cite{bender_brody_hook,bender_hook_kooner} for other models. In case of figure \ref{figbohm16}, the width of the potential sink being relatively wider and the initial values being close to the region of potential sink, a kink is naturally expected because of the strong attraction of the particle towards the potential sink during tunnelling.

\newpage

\section{Noncommutative Harmonic Oscillator}
So far we have examined the quality of the Klauder coherent states utilising the standard formulations of Bohmian mechanics for the models in the usual space. Certainly our original motivation is to inspect the quality of the noncommutative coherent states that we developed in chapter \ref{chapter_coherent}. Indeed admitting the fact that the computations of Bohmian trajectories are more difficult in noncommutative space than that of the usual space, we first developed the method for the usual case so that we can obtain sufficient knowledge on the subject and move on to the computations in noncommutative space. However, it is quite astonishing that we have not found a single model available in noncommutative space which has been solved in position space. Therefore it becomes essential to look at the momentum space formulation of Bohmian mechanics.

However, we noticed that the standard Bohmian mechanics has only been explored in position space, not in momentum space. Due to interpretational difficulties of Bohmian mechanics, which is itself contradictory even in position space, it is fairly complicated to formulate the structure in momentum space and we observed that there are very few people who have concentrated \cite{brown_hiley,hiley1} on momentum space formulations and have not succeeded yet. They argued that instead of developing the formulation of Bohmian mechanics in standard momentum space it is convenient to start with the algebra of functions and then deduce the properties of the underlying space. Therefore the results they obtain certainly is not in the standard momentum space but in some other phase space, which they call "Shadow phase space".

However, we do not want to be involved in the interpretational difficulties, rather we focussed on the position space models in noncommutative space and noticed that it is indeed difficult to obtain models in position space. Nevertheless we picked up a noncommutative harmonic oscillator wave function in momentum space and albeit the standard Fourier transform makes the task easier. Though we have not found a compact form of the wave function together with the normalisation constant, but we can certainly compute them separately corresponding to each quantum numbers, $n=0,1,2,3$ and so on. It is worth mentioning that few excited state wave functions together with the ground state may solve our purpose because we already noticed that the infinite sum that computes the coherent states (\ref{klaudercoherent}) converges fast and one hardly requires the knowledge of all the wavefunctions. We choose the noncommutative harmonic oscillator wave function in momentum space (\ref{310}) 
\begin{equation}
\psi_n(p)=\frac{1}{\sqrt{N_n}}\frac{1}{(1+\check{\tau}p^2)^{1/4}} P_{n-\mu_-}^{\mu_-}\left(\frac{\sqrt{\check{\tau}}p}{\sqrt{1+\check{\tau}p^2}}\right),\quad \text{with}\quad \mu_-=-\frac{\sqrt{\tau ^2+4}}{2 \tau },
\end{equation}
and compute some of the Fourier transformed versions in position space
\begin{eqnarray}
\psi_0(x) &=& \frac{3^{7/16} \left| x\right| ^{3/4} K_{\frac{3}{4}}\left(\frac{\sqrt[4]{3} \left| x\right| }{\sqrt{2}}\right)}{4 \sqrt[8]{2}~\Gamma \left(\frac{5}{4}\right)}, \\
\psi_1(x) &=& \frac{3^{9/16} x \sqrt[4]{\left| x\right| } K_{\frac{1}{4}}\left(\frac{\sqrt[4]{3} \left|x\right| }{\sqrt{2}}\right)}{4\ 2^{7/8}~\Gamma \left(\frac{7}{4}\right)}, \\
\psi_2(x) &=& \frac{5\ 3^{7/16} \left| x\right| ^{3/4} K_{\frac{3}{4}}\left(\frac{\sqrt[4]{3} \left|
   x\right| }{\sqrt{2}}\right)}{16 \sqrt[8]{2}~\Gamma \left(\frac{9}{4}\right)}-\frac{5\
   3^{11/16} \left| x\right| ^{7/4} K_{\frac{7}{4}}\left(\frac{\sqrt[4]{3} \left| x\right|
   }{\sqrt{2}}\right)}{32\ 2^{5/8}~\Gamma \left(\frac{9}{4}\right)}, \\
\psi_3(x) &=& \frac{7\ 3^{9/16} x \sqrt[4]{\left| x\right| } K_{\frac{1}{4}}\left(\frac{\sqrt[4]{3} \left|
   x\right| }{\sqrt{2}}\right)}{16\ 2^{7/8}~\Gamma \left(\frac{11}{4}\right)}-\frac{7\
   3^{13/16} x \left| x\right| ^{5/4} K_{\frac{5}{4}}\left(\frac{\sqrt[4]{3} \left| x\right|
   }{\sqrt{2}}\right)}{64\ 2^{3/8}~\Gamma \left(\frac{11}{4}\right)}, \\
\psi_4(x) &=& \frac{21\ 3^{15/16} \left| x\right| ^{11/4} K_{\frac{11}{4}}\left(\frac{\sqrt[4]{3} \left|
   x\right| }{\sqrt{2}}\right)}{256 \sqrt[8]{2} \Gamma \left(\frac{13}{4}\right)}-\frac{63\
   3^{11/16} \left| x\right| ^{7/4} K_{\frac{7}{4}}\left(\frac{\sqrt[4]{3} \left| x\right|
   }{\sqrt{2}}\right)}{64\ 2^{5/8} \Gamma \left(\frac{13}{4}\right)} \nonumber \\
   &&+\frac{45\ 3^{7/16} \left|
   x\right| ^{3/4} K_{\frac{3}{4}}\left(\frac{\sqrt[4]{3} \left| x\right|
   }{\sqrt{2}}\right)}{64 \sqrt[8]{2} \Gamma \left(\frac{13}{4}\right)}, \\
\psi_5(x) &=& \frac{99 \sqrt[16]{3} x \left| x\right| ^{9/4} K_{\frac{9}{4}}\left(\frac{\sqrt[4]{3} \left|
   x\right| }{\sqrt{2}}\right)}{256\ 2^{7/8} \Gamma \left(\frac{15}{4}\right)}-\frac{99\
   3^{13/16} x \left| x\right| ^{5/4} K_{\frac{5}{4}}\left(\frac{\sqrt[4]{3} \left| x\right|
   }{\sqrt{2}}\right)}{128\ 2^{3/8} \Gamma \left(\frac{15}{4}\right)} \nonumber \\
   && +\frac{77\ 3^{9/16} x
   \sqrt[4]{\left| x\right| } K_{\frac{1}{4}}\left(\frac{\sqrt[4]{3} \left| x\right|
   }{\sqrt{2}}\right)}{64\ 2^{7/8} \Gamma \left(\frac{15}{4}\right)}, \\
\psi_6(x) &=& -\frac{1287\ 3^{3/16} \left| x\right| ^{15/4} K_{\frac{15}{4}}\left(\frac{\sqrt[4]{3} \left|
   x\right| }{\sqrt{2}}\right)}{4096\ 2^{5/8} \Gamma \left(\frac{17}{4}\right)}+\frac{1287\
   3^{15/16} \left| x\right| ^{11/4} K_{\frac{11}{4}}\left(\frac{\sqrt[4]{3} \left| x\right|
   }{\sqrt{2}}\right)}{1024 \sqrt[8]{2} \Gamma \left(\frac{17}{4}\right)} \nonumber \\
   && -\frac{3159\
   3^{11/16} \left| x\right| ^{7/4} K_{\frac{7}{4}}\left(\frac{\sqrt[4]{3} \left| x\right|
   }{\sqrt{2}}\right)}{512\ 2^{5/8} \Gamma \left(\frac{17}{4}\right)}+\frac{585\ 3^{7/16}
   \left| x\right| ^{3/4} K_{\frac{3}{4}}\left(\frac{\sqrt[4]{3} \left| x\right|
   }{\sqrt{2}}\right)}{256 \sqrt[8]{2} \Gamma \left(\frac{17}{4}\right)}, \\
\psi_7(x) &=& -\frac{2145\ 3^{5/16} x \left| x\right| ^{13/4} K_{\frac{13}{4}}\left(\frac{\sqrt[4]{3} \left|
   x\right| }{\sqrt{2}}\right)}{8192\ 2^{3/8} \Gamma \left(\frac{19}{4}\right)}+\frac{6435
   \sqrt[16]{3} x \left| x\right| ^{9/4} K_{\frac{9}{4}}\left(\frac{\sqrt[4]{3} \left|
   x\right| }{\sqrt{2}}\right)}{1024\ 2^{7/8} \Gamma \left(\frac{19}{4}\right)} \nonumber \\
&& -\frac{5445\
   3^{13/16} x \left| x\right| ^{5/4} K_{\frac{5}{4}}\left(\frac{\sqrt[4]{3} \left| x\right|
   }{\sqrt{2}}\right)}{1024\ 2^{3/8} \Gamma \left(\frac{19}{4}\right)}+\frac{1155\ 3^{9/16} x
   \sqrt[4]{\left| x\right| } K_{\frac{1}{4}}\left(\frac{\sqrt[4]{3} \left| x\right|
   }{\sqrt{2}}\right)}{256\ 2^{7/8} \Gamma \left(\frac{19}{4}\right)}, \\
\psi_8(x) &=& \frac{7293\ 3^{7/16} \left| x\right| ^{19/4} K_{\frac{19}{4}}\left(\frac{\sqrt[4]{3} \left|
   x\right| }{\sqrt{2}}\right)}{32768 \sqrt[8]{2} \Gamma
   \left(\frac{21}{4}\right)}-\frac{36465\ 3^{3/16} \left| x\right| ^{15/4}
   K_{\frac{15}{4}}\left(\frac{\sqrt[4]{3} \left| x\right| }{\sqrt{2}}\right)}{4096\ 2^{5/8}
   \Gamma \left(\frac{21}{4}\right)} \nonumber \\
&& +\frac{31603\ 3^{15/16} \left| x\right| ^{11/4}
   K_{\frac{11}{4}}\left(\frac{\sqrt[4]{3} \left| x\right| }{\sqrt{2}}\right)}{2048
   \sqrt[8]{2} \Gamma \left(\frac{21}{4}\right)}-\frac{21879\ 3^{11/16} \left| x\right| ^{7/4}
   K_{\frac{7}{4}}\left(\frac{\sqrt[4]{3} \left| x\right| }{\sqrt{2}}\right)}{512\ 2^{5/8}
   \Gamma \left(\frac{21}{4}\right)} \nonumber \\
&& +\frac{9945\ 3^{7/16} \left| x\right| ^{3/4}
   K_{\frac{3}{4}}\left(\frac{\sqrt[4]{3} \left| x\right| }{\sqrt{2}}\right)}{1024 \sqrt[8]{2}
   \Gamma \left(\frac{21}{4}\right)},
\end{eqnarray} 
where $K$ denotes the modified Bessel function of second kind. For our convenience we choose the natural units $\hbar=\omega=m=1$ so that $\tau=\check{\tau}=2/\sqrt{3}$. Therefore $\mu_-$ is specified to $-1$. The wave functions computed above are all physically well behaved as they satisfy the proper boundary conditions as depicted in figure \ref{figbohm18}.

\begin{figure}[H]
\centering   \includegraphics[width=7.5cm,height=6.0cm]{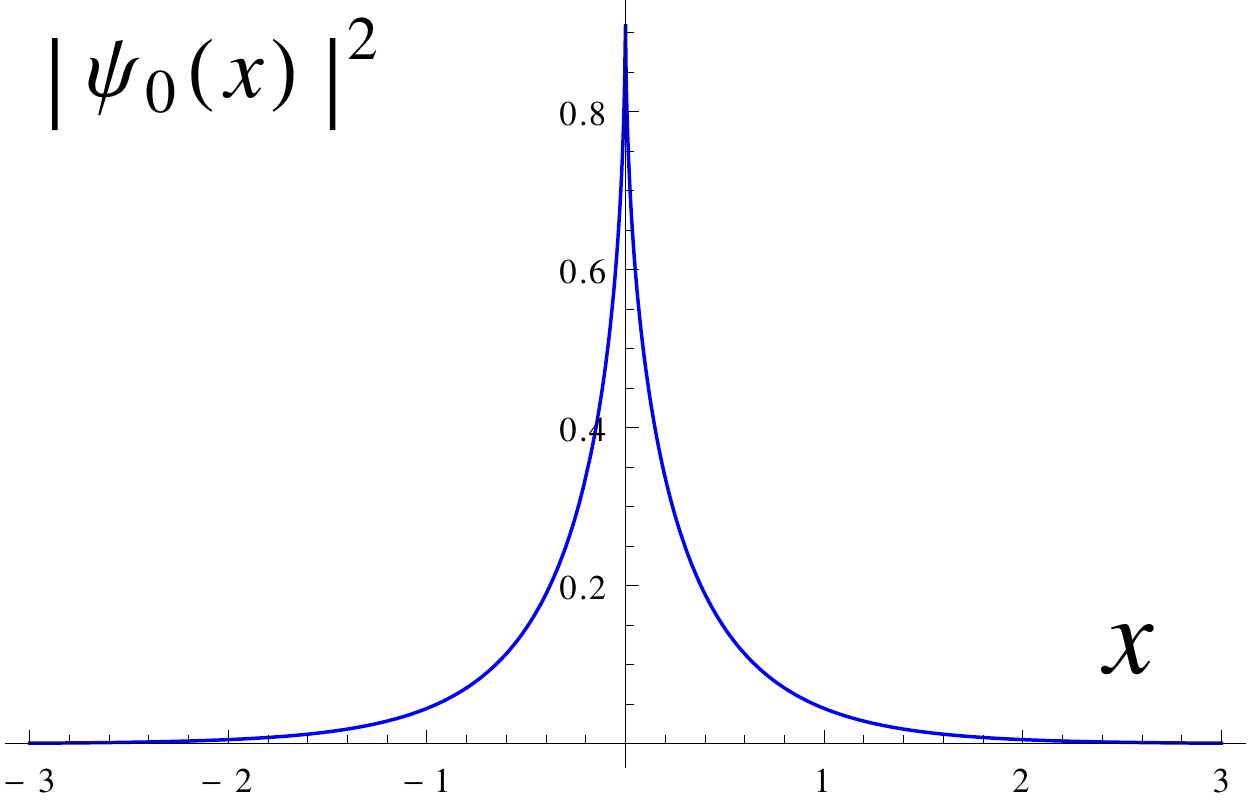}
\includegraphics[width=7.5cm,height=6.0cm]{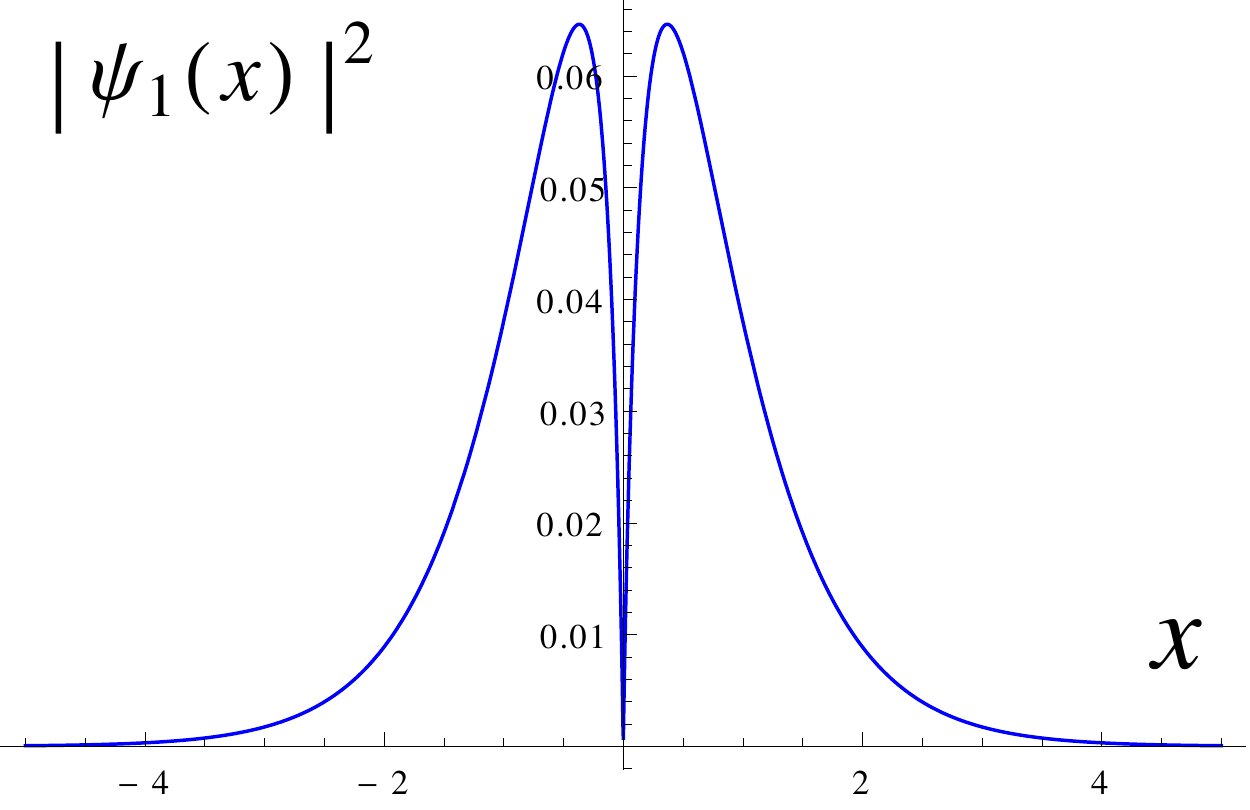}
\includegraphics[width=7.5cm,height=6.0cm]{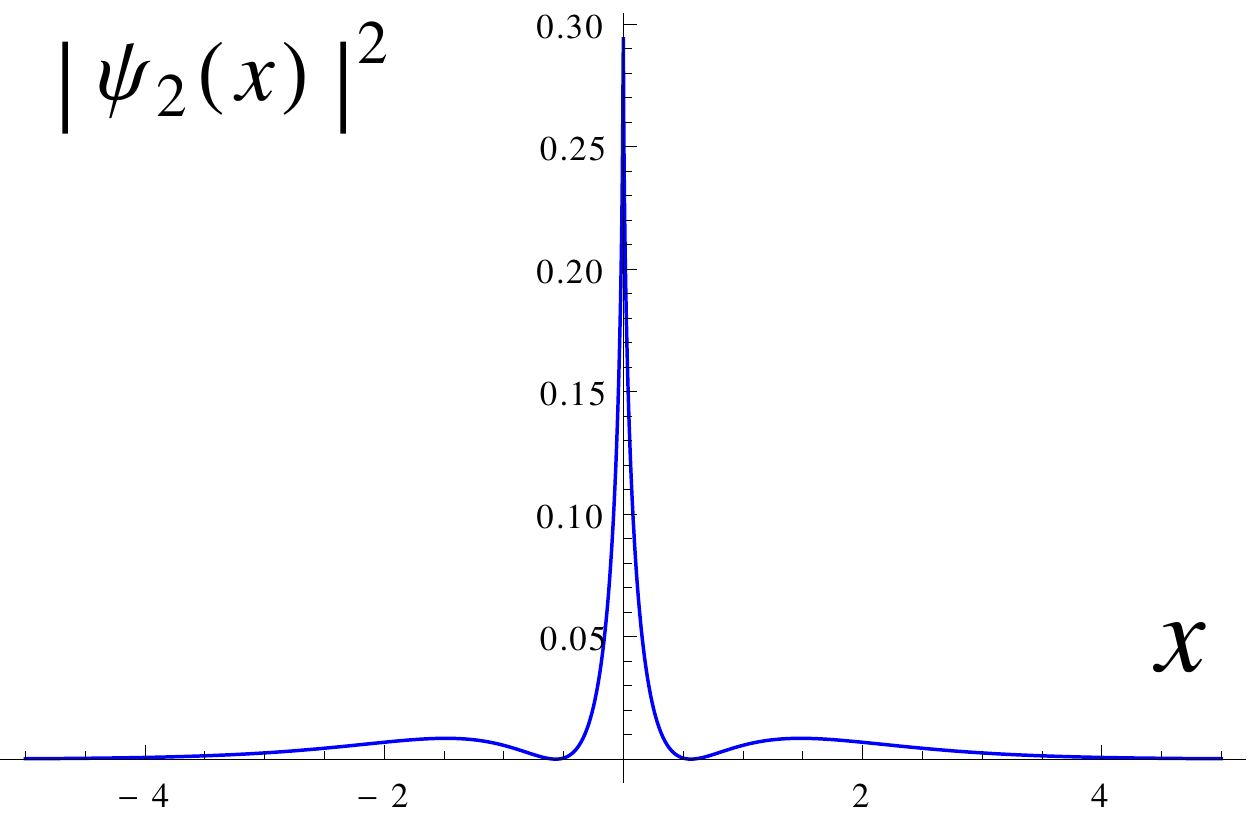}
\includegraphics[width=7.5cm,height=6.0cm]{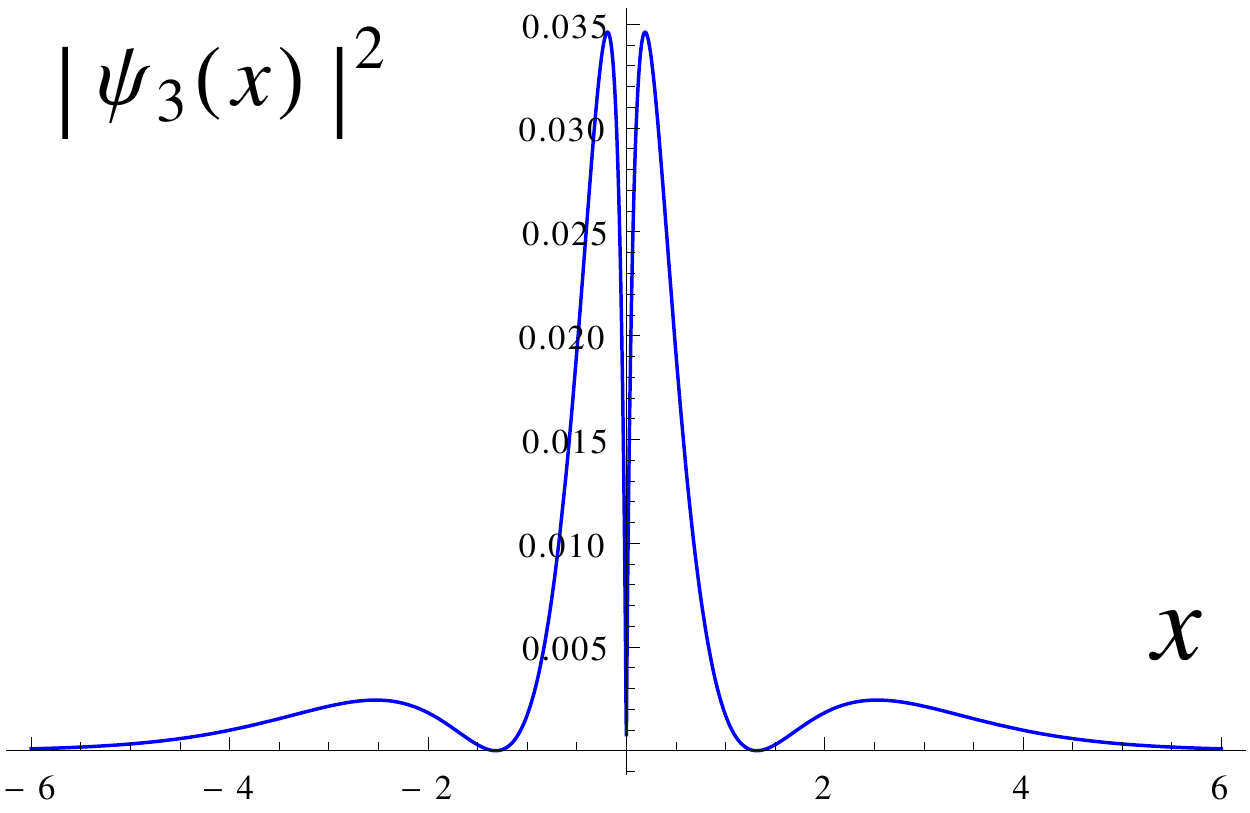} \centering   
\caption{\small{Probability amplitudes of ground state and first three excited states of the noncommutative harmonic oscillator.}}
\label{figbohm18}
\end{figure}

However, we have not computed the Bohmian trajectories to examine the qualities of coherent states yet. Nevertheless, one can utilize the above wave functions to compute them easily. We think this is clearly a new contribution in the literature as we mentioned earlier that no one has solved the noncommutative models in position space yet. 

\section{Discussions}
We have computed real and complex quantum trajectories in two alternative ways, either by solving the associated equation for the velocity or by solving the Hamilton-Jacobi equations taking the quantum potential as a starting point. In all cases considered we found perfect agreement for the same initial values in the position.

Our main concern in this chapter has been to investigate the quality of the Klauder coherent states and test how close they can mimic a purely classical description. This line of enquiry continues our previous investigations \cite{dey_fring_squeezed,dey_fring_gouba_castro} for these type of states in a different context. We have demonstrated that in the quasi-Poissonian regime well
localized Klauder coherent states produce the same qualitative behaviour as a purely classical analysis. We found these features in the real as well as in the complex scenario. For the real trajectories we conjectured some analytical expressions reproducing the numerically obtained results. Whereas the real case required always some adjustments, we found for the complex analysis of the harmonic oscillator, the P\"{o}schl-Teller potential and a Calogero type potential a precise match with the purely classical treatment. While investigating the Calogero type potential we found an interesting result in the context of complex classical mechanics. In \cite{bender_holm_hook,bender_holm_hook_2,bender_brody_hook}, the authors claimed that the trajectories of classical particles moving in a complex plane are always real and periodic if their energy is real, whereas they are open if the energy is complex. We argue that in general this is not true and demand that they might not have any connection of $\mathcal{PT}$-symmetry with the trajectories of complex classical particle.

Naturally there are a number of open problems left: Clearly it would be interesting to produce more sample computations for different types of potentials, especially for the less well explored complex case. In that case it would also be very interesting to explore further how the conventional quantum mechanical description can be reproduced. Since Bohmian quantum trajectories allow to establish that link, there would be no need to guess any rules in the classical picture mimicking some quantum behaviour as done in the literature. Moreover, we have only provided a general procedure to test the qualities of the coherent states and we have not finished our line of investigation in the noncommutative space. It would be very interesting to utilize the wave functions provided in the position space and apply the procedure to the case of noncommutative models. Complex classical mechanics is an interesting subject in itself and would be very interesting to study for some other models. As it turned out that there might not be a connection of classical trajectories with $\mathcal{PT}$-symmetry property of the Hamiltonians, it would be fantastic to provide some exact explanations behind the nature of open and closed trajectories. 


\chapter{Conclusions} \label{conclusions}
Although we have provided each of the chapters of this thesis with a summary of the main results and open problems, in this chapter our aim is now to provide the reader with a more general and concise review of all the original results investigated as well as the open problems.

In this thesis, we carried out our investigations on noncommutative space-time structure chiefly in a couple of different directions. In the former part, in chapters \ref{noncommutativemodelsin3D} and \ref{chapter_solvable} \cite{dey_fring_gouba,dey_fring_2DHO,dey_fring_khantoul}, we focussed on the construction of mathematical models in $q$-deformed noncommutative space-time structure in the context of minimal length and in the later part, in chapters \ref{chapter_coherent} and \ref{chapterBohmian} \cite{dey_fring_squeezed,dey_fring_gouba_castro,dey_fring_bohmian,dey_fring_bohmian_2}, we explored the physical behaviour of these kind of models in such spaces. 

Models were constructed in a couple of different independent approaches. In the first approach, we have provided a systematic procedure to relate a three dimensional $q$-deformed oscillator algebra to the corresponding algebra satisfied by canonical variables describing non-commutative spaces. We started from a generic Ansatz for the canonical variables obeying a $q$-deformed oscillator algebra and employed $\mathcal{PT}$-symmetry to limit the amount of free parameters to a reasonable number, so that the calculations became easier and the equations turned out to be solvable. We presented three different versions of $\mathcal{PT}$-symmetry and provided a specific solution for one of them. Of course it is possible to employ many other symmetries and consequently there exist many other solutions, which we leave as open problems. However, we concentrated on the specific solution and constructed an explicit representation for the algebra obtained in the non-trivial limit $q\rightarrow 1$ in terms of the generators of a flat noncommutative space. Therefore, it became quite straight-forward to compute different possible models with the observables obtained. We presented the solution of a three-dimensional harmonic oscillator as an example. Subsequently, we decoupled the three-dimensional representation into lower dimensions and therefore computed the harmonic oscillator models in lower dimensions and checked the consistencies with the results obtained previously in the literature \cite{kempf,kempf_mangano_mann,bagchi_fring,fring_gouba_scholtz, fring_gouba_bagchi_area}. In all the above cases, namely for the harmonic oscillators in one, two and three dimensions, despite of being the observables non-Hermitian, we obtained real eigenvalue spectrums and thus self-consistent theories. The reason can be traced back to the built-in $\mathcal{PT}$-symmetry of the systems. The perturbative computation of the energy eigenvalues suggests that there exists a parameter regime for which the $\mathcal{PT}$-symmetry is broken. Therefore it is quite obvious why we studied the theory of $\mathcal{PT}$-symmetry parallelly with the noncommutative space-time structure. In this context, in chapter \ref{chapterPT} \cite{dey_fring_mathanaranjan}, we computed an exact form of the metric operator that transforms the Euclidean Lie algebraic type non-Hermitian Hamiltonians into their Hermitian counterparts, which we believe to be one of the excellent results in this thesis as this can be found rarely in the literature. We pointed out the region of broken and unbroken $\mathcal{PT}$-symmetry together with a theoretical observation of the gain/loss behaviour in the unbroken $\mathcal{PT}$-regime which are of interest nowadays of the experimental physicists. However, apart from the construction of models in the noncommutative spaces, we also computed the minimal lengths and momenta resulting from generalised uncertainty relations, which overall give rise to minimal areas and minimal volumes in phase space and provided an understanding of the Planck scale physics in the context of quantum gravity. Again we believe that the existence of minimal volume has not been found notably in the literature.

\rhead{Conclusions}
\lhead{}
\chead{}

The theoretical models constructed above were mainly based on the standard perturbation theory. The other approach in the process of constructing models that we explored, is the formulation of exactly solvable models in noncommutative spaces. We provided a general construction procedure for solvable non-Hermitian potentials and implemented it in the case of noncommutative harmonic oscillator in four different representations and later to a couple of other characteristically different models, the non-Hermitian Swanson model and an intrinsically noncommutative model with P{\"o}schl-Teller type potential. In all cases we showed that an appropriate metric can be found in an analytical fashion, such that expectation values result to be representation independent. We presented an explicit formula for this metric, involving the quantities computed in the steps of the general procedure. We also found that the representation presented in \cite{castro_kullock_toppan} does not lead to the proper uncertainty relation as claimed. Moreover, the models investigated in \cite{castro_kullock_toppan} always give rise to non-physical spectra which are not bounded from below, which suggests that the general procedure requires a mild modification as outlined in this thesis.

In the later part of the thesis we focussed more on the physical implications of the noncommutative theories rather than building up the mathematical models. We constructed the coherent states for the noncommutative harmonic oscillators that were developed in the previous part. We preferred to choose the Klauder coherent states instead of the well-known Glauber states because the Klauder states were constructed in such a way that they can be applied to any generalised setting, whereas the Glauber states require a formulation of the Hamiltonian in terms of creation and annihilation operators, which is not always possible. We implemented the general formulations of Klauder states to a couple of versions of the harmonic oscillator, firstly, to the perturbative one and later to the exact case. In both the cases we showed that the states are squeezed, in the perturbative case for all parameter values, whereas in the exact case for a specific value. It was shown that both of them obey all the Gazeau-Klauder axioms \cite{gazeau_klauder}, especially two of them which are nontrivial. Firstly, the states are shown to be temporarily stable, i.e., they remain coherent under time evolution and secondly, the states satisfy the action identity allowing for a close relation to a classical description in terms of action-angle variables. We also demonstrated that the Ehrenfest's theorem is satisfied in both cases for the observables $X$ and $P$. The desired resemblance of the coherent states with a classical description was further underpinned by an analysis of the revival structure exhibiting the typical quasi-classical evolution of the original wave packet. The revival structure being associated with the energy eigenvalues $E_n$ on the quantum number $n$, the perturbative case produced only the revival structure, whereas the exact case manifested infinitely many such structures. We computed the structures up to the super-revival time,  but in principle, one can compute the super-super-revival and so on, which we leave as an open problem. 

Further insights into the comparison between the classical and coherent states were carried out mainly based on numerical technique. This approach is quite different from the previous one and certainly more promising as one can directly compare the dynamics coming out of both sides. In the process, we computed the dynamics of the coherent states numerically by using the general formulation of the Bohmian mechanics and compared them with the dynamics of the classical system by solving the standard canonical equations. We implemented the procedure in quite a number of systems, for example, for the real and complex potential of the harmonic oscillator, the P{\"o}schl-Teller model and a Calogero-type model with singularity at the origin. Quite convincingly, we have found exact resemblances of the classical trajectories with the trajectories of the coherent states in all the above cases suggesting that the Klauder states produce qualitatively good coherent states, but this includes a restriction. We pointed out the region on which one must operate to obtain qualitatively good coherent states and that is the restriction on the Mandel parameter. We are convinced that this method is very powerful for the situations where one requires qualitatively better coherent states, for instance, in certain experimental situations. At first we applied the procedure to the simple models in usual space, however, we intended to apply this procedure rigorously to the case of noncommutative models to examine the qualities of the noncommutative coherent states. Quite surprisingly there is no noncommutative model available in the position space while the Bohmian mechanics has only been formulated in the position space. Therefore one either requires to solve the noncommutative models in position space or one needs to formulate the Bohmian mechanics in momentum space, which is yet under construction and does not provide any satisfactory progress. That is why we computed the position space wave functions of the noncommutative harmonic oscillator, which we have not been able to apply to the Bohmian mechanics to inspect the  qualities of the noncommutative coherent states further. But we intend to investigate this further which we also leave as open problem to the readers.

Meanwhile we also found the classical mechanics in complex plane to be another interesting research topic. The subject is immensely rich and interesting in itself as one may experience many appealing phenomena while investigating. In the literature various authors claimed \cite{bender_holm_hook,bender_holm_hook_2,bender_brody_hook,anderson_bender_morone} that classical particles having real energy while moving in a complex plane produce closed and periodic orbit, whereas those which possess complex energy, create open trajectories and therefore they demand that there might have a close connection of complex classical trajectories with the $\mathcal{PT}$-symmetry of the Hamiltonian. However, we found that this statement is not always true and one can produce closed trajectories for a classical particle having complex energy and vice-versa. Thus we claim that there might not have any connection of the topic with the $\mathcal{PT}$-symmetry property of the Hamiltonian. This line of thought would be interesting to investigated further. To provide an exact explanation of the behaviour would surely be compelling, which we leave as an open problem.

We have been able to  provide a large amount of physical motivation for noncommutative quantum mechanics in this thesis. Our central goal was to develop the framework of noncommutative quantum mechanics towards the more challenging physical consideration as well as to construct the mathematical framework wherever needed. Besides those open problems which are directly related to our research as we mentioned earlier, there is a multitude of ways in which one can extend our ideas. For instance, we have not explored the analysis of our noncommutative algebra to the realm of quantum field theory. The key ingredients that one needs for the study of noncommutative quantum field theories is either the path integral formulation or replacement of products of fields by the Gronewald-Moyal star product. It would be interesting to compare the outcomes of the noncommutative field theoretical results with the standard results of field theory, which is where one may obtain plenty of new results and ideas which might be useful to explain the hidden structure of our universe. One may also extend the understanding of the deformed algebra towards the relativistic quantum mechanical framework in the noncommutative space time preserving the Lorentz symmetry, which will ultimately lead to the relativistic wave equation describing the spinning particle on co-ordinate dependent noncommutative space time (noncommutative Dirac equation).  

Noncommutative space has a wide range of success in many areas of Physics and Mathematics, for instance in quantum gravity, string theory etc. However it has not been studied considerably in the field of quantum information, condensed matter or optical Physics, in particular quantum entanglement which is itself a very widely studied subject having a wide range of application to the quantum cryptography, teleportation and many other fields. To be precise, quantum entanglement has undoubtedly proven to be a successful theory, however it will be very exciting to study the subject in the field of noncommutative space which might provide us a new insight into the field. One could investigate, whether the noncommutative coherent states that we have explored are quantum mechanically entangled or not. In the later case one may construct the entangled state representation of the coherent states and to examine if the Bell's inequality is violated using the idea of Sanders and his collaborators \cite{mann_sanders_munro}.

Time dependent Hamiltonians have various applications in modern physics including non-equilibrium transport problem, high frequency electronic response, molecular dynamics and many more. Thus another interesting aspect that one could think about is to explore the noncommutative structures in time dependent backgrounds and subsequently solve them to obtain concrete models in the time dependent framework.
    
}

\small{
    
}


\begin{thebibliography}{100}

\addcontentsline{toc}{chapter}{Bibliography}

\bibitem{polchinski_volI}
J.~Polchinski,
\newblock {\em String theory}, volume {I: An introduction to the bosonic string},
\newblock Cambridge University Press (2005).

\bibitem{polchinski_volII}
J.~Polchinski,
\newblock {\em String theory}, volume {II: Superstring theory and beyond},
\newblock Cambridge University Press (2005).

\bibitem{green_schwarz_witten}
M.~B. Green, J.~H. Schwarz and E.~Witten,
\newblock {\em Superstring theory}, volume~{II},
\newblock Cambridge University Press (2012).


\rhead{Bibliography}
\lhead{}
\chead{}


\bibitem{rovelli}
C.~Rovelli,
\newblock Loop quantum gravity,
\newblock {\em Living Rev. Rel.} \textbf{1} (1998).

\bibitem{reuter}
M.~Reuter,
\newblock Nonperturbative evolution equation for quantum gravity,
\newblock {\em Phys. Rev.} \textbf{D57}, 971 (1998).

\bibitem{niedermaier_reuter}
M.~Niedermaier and M.~Reuter,
\newblock The asymptotic safety scenario in quantum gravity,
\newblock {\em Living Rev. Rel.} \textbf{9}, 173 (2006).

\bibitem{ambjorn_loll}
J.~Ambj{\o}rn and R.~Loll,
\newblock {Non-perturbative Lorentzian quantum gravity, causality and topology change},
\newblock {\em Nuc. Phys.} \textbf{B536}, 407--434 (1998).

\bibitem{amelino1}
G.~Amelino-Camelia,
\newblock Doubly special relativity,
\newblock {\em Nature} \textbf{418}, 34--35 (2002).

\bibitem{dewitt}
B.~S. DeWitt,
\newblock Quantum theory of {gravity-I, The canonical} theory,
\newblock {\em Phys. Rev.} \textbf{160}, 1113 (1967).

\bibitem{hartle_hawking}
J.~B. Hartle and S.~W. Hawking,
\newblock Wave function of the universe,
\newblock {\em Phys. Rev.} \textbf{D28}, 2960 (1983).

\bibitem{hawking}
S.~W. Hawking,
\newblock The quantum state of the universe,
\newblock {\em Nuc. Phys.} \textbf{B239}, 257--276 (1984).

\bibitem{snyder}
H.~S. Snyder,
\newblock Quantized space-time,
\newblock {\em Phys. Rev.} \textbf{71}, 38--41 (1947).

\bibitem{yang}
C.~N. Yang,
\newblock On quantized space-time,
\newblock {\em Phys. Rev.} \textbf{72}, 874 (1947).

\bibitem{seiberg_witten}
N.~Seiberg and E.~Witten,
\newblock String theory and noncommutative geometry,
\newblock {\em JHEP} \textbf{1999}, 032 (1999).

\bibitem{connes1985}
A.~Connes,
\newblock Non-commutative differential geometry,
\newblock {\em Pub. Math. de l'IHES} \textbf{62}, 41--144 (1985).

\bibitem{woronowicz}
S.~L. Woronowicz,
\newblock Compact matrix pseudogroups,
\newblock {\em Comm. Math. Phys.} \textbf{111}, 613--665 (1987).

\bibitem{garay}
L.~J. Garay,
\newblock Quantum gravity and minimum length,
\newblock {\em Int. J. Mod. Phys.} \textbf{A10}, 145--165 (1995).

\bibitem{connes}
A.~Connes,
\newblock {\em Noncommutative geometry},
\newblock Academic Press (1995).

\bibitem{madore}
J.~Madore,
\newblock {\em An introduction to noncommutative differential geometry and its physical applications},
\newblock Cambridge University Press (1999).

\bibitem{douglas_nekrasov}
M.~R. Douglas and N.~A. Nekrasov,
\newblock Noncommutative field theory,
\newblock {\em Rev. Mod. Phys.} \textbf{73}, 977 (2001).

\bibitem{szabo}
R.~J. Szabo,
\newblock Quantum field theory on noncommutative spaces,
\newblock {\em Phys. Rep.} \textbf{378}, 207--299 (2003).

\bibitem{doplicher_fredenhagen_roberts}
S.~Doplicher, K.~Fredenhagen and J.~E. Roberts,
\newblock Spacetime quantization induced by classical gravity,
\newblock {\em Phys. Lett.} \textbf{B331}, 39--44 (1994).

\bibitem{doplicher_fredenhagen_roberts2}
S.~Doplicher, K.~Fredenhagen and J.~E. Roberts,
\newblock The quantum structure of spacetime at the {Planck} scale and quantum fields,
\newblock {\em Comm. Math. Phys.} \textbf{172}, 187--220 (1995).

\bibitem{veneziano}
G.~Veneziano,
\newblock A stringy nature needs just two constants,
\newblock {\em Europhys. Lett.} \textbf{2}, 199 (1986).

\bibitem{gross_mende}
D.~J. Gross and P.~F. Mende,
\newblock String theory beyond the {Planck} scale,
\newblock {\em Nuc. Phys.} \textbf{B303}, 407--454 (1988).

\bibitem{yoneya}
T.~Yoneya,
\newblock On the interpretation of minimal length in string theories,
\newblock {\em Mod. Phys. Lett.} \textbf{A4}, 1587--1595 (1989).

\bibitem{amati_cliafaloni_veneziano}
D.~Amati, M.~Ciafaloni and G.~Veneziano,
\newblock Can spacetime be probed below the string size?
\newblock {\em Phys. Lett.} \textbf{B216}, 41--47 (1989).

\bibitem{guida_konishi_provero}
R.~Guida, K.~Konishi and P.~Provero,
\newblock On the short distance behavior of string theories,
\newblock {\em Mod. Phys. Lett.} \textbf{A6}, 1487--1503 (1991).

\bibitem{padmanabhan1}
T.~Padmanabhan,
\newblock Physical significance of {Planck} length,
\newblock {\em Ann. Phys.} \textbf{165}, 38--58 (1985).

\bibitem{padmanabhan2}
T.~Padmanabhan,
\newblock Planck length as the lower bound to all physical length scales,
\newblock {\em Gen. Rel. Grav.} \textbf{17}, 215--221 (1985).

\bibitem{padmanabhan3}
T.~Padmanabhan,
\newblock The role of general relativity in the uncertainty principle,
\newblock {\em Class. Quant. Grav.} \textbf{3}, 911 (1986).

\bibitem{padmanabhan4}
T.~Padmanabhan,
\newblock Limitations on the operational definition of spacetime events and quantum gravity,
\newblock {\em Class. Quant. Grav.} \textbf{4}, L107 (1987).

\bibitem{greensite}
J.~Greensite,
\newblock Is there a minimum length {in $D=4$ lattice} quantum gravity?
\newblock {\em Phys. Lett.} \textbf{B255}, 375--380 (1991).

\bibitem{magueijo_smolin}
J.~Magueijo and L.~Smolin,
\newblock Lorentz invariance with an invariant energy scale,
\newblock {\em Phys. Rev. Lett.} \textbf{88}, 190403 (2002).

\bibitem{magueijo_smolin2}
J.~Magueijo and L.~Smolin,
\newblock String theories with deformed energy-momentum relations and a possible nontachyonic bosonic string,
\newblock {\em Phys. Rev.} \textbf{D71}, 026010 (2005).

\bibitem{girelli_livine_oriti}
F.~Girelli, E.~R. Livine and D.~Oriti,
\newblock Deformed special relativity as an effective flat limit of quantum gravity,
\newblock {\em Nuc. Phys.} \textbf{B708}, 411--433 (2005).

\bibitem{cortes_gamboa}
J.~L. Cortes and J.~Gamboa, 
\newblock Quantum uncertainty in doubly special relativity,
\newblock {\em Phys. Rev.} \textbf{D71}, 065015 (2005).

\bibitem{ghosh_subir}
S.~Ghosh,
\newblock Lagrangian for doubly special relativity particle and the role of noncommutativity,
\newblock {\em Phys. Rev.} \textbf{D74}, 084019 (2006).

\bibitem{ali_das_vagenas}
A.~F. Ali, S.~Das and E.~C. Vagenas,
\newblock Discreteness of space from the generalized uncertainty principle,
\newblock {\em Phys. Lett.} \textbf{B678}, 497--499 (2009).

\bibitem{nozari_etemadi}
K.~Nozari and A.~Etemadi,
\newblock Minimal length, maximal momentum, and {Hilbert} space representation of quantum mechanics,
\newblock {\em Phys. Rev.} \textbf{D85}, 104029 (2012).

\bibitem{kato}
M.~Kato, 
\newblock Particle theories with minimum observable length and open string theory,
\newblock {\em Phys. Lett.} \textbf{B245}, 43--47 (1990).

\bibitem{konishi_paffuti_provero}
K.~Konishi, G.~Paffuti and P.~Provero,
\newblock Minimum physical length and the generalized uncertainty principle in string theory,
\newblock {\em Phys. Lett.} \textbf{B234}, 276--284 (1990).

\bibitem{maggiore}
M.~Maggiore,
\newblock A generalized uncertainty principle in quantum gravity,
\newblock {\em Phys. Lett.} \textbf{B304}, 65--69 (1993).

\bibitem{scardigli}
F.~Scardigli,
\newblock Generalized uncertainty principle in quantum gravity from micro-black hole gedanken experiment,
\newblock {\em Phys. Lett.} \textbf{B452}, 39--44 (1999).

\bibitem{ng_dam}
Y.~J. Ng and H.~Van~Dam,
\newblock {Spacetime Foam, Holographic Principle and Black Hole Quantum Computers},
\newblock {\em Int. J. Mod. Phys.} \textbf{A20}, 1328--1335 (2005).

\bibitem{mead}
C.~A. Mead,
\newblock Possible connection between gravitation and fundamental length,
\newblock {\em Phys. Rev.} \textbf{135}, B849 (1964).

\bibitem{ng}
Y.~J. Ng,
\newblock Selected topics in {Planck}-scale physics,
\newblock {\em Mod. Phys. Lett.} \textbf{A18}, 1073--1097 (2003).

\bibitem{hossenfelder}
S.~Hossenfelder,
\newblock The minimal length and large extra dimensions,
\newblock {\em Mod. Phys. Lett.} \textbf{A19}, 2727--2744 (2004).

\bibitem{kempf}
A.~Kempf,
\newblock Uncertainty relation in quantum mechanics with quantum group symmetry,
\newblock {\em J. Math. Phys.} \textbf{35}, 4483--4496 (1994).

\bibitem{kempf_mangano_mann}
A.~Kempf, G.~Mangano and R.~B. Mann, 
\newblock Hilbert space representation of the minimal length uncertainty relation,
\newblock {\em Phys. Rev.} \textbf{D52}, 1108--1118 (1995).

\bibitem{sun_fu}
C.~Sun and H.~Fu,
\newblock The q-deformed boson realisation of the quantum group {$SU(n)_q$} and its representations,
\newblock {\em J. Phys. A: Math. Gen.} \textbf{22}, L983 (1989).

\bibitem{macfarlane}
A.~J. Macfarlane,
\newblock On q-analogues of the quantum harmonic oscillator and the quantum group {$SU(2)_q$},
\newblock {\em J. Phys. A: Math. Gen.} \textbf{22}, 4581 (1989).

\bibitem{biedenharn}
L.~C. Biedenharn,
\newblock The quantum group {$SU_q(2)$} and a q-analogue of the boson operators,
\newblock {\em J. Phys. A: Math. Gen.} \textbf{22}, L873 (1989).

\bibitem{brodimas_jannussis_mignani}
G.~Brodimas, A.~Jannussis and R.~Mignani,
\newblock Bose realization of a non-canonical {Heisenberg} algebra,
\newblock {\em J. Phys. A: Math. Gen.} \textbf{25}, L329 (1992).

\bibitem{quesne_tkachuk}
C.~Quesne and V.~M. Tkachuk,
\newblock Generalized deformed commutation relations with nonzero minimal uncertainties in position and/or momentum and applications to quantum mechanics,
\newblock {\em SIGMA} \textbf{3}, 16--18 (2007).

\bibitem{bagchi_fring}
B.~Bagchi and A.~Fring,
\newblock Minimal length in quantum mechanics and {non-Hermitian Hamiltonian systems},
\newblock {\em Phys. Lett.} \textbf{A373}, 4307--4310 (2009).

\bibitem{fring_gouba_scholtz}
A.~Fring, L.~Gouba and F.~G. Scholtz,
\newblock Strings from position-dependent noncommutativity,
\newblock {\em J. Phys. A: Math. Theor.} \textbf{43}, 345401 (2010).

\bibitem{fring_gouba_bagchi_area}
A.~Fring, L.~Gouba and B.~Bagchi,
\newblock Minimal areas from q-deformed oscillator algebras,
\newblock {\em J. Phys. A: Math. Theor.} \textbf{43}, 425202 (2010).

\bibitem{hossenfelder1}
S.~Hossenfelder,
\newblock A note on theories with a minimal length,
\newblock {\em Class. Quant. Grav.} \textbf{23}, 1815 (2006).

\bibitem{chang_lewis_minic_takeuchi}
L.~N. Chang, Z.~Lewis, D.~Minic and T.~Takeuchi,
\newblock On the minimal length uncertainty relation and the foundations of string theory,
\newblock {\em Adv. High Energy Phys.} \textbf{2011}, 493514 (2011).

\bibitem{lewis_takeuchi}
Z.~Lewis and T.~Takeuchi,
\newblock Position and momentum uncertainties of the normal and inverted harmonic oscillators under the minimal length uncertainty relation,
\newblock {\em Phys. Rev.} \textbf{D84}, 105029 (2011).

\bibitem{pedram_nozari_taheri}
P.~Pedram, K.~Nozari and S.~H. Taheri,
\newblock The effects of minimal length and maximal momentum on the transition rate of ultra cold neutrons in gravitational field,
\newblock {\em JHEP} \textbf{2011}, 1--11 (2011).

\bibitem{dorsch_nogueira}
G.~Dorsch and J.~A. Nogueira, 
\newblock Minimal length in quantum mechanics via modified {Heisenberg} algebra,
\newblock Technical report CERN (2011).

\bibitem{gavrilik_kachurik}
A.~M. Gavrilik and I.~I. Kachurik,
\newblock Three-parameter (two-sided) deformation of {Heisenberg} algebra,
\newblock {\em Mod. Phys. Lett.} \textbf{A27} (2012).

\bibitem{dey_fring_mathanaranjan}
S.~Dey, A.~Fring and T.~Mathanaranjan,
\newblock {Non-Hermitian systems of Euclidean Lie algebraic type with real energy spectra},
\newblock {\em Ann. Phys.} \textbf{346}, 28--41 (2014).

\bibitem{dey_fring_gouba}
S.~Dey, A.~Fring and L.~Gouba,
\newblock $\mathcal{PT}$-symmetric non-commutative spaces with minimal volume uncertainty relations,
\newblock {\em J. Phys. A: Math. Theor.} \textbf{45}, 385302 (2012).

\bibitem{dey_fring_2DHO}
S.~Dey and A.~Fring,
\newblock The two dimensional harmonic oscillator on a noncommutative space with minimal uncertainties,
\newblock {\em Acta Polytechnica} \textbf{53}, 268--270 (2013).

\bibitem{dey_fring_khantoul}
S.~Dey, A.~Fring and B.~Khantoul,
\newblock Hermitian versus {non-Hermitian} representations for minimal length uncertainty relations,
\newblock {\em J. Phys. A: Math. Theor.} \textbf{46}, 335304 (2013).

\bibitem{dey_fring_squeezed}
S.~Dey and A.~Fring,
\newblock Squeezed coherent states for noncommutative spaces with minimal length uncertainty relations,
\newblock {\em Phys. Rev.} \textbf{D86}, 064038 (2012).

\bibitem{dey_fring_gouba_castro}
S.~Dey, A.~Fring, L.~Gouba and P.~G. Castro,
\newblock Time-dependent q-deformed coherent states for generalized uncertainty relations,
\newblock {\em Phys. Rev.} \textbf{D87}, 084033 (2013).

\bibitem{dey_fring_bohmian}
S.~Dey and A.~Fring,
\newblock Bohmian quantum trajectories from coherent states,
\newblock {\em Phys. Rev.} \textbf{A88}, 022116 (2013).

\bibitem{dey_fring_bohmian_2}
S.~Dey and A.~Fring,
\newblock {Quantum versus classical $\mathcal{PT}$-symmetry},
\newblock {\em in preparation}.

\bibitem{bender_holm_hook}
C.~M. Bender, D.~D. Holm and D.~W. Hook,
\newblock Complex trajectories of a simple pendulum,
\newblock {\em J. Phys. A: Math. Theor.} \textbf{40}, F81 (2007).

\bibitem{bender_holm_hook_2}
C.~M. Bender, D.~D. Holm and D.~W. Hook,
\newblock Complexified dynamical systems,
\newblock {\em J. Phys. A: Math. Theor.} \textbf{40}, F793 (2007).

\bibitem{birkhoff_neumann}
G.~Birkhoff and J.~Von~Neumann,
\newblock The logic of quantum mechanics,
\newblock In {\em The logico-algebraic approach to quantum mechanics}, 1--26, Springer (1975).

\bibitem{landi}
G.~Landi,
\newblock {\em An introduction to noncommutative spaces and their geometries},
\newblock Springer Berlin (1997).

\bibitem{carroll_harvey_kostelecky_lane_okamoto}
S.~M. Carroll, J.~A. Harvey, V.~A. Kostelecky, C.~D. Lane and T.~Okamoto,
\newblock Noncommutative field theory and {Lorentz} violation,
\newblock {\em Phys. Rev. Lett.} \textbf{87}, 141601 (2001).

\bibitem{banerjee_chakraborty_kumar}
R.~Banerjee, B.~Chakraborty and K.~Kumar,
\newblock Noncommutative gauge theories and {Lorentz} symmetry,
\newblock {\em Phys. Rev.} \textbf{D70}, 125004 (2004).

\bibitem{gomis_mehen}
J.~Gomis and T.~Mehen, 
\newblock Space--time noncommutative field theories and unitarity,
\newblock {\em Nuc. Phys.} \textbf{B591}, 265--276 (2000).

\bibitem{seiberg_susskind_toumbas}
N.~Seiberg, L.~Susskind and N.~Toumbas,
\newblock Space/time non-commutativity and causality,
\newblock {\em JHEP} \textbf{2000}, 044 (2000).

\bibitem{minwalla_van_seiberg}
S.~Minwalla, M.~Van~Raamsdonk and N.~Seiberg, 
\newblock Noncommutative perturbative dynamics,
\newblock {\em JHEP} \textbf{2000}, 020 (2000).

\bibitem{ardalan_sadooghi}
F.~Ardalan and N.~Sadooghi, 
\newblock Axial anomaly in noncommutative {QED on R4},
\newblock {\em Int. J. Mod. Phys.} \textbf{A16}, 3151--3177 (2001).

\bibitem{banerjee_ghosh}
R.~Banerjee and S.~Ghosh,
\newblock {Seiberg--Witten map and the axial anomaly in noncommutative field theory},
\newblock {\em Phys. Lett.} \textbf{B533}, 162--167 (2002).

\bibitem{banerjee1}
R.~Banerjee,
\newblock Anomalies in noncommutative gauge theories, {Seiberg--Witten transformation and Ramond--Ramond couplings},
\newblock {\em Int. J. Mod. Phys.} \textbf{A19}, 613--630 (2004).

\bibitem{moyal}
J.~E. Moyal,
\newblock Quantum mechanics as a statistical theory,
\newblock In {\em Math. Proc. Cambridge Phil. Soc.} \textbf{45}, 99--124, Cambridge University Press (1949).

\bibitem{fairlie}
D.~B. Fairlie,
\newblock Moyal brackets in {M-theory},
\newblock {\em Mod. Phys. Lett.} \textbf{A13}, 263--274 (1998).

\bibitem{dimakis}
A.~Dimakis and F.~M{\"u}ller-Hoissen,
\newblock Bicomplexes, integrable models, and noncommutative geometry,
\newblock {\em Int. J. Mod. Phys.} \textbf{B14}, 2455--2460 (2000).

\bibitem{grisaru}
M.~T. Grisaru and S.~Penati,
\newblock An integrable noncommutative version of the {sine-Gordon} system,
\newblock {\em Nuc. Phys.} \textbf{B655}, 250--276 (2003).

\bibitem{moriconi}
I.~Cabrera-Carnero and M.~Moriconi,
\newblock Noncommutative integrable field theories in 2d,
\newblock {\em Nuc. Phys.} \textbf{B673}, 437--454 (2003).

\bibitem{lechtenfeld}
O.~Lechtenfeld, L.~Mazzanti, S.~Penati, A.~D. Popov and L.~Tamassia,
\newblock {Integrable noncommutative sine-Gordon model},
\newblock {\em Nuc. Phys.} \textbf{B705}, 477--503 (2005).

\bibitem{fairlie1}
D.~B Fairlie,
\newblock Moyal brackets, star products and the generalised {Wigner} function,
\newblock {\em Chaos, Solitons \& Fractals} \textbf{10}, 365--371 (1999).

\bibitem{carroll}
R.~Carroll,
\newblock {Quantum Therory, Deformation and Integrability},
\newblock {\em North-Holland Math. Studies} \textbf{186}, 365--371 (2000).

\bibitem{banerjee_kulkarni_samanta}
R.~Banerjee, S.~Kulkarni and S.~Samanta,
\newblock Deformed symmetry in {Snyder space and relativistic particle dynamics},
\newblock {\em JHEP} \textbf{2006}, 077 (2006).

\bibitem{wess}
J.~Wess,
\newblock Deformed coordinate spaces derivatives,
\newblock In {\em Math. Theo. Pheno. Challenges Beyond the Standard Model} \textbf{1}, 122--128 (2005).

\bibitem{aschieri_blohmann_dimitrijevic_meyer_schupp_wess}
P.~Aschieri, C.~Blohmann, M.~Dimitrijevi{\'c}, F.~Meyer, P.~Schupp and J.~Wess,
\newblock A gravity theory on noncommutative spaces,
\newblock {\em Class. Quant. Grav.} \textbf{22}, 3511 (2005).

\bibitem{koch_tsouchnika}
F.~Koch and E.~Tsouchnika,
\newblock {Construction of $\theta$-Poincar{\'e} algebras and their invariants on $\mathcal{M}_\theta$},
\newblock {\em Nuc. Phys.} \textbf{B717}, 387--403 (2005).

\bibitem{banerjee2}
R.~Banerjee,
\newblock Deformed {Schr{\"o}dinger symmetry on noncommutative space},
\newblock {\em Euro. Phys. J.} \textbf{C47}, 541--545 (2006).

\bibitem{chaichian_kulish_nishijima_tureanu}
M.~Chaichian, P.~P. Kulish, K.~Nishijima and A.~Tureanu,
\newblock On a {Lorentz-invariant interpretation of noncommutative space--time and its implications on noncommutative QFT},
\newblock {\em Phys. Lett.} \textbf{B604}, 98--102 (2004).

\bibitem{chaichian_presnajder_tureanu}
M.~Chaichian, P.~Pre{\v{s}}najder and A.~Tureanu,
\newblock New concept of relativistic invariance in noncommutative space-time: twisted {Poincar{\'e} symmetry} and its implications,
\newblock {\em Phys. Rev. Lett.} \textbf{94}, 151602 (2005).

\bibitem{aschieri_dimitrijevic_meyer_wess}
P.~Aschieri, M.~Dimitrijevi{\'c}, F.~Meyer and J.~Wess,
\newblock Noncommutative geometry and gravity,
\newblock {\em Class. Quant. Grav.} \textbf{23}, 1883 (2006).

\bibitem{castro_kullock_toppan}
P.~G. Castro, R.~Kullock and F.~Toppan,
\newblock Snyder noncommutativity and {pseudo-Hermitian Hamiltonians from a Jordanian twist},
\newblock {\em J. Math. Phys.} \textbf{52}, 062105 (2011).

\bibitem{lukierski_woronowicz}
J.~Lukierski and M.~Woronowicz,
\newblock New {Lie-algebraic and quadratic deformations of Minkowski space from twisted Poincare symmetries},
\newblock {\em Phys. Lett.} \textbf{B633}, 116--124 (2006).

\bibitem{gomes_kupriyanov}
M.~Gomes and V.~G. Kupriyanov,
\newblock Position-dependent noncommutativity in quantum mechanics,
\newblock {\em Phys. Rev.} \textbf{D79}, 125011 (2009).

\bibitem{gomes_kupriyanov_silva}
M.~Gomes, V.~G. Kupriyanov and A.~J. Da~Silva,
\newblock Dynamical noncommutativity,
\newblock {\em J. Phys. A: Math. Theor.} \textbf{43}, 285301 (2010).

\bibitem{takhtadzhan_faddeev}
L.~A. Takhtadzhan and L.~D. Faddeev,
\newblock The quantum method of the inverse problem and the heisenberg {XYZ} model,
\newblock {\em Russian Math. Surveys} \textbf{34}, 11--68 (1979).

\bibitem{kulish_sklyanin}
P.~P. Kulish and E~.K. Sklyanin,
\newblock Lecture notes in physics (Springer, Berlin),
\newblock \textbf{151}, 61 (1982).

\bibitem{jimbo1}
M.~Jimbo,
\newblock Aq-difference analogue of {U(g) and the Yang-Baxter equation},
\newblock {\em Lett. Math. Phys.} \textbf{10}, 63--69 (1985).

\bibitem{jimbo2}
M.~Jimbo,
\newblock A q-analogue of {U(g(N+ 1)), Hecke algebra, and the Yang-Baxter equation},
\newblock {\em Lett. Math. Phys.} \textbf{11}, 247--252 (1986).

\bibitem{drinfeld}
VG~Drinfeld,
\newblock Quantum groups,
\newblock {\em Proc. ICM (Berkeley)}, 798--819 (1986).

\bibitem{kulish_damaskinsky}
P.~P. Kulish and E.~V. Damaskinsky,
\newblock On the q oscillator and the quantum algebra $su_q(1, 1)$,
\newblock {\em J. Phys. A: Math. Gen.} \textbf{23}, L415 (1990).

\bibitem{arik_coon}
M.~Arik and D.~D. Coon,
\newblock Hilbert spaces of analytic functions and generalized coherent states,
\newblock {\em J. Math. Phys.} \textbf{17}, 524--527 (1976).

\bibitem{schwenk}
J.~Schwenk and J.~Wess,
\newblock A q-deformed quantum mechanical toy model,
\newblock {\em Phys. Lett.} \textbf{B291}, 273--277 (1992).

\bibitem{bird}
D.~J. Bird~et al,
\newblock Detection of a cosmic ray with measured energy well beyond the expected spectral cutoff due to cosmic microwave radiation,
\newblock {\em Astrophys. J.} \textbf{441}, 144--150 (1995).

\bibitem{amelino3}
G.~Amelino-Camelia,
\newblock Doubly-special relativity: {First} results and key open problems,
\newblock {\em Int. J. Mod. Phys.} \textbf{D11}, 1643--1669 (2002).

\bibitem{amelino2}
G.~Amelino-Camelia,
\newblock Testable scenario for relativity with minimum length,
\newblock {\em Phys. Lett.} \textbf{B510}, 255--263 (2001).

\bibitem{kowalski}
J.~Kowalski-Glikman,
\newblock Introduction to doubly special relativity,
\newblock {\em Lect. Notes Phys.} \textbf{669}, 131 (2005).

\bibitem{lukierski_ruegg_nowicki_tolstoy}
J.~Lukierski, H.~Ruegg, A.~Nowicki and V.~N. Tolstoy,
\newblock q--deformation of {Poincar{\'e} algebra},
\newblock {\em Phys. Lett.} \textbf{B264}, 331--338 (1991).

\bibitem{majid_ruegg}
S.~Majid and H.~Ruegg,
\newblock Bicrossproduct structure of $\kappa$--{Poincar{\'e}} group and non-commutative geometry,
\newblock {\em Phys. Lett.} \textbf{B334}, 348--354 (1994).

\bibitem{lukierski_ruegg_zakrzewski}
J.~Lukierski, H.~Ruegg and W.~J. Zakrzewski,
\newblock Classical and quantum mechanics of free $\kappa$--relativistic systems,
\newblock {\em Ann. Phys.} \textbf{243}, 90--116 (1995).

\bibitem{kowalski2}
J.~Kowalski-Glikman,
\newblock Observer-independent quantum of mass,
\newblock {\em Phys. Lett.} \textbf{A286}, 391--394 (2001).

\bibitem{kowalski_nowak}
J.~Kowalski-Glikman and S.~Nowak,
\newblock Non-commutative space--time of doubly special relativity theories,
\newblock {\em Int. J. Mod. Phys.} \textbf{D12}, 299--315 (2003).

\bibitem{kowalski_nowak2}
J.~Kowalski-Glikman and S.~Nowak,
\newblock Doubly special relativity theories as different bases of $\kappa$-{Poincar{\'e}} algebra,
\newblock {\em Phys. Lett.} \textbf{B539}, 126--132 (2002).

\bibitem{barton}
G.~Barton,
\newblock {\em Introduction to advanced field theory},
\newblock John Wiley \& Sons, chap. 12 (1963).

\bibitem{neumann}
J.~Von~Neumann and E.~Wigner,
\newblock {{\"U}ber merkw{\"u}rdige diskrete Eigenwerte. {\"U}ber das Verhalten von Eigenwerten bei adiabatischen Prozessen},
\newblock {\em Zhurnal Physik} \textbf{30}, 467--470 (1929).

\bibitem{friedrich1}
H.~Friedrich and D.~Wintgen,
\newblock Interfering resonances and bound states in the continuum,
\newblock {\em Phys. Rev.} \textbf{A32}, 3231 (1985).

\bibitem{magunov}
A.~I. Magunov, I.~Rotter and S.~I. Strakhova,
\newblock Laser induced resonance trapping in atoms,
\newblock {\em J. Phys.} \textbf{B32}, 1669 (1999).

\bibitem{rotter1}
I.~Rotter and A.~F. Sadreev,
\newblock Zeros in single-channel transmission through double quantum dots,
\newblock {\em Phys. Rev.} \textbf{E71}, 046204 (2005).

\bibitem{persson}
E.~Persson, T.~Gorin and I.~Rotter,
\newblock Decay rates of resonance states at high level density,
\newblock {\em Phys. Rev.}, \textbf{E54}, 3339 (1996).

\bibitem{moiseyev}
N.~Moiseyev,
\newblock Quantum theory of resonances: calculating energies, widths and cross-sections by complex scaling,
\newblock {\em Phys. Rep.} \textbf{302}, 212--293 (1998).

\bibitem{bender_boettcher}
C.~M. Bender and S.~Boettcher,
\newblock Real spectra in {non-Hermitian Hamiltonians} having
  $\mathcal{PT}$-symmetry,
\newblock {\em Phys. Rev. Lett.} \textbf{80}, 5243 (1998).

\bibitem{wigner}
E.~P. Wigner,
\newblock Normal form of antiunitary operators,
\newblock {\em J. Math. Phys.} \textbf{1}, 409 (1960).

\bibitem{bessis}
D.~Bessis and Z.~Justin,
\newblock Private communication with {C. M. Bender}
\newblock (1998)

\bibitem{bender_making_sense}
C.~M. Bender,
\newblock Making sense of {non-Hermitian Hamiltonians},
\newblock {\em Rep. Prog. Phys.} \textbf{70}, 947 (2007).

\bibitem{bender_brody_jones}
C.~M. Bender, D.~C. Brody and H.~F. Jones,
\newblock Complex extension of quantum mechanics,
\newblock {\em Phys. Rev. Lett.} \textbf{89}, 270401 (2002).

\bibitem{weigert}
S.~Weigert,
\newblock $\mathcal{PT}$-symmetry and its spontaneous breakdown explained by anti-linearity,
\newblock {\em J. Phys.} \textbf{B5}, S416 (2003).

\bibitem{dorey}
P.~Dorey, C.~Dunning and R.~Tateo,
\newblock A reality proof in $\mathcal{PT}$-symmetric quantum mechanics,
\newblock {\em Czech. J. Phys.} \textbf{54}, 35--41 (2004).

\bibitem{weigert_detecting}
S.~Weigert,
\newblock Detecting broken $\mathcal{PT}$-symmetry,
\newblock {\em J. Phys. A: Math. Gen.} \textbf{39}, 10239 (2006).

\bibitem{weigert_completeness}
S.~Weigert,
\newblock Completeness and orthonormality in $\mathcal{PT}$-symmetric quantum systems,
\newblock {\em Phys. Rev.} \textbf{A68}, 062111 (2003).

\bibitem{bender_wang}
C.~M. Bender, P.~N. Meisinger and Q.~Wang,
\newblock Calculation of the hidden symmetry operator in
  $\mathcal{PT}$-symmetric quantum mechanics,
\newblock {\em J. Phys. A: Math. Gen.} \textbf{36}, 1973 (2003).

\bibitem{bender_jones}
C.~M. Bender and H.~F. Jones,
\newblock Semiclassical calculation of the $\mathcal{C}$ operator in $\mathcal{PT}$-symmetric quantum mechanics,
\newblock {\em Phys. Lett.} \textbf{A328}, 102--109 (2004).

\bibitem{bender_tan}
C.~M. Bender and B.~Tan,
\newblock Calculation of the hidden symmetry operator for a
  $\mathcal{PT}$-symmetric square well,
\newblock {\em J. Phys. A: Math. Gen.} \textbf{39}, 1945 (2006).

\bibitem{roychoudhury_roy}
R.~Roychoudhury and P.~Roy,
\newblock Construction of the $\mathcal{C}$ operator for a
  $\mathcal{PT}$-symmetric model,
\newblock {\em J. Phys. A, Math. Theor.} \textbf{40} (2007).

\bibitem{mostafazadeh4}
A.~Mostafazadeh,
\newblock {Exact $\mathcal{PT}$-symmetry is equivalent to Hermiticity},
\newblock {\em J. Phys. A: Math. Gen.} \textbf{36}, 7081 (2003).

\bibitem{mostafazadeh6}
A.~Mostafazadeh,
\newblock A critique of $\mathcal{PT}$-symmetric quantum mechanics,
\newblock {\em arXiv:} 0310164 (2003).

\bibitem{mostafazadeh_time_dependent}
A.~Mostafazadeh,
\newblock Time-dependent {pseudo-Hermitian Hamiltonians defining a unitary quantum system and uniqueness of the metric operator},
\newblock {\em Phys. Lett.} \textbf{B650}, 208--212 (2007).

\bibitem{bender_brody_jones_prd}
C.~M. Bender, D.~C. Brody and H.~F. Jones,
\newblock Extension of $\mathcal{PT}$-symmetric quantum mechanics to quantum field theory with cubic interaction,
\newblock {\em Phys. Rev.} \textbf{D70}, 025001 (2004).

\bibitem{pauli}
W.~Pauli,
\newblock {On Dirac's new method of field quantization},
\newblock {\em Rev. Mod. Phys.} \textbf{15}, 175 (1943).

\bibitem{gupta}
S.~N. Gupta,
\newblock Theory of longitudinal photons in quantum electrodynamics,
\newblock {\em Proc. Phys. Soc. London} \textbf{63}, 681 (1950).

\bibitem{bleuler}
K.~Bleuler,
\newblock A new method for the treatment of longitudinal and scalar photons,
\newblock {\em Helv. Phys. Acta} \textbf{23}, 567 (1950).

\bibitem{sudarshan}
E.~C.~G. Sudarshan,
\newblock {Quantum mechanical systems with indefinite metric I},
\newblock {\em Phys. Rev.} \textbf{123}, 2183 (1961).

\bibitem{lee}
T.~D. Lee and G.~C. Wick,
\newblock Negative metric and the unitarity of the $s$-matrix,
\newblock {\em Nuc. Phys.}, \textbf{B9}, 209--243 (1969).

\bibitem{dieudonne}
J.~Dieudonn{\'e},
\newblock {Quasi-Hermitian operators},
\newblock {\em Proc. Int. Symp. on Linear Spaces (Jerusalem, 1960), Pergamon, Oxford}, 115--122 (1961).

\bibitem{scholtz_geyer_hahne}
F.~G. Scholtz, H.~B. Geyer and F.~J.~W. Hahne,
\newblock {Quasi-Hermitian operators in quantum mechanics and the variational principle},
\newblock {\em Ann. Phys.} \textbf{213}, 74--101 (1992).

\bibitem{mostafazadeh1}
A.~Mostafazadeh,
\newblock {Pseudo-Hermiticity versus $\mathcal{PT}$ symmetry: the necessary condition for the reality of the spectrum of a non-Hermitian Hamiltonian},
\newblock {\em J. Math. Phys.} \textbf{43}, 205 (2002).

\bibitem{mostafazadeh2}
A.~Mostafazadeh,
\newblock {Pseudo-Hermiticity versus $\mathcal{PT}$-symmetry II: A complete characterization of non-Hermitian Hamiltonians with a real spectrum},
\newblock {\em J. Math. Phys.} \textbf{43}, 2814 (2002).

\bibitem{mostafazadeh3}
A.~Mostafazadeh,
\newblock {Pseudo-Hermiticity versus $\mathcal{PT}$-symmetry III: Equivalence of pseudo-Hermiticity and the presence of antilinear symmetries},
\newblock {\em J. Math. Phys.} \textbf{43}, 3944 (2002).

\bibitem{dyson}
F.~J. Dyson,
\newblock General theory of spin-wave interactions,
\newblock {\em Phys. Rev.} \textbf{102}, 1217 (1956).

\bibitem{mostafazadeh5}
A.~Mostafazadeh, 
\newblock {Pseudo-Hermitian representation of quantum mechanics}, 
\newblock {\em Int. J. Geo. Meth. Mod. Phys.} \textbf{7}, 1191--1306 (2010).

\bibitem{williams}
J.~P. Williams,
\newblock Operators similar to their adjoints,
\newblock {\em Proc. Am. Math. Soc.} \textbf{20}, 121--123 (1969).

\bibitem{bagchi_quesne_roychoudhury}
B.~Bagchi, C.~Quesne and R.~Roychoudhury,
\newblock {Pseudo-Hermiticity} and some consequences of a generalized quantum condition,
\newblock {\em J. Phys. A: Math. Gen.} \textbf{38}, L647--L652 (2005).

\bibitem{znojil_geyer}
M.~Znojil and H.~B. Geyer,
\newblock Construction of a unique metric in {quasi-Hermitian} quantum mechanics: nonexistence of the charge operator in a 2 $\times$ 2 matrix model,
\newblock {\em Phys. Lett.} \textbf{B640}, 52--56 (2006).

\bibitem{musumbu_geyer_heiss}
D.~P. Musumbu, H.~B. Geyer and W.~D. Heiss,
\newblock Choice of a metric for the {non-Hermitian} oscillator,
\newblock {\em J. Phys. A: Math. Theor.} \textbf{40}, F75 (2007).

\bibitem{assis_fring_lie}
P.~E.~G. Assis and A.~Fring,
\newblock {Non-Hermitian Hamiltonians of Lie algebraic type},
\newblock {\em J. Phys. A: Math. Theor.} \textbf{42}, 015203 (2009).

\bibitem{mostafazadeh_batal}
A.~Mostafazadeh and A.~Batal,
\newblock Physical aspects of {pseudo-Hermitian and $\mathcal{PT}$-symmetric} quantum mechanics,
\newblock {\em J. Phys. A: Math. Gen.} \textbf{37}, 11645 (2004).

\bibitem{jones}
H.~F. Jones,
\newblock {On pseudo-Hermitian Hamiltonians and their Hermitian counterparts},
\newblock {\em J. Phys. A: Math. Gen.} \textbf{38}, 1741 (2005).

\bibitem{mostafazadeh8}
A.~Mostafazadeh, 
\newblock $\mathcal{PT}$-symmetric cubic anharmonic oscillator as a physical model, 
\newblock {\em J. Phys. A: Math. Gen.} \textbf{38}, 6557 (2005).

\bibitem{krejvcivrik_bila_znojil}
D.~Krej{\v{c}}i{\v{r}}{\'\i}k, H.~B{\'\i}la and M.~Znojil,
\newblock Closed formula for the metric in the {Hilbert space of a $\mathcal{PT}$-symmetric} model,
\newblock {\em J. Phys. A: Math. Gen.} \textbf{39}, 10143 (2006).

\bibitem{mostafazadeh9}
A.~Mostafazadeh,
\newblock Metric operator in {pseudo-Hermitian} quantum mechanics and the imaginary cubic potential,
\newblock {\em J. Phys. A: Math. Gen.} \textbf{39}, 10171 (2006).

\bibitem{ghatak_mandal}
A.~Ghatak and B.~P. Mandal,
\newblock Comparison of different approaches of finding the positive definite metric in {pseudo-Hermitian} theories,
\newblock {\em Comm. Theor. Phys.} \textbf{59}, 533 (2013).

\bibitem{faria_fring}
C.~F. de~Morisson~Faria and A.~Fring,
\newblock {Time evolution of non-Hermitian Hamiltonian systems},
\newblock {\em J. Phys. A: Math. and Gen.} \textbf{39}, 9269 (2006).

\bibitem{scholtz_geyer1}
F.~G. Scholtz and H.~B. Geyer,
\newblock {Operator equations and Moyal products-metrics in quasi-Hermitian quantum mechanics},
\newblock {\em Phys. Lett.} \textbf{B634}, 84--92 (2006).

\bibitem{scholtz_geyer2}
F.~G. Scholtz and H.~B. Geyer,
\newblock {Moyal products a new perspective on quasi-Hermitian quantum mechanics},
\newblock {\em J. Phys. A: Math. Gen.} \textbf{39}, 10189 (2006).

\bibitem{faria_fring1}
C.~F. de~Morisson~Faria and A.~Fring,
\newblock {Isospectral Hamiltonians from Moyal products},
\newblock {\em Czech. J. Phys.} \textbf{56}, 899--908 (2006).

\bibitem{bender_chen_milton}
C.~M. Bender, J.-H. Chen and K.~A. Milton,
\newblock $\mathcal{PT}$-symmetric versus {Hermitian} formulations of quantum mechanics,
\newblock {\em J. Phys. A: Math. Gen.} \textbf{39}, 1657 (2006).

\bibitem{mostafazadeh7}
A.~Mostafazadeh,
\newblock Comment on $\mathcal{PT}$-symmetric versus hermitian formulations of quantum mechanics,
\newblock {\em Los alamos preprint hep-th}/0603059 (2006).

\bibitem{mallick_kundu}
B.~Basu-Mallick and A.~Kundu,
\newblock {Exact solution of Calogero model with competing long-range interactions},
\newblock {\em Phys. Rev.} \textbf{B62}, 9927 (2000).

\bibitem{fring1}
A.~Fring,
\newblock {A note on the integrability of non-Hermitian extensions of Calogero--Moser--Sutherland models},
\newblock {\em Mod. Phys. Lett.} \textbf{A21}, 691--699 (2006).

\bibitem{el_makris}
R.~El-Ganainy, K.~G. Makris, D.~N. Christodoulides and Z.~H. Musslimani,
\newblock Theory of coupled {optical $\mathcal{PT}$-symmetric} structures,
\newblock {\em Opt. Lett.} \textbf{32}, 2632 (2007).

\bibitem{musslimani}
Z.~H. Musslimani, K.~G. Makris, R.~El-Ganainy and D.~N. Christodoulides,
\newblock Optical solitons in $\mathcal{PT}$ periodic potentials,
\newblock {\em Phys. Rev. Lett.} \textbf{100}, 030402 (2008).

\bibitem{makris_el_chris}
K.~G. Makris, R.~El-Ganainy, D.~N. Christodoulides and Z.~H. Musslimani,
\newblock {Beam Dynamics in $\mathcal{PT}$-symmetric Optical Lattices},
\newblock {\em Phys. Rev. Lett.} \textbf{100}, 103904 (2008).

\bibitem{guo_salamo}
A.~Guo, G.~J. Salamo, D.~Duchesne, R.~Morandotti, M.~Volatier-Ravat, V.~Aimez, G.~A. Siviloglou and D.~N. Christodoulides,
\newblock Observation of $\mathcal{PT}$-symmetry breaking in complex optical potentials,
\newblock {\em Phys. Rev. Lett.} \textbf{103}, 093902 (2009).

\bibitem{makris}
K.~G. Makris, R.~El-Ganainy, D.~N. Christodoulides and Z.~H. Musslimani,
\newblock $\mathcal{PT}$-symmetric optical lattices,
\newblock {\em Phys. Rev.} \textbf{A81}, 063807 (2010).

\bibitem{ruter}
C.~E. R{\"u}ter, K.~G. Makris, R.~El-Ganainy, D.~N. Christodoulides, M.~Segev and D.~Kip,
\newblock Observation of parity--time symmetry in optics,
\newblock {\em Nat. Phys.} \textbf{6}, 192--195 (2010).

\bibitem{midya_roy_roychoudhury}
B.~Midya, B.~Roy, and R.~Roychoudhury,
\newblock A note on the $\mathcal{PT}$ invariant periodic potential $V(x)=4\cos^2 x+4 i V_0\sin 2x$,
\newblock {\em Phys. Lett.} \textbf{A374}, 2605--2607 (2010).

\bibitem{jones1}
H.~F. Jones,
\newblock {Use of equivalent Hermitian Hamiltonian for $\mathcal{PT}$-symmetric sinusoidal optical lattices},
\newblock {\em J. Phys. A: Math. Theor.} \textbf{44}, 345302 (2011).

\bibitem{graefe_jones}
E.~M. Graefe and H.~F. Jones,
\newblock $\mathcal{PT}$-symmetric sinusoidal optical lattices at the symmetry-breaking threshold,
\newblock {\em Phys. Rev.} \textbf{A84}, 013818 (2011).

\bibitem{longhi_valle}
S.~Longhi and G.~D. Valle,
\newblock Invisible defects in complex crystals,
\newblock {\em Ann. Phys.} \textbf{334}, 35--46 (2013).

\bibitem{benisty_degiron}
H.~Benisty, A.~Degiron, A.~Lupu, A.~De~Lustrac, S.~Ch{\'e}nais, S.~Forget, M.~Besbes, G.~Barbillon, A.~Bruyant and S.~Blaize,
\newblock Implementation of {$\mathcal{PT}$-symmetric devices using plasmonics: principle and applications},
\newblock {\em Opt. Exp.} \textbf{19}, 18004--18019 (2011).

\bibitem{lupu_benisty}
A.~Lupu, H.~Benisty and A.~Degiron, 
\newblock Switching {using $\mathcal{PT}$-symmetry in plasmonic systems: positive role of the losses}, 
\newblock {\em Opt. Exp.} \textbf{21}, 21651--21668 (2013).

\bibitem{hang_huang_konotop}
C.~Hang, G.~Huang, and V.~V. Konotop,
\newblock {$\mathcal{PT}$-symmetry with a System of Three-Level Atoms},
\newblock {\em Phys. Rev. Lett.} \textbf{110}, 083604 (2013).

\bibitem{sheng_miri}
J.~Sheng, M.~A. Miri, D.~N. Christodoulides and M.~Xiao,
\newblock {$\mathcal{PT}$-symmetric optical potentials in a coherent atomic medium},
\newblock {\em Phys. Rev.} \textbf{A88}, 059904 (2013).

\bibitem{graefe_korsch_niederle}
E.~M. Graefe, H.~J. Korsch and A.~E. Niederle,
\newblock Mean-field dynamics of a {non-Hermitian Bose-Hubbard dimer},
\newblock {\em Phys. Rev. Lett.} \textbf{101}, 150408 (2008).

\bibitem{klaiman_gunther_moiseyev}
S.~Klaiman, U.~G{\"u}nther and N.~Moiseyev,
\newblock {Visualization of Branch Points in $\mathcal{PT}$-Symmetric Waveguides},
\newblock {\em Phys. Rev. Lett.} \textbf{101}, 080402 (2008).

\bibitem{graefe_korsch_niederle_2}
E.~M. Graefe, H.~J. Korsch and A.~E. Niederle,
\newblock Quantum-classical correspondence for a {non-Hermitian Bose-Hubbard dimer},
\newblock {\em Phys. Rev.} \textbf{A82}, 013629 (2010).

\bibitem{cartarius_wunner}
H.~Cartarius and G.~Wunner, 
\newblock {Model of a $\mathcal{PT}$-symmetric Bose-Einstein} condensate in a $\delta$-function double-well potential,
\newblock {\em Phys. Rev.} \textbf{A86}, 013612 (2012).

\bibitem{graefe_JPA_2012}
E.~M. Graefe,
\newblock {Stationary states of a $\mathcal{PT}$-symmetric two-mode Bose--Einstein condensate},
\newblock {\em J. Phys. A: Math. Theor.} \textbf{45}, 444015 (2012).

\bibitem{dast_haag}
D.~Dast, D.~Haag, H.~Cartarius, J.~Main and G.~Wunner,
\newblock Eigenvalue structure of a {Bose--Einstein condensate in a $\mathcal{PT}$-symmetric double well},
\newblock {\em J. Phys. A: Math. Theor.} \textbf{46}, 375301 (2013).

\bibitem{heiss_cartarius}
W.~D. Heiss, H.~Cartarius, G.~Wunner and J.~Main,
\newblock {Spectral singularities in $\mathcal{PT}$-symmetric Bose--Einstein condensates},
\newblock {\em J. Phys. A: Math. Theor.} \textbf{46}, 275307 (2013).

\bibitem{schindler_li}
J.~Schindler, A.~Li, M.~Zheng, F.~Ellis and T.~Kottos,
\newblock Experimental study of active {LRC circuits with $\mathcal{PT}$ symmetries},
\newblock {\em Phys. Rev.} \textbf{A84}, 040101 (2011).

\bibitem{schindler_lin}
J.~Schindler, Z.~Lin, J.~Lee, H.~Ramezani, F.~Ellis and T.~Kottos,
\newblock {$\mathcal{PT}$-symmetric electronics},
\newblock {\em J. Phys. A: Math. Theor.} \textbf{45}, 444029 (2012).

\bibitem{kato_book}
T.~Kato,
\newblock {\em Perturbation theory for linear operators} \textbf{132},
\newblock springer (1995).

\bibitem{cartarius_main_wunner}
H.~Cartarius, J.~Main and G.~Wunner,
\newblock Exceptional points in atomic spectra,
\newblock {\em Phys. Rev. Lett.} \textbf{99}, 173003 (2007).

\bibitem{cejnar_heinze_macek}
P.~Cejnar, S.~Heinze and M.~Macek,
\newblock Coulomb analogy for nonhermitian degeneracies near quantum phase transitions,
\newblock {\em Phys. Rev. Lett.} \textbf{99}, 100601 (2007).

\bibitem{berry}
M.~V. Berry,
\newblock Lop-sided diffraction by absorbing crystals,
\newblock {\em J. Phys. A: Math. Gen.} \textbf{31}, 3493 (1998).

\bibitem{chong}
Y.~D. Chong, L.~Ge, H.~Cao and A.~D. Stone,
\newblock Coherent perfect absorbers: time-reversed lasers,
\newblock {\em Phys. Rev. Lett.} \textbf{105}, 053901 (2010).

\bibitem{longhi}
S.~Longhi,
\newblock {$\mathcal{PT}$-symmetric} laser absorber,
\newblock {\em Phys. Rev.} \textbf{A82}, 031801 (2010).

\bibitem{turbiner}
A.~Turbiner,
\newblock Lie-algebras and linear operators with invariant subspaces,
\newblock {\em Contemp. Math.} \textbf{160}, 263 (1994).

\bibitem{kamran}
N.~Kamran and P.~J. Olver,
\newblock {Lie algebras of differential operators and Lie-algebraic potentials},
\newblock {\em J. Math. Anal. Appl.} \textbf{145}, 342--356 (1990).

\bibitem{humphreys}
J.~E. Humphreys,
\newblock {\em {Introduction to Lie algebras and representation theory}} \textbf{1980},
\newblock Springer New York (1972).

\bibitem{assis}
P.~E.~G. Assis,
\newblock {Metric operators for non-Hermitian quadratic $su (2)$ Hamiltonians},
\newblock {\em J. Phys. A: Math. Theor.} \textbf{44}, 265303 (2011).

\bibitem{bender_kalveks}
C.~M. Bender and R.~J. Kalveks,
\newblock {Extending $\mathcal{PT}$-Symmetry from Heisenberg Algebra to $E2$ Algebra},
\newblock {\em Int. J. Theor. Phys.} \textbf{50}, 955--962 (2011).

\bibitem{jones_kalveks}
K.~Jones-Smith and R.~Kalveks,
\newblock {Vector Models in $\mathcal{PT}$ Quantum Mechanics},
\newblock {\em Int. J. Theor. Phys.} \textbf{52}, 2187--2195 (2013).

\bibitem{isham}
C.~J. Isham and N.~Linden,
\newblock Group theoretic quantisation of strings on tori,
\newblock {\em Cl. and Quan. Gravity} \textbf{5}, 71 (1988).

\bibitem{bender_berry_mandilara}
C.~M. Bender, M.~V. Berry and A.~Mandilara,
\newblock Generalized $\mathcal{PT}$-symmetry and real spectra,
\newblock {\em J. Phys. A: Math. Gen.} \textbf{35}, L467 (2002).

\bibitem{fring_smith}
A.~Fring and M.~Smith,
\newblock Antilinear deformations of {Coxeter groups, an application to Calogero models},
\newblock {\em J. Phys. A: Math. Theor.} \textbf{43}, 325201 (2010).

\bibitem{bagarello}
F.~Bagarello,
\newblock Pseudo-bosons, so far,
\newblock {\em Rep. Math. Phys.} \textbf{68}, 175-210 (2011).

\bibitem{bagarello_lattuca}
F.~Bagarello and M.~Lattuca,
\newblock $\mathcal{D}$ pseudo-bosons in quantum models,
\newblock {\em Phys. Lett. A} \textbf{377}, 3199-3204 (2013).

\bibitem{bagarello_fring}
F.~Bagarello and A.~Fring,
\newblock Non-self-adjoint model of a two-dimensional noncommutative space with an unbound metric,
\newblock {\em Phys. Rev. A} \textbf{88}, 042119 (2013).

\bibitem{gradshteyn}
I.~Gradshteyn and I.~Ryzhik,
\newblock {\em Table of Integrals, Series, and Products},
\newblock (Academic Press London) (2007).

\bibitem{douglas_guise}
A.~Douglas and H.~de~Guise,
\newblock Some nonunitary, indecomposable representations of the euclidean algebra $e(3)$,
\newblock {\em J. Phys. A: Math. Theor.} \textbf{43}, 085204 (2010).

\bibitem{giri_roy}
P.~R. Giri and P.~Roy,
\newblock {Non-Hermitian} quantum mechanics in non-commutative space,
\newblock {\em Euro. Phys. J.} \textbf{C60}, 157--161 (2009).

\bibitem{alvaredo_fring}
O.~A. Castro-Alvaredo and A.~Fring,
\newblock {A spin chain model with non-Hermitian interaction: the Ising quantum spin chain in an imaginary field},
\newblock {\em J. Phys. A: Math. Theor.} \textbf{42}, 465211 (2009).

\bibitem{graefe_gunther_korsch_niederle}
E.~M. Graefe, U.~G{\"u}nther, H.~J. Korsch and A.~E. Niederle,
\newblock {A non-Hermitian $\mathcal{PT}$-symmetric Bose--Hubbard model: eigenvalue rings from unfolding higher order exceptional points},
\newblock {\em J. Phys. A: Math. Theor.} \textbf{41}, 255206 (2008).

\bibitem{curtright_fairlie_zachos}
T.~Curtright, D.~Fairlie and C.~Zachos,
\newblock Features of time-independent {Wigner functions},
\newblock {\em Phys. Rev.} \textbf{D58}, 025002 (1998).

\bibitem{li_wang}
K.~Li and J.~Wang,
\newblock The topological {AC effect on non-commutative phase space},
\newblock {\em Euro. Phys. J.} \textbf{C50}, 1007--1011 (2007).

\bibitem{bender_wu}
C.~M. Bender and T.~T. Wu,
\newblock Anharmonic oscillator,
\newblock {\em Phys. Rev.} \textbf{184}, 1231 (1969).

\bibitem{pierreaugercollaboration1}
Pierre~Auger Collaboration,
\newblock Update on the correlation of the highest energy cosmic rays with nearby extragalactic matter,
\newblock {\em Astropart. Phys.} \textbf{34}, 314--326 (2010).

\bibitem{pierreaugercollaboration2}
Pierre~Auger Collaboration,
\newblock {Measurement of the energy spectrum of cosmic rays above $10^{18} eV$ using the Pierre Auger Observatory},
\newblock {\em Phys. Lett.} \textbf{B685}, 239--246 (2010).

\bibitem{HiRes_collaboration}
The~HiRes Collaboration,
\newblock Final results from the high resolution {Fly's eye (HiRes)} experiment,
\newblock {\em arXiv}:1010.2690 (2010).

\bibitem{bhattacharjie_sudarshan}
A.~Bhattacharjie and E.~C.~G. Sudarshan,
\newblock A class of solvable potentials,
\newblock {\em Nuovo Cim.} \textbf{25}, 864--879 (1962).

\bibitem{jana_roy}
T.~K. Jana and P.~Roy,
\newblock {Non-Hermitian} quantum mechanics with minimal length uncertainty,
\newblock {\em SIGMA} \textbf{5}, 83 (2009).

\bibitem{levai}
G.~L{\'e}vai,
\newblock A search for shape-invariant solvable potentials,
\newblock {\em J. Phys. A: Math. Gen.} \textbf{22}, 689--702 (1989).

\bibitem{swanson}
M.~S. Swanson,
\newblock Transition elements for a {non-Hermitian quadratic Hamiltonian},
\newblock {\em J. Math. Phys.} \textbf{45}, 585--601 (2004).

\bibitem{poschl_teller}
G.~P{\"o}schl and E.~Teller,
\newblock Bemerkungen zur {Quantenmechanik des anharmonischen Oszillators},
\newblock {\em Z. f{\"u}r Physik} \textbf{83}, 143--151 (1933).

\bibitem{schrodinger}
E.~Schr{\"o}dinger,
\newblock {Der stetige {\"U}bergang von der Mikro-zur Makromechanik},
\newblock {\em Naturwissenschaften} \textbf{14}, 664--666 (1926).

\bibitem{iwata}
G.~Iwata,
\newblock {Non-Hermitian operators and eigenfunction expansions},
\newblock {\em Prog. Theor. Phys.} \textbf{6}, 216--226 (1951).

\bibitem{glauber}
R.~J. Glauber,
\newblock Coherent and incoherent states of the radiation field,
\newblock {\em Phys. Rev.} \textbf{131}, 2766 (1963).

\bibitem{feynman}
R.~P. Feynman,
\newblock An operator calculus having applications in quantum electrodynamics,
\newblock {\em Phys. Rev.} \textbf{84}, 108 (1951).

\bibitem{glauber1951}
R.~J. Glauber,
\newblock Some notes on multiple-boson processes,
\newblock {\em Phys. Rev.} \textbf{84}, 395 (1951).

\bibitem{dodonov}
V.~V. Dodonov,
\newblock 'nonclassical'states in quantum optics: a 'squeezed' review of the first 75 years,
\newblock {\em J. Opt. B: Quantum and Semiclas. Opt.} \textbf{4}, R1 (2002).

\bibitem{klauder1995}
J.~R. Klauder,
\newblock Quantization without quantization,
\newblock {\em Ann. Phys.} \textbf{237}, 147--160 (1995).

\bibitem{klauder1996}
J.~R. Klauder,
\newblock Coherent states for the hydrogen atom,
\newblock {\em J. Phys. A: Math. Gen.} \textbf{29}, L293--L298 (1996).

\bibitem{gazeau_klauder}
J-P. Gazeau and J.~R. Klauder,
\newblock Coherent states for systems with discrete and continuous spectrum,
\newblock {\em J. Phys. A: Math. Gen.} \textbf{32}, 123 (1999).

\bibitem{ali_bagarello}
S.~T. Ali, and F. Bagarello,
\newblock Some physical appearances of vector coherent states and coherent states related to degenerate {Hamiltonians},
\newblock {\em J. Math. Phys.} \textbf{46}, 053518 (2005).

\bibitem{gazeau_monceau}
J-P. Gazeau and P.~Monceau,
\newblock Generalized coherent states for arbitrary quantum systems,
\newblock In {\em Conf{\'e}rence Mosh{\'e} Flato 1999}, 131--144, Springer (2000).

\bibitem{jackson}
F.~H. Jackson,
\newblock A q-form of {Taylor’s theorem},
\newblock {\em Messenger Math.} \textbf{38}, 62--64 (1909).

\bibitem{ali_antoine_gazeau}
S.~T. Ali, J-P. Antoine and J-P. Gazeau,
\newblock {\em Coherent states, wavelets and their generalizations},
\newblock Springer (2000).

\bibitem{antoine_gazeau_monceau_klauder_penson}
J-P Antoine, J-P Gazeau, P.~Monceau, J.~R. Klauder and K.~A. Penson,
\newblock Temporally stable coherent states for infinite well {and P{\"o}schl--Teller potentials},
\newblock {\em J. Math. Phys.} \textbf{42}, 2349--2387 (2001).

\bibitem{ghosh_roy}
S.~Ghosh and P.~Roy,
\newblock "Stringy" coherent states inspired by generalized uncertainty principle,
\newblock {\em Phys. Lett.} \textbf{B711}, 423--427 (2012).

\bibitem{atakishiyev_nagiyev}
N.~M. Atakishiyev and S.~M. Nagiyev,
\newblock On the {Rogers-Szego polynomials},
\newblock {\em J. Phys. A: Math. Gen.} \textbf{27}, L611 (1994).

\bibitem{andrews}
G.~E. Andrews,
\newblock {\em The theory of Partitions},
\newblock Cambridge University Press, Cambridge, England (1984).

\bibitem{burban_klimyk}
I.~M. Burban and A.~U. Klimyk,
\newblock On spectral properties of q-oscillator operators,
\newblock {\em Lett. Math. Phys.} \textbf{29}, 13--18 (1993).

\bibitem{jackson1}
F.~H. Jackson,
\newblock On {q-Functions and a certain Difference Operator},
\newblock {\em Trans. R. Soc. Edinburgh} \textbf{46}, 253--281 (1909).

\bibitem{littlejohn}
R.~G. Littlejohn,
\newblock The semiclassical evolution of wave packets,
\newblock {\em Phys. Rep.} \textbf{138}, 193--291 (1986).

\bibitem{averbukh_kovarsky_perelman}
I.~S. Averbukh, V.~A. Kovarsky and N.~F. Perelman,
\newblock The diffractional grating effect in the luminescence of a pulse excited multilevel system,
\newblock {\em Phys. Lett.} \textbf{A70}, 289--291 (1979).

\bibitem{parker_stroud}
J.~Parker and C.~R. Stroud,
\newblock Coherence and decay of {Rydberg wave packets},
\newblock {\em Phys. Rev. Lett.} \textbf{56}, 716--719 (1986).

\bibitem{alber_ritsch_zoller}
G.~Alber, H.~Ritsch and P.~Zoller,
\newblock Generation and detection of {Rydberg} wave packets by short laser pulses,
\newblock {\em Phys. Rev.} \textbf{A34}, 1058 (1986).

\bibitem{milburn}
G.~J. Milburn,
\newblock Quantum and classical liouville dynamics of the anharmonic oscillator,
\newblock {\em Phys. Rev.} \textbf{A33}, 674 (1986).

\bibitem{yurke_stoler}
B.~Yurke and D.~Stoler,
\newblock Generating quantum mechanical superpositions of macroscopically distinguishable states via amplitude dispersion,
\newblock {\em Phys. Rev. Lett.} \textbf{57}, 13 (1986).

\bibitem{mecozzi_tombesi}
A.~Mecozzi and P.~Tombesi,
\newblock Distinguishable quantum states generated via nonlinear birefringence,
\newblock {\em Phys. Rev. Lett.} \textbf{58}, 1055--1058 (1987).

\bibitem{averbukh_perelman}
I.~S. Averbukh and N.~F. Perelman,
\newblock Fractional revivals: {Universality} in the long-term evolution of quantum wave packets beyond the correspondence principle dynamics,
\newblock {\em Phys. Lett.} \textbf{A139}, 449--453 (1989).

\bibitem{mandel}
L.~Mandel,
\newblock Sub-{Poissonian} photon statistics in resonance fluorescence,
\newblock {\em Opt. Lett.} \textbf{4}, 205--207 (1979).

\bibitem{mandel1}
L.~Mandel,
\newblock Squeezed states and sub-Poissonian photon statistics,
\newblock {\em Phys. Rev. Lett.} \textbf{49}, 136 (1982).

\bibitem{short_mandel}
R.~Short and L.~Mandel,
\newblock Observation of {sub-Poissonian} photon statistics,
\newblock {\em Phys. Rev. Lett.} \textbf{51}, 384 (1983).

\bibitem{bluhm_kostelecky_porter}
R.~Bluhm, A.~Kostelecky and J.~Porter,
\newblock The evolution and revival structure of localized quantum wave packets,
\newblock {\em Am. J. Phys.} \textbf{64}, 944--953 (1996).

\bibitem{vrakking_villeneuve_stolow}
M.~Vrakking, D.~Villeneuve and A.~Stolow, 
\newblock Observation of fractional revivals of a molecular wave packet,
\newblock {\em Phys. Rev.} \textbf{A54}, R37 (1996).

\bibitem{graefe_schubert}
E.-M. Graefe and R.~Schubert,
\newblock Complexified coherent states and quantum evolution with {non-Hermitian Hamiltonians},
\newblock {\em J Phys. A: Math. Theor.} \textbf{45}, 244033 (2012).

\bibitem{bohm1}
D.~Bohm,
\newblock A suggested interpretation of the quantum theory in terms of "hidden" {variables. I},
\newblock {\em Phys. Rev.} \textbf{85}, 166 (1952).

\bibitem{bohm2}
D.~Bohm,
\newblock A suggested interpretation of the quantum theory in terms of "hidden" {variables. II},
\newblock {\em Phys. Rev.} \textbf{85}, 180 (1952).

\bibitem{holland}
P.~R. Holland,
\newblock {\em {The Quantum Theory of Motion: An Account of the Broglie-Bohm Causal Interpretation of Quantum Mechanics}},
\newblock Cambridge University Press (1993).

\bibitem{hiley}
B.~J. Hiley,
\newblock {Bohmian Non-commutative Dynamics: History and New Developments},
\newblock {\em arXiv: }1303.6057 (2013).

\bibitem{bohm_hiley_book}
D.~Bohm and B.~J. Hiley,
\newblock {\em {The undivided universe: An ontological interpretation of quantum theory}},
\newblock Routledge, London (2006).

\bibitem{mayor_askar_rabitz}
F.~S. Mayor, A.~Askar and H.~A. Rabitz,
\newblock {Quantum fluid dynamics in the Lagrangian representation and applications to photodissociation problems},
\newblock {\em J. Chem. Phys.} \textbf{111}, 2423--2435 (1999).

\bibitem{chou_wyatt_scattering}
C.~Chou and R.~E. Wyatt,
\newblock {Computational method for the quantum Hamilton-Jacobi equation: One-dimensional scattering problems},
\newblock {\em Phys. Rev.} \textbf{E74}, 066702 (2006).

\bibitem{lopreore_wyatt}
C.~L. Lopreore and R.~E. Wyatt,
\newblock Quantum wave packet dynamics with trajectories,
\newblock {\em Phys. Rev. Lett.} \textbf{82}, 5190 (1999).

\bibitem{sanz_borondo_miret}
A.~S. Sanz, F.~Borondo and S.~Miret-Art{\'e}s,
\newblock Causal trajectories description of atom diffraction by surfaces,
\newblock {\em Phys. Rev.} \textbf{B61}, 7743 (2000).

\bibitem{wang_darling_holloway}
Z.~S. Wang, G.~R. Darling and S.~Holloway,
\newblock {Dissociation dynamics from a de Broglie--Bohm perspective},
\newblock {\em J. Chem. Phys.} \textbf{115}, 10373--10381 (2001).

\bibitem{guantes_sanz}
R.~Guantes, A.~S. Sanz, J.~Margalef-Roig and S.~Miret-Art{\'e}s,
\newblock Atom--surface diffraction: a trajectory description,
\newblock {\em Surface Science Rep.} \textbf{53}, 199--330 (2004).

\bibitem{wu_augstein_faria}
J.~Wu, B.~B. Augstein and C.~F. de~Morisson~Faria,
\newblock {Bohmian-trajectory analysis of high-order-harmonic generation: Ensemble averages, nonlocality, and quantitative aspects},
\newblock {\em Phys. Rev.} \textbf{A88}, 063416 (2013).

\bibitem{leacock_padgett}
R.~A. Leacock and M.~J. Padgett,
\newblock {Hamilton-Jacobi theory and the quantum action variable},
\newblock {\em Phys. Rev. Lett.} \textbf{50}, 3--6 (1983).

\bibitem{leacock_padgett1}
R.~A. Leacock and M.~J. Padgett,
\newblock {Hamilton-Jacobi/action-angle quantum mechanics},
\newblock {\em Phys. Rev.} \textbf{D28}, 2491 (1983).

\bibitem{goldfarb_degani_tannor}
Y.~Goldfarb, I.~Degani and D.~J. Tannor,
\newblock {Bohmian mechanics with complex action: A new trajectory-based formulation of quantum mechanics},
\newblock {\em J. Chem. Phys.} \textbf{125}, 231103 (2006).

\bibitem{chou_wyatt_complex}
C.~Chou and R.~E. Wyatt,
\newblock Quantum trajectories in complex space,
\newblock {\em Phys. Rev.} \textbf{A76}, 012115 (2007).

\bibitem{cdyang}
C.~D. Yang,
\newblock {Trajectory interpretation of the uncertainty principle in 1D systems using complex Bohmian mechanics},
\newblock {\em Phys. Lett.} \textbf{A372}, 6240--6253 (2008).

\bibitem{mvjohn}
M.~V. John,
\newblock Probability and complex quantum trajectories,
\newblock {\em Ann. Phys.} \textbf{324}, 220--231 (2009).

\bibitem{mvjohn1}
M.~V. John,
\newblock Modified de {Broglie-Bohm} approach to quantum mechanics,
\newblock {\em Found. Phys. Lett.} \textbf{15}, 329--343 (2002).

\bibitem{cdyang1}
C.~D. Yang,
\newblock Modeling quantum harmonic oscillator in complex domain,
\newblock {\em Chaos, Solitons and Fractals} \textbf{30}, 342--362 (2006).

\bibitem{kleinert_mustapic}
H.~Kleinert and I.~Mustapic,
\newblock Summing the spectral {representations of P{\"o}schl--Teller and Rosen--Morse fixed-energy amplitudes},
\newblock {\em J. Math. Phys.} \textbf{33}, 643--662 (1992).

\bibitem{john_mathew}
M.~V. John and K.~Mathew,
\newblock {Coherent States and Modified de Broglie-Bohm Complex Quantum Trajectories},
\newblock {\em Found. Phys.} \textbf{43}, 859--871 (2013).

\bibitem{bender_hook_kooner}
C.~M. Bender, D.~W. Hook and K.~S. Kooner,
\newblock Classical particle in a complex elliptic potential,
\newblock {\em J. Phys. A: Math. Theo.} \textbf{43}, 165201 (2010).

\bibitem{haar}
D.~T. Haar,
\newblock {\em Problems in quantum mechanics}
\newblock (1975).

\bibitem{nanayakkara}
A.~Nanayakkara,
\newblock Classical trajectories of {1D complex non-Hermitian Hamiltonian systems},
\newblock {\em J. Phys. A: Math. Gen.} \textbf{37}, 4321 (2004).

\bibitem{bender_brody_hook}
C.~M. Bender, D.~C. Brody and D.~W. Hook,
\newblock Quantum effects in classical systems having complex energy,
\newblock {\em J. Phys. A: Math. Theor.} \textbf{41}, 352003 (2008).

\bibitem{arpornthip_bender}
T.~Arpornthip and C.~M. Bender,
\newblock Conduction bands in classical periodic potentials,
\newblock {\em Pramana} \textbf{73}, 259--267 (2009).

\bibitem{bender_feinberg_hook_weir}
C.~M. Bender, J.~Feinberg, D.~W. Hook and D.~J. Weir,
\newblock Chaotic systems in complex phase space,
\newblock {\em Pramana} \textbf{73}, 453--470 (2009).

\bibitem{anderson_bender_morone}
A.~G. Anderson, C.~M. Bender and U.~I. Morone,
\newblock Periodic orbits for classical particles having complex energy,
\newblock {\em Phys. Lett.} \textbf{A375}, 3399--3404 (2011).

\bibitem{cavaglia_fring_bagchi}
A.~Cavaglia, A.~Fring and B.~Bagchi,
\newblock $\mathcal{PT}$-symmetry breaking in complex nonlinear wave equations and their deformations,
\newblock {\em J. Phys. A: Math. Theor.} \textbf{44}, 325201 (2011).

\bibitem{brown_hiley}
M.~R. Brown and B.~J. Hiley,
\newblock Schrodinger revisited: an algebraic approach,
\newblock {\em arXiv: }0005026 (2000).

\bibitem{hiley1}
B.~J. Hiley,
\newblock Non-commutative quantum geometry: a reappraisal of the {Bohm} approach to quantum theory,
\newblock in {\em Quo Vadis Quantum Mechanics?}, 299--324, Springer (2005).

\bibitem{mann_sanders_munro}
A.~Mann, B.~C. Sanders and W.~J. Munro,
\newblock Bell’s inequality for an entanglement of nonorthogonal states,
\newblock {\em Phys. Rev.} \textbf{A51}, 989 (1995).

\end{thebibliography}
\end{document}